\documentclass[11pt]{article}
\usepackage{authblk}
\title{Competition of Core-Shell and Janus Morphology in Alloy Nanoparticles: Insights From a
Phase-Field Model}
\author{P. Pankaj} 
\author{Saswata Bhattacharya\thanks{Corresponding author.}}
\author{Subhradeep Chatterjee\thanks{Corresponding author.}}

\affil{Department of Materials Science and Metallurgical Engineering, \\
Indian Institute of Technology, Hyderabad, 502285, India.\\
Emails: ms15resch11005@iith.ac.in, saswata@msme.iith.ac.in, subhradeep@msme.iith.ac.in}

\usepackage[super,sort&compress,comma]{natbib} 
\usepackage[version=3]{mhchem}
\usepackage[left=1.5cm, right=1.5cm, top=1.785cm, bottom=2.0cm]{geometry}
\usepackage{balance}
\usepackage{mathptmx}
\usepackage{sectsty}
\usepackage{graphicx} 
\usepackage{lastpage}
\usepackage[format=plain,justification=justified,singlelinecheck=false,font={stretch=1.125,small,sf},labelfont=bf,labelsep=space]{caption}
\usepackage{float}
\usepackage[english]{babel}
\usepackage{scalerel}
\addto{\captionsenglish}{%
  
}
\usepackage{array}
\usepackage[T1]{fontenc}
\usepackage[usenames,dvipsnames]{xcolor}
\usepackage{setspace}
\usepackage[compact]{titlesec}
\usepackage{hyperref}
\usepackage{epstopdf}%
\usepackage{amsfonts}
\usepackage{amssymb}
\usepackage{amsmath}
\usepackage{miller}
\usepackage{lineno}
\usepackage{booktabs}
\usepackage{subcaption}
\usepackage{wrapfig}
\usepackage{siunitx}
\usepackage{mathrsfs}
\usepackage{mathtools}
\usepackage[abs]{overpic}
\usepackage[export]{adjustbox}
\usepackage{tikz}
\usetikzlibrary{arrows}
\DeclareGraphicsExtensions{.pdf,.png,.jpg,.jpeg, .eps}
\graphicspath{{figures/}}

\begin{document}

\maketitle

\begin{abstract}
Bimetallic nanoparticles (BNPs) exhibit diverse morphologies such as core-shell, Janus, onion-like, quasi-Janus, and homogeneous structures. Although extensive effort has been directed towards understanding the equilibrium configurations of BNPs, kinetic mechanisms involved in their development have not been explored systematically. Since these systems often contain a miscibility gap, experimental studies have alluded to spinodal decomposition (SD) as a likely mechanism for the formation of such structures. We present a novel phase-field model for confined (embedded) systems to study SD-induced morphological evolution within a BNP. It initiates with the formation of compositionally modulated rings as a result of surface directed SD, and eventually develops into core-shell or Janus structures due to coarsening/breakdown of the rings. The final configuration depends crucially on contact angle and particle size - Janus is favored at smaller sizes and higher contact angles. Our simulations also illustrate the formation of metastable, kinetically trapped structures as a result of competition between capillarity and diffusion.    
\end{abstract}

\section{Introduction}

Bimetallic nanoparticles (BNPs) are an important class of materials
that exhibit novel chemical, optical, electrical, electronic, magnetic 
phenomena. A general overview of their methods of synthesis, structure
and its underlying theoretical aspects, properties, and potential 
applications can be found, for example, in Ferrando et 
al.~\cite{Ferrando2008} and Gilroy et al.~\cite{Gilroy2016}, while more
specific reviews concerning properties targeted to their 
catalytic~\cite{Fang2018}, plasmonic~\cite{Cortie2011}, or 
magnetic~\cite{Zeng2008} applications are also available. These 
properties in turn depend on the internal configuration of the 
nanoparticles (NPs), with core-shell (CS), onion-like rings and Janus
(side-segregated/dumbbell-like) being the most frequently observed
morphologies.
BNPs have been synthesized in a wide range of systems such as 
Ag-Cu~\cite{Langlois2012,Malviya2014,Lee2015,Sopousek2015},
Ag-Ni~\cite{Srivastava2011}, 
Au-Co~\cite{Wang2010,Mayoral2015}, 
Au-Ni~\cite{Schnedlitz2018,Wang2010}, Au-Pd~\cite{Lee2009}, 
Co-Cu~\cite{McKeown2015}, Co-Pt~\cite{Jun2006}, 
Cu-Mo~\cite{Krishnan2013} and Ni-Pt~\cite{Lim2018}. Most of these systems
exhibit immiscibility or very limited solid solubility over a wide range of 
temperatures. Thus, bulk thermodynamics favors phase separation in the NPs
and suppresses formation of homogeneous alloyed phase. However, capillary
forces come into dominance at small sizes, and depending on the composition,
size, and method of preparation, NPs are sometimes found in the homogeneous
alloyed state as well~\cite{Krishnan2013,Malviya2016,Radnoczi2017}. 
In the same alloy system, variation of processing variables or exposure to
moderately high temperatures can lead to the formation of thermodynamically 
more stable phase separated configurations.

For example, through a variation of process parameters of magnetron sputtering,
Krishnan et al.~\cite{Krishnan2013} synthesized Cu-Mo NPs in a range of sizes
(16-59 nm) and compositions (14-30 at.\%Cu). They observed single phase, 
Cu@Mo\footnote[2]{we use A@B to refer to a CS configuration where A is the shell
and B is the core} CS, and Janus morphologies in these NPs. They also reported an
onion-like concentric ring configuration where a Cu middle ring was sandwiched
between the outer shell and inner core made of Mo. This is surprising, because,
based on surface energy considerations, an outer ring of Cu would have been more 
likely ($\sigma_\text{Cu}<\sigma_\text{Mo}$, $\sigma$ being the surface energy).

\begin{figure}[htbp]
\centering
     \begin{subfigure}[b]{.24\textwidth}%
     \centering
        {\includegraphics[scale=0.2]{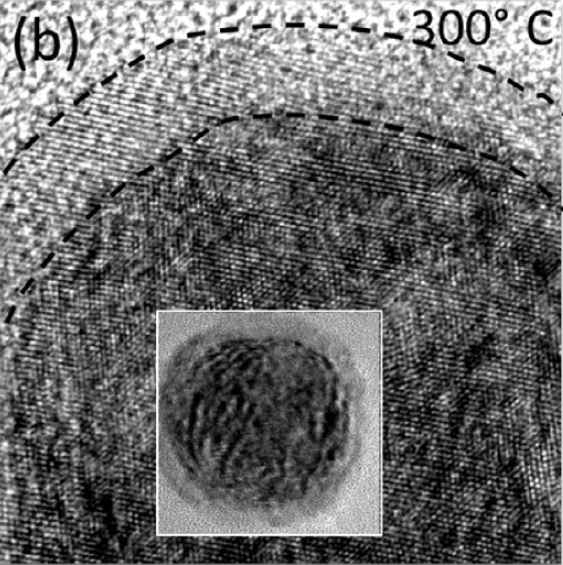}}
     \caption{}\label{fig:Malviya-a}
     \end{subfigure}%
    \begin{subfigure}[b]{.24\textwidth}%
     \centering
        {\includegraphics[scale=0.9]{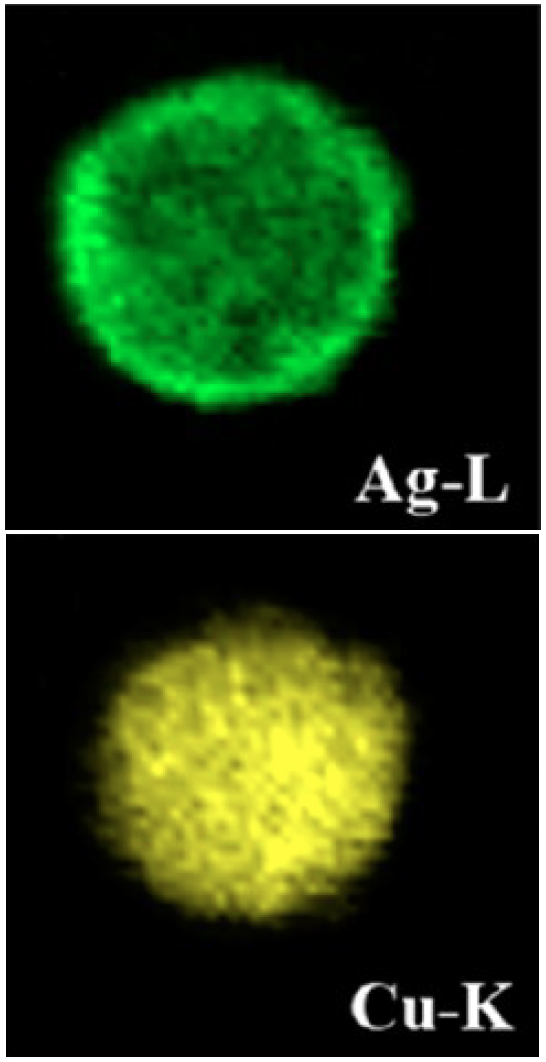}}
     \caption{}\label{fig:Malviya-b}
     \end{subfigure}\\%
     \begin{subfigure}[b]{.24\textwidth}%
     \centering
        {\includegraphics[scale=0.2]{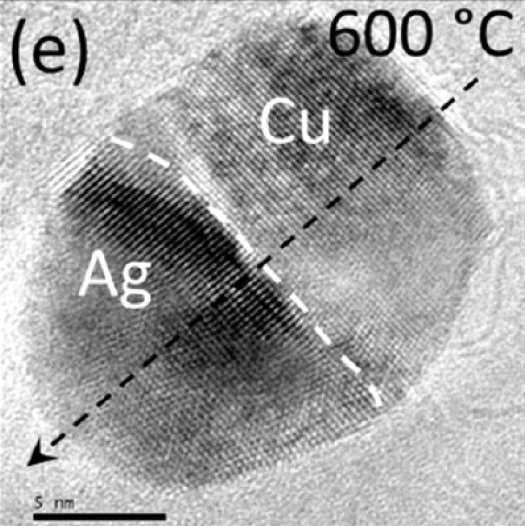}}
     \caption{}\label{fig:Malviya-c}
     \end{subfigure}%
    \begin{subfigure}[b]{.24\textwidth}%
     \centering
        {\includegraphics[scale=0.2]{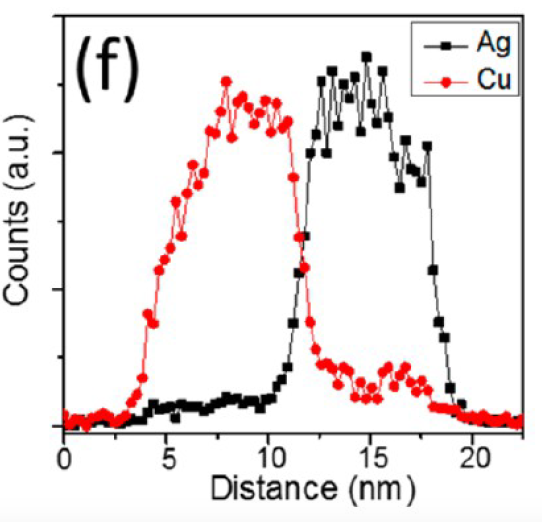}}
     \caption{}\label{fig:Malviya-d}
     \end{subfigure}%
\caption{Core-shell and Janus structures of pulsed laser deposited 
near equiatomic Ag-Cu NPs as a function of  particle size~\cite{Malviya2016}:
(a, b)  high resolution transmission electron micrograph (HRTEM) of a
\SI{40}{\nano\meter} core-shell particle and the corresponding elemental composition map;
(c, d)  HRTEM of a \SI{20}{\nano\meter} Janus particle and the corresponding elemental 
composition map. Reprinted with permission from Kirtiman Deo Malviya and K. Chattopadhyay, 
\emph{Journal of Physical Chemistry C}, 2016, 120, 27699-27706. Copyright $\copyright$
2016 American Chemical Society.
}
\label{fig:Malviya}
\end{figure}

Exposure to high temperature too can lead to a change in the configuration. When 
Ag@Cu CS NPs of diameter $\sim$25 nm prepared by sequential chemical 
synthesis~\cite{Tsai2013} were  heated above 500 \textdegree{C}, the thin Ag shells
($\sim$0.7 nm thick) dewetted and produced Ag nodules on the surface  of Cu cores. 
During continued heating, the NPs eventually coarsened to several hundred nanometers
in size through coalescence. In contrast, when Cu-Ag NPs of near-equiatomic composition 
produced by pulsed laser deposition~\cite{Malviya2016} were exposed to temperatures up
to 600 \textdegree{C}, the initially homogeneous solid solution phase decomposed to 
form Ag- and Cu-rich phases. Figure~\ref{fig:Malviya}, which is reproduced from
this work~\cite{Malviya2016}, shows that particles below $\sim$20 nm in size
developed into a Janus configuration, while the larger ones (size $\sim$40 nm)
exhibited an Ag@Cu CS morphology.

Radnoczi et al.~\cite{Radnoczi2017} identified spinodal decomposition (SD) of the
solid solution phase to be the underlying the mechanism of phase separation in Ag-Cu
NPs prepared by DC magnetron sputtering. They found that irrespective of particle 
composition, there was no phase separation if the particle size was below a critical 
value (5 nm), and it matched well with the critical spinodal wavelength estimated
using the Cahn-Hilliard (CH) theory~\cite{cahn1961spinodal}. They observed phase
separation in larger Cu-rich NPs (30 at.\%Ag) that resulted in the formation of 
Janus and off-centered Ag@Cu CS morphologies. Formation of other configurations such
as `crescent'~\cite{Osowiecki2018,Tang2019}, \emph{i.e.}, side segregated with a curved
interface between the two halves, and inverse CS~\cite{Schnedlitz2018} where the higher
surface energy phase forms the shell, have also been reported. 

Although experiments report a myriad of morphologies of BNPs as a function of processing, 
one requires to supplement them with theoretical insights to understand the role of
thermodynamic and kinetic factors leading to these structures.
Computational studies aiming to understand the role of different factors
influencing BNP configurations fall into two broad categories: (i) bulk 
and surface thermodynamics based models and (ii) electronic and atomistic
simulations. The former essentially follows the CALPHAD 
approach~\cite{Saunders1998}, but additionally incorporates finite-size
corrections for surface energy into the Gibbs free energy description of
the corresponding bulk alloy. This enables a re-assessment of thermal 
stability of nanoalloys and determination of phase diagrams as a function
of particle size%
~\cite{ParkLee2008, Garzel2012, CHU2021,Hajra2004,JABBAREH2018,Peng2015}. 
Although these models have been successful in predicting a shift in 
miscibility gap in nanoalloy systems, they do not provide sufficient insight
into the internal morphological details of BNPs. 

Electronic or atomistic models of nanoalloy clusters%
~\cite{Bochicchio2013,Polak2014,Chandross2014,Peng2015modeling,Ferrando2018,FevreBouarFinel2018}, 
on the other hand, explore the relative stability of the different morphologies in BNPs
using density functional theory, molecular dynamics or Monte Carlo (MC) simulations.
Details of the extensive research carried out using these methods can be found in the 
monograph by Ferrando~\cite{Ferrandobook}. The primary goal of most, if not all, of these
studies, however, is to predict the \emph{equilibrium} or lowest energy configuration in
specific systems as a function of particle size. 
A notable exception is the recent study by Li et al.~\cite{Li2019Nanoscale} 
who performed kinetic MC simulations. It indicated the 
possibility of metastable, `kinetically trapped' configurations due to partial 
redistribution of atoms in BNPs as small as \SI{2}{\nano\meter}. 
Their study brought out the limitations of equilibrium studies which fail to 
account for diffusion-limited processes. In particular, they show that the 
diffusion process becomes extremely sluggish after initial phase separation, leading
to such metastable configurations. 

Barring the last one, most studies have focused on equilibrium structures of BNPs.  
Although they reveal several important aspects of BNP morphology at equilibrium, 
one needs a systematic investigation of kinetic pathways leading to the formation of
CS or Janus configurations. Such an investigation will be useful to identify the key
mechanisms of evolution of these morphologies as a function of particle size, contact
angle, alloy chemistry and other processing conditions.

Majority of the bimetallic systems constituting BNPs exhibit a miscibility gap in their
bulk phase diagrams and their microstructures often consist of two compositionally-distinct, 
isostructural phases. This suggests spinodal to be a viable mechanism by which an initially 
homogeneous, alloyed BNP with equiatomic or near-equiatomic composition can phase separate and
form CS/Janus structures~\cite{Sopousek2015,Malviya2016,Radnoczi2017}.

Although SD in bulk systems has been investigated extensively since Cahn~\cite{cahn1961spinodal},
there have been fewer studies on phase separation in confined geometries where external surfaces 
influence the decomposition. In a finite system, the constituent phase having a lower surface energy
may prefer to form at the surface and trigger concentration waves propagating from surface to the
interior. Compositionally modulated structures resulting from this surface-directed spinodal 
decomposition (SDSD) appear different from bulk spinodal morphology~\cite{Jones1991,PuriBinderPhysRevA1992}.

In the past three decades, phase field models~\cite{chen2002Review} have emerged as a very powerful
continuum, mesoscale computational tool. They incorporate thermodynamics and kinetics of a system 
seamlessly, and are ideal for studying microstructures resulting from phase transformations such as
SD. In addition to their extensive application in understanding bulk SD, they have also been
used to study SD in finite systems such as thin films~\cite{Johnson2002,Puri2020}, polycrystalline 
materials containing multiple internal surfaces~\cite{Ramanarayan2003,RAMANARAYAN2004}, systems 
containing inert dispersoids~\cite{supriyo2017}, and isolated, radially-stressed 
spheres~\cite{Johnson2001}. 

Using a CH~\cite{cahn1958free} formalism with an auxiliary non-conserved phase field 
variable~\cite{AC1976}, we present a novel phase-field model of phase separation in confined systems.
We report results of two-dimensional simulations using this model, and investigate morphological 
evolution driven by SD in an initially homogeneous, alloyed BNP. In particular, we address the 
following questions:

\begin{enumerate}

  \item 
  For a circular BNP of diameter $d$, how does the contact angle $\theta$, which represents the 
  balance  between surface and interfacial energies, influence the morphological evolution and final 
  configuration?
  \item
   For a given contact angle, how does the particle size influence the kinetic pathways and final morphology?
\end{enumerate}

Further, we systematically explore the $\theta-d$ space to study how both these factors interact, and 
summarize our findings through a morphology map. Finally, we also show how the shape of a BNP can be
another important factor in determining its internal morphology.

\section{Model Formulation}

When we study phase transformations in bulk systems, the governing CH
equations employ periodic or zero-flux boundary conditions that arise
naturally. However, since nanoparticles are finite systems by definition,
we can no longer apply periodic boundary conditions, and zero-flux
boundary conditions too fail to correctly incorporate the three-phase 
contact angle. Even the so-called `contact angle boundary 
condition'~\cite{Kim2011,Kim2016} is inappropriate and ill-imposed
when contact angle develops dynamically, e.g., when an initial two-phase
contact evolves to a three-phase contact or vice versa. As pointed out by
Cahn in his pioneering work on `critical point wetting', the equilibrium
contact angle $\theta$ cannot be defined~\cite{Cahn1977} beyond complete
wetting transition.

Thus, one of the critical challenges is to address the boundary condition
in our problem. Immersed boundary (IB) method  has emerged as an efficient
technique in computational fluid dynamics to solve Navier-Stokes and other
transport equations in complex geometries~\cite{MittalAnnRevFluidMech2005}. 
IB method uses an auxiliary function to capture irregular boundaries of a 
complex geometry inside a regular computational domain such that regular 
solvers can be employed without special meshing techniques. Bueno-Orovio
et al.~\cite{BuenoOrovioSIAM2006} implemented an IB technique in the Fourier
space to solve partial differential equations in irregular geometries with 
zero-flux boundary conditions. They used a smoothly varying phase-field 
parameter to inscribe the physical region of interest in the extended 
computational domain. Li et al.~\cite{LiCommMathSci2009} and Yu et 
al.~\cite{Yu2012} extended this method to include Dirichlet, Neumann and 
Robin boundary conditions. Poulsen and Voorhees~\cite{Voorhees2018} 
introduced an alternative energy-penalty based formulation which integrated
IB with the variational phase-field framework. They demonstrated the efficacy
of their technique by solving the heat diffusion problem during annealing.   

We introduce a method similar in spirit to these smoothed IB techniques. However, since microstructural evolution inside a BNP can lead to a dynamical
transition in the contact angle at the particle boundary, contact angle boundary 
condition on the particle surface cannot be imposed explicitly. As contact angle
results from a balance of surface energies and interfacial energy at the particle
surface, we implicitly incorporate it by first constructing an extended system 
containing a particle embedded in a matrix, and then choosing an appropriate free
energy functional for the system. Through benchmark simulations, we demonstrate
the capability of the model to successfully recover the correct contact angle.

\begin{figure}[htbp]
  \center
  \includegraphics[width=5.5cm]{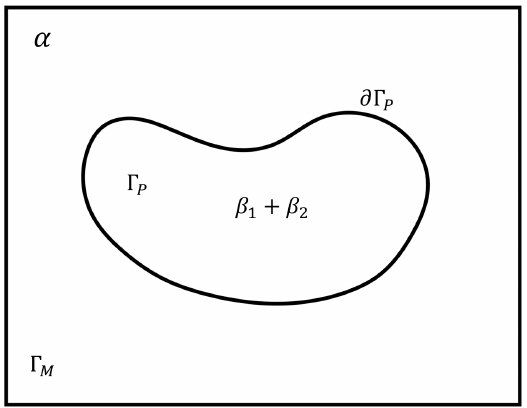}
  \caption{A spinodally decomposing particle of volume $\Gamma_P$ embedded in a matrix of
  volume $\Gamma_M$. $\Gamma = \Gamma_M \cup \Gamma_P$ is the volume of the entire domain.}
  \label{method}
\end{figure}

Our model, shown schematically in Fig.~\ref{method}, consists of an alloyed BNP
immersed in an inert matrix phase $\alpha$ that forms the surrounding environment
of the particle. Chemically, it is a model binary system consisting of two atomic
species A and B. When an initially homogeneous particle of composition $c_0$ is
brought below a critical temperature, it undergoes phase separation that leads to
the formation of two isostructural terminal phases $\beta_1$ and $\beta_2$. We use
a non-conserved order parameter field $\phi(\mathbf{r})$ to distinguish the particle
($\phi=1$) from the matrix ($\phi=0$), and a conserved composition field 
$c(\mathbf{r},t)$ to describe phase separation within the particle. For numerical 
convenience, we scale $c(\mathbf{r},t)$ using the expression: 
$c=(c^{\prime} - c_{\beta_1}^{\textrm{eq}})/(c_{\beta_2}^{\textrm{eq}} - %
c_{\beta_1}^{\textrm{eq}})$ where $c^{\prime}$ denotes the unscaled composition
variable, and $c_{\beta_1}^{\textrm{eq}}$, $c_{\beta_2}^{\textrm{eq}}$ are the
unscaled equilibrium compositions of $\beta_1$ and $\beta_2$, respectively.

\subsection{Energetics}
\label{Energetics}

Three energy contributions (expressed in per atom basis) constitute the total free energy density
of the phase-separating system:

\begin{enumerate}
   \item 
   The bulk \emph{chemical} free energy density $f_{\textrm{ch}}$, which is the weighted average of the 
   particle and the matrix free energies: 
   $f_{\textrm{ch}}(c,\phi) = h(\phi)f^{\beta} + (1 - h(\phi))f^{\alpha}$, where 
   $f^\beta=f_0^p c^2 (1-c)^2$ is a double well potential denoting the particle free energy and
   $f^{\alpha}=f_0^m(c-0.5)^2$ is the free energy of the matrix. Here $h(\phi)=\phi^3(10 - 15\phi + 6\phi^2)$
   is a weighing function~\cite{Wang1993} varying smoothly between $0$ and $1$, and $f_0^i (i=m,p)$ are 
   temperature-dependent phenomenological constants that set the scale of free energy functions of the matrix 
   and particle, respectively. This particular choice of $f^\beta$ makes the scaled equilibrium
   $\beta_1$ and $\beta_2$ compositions $0$ and $1$, respectively. Without any loss of generality, we choose 
   the minimum of the matrix free energy function at $c=0.5$ (midway between equilibrium compositions of 
   $\beta_1$ and $\beta_2$) describing a three-phase equilibrium between the matrix ($\alpha$) and the particle 
   ($\beta_1$ and $\beta_2$). The equilibrium values of $c$ and $\phi$ in different phases are summarized in 
   Table~\ref{eq_value}. 
  
Note that the double-well potential $f^\beta$ used in our model can be compared to the regular solution 
potential through the relation $\dfrac{f^\beta}{k_BT}= \Omega c(1-c) + c \ln c + (1-c) \ln (1-c)$ where $k_B$
is Boltzmann constant, $T$ is temperature in kelvin, and $\Omega$ is the regular solution parameter 
normalized with $k_B T$. Free energy barrier heights in these two potentials at equiatomic composition
are related by $\dfrac{f_0^p}{16k_BT} = \Omega/4 - \ln 2$.

 \begin{table}[hb]
   \centering
   \small\caption{Equilibrium values of $c$ and $\phi$ for all the phases}
   \label{eq_value}
   \begin{tabular*}{0.15\textwidth}{lll}
   \hline
   Phase & $c^\textrm{eq}$ & $\phi^\textrm{eq}$ \\
   \hline
   $\beta_1$ & $0$ & $1$\\ 
   $\beta_2$ & $1$ & $1$\\ 
   $\alpha$ & $0.5$ & $0$\\ 
   \hline
 \end{tabular*}
 \end{table}  

  \item 
  A double-well free energy function $f_{\textrm{br}}$ that introduces a potential \emph{barrier} 
  between the particle and the matrix: $f_{\textrm{br}}(c,\phi)=\omega(c) \phi^2(1-\phi)^2$, where the
  composition-dependent variable $\omega(c)$ sets the scale for the barrier. We choose 
  $\omega(c)=\omega_0(1-\chi c)$ where $\chi$ is a parameter that governs solute segregation to the 
  particle-matrix interface, similar to the one used by {\AA}gren and coworkers~\cite{Strandlund2008}
  to study  solute drag effects on grain boundary migration.

  \item
  \emph A {gradient} energy contribution $f_{\textrm{gr}}$ that accounts for the surface excess
  associated with the matrix-particle interface as well as interfaces between $\beta_1$ and $\beta_2$ 
  phases: $f_{\textrm{gr}} = \kappa_\phi|\nabla\phi|^2 + \kappa_c|\nabla c|^2 $, where $\kappa_{\phi}$
  and $\kappa_c$ are the positive gradient energy coefficients associated with composition and phase field
  variables, respectively. For a given alloy particle, $\kappa_c$ controls the length scale of SD leading
  to CS/Janus structures.
  
 \end{enumerate}
 
The total free energy functional of the system, $\mathcal{F}$, is given as:
\begin{equation}
 \mathcal{F} = N_v \int_{\Gamma} \Big(f_{\textrm{ch}} + f_{\textrm{br}} + f_{\textrm{gr}}\Big) d\Gamma,
\end{equation}
where $N_v$ denotes number of atoms per unit volume. Since we use $\phi$ only as an auxiliary parameter 
to distinguish the particle from the matrix, any continuous $\phi$ profile that satisfies $\phi=0$ and 
$\phi=1$ for matrix and particle, respectively, would suffice. We have chosen a smooth \texttt{tanh}
function for the variation of $\phi$ across the particle-matrix interface:
\begin{equation}
\phi(\mathbf{r})=\frac{1}{2}\left[1 - \tanh\left(\frac{|\mathbf{r}|- (d/2)}{W_{\phi}}\right)\right].  
\label{tanh}
\end{equation}

Here $W_\phi$ sets the width of the matrix-particle interface (kept constant for all our simulations), 
$|\mathbf{r}|$ is the magnitude of the position vector measured from the centre of the particle, and 
$d$ is the particle diameter. Alternatively, one could additionally solve Allen-Cahn equation~\cite{AC1976}
for $\phi$ for a specified number of time steps to achieve regularization of matrix-particle interface
of any arbitrary geometry. For the chosen $\phi$ profile, we obtain the equilibrium composition profile $c^{\textrm{eq}}(\mathbf{r}$) by solving the Euler-Lagrange equation for the variational problem:

\begin{align}
    \frac{\delta \mathcal{F}}{\delta c} = \frac{\partial f}{\partial c} - 2 \kappa_{c} \nabla^2 c = 0.
\end{align}

Using the $c^{\textrm{eq}}(\mathbf{r})$ profile between any two of $\alpha$, $\beta_1$ and 
$\beta_2$ phases at a time, we compute the corresponding interfacial energy $\sigma$ as: 
\begin{align}
    \sigma &=
    \int_{-\infty}^{\infty}{\left( \Delta f(c^{\textrm{eq}}, \phi)  
    + \kappa_c |\nabla{c^{\textrm{eq}}}|^2 + \kappa_{\phi} |\nabla{\phi}|^2  \right)}d\Gamma.
     \label{IEexpr}
\end{align}
Here, $\Delta f(c^{\textrm{eq}},\phi)=f(c^{\textrm{eq}},\phi) - (1-c)\mu_A^{\textrm{eq}} - c\mu_B^{\textrm{eq}}$,
where $\mu_{A/B}^{\textrm{eq}}$ denotes the chemical potential of the corresponding species. In what follows, 
$\sigma_1$ and $\sigma_2$ refer to the energies of $\alpha$-$\beta_1$ and $\alpha$-$\beta_2$ interfaces (the 
`surface energies'), respectively, while $\sigma_{12}$ denotes the $\beta_1$-$\beta_2$ interfacial energy. 

\begin{figure*}[htbp]
     \begin{subfigure}[b]{.5\linewidth}%
     \centering
        {\includegraphics[scale=0.25]{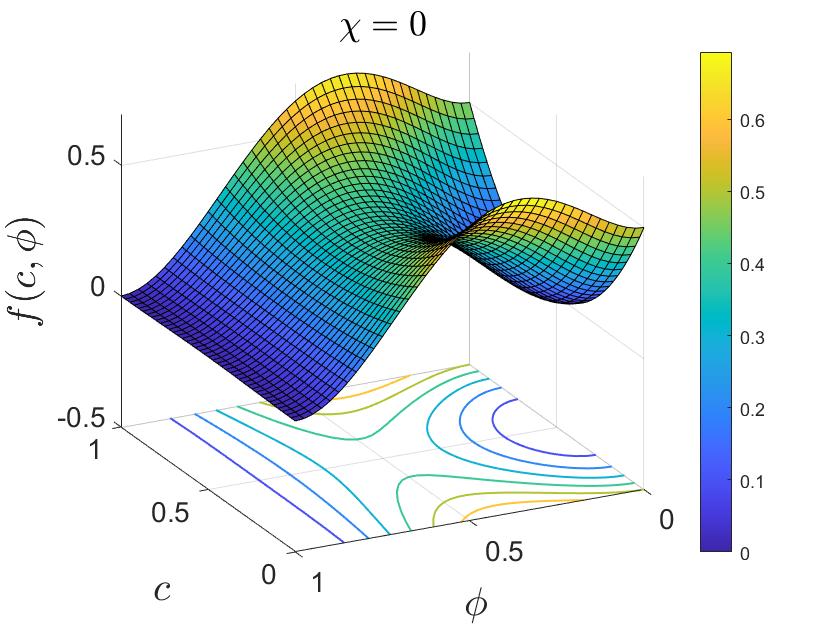}}
     \caption{}\label{fig:cont-chi0}
     \end{subfigure}%
    \begin{subfigure}[b]{.5\linewidth}%
     \centering
        {\includegraphics[scale=0.25]{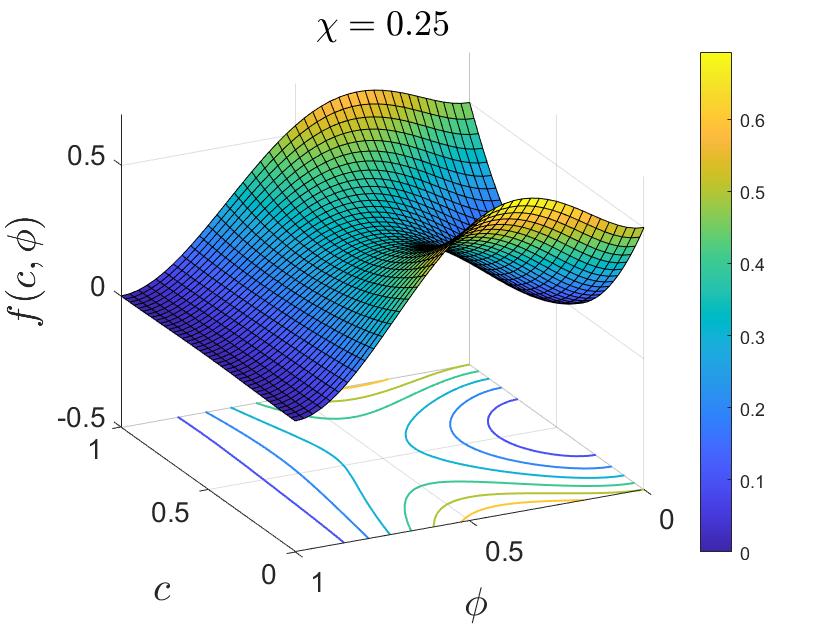}}
     \caption{}\label{fig:cont-chi25}
     \end{subfigure}\\%
    \begin{subfigure}[b]{.5\linewidth}%
     \centering
        {\includegraphics[scale=0.25]{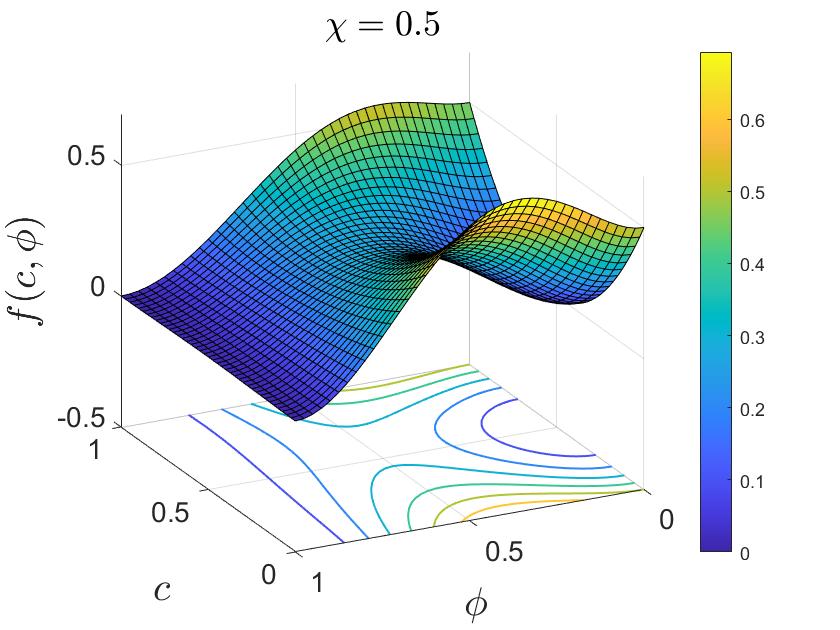}}
    \caption{} \label{fig:cont-chi50}
     \end{subfigure}%
    \begin{subfigure}[b]{.5\linewidth}%
     \centering
        {\includegraphics[scale=0.25]{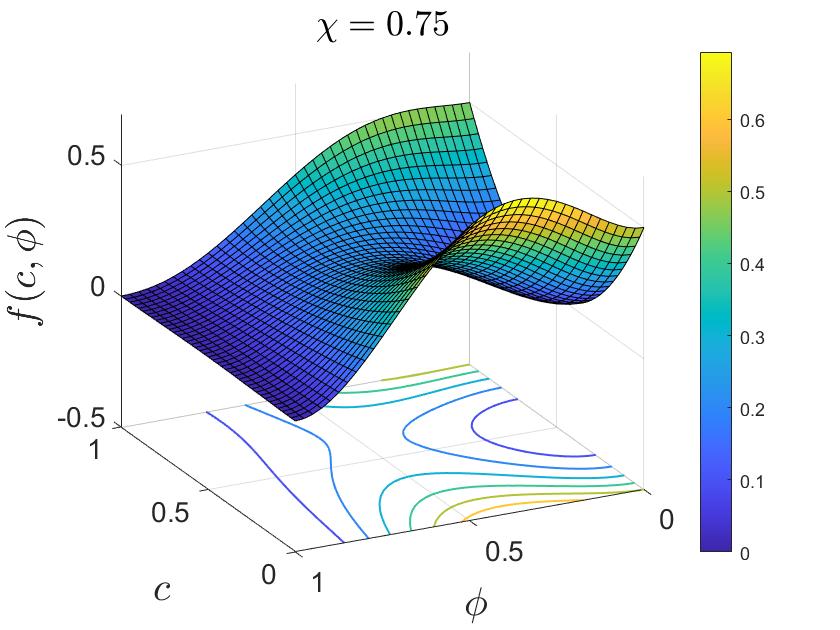}}
     \caption{}\label{fig:cont-chi75}
     \end{subfigure}\\%
     \begin{subfigure}[b]{.33\linewidth}%
     \centering
        {\includegraphics[scale=0.16]{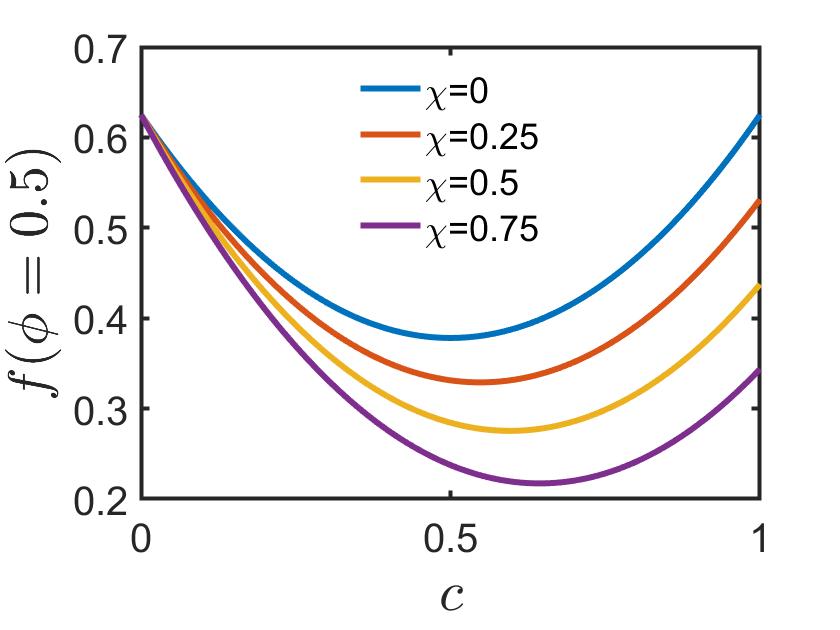}}
     \caption{}\label{fig:fmiddle}
     \end{subfigure}%
     \begin{subfigure}[b]{.33\linewidth}%
     \centering
        {\includegraphics[scale=0.16]{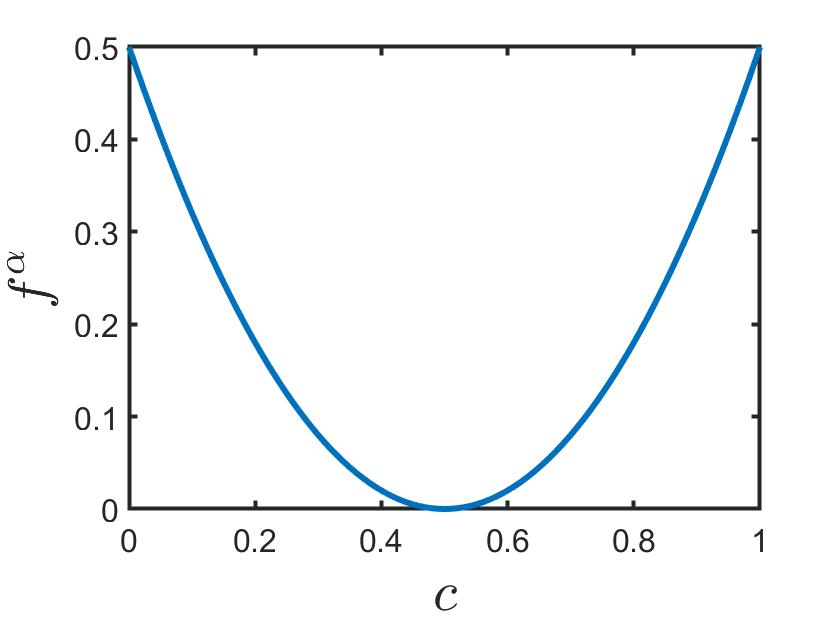}}
     \caption{}\label{fig:falpha}
     \end{subfigure}%
     \begin{subfigure}[b]{.33\linewidth}%
     \centering
        {\includegraphics[scale=0.16]{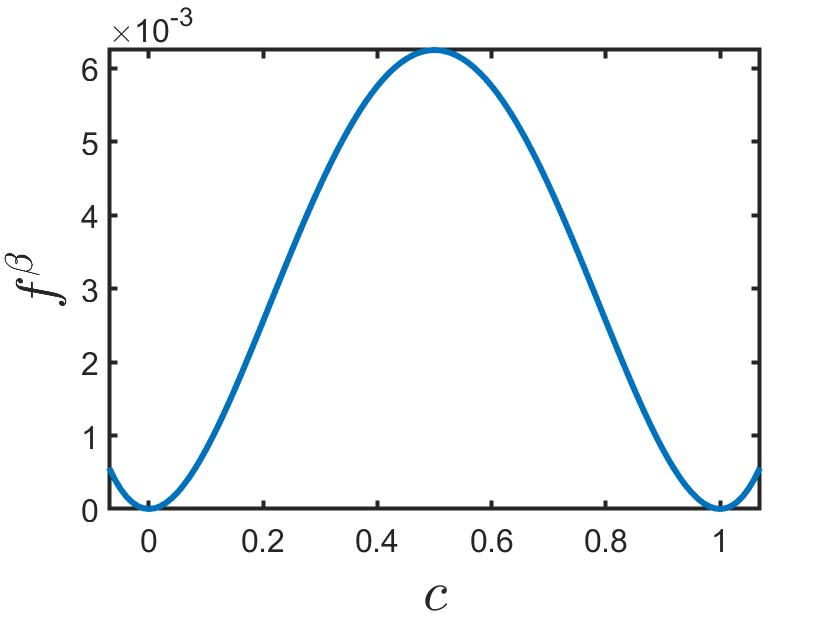}}
     \caption{}\label{fig:fbeta}
     \end{subfigure}

  \caption{Free energy density ($f_{ch}+f_{br}$) of the matrix-particle system as a function of
  $c$ and $\phi$. (a--d): Surface plots along with iso-energy contours for different values of
  the segregation parameter $\chi$. (e) Free energy-composition curves drawn at $\phi=0.5$ for 
  different values of $\chi$. (f--g): Bulk free energy-composition curves of the matrix ($\phi=0$) 
  and particle ($\phi=1$), respectively; only one curve suffices as $\chi$ does not
  affect free energy of bulk regions.}
  \label{FE_Surface}
 \end{figure*}

The segregation parameter $\chi$ modifies the free energy landscape in regions of non-uniform 
$\phi$ (\emph{viz.}, within the matrix-particle interface). In doing so, a non-zero $\chi$ makes
$\alpha$-$\beta_1$ and $\alpha$-$\beta_2$ interfaces non-equivalent, which is illustrated in
Fig.~\ref{FE_Surface} using surface and contour plots of $f_{ch}+f_{br}$ in the $c$-$\phi$ space.
While the free energy is perfectly symmetric about $c=0.5$ in Fig.~\ref{fig:cont-chi0} ($\chi=0$), it becomes
increasingly more asymmetric  as $\chi$ continues to increase in Fig.~\ref{fig:cont-chi25} through 
Fig.~\ref{fig:cont-chi75}. This asymmetry is revealed most strikingly in Fig.~\ref{fig:fmiddle} through the free
energy -- composition ($f$-$c$) curves drawn at $\phi=0.5$. In this figure, $f(c,0.5)$ is symmetric about $c=0.5$
for $\chi=0$, but non-zero $\chi$'s create asymmetry by shifting the minimum of $f$ towards right. 
Figs.~\ref{fig:falpha} and~\ref{fig:fbeta} present the $f$-$c$ curves corresponding to matrix ($\phi=0$) and 
particle ($\phi=1$), respectively. The free energy function for the matrix has a single minimum, whereas that for
the particle is a double-well potential corresponding to the two equilibrium phases $\beta_1$ and $\beta_2$.

Since the choice of $\chi$ influences interface profiles and thereby energies associated with the
interfaces, it provides us a useful means to tailor the surface energies of the $\beta_1$ and 
$\beta_2$ phases. Young's equation for the contact angle $\theta$ at a three-phase coexistence 
point connects these surface energies and the $\beta_1$-$\beta_2$ interfacial energy:

\begin{equation}{\label{eq-Young}}
    \cos\theta = \frac{\sigma_1 - \sigma_2}{\sigma_{12}}.
\end{equation}

Hence, we use $\chi$ as a parameter to systematically vary the contact angle and study how the
balance between surface and interfacial energies influences the morphological evolution in BNPs.
As an extension to Young's equation, Cahn has discussed the condition for spontaneous or complete
wetting~\cite{Cahn1977} when the following condition is satisfied:
\begin{equation}
 \sigma_1 \geq\sigma_{2} +\sigma_{12}.
\label{eq_Cahn_wet}
\end{equation}
In this case, the lower surface energy phase $\beta_2$ perfectly wets the high-energy 
$\alpha\text{-}\beta_1$ interface and replaces it with two low-energy interfaces, \emph{viz.},
$\alpha\text{-}\beta_2$ and $\beta_1\text{-}\beta_2$. The inequality condition in 
Eq.~\eqref{eq_Cahn_wet} sets the limit of applicability of Eq.~\eqref{eq-Young}.
Note that the chosen values of $\chi$ for BNP simulations result in a variation of the contact 
angle $\theta$ from \ang{90} to \ang{0}.

\subsection{Kinetics}

In our simulations, a  BNP is initially in the homogeneous alloyed configuration. However, this 
state is thermodynamically unstable, and it undergoes phase separation by SD. Phase separation and 
subsequent establishment of the proper three-phase equilibrium at the particle surface are achieved 
by solute transport inside the BNP. Cahn-Hilliard equation~\cite{cahn1961spinodal}, which describes
the evolution of composition field over time, governs the kinetics of solute transport in the system:

\begin{align}\label{CahnHilliard}
    \frac{\partial c}{\partial t}&=\nabla\cdot M(\mathbf{r})\nabla \mu \nonumber \\
                                 &=\nabla\cdot M(\mathbf{r}) \nabla\left(\frac{\partial f}{\partial c} -
        2\kappa_c\nabla^2c\right),
\end{align}
where $\mu \equiv \dfrac{\delta\mathcal{F}}{\delta c}=\partial f/\partial c - 2\kappa_c\nabla^2c$ 
is the generalized diffusion potential. $M(\mathbf{r})=M_c\phi$ defines the position-dependent 
atomic mobility of the solute, $M_c$ being the constant solute mobility within the particle 
($\phi=1$). This functional form of $M(\mathbf{r})$ ensures that no diffusion takes place in the
matrix and diffusional mobility is  constant within the particle. Thus, SD and wetting dynamics
are controlled solely by diffusion within the particle and precludes interactions with neighbouring 
particles due to periodic computational domain $\Gamma$. Although here we do not report the effects
of matrix diffusion or coalescence of particles on wetting, our formulation is general enough to 
include such effects through an appropriate choice of $M(\mathbf{r})$. We nondimensionalize all
parameters in the governing equation (Eq.~\eqref{CahnHilliard}) using a characteristic energy
$E_c=k_B T$, a characteristic length $L_c = (\kappa_c/k_BT)^{1/2}$, and a characteristic time
$\tau_c=L_c^2/M_c E_c$.  Nondimensional values of all physical and numerical parameters used 
in the simulations are listed in Table~\ref{SimulationParameters}.

We solve Eq.~\eqref{CahnHilliard} numerically in the extended computational domain $\Gamma$ using 
a semi-implicit Fourier spectral method~\cite{Zhu1999}. An operator splitting approach using 
a constant weighing parameter $Q$ ($0 \leq Q \leq 1$) introduces implicitness to the linear terms 
in the discretized equation in the reciprocal space and thereby reduces the computational
time~\cite{cogs-thesis}:

\begin{align}
    \Tilde{c}^{n+1} = \frac{\Tilde{c}^{n}(1 + 2Q \kappa_c k^4  \Delta t ) + 
    					\mathrm{i}\mathbf{k} \Delta t \cdot \left\{ M\phi(\mathbf{r}) 
    					\left(\left[ \mathrm{i}\mathbf{k}^{\prime}
    					\Tilde{\mu} \right]^n_{\mathbf k^{\prime}}\right)_{\mathbf{r}} 
    					\right\}_{\mathbf{k}}}{1+2Q\kappa_c k^4 \Delta t }.
    					\label{Eq:discretization}
\end{align}

Here $\mathbf{k}$ is the reciprocal (Fourier) space vector and $k=|\mathbf{k}|$
denotes its magnitude, $\Tilde{\{.\}}$ over a field variable represents it in the
Fourier space, $\{.\}_\mathbf{k}$ or $[.]_\mathbf{k'}$ denotes Fourier transform of
the quantity inside the brackets, $(.)_\mathbf{r}$ represents the inverse Fourier
transform, $\mathrm{i}=\sqrt{-1}$, $\Delta{t}$ is and the value of a discrete time
step, and superscripts $n$ and $n+1$ denote the current and next time-steps, 
respectively. We use the forward Euler method to advance the composition field in 
time, and carry out a sensitivity analysis to choose an appropriately small time
step that ensures temporal stability and accuracy.

The discretized equation (Eq.~\ref{Eq:discretization}) is solved on a Xeon Phi vector processor
with 72 CPU cores using OpenMP parallelization or an NVIDIA Tesla V100 GPU card having 5120 cores.
The Fourier transforms are implemented using the open source \texttt{fftw}~\cite{FFTW05} or 
NVIDIA \texttt{cufft} library. We systematically investigate the effects of contact angle and 
particle size on SD and wetting behaviour of an isolated, circular NP embedded in an inert matrix.
The simulations are run till equilibrium, measured computationally by comparing the largest 
difference in chemical potential of the system between successive steps with a tolerance of
$10^{-10}$.

\begin{table}[htbp]
 \small\caption{Simulation Parameters}
 \begin{tabular*}{0.48\textwidth}{ll}
 \hline
  Parameter & Value \\
  \hline
  \small{Average particle composition $c_0$} & $0.5$  \\
  \small{Number density of atoms $N_v$} & $0.37$ \\
  \small{Matrix free energy coefficient $f_0^m$} & $2$ \\
  \small{Particle free energy coefficient $f_0^p$} & $0.1$\\
  \small{Barrier height $\omega_0$} & $6$ \\
  \small{Gradient energy coefficient $\kappa_c$} & $1$ \\
  \small{Gradient energy coefficient $\kappa_{\phi}$} & $15$ \\
  \small{Segregation parameter $\chi$} & $0.0, 0.25, 0.5, 0.75$ \\
  \small{Solute mobility in particle $M_c$} & 1.0\\
  \small{Grid size $\Delta x$, $\Delta y$} & 1.0, 1.0 \\
  \small{Time step $\Delta t$} & 0.1 \\
  \small{Particle size $d$} & 140, 160, 180, 240 \\
  \small{Amplitude of initial noise $\eta$} & $1\%$ \\
  \hline
 \end{tabular*}
 \label{SimulationParameters}
\end{table}

\section{Results and Discussion}

We first discuss the role of the parameter $\chi$ on solute segregation at the particle surface
and consequently on surface energies. Through a series of drop-on-substrate simulations, we 
demonstrate how $\chi$ can be used to systematically vary the three-phase 
($\alpha\text{-}\beta_1\text{-}\beta_2$) contact angle $\theta$. Following this, we present
simulations of phase-separating particles embedded in a matrix as a function of their size and 
the contact angle, and discuss underlying mechanisms for the morphological evolution. These are
summarized through a morphology map for CS and Janus configurations. Finally, through 
simulations of elliptical BNPs, we illustrate how the shape of the NPs can influence
morphological evolution and even give rise to novel configurations.

\subsection{Surface segregation, surface energy and contact angle}

\begin{figure*}[ht]
 \captionsetup[subfigure]{justification=centering}
  \begin{subfigure}[b]{.25\linewidth}%
    \centering
    \includegraphics[scale=0.18]{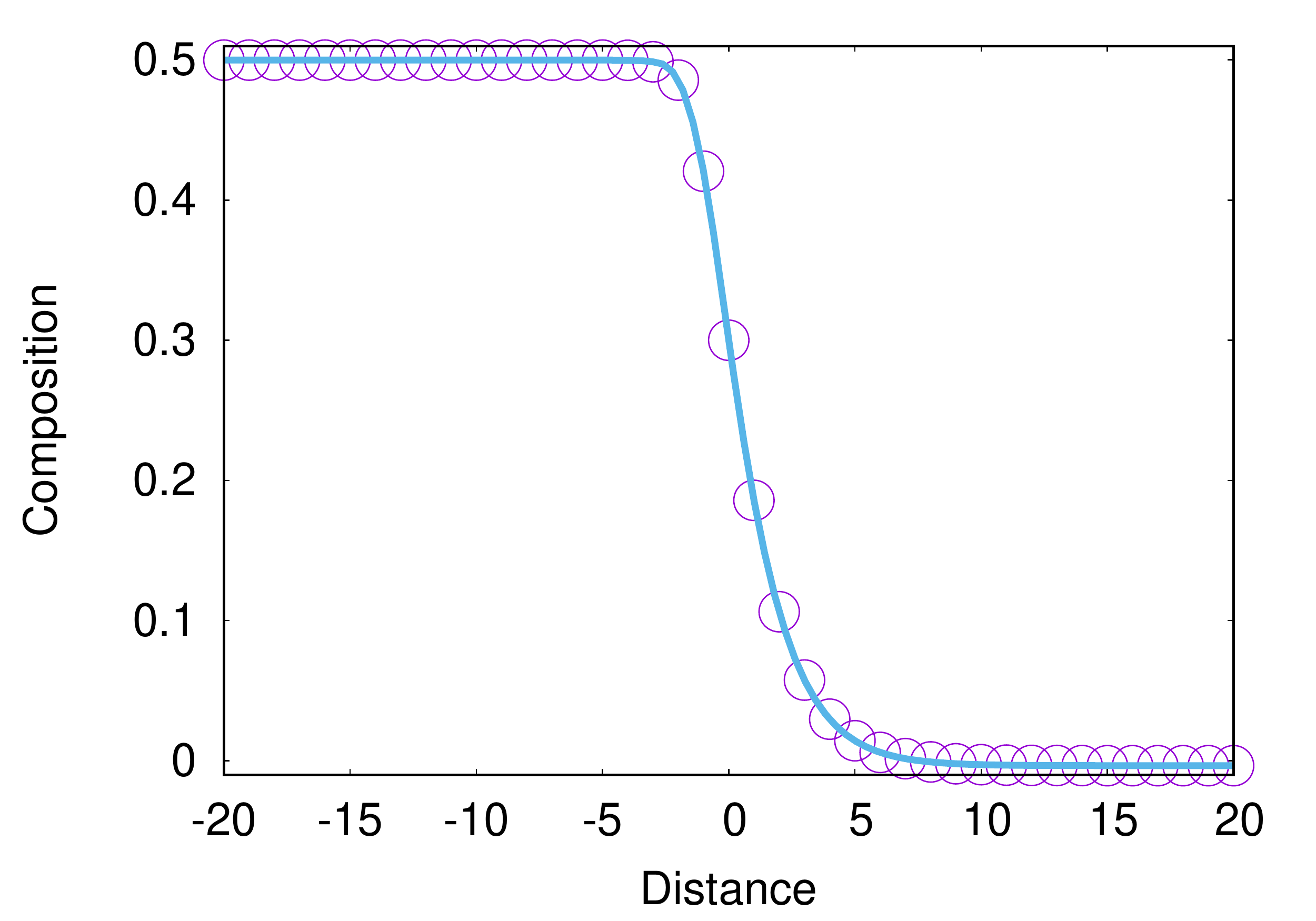}
    \caption{$\chi=0$, $\alpha\text{-}\beta_1$}
    \label{chi0_AB1}
  \end{subfigure}%
  \begin{subfigure}[b]{.25\linewidth}%
    \centering
    \includegraphics[scale=0.18]{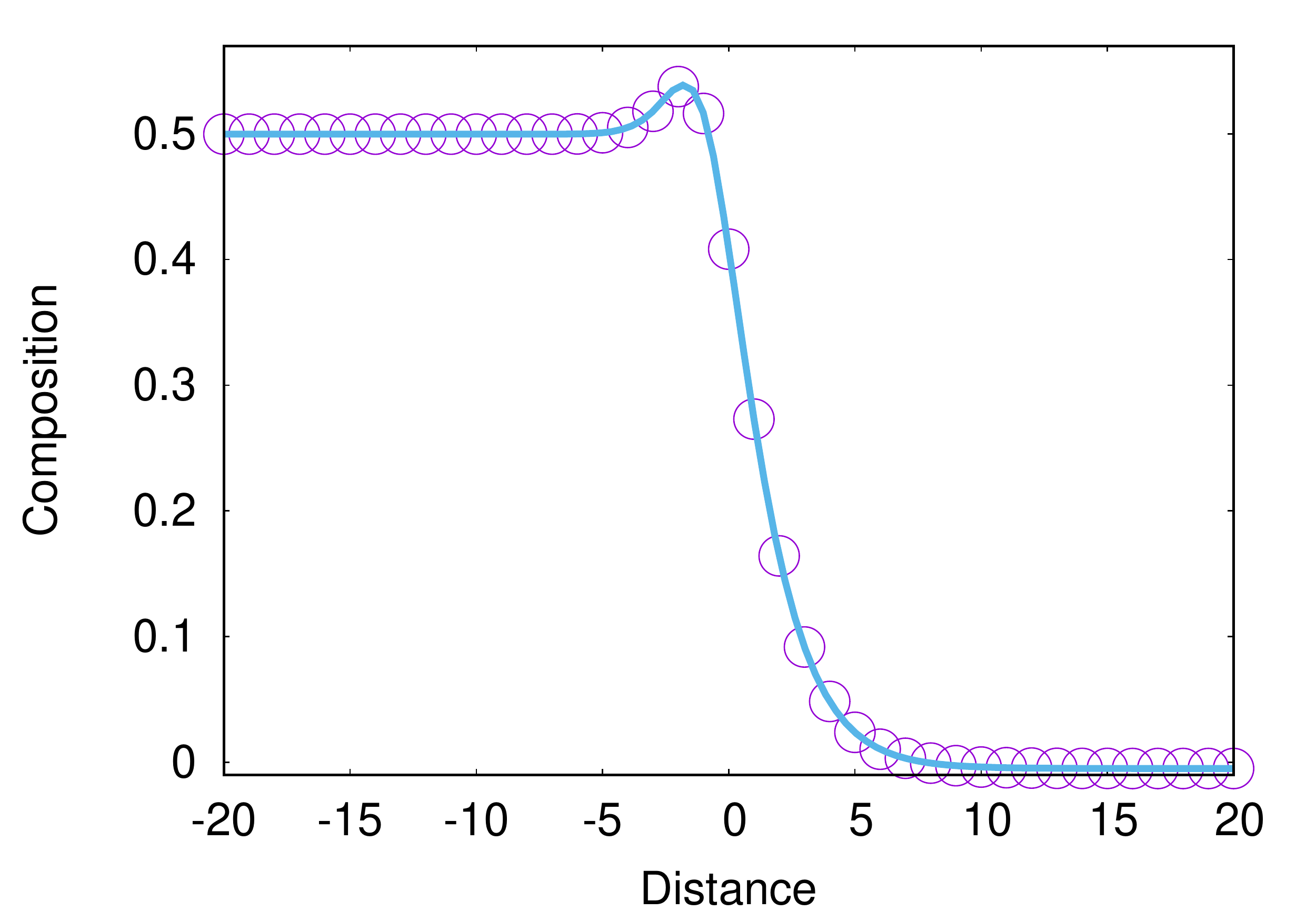}
    \caption{$\chi=0.25$, $\alpha\text{-}\beta_1$}
    \label{chi025_AB1}
  \end{subfigure}%
  \begin{subfigure}[b]{.25\linewidth}%
    \centering
    \includegraphics[scale=0.18]{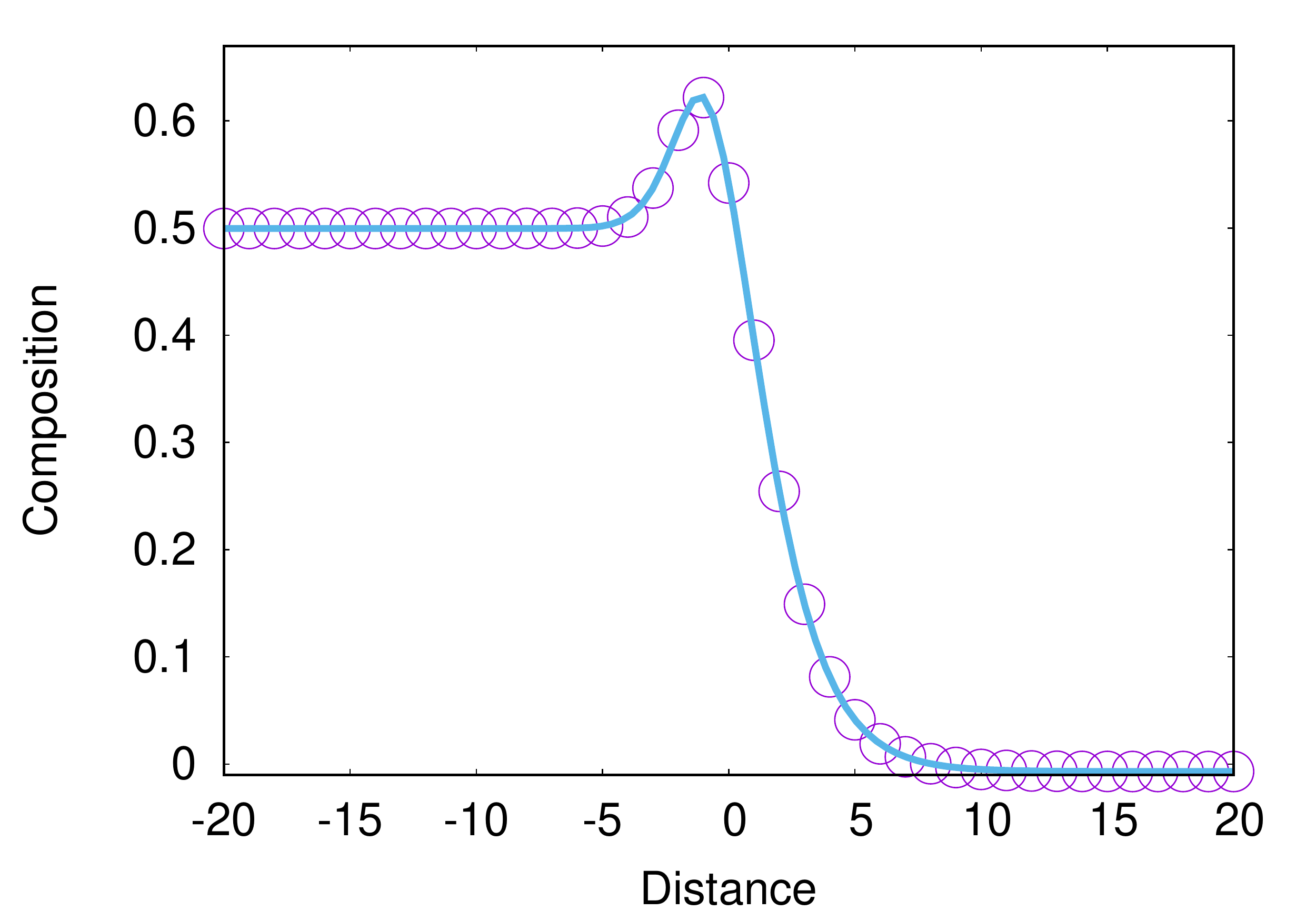}
    \caption{$\chi=0.5$, $\alpha\text{-}\beta_1$}
    \label{chi05_AB1}
  \end{subfigure}%
  \begin{subfigure}[b]{.25\linewidth}%
    \centering
    \includegraphics[scale=0.18]{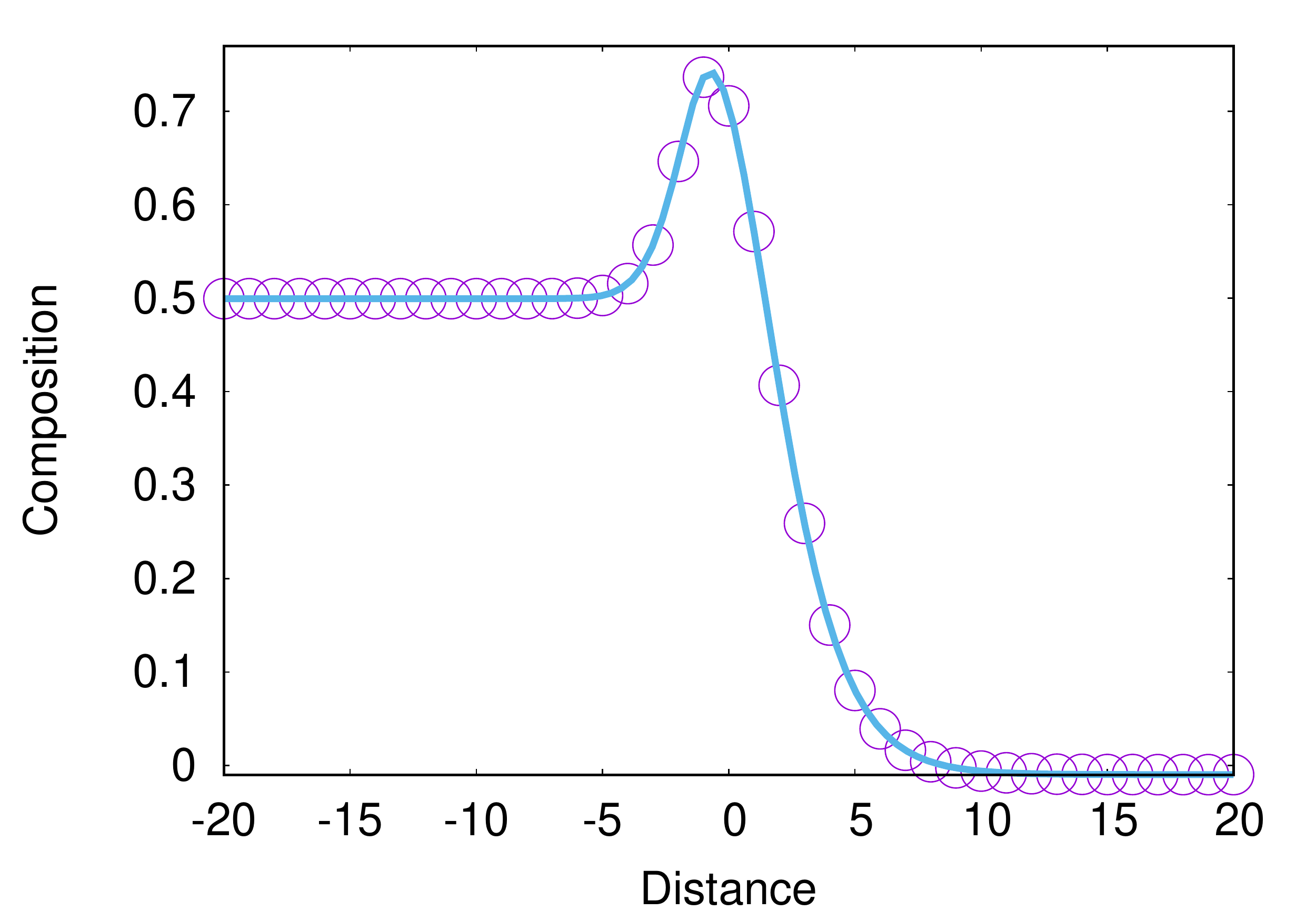}
    \caption{$\chi=0.75$, $\alpha\text{-}\beta_1$}
    \label{chi075_AB1}
  \end{subfigure}\\%
  \begin{subfigure}[b]{.25\linewidth}%
  \centering
  \includegraphics[scale=0.18]{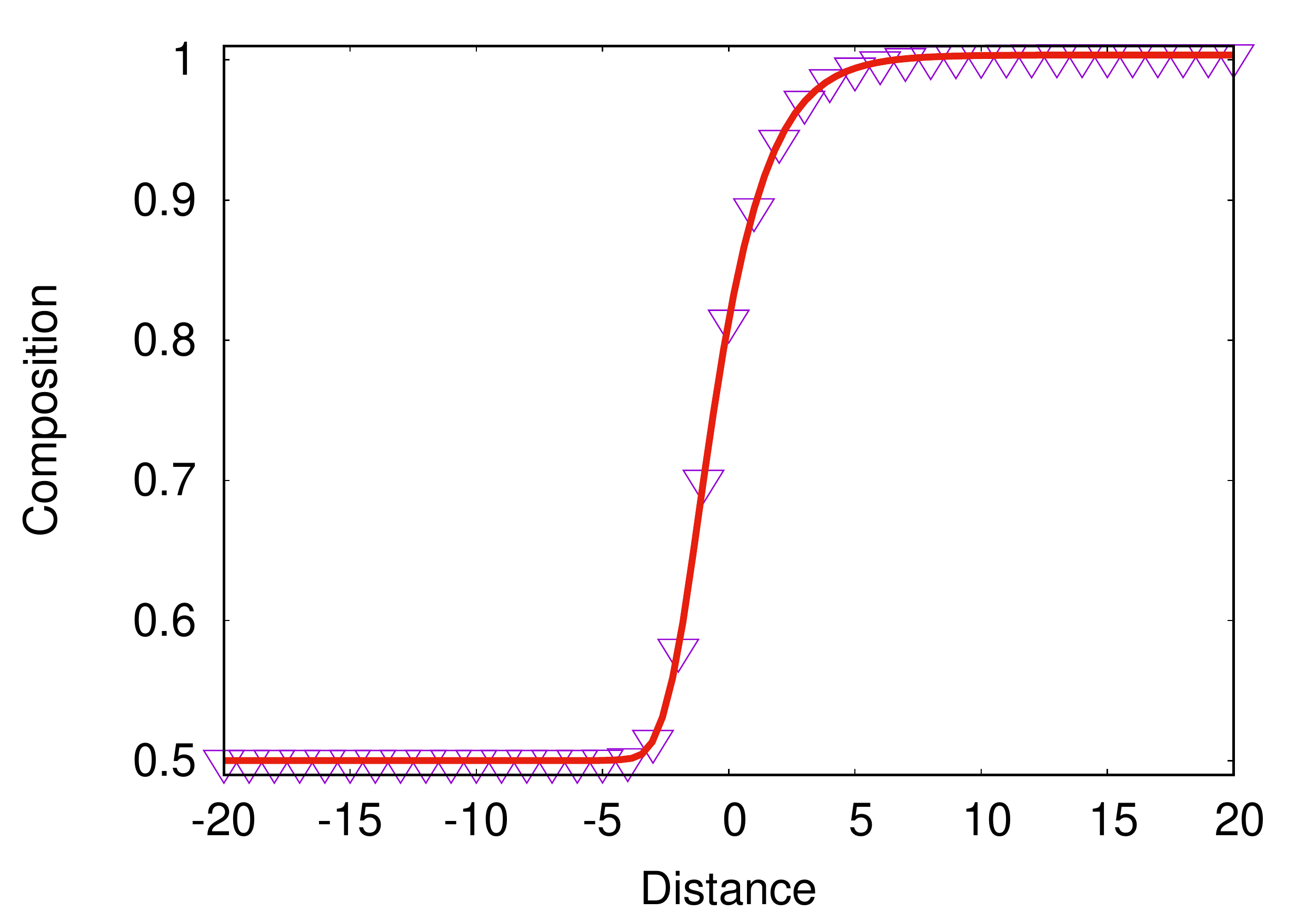}
  \caption{$\chi=0$, $\alpha\text{-}\beta_2$}
  \label{chi0_AB2}
\end{subfigure}%
\begin{subfigure}[b]{.25\linewidth}%
  \centering
  {\includegraphics[scale=0.18]{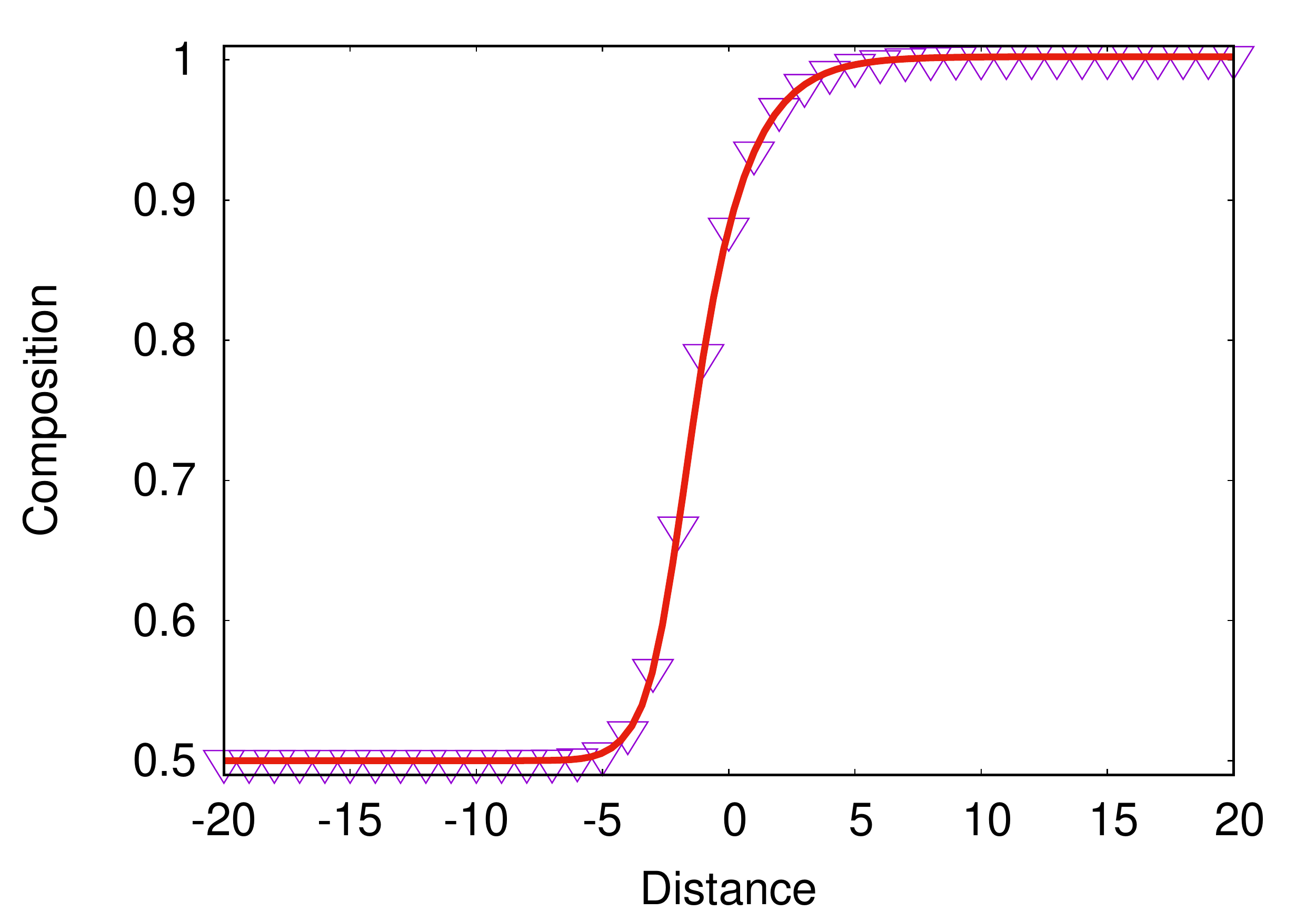}}
  \caption{$\chi=0.25$, $\alpha\text{-}\beta_2$}
  \label{chi025_AB2}
\end{subfigure}%
\begin{subfigure}[b]{.25\linewidth}%
  \centering
  \includegraphics[scale=0.18]{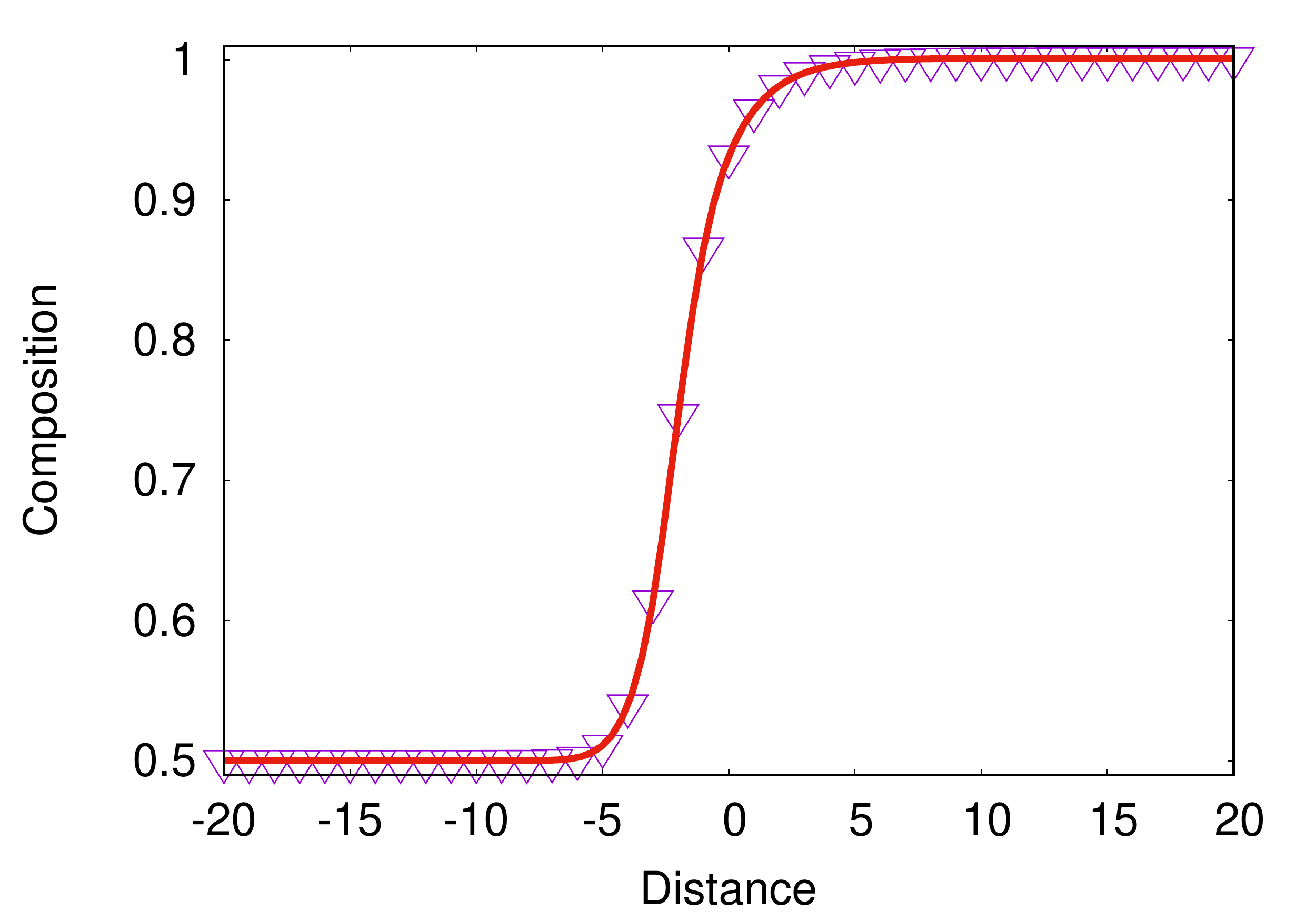}
  \caption{$\chi=0.5$, $\alpha\text{-}\beta_2$}
  \label{chi05_AB2}
\end{subfigure}%
\begin{subfigure}[b]{.25\linewidth}%
  \centering
  \includegraphics[scale=0.18]{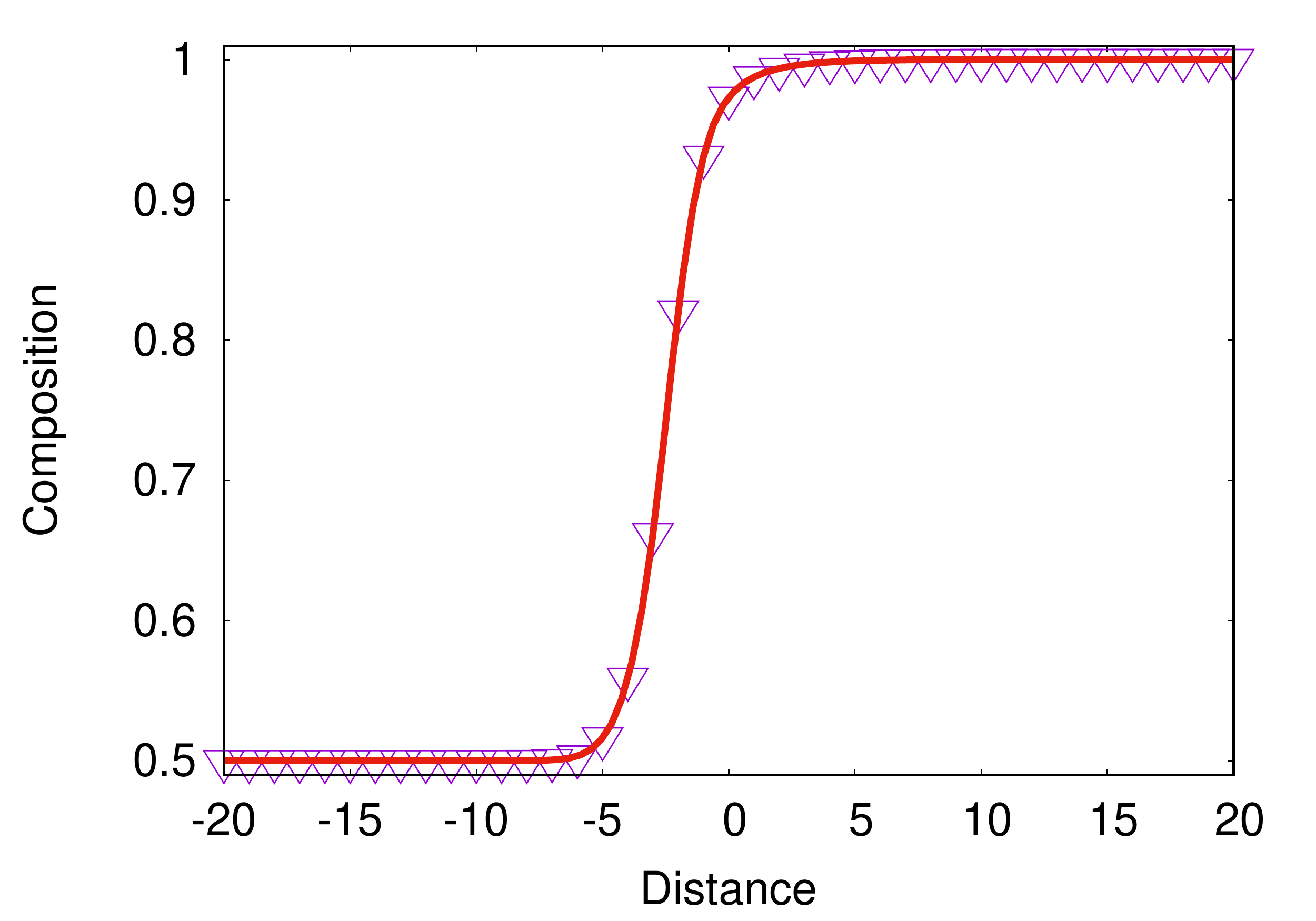}
  \caption{$\chi=0.75$, $\alpha\text{-}\beta_2$}
  \label{chi075_AB2}
\end{subfigure}%
\caption{Equilibrium composition profiles across $\alpha\text{-}\beta_1$ (top row)  and 
$\alpha\text{-}\beta_2$ (bottom row) interfaces for (a, e) $\chi=0$, (b, f) $\chi=0.25$, (c, g) 
$\chi=0.5$, (d, h) $\chi=0.75$%
\label{fig:1d-eqm-profiles}
}
  \label{1deq}
\end{figure*}

In our model, the parameter $\chi$ controls solute segregation at the particle surface, and 
consequently, influences the surface energies of the phases. To compute interfacial energies 
between different phases using Eq.~\ref{1deq}, we first obtain equilibrium composition profiles 
across them by carrying out one-dimensional (1D) simulations.  Figs.~\ref{1deq}(a--d) present
these profiles across the $\alpha\text{-}\beta_1$ interface for different values of $\chi$. 
These profiles have peaks in solute concentration within the interface region, and the peak
compositions increase with increasing $\chi$. The profiles for the $\alpha\text{-}\beta_2$
interface shown in Figs.~\ref{1deq}(e--h), on the other hand, do not exhibit any intermediate
peak, but increased $\chi$ appears to reduce the width of the compositional interface. The
composition profile for the $\beta_1\text{-}\beta_2$ interface remains completely unaffected 
by $\chi$ and is not presented here.  

Table~\ref{IE_CA} lists interfacial energies computed from the equilibrium composition profiles
and corresponding contact angles estimated using Eq.~\eqref{eq-Young}. 

\begin{table}[htbp]
 \centering
 \small\caption{Variation of interfacial energies and contact angle with $\chi$}
 \label{IE_CA}
 \begin{tabular*}{0.45\textwidth}{@{\extracolsep{\fill}}*{5}l}
 \hline
  $\chi$& $\sigma_1$ & $\sigma_2$ &$\sigma_{12}$ & $\theta$\\
  \hline
  $0$ & $3.31$ & $3.31$ & $0.11$ & \ang{90} \\
  $0.25$ & $3.07$ & $3.03$ & $0.11$ & \ang{69} \\
  $0.5$ & $2.80$ & $2.72$ & $0.11$ & \ang{43} \\
  $0.75$ & $2.50$ & $2.39$ & $0.11$ & \ang{0}\\
  \hline
 \end{tabular*}
\end{table}

\begin{figure}[htbp]
 \centering%
 \includegraphics[scale=0.3]{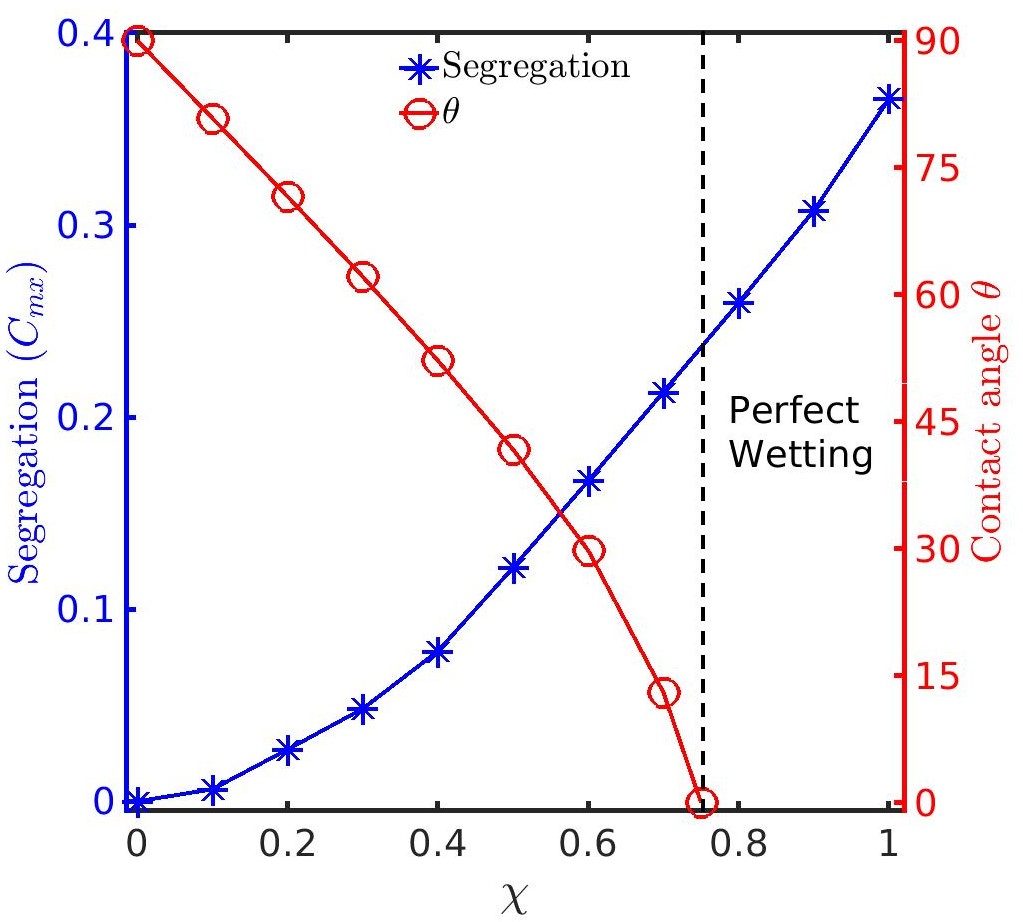}
 \caption{Variation of equilibrium segregation and contact angle with $\chi$}
 \label{fig:Seg_CA}
\end{figure}

Fig.~\ref{fig:Seg_CA} shows that the equilibrium solute segregation at the high-energy 
$\alpha\text{-}\beta_1$ interface increases with increasing $\chi$. The extent of segregation
is represented here by the maximum in interface composition, $C_{\textrm{mx}}$, as revealed in
Fig.~\ref{1deq}. We note that $C_{\textrm{mx}}$ is similar to the limiting surface composition
$C_s$ defined by Cahn in his theory of wetting~\cite{Cahn1977}; however, a direct correspondence
is not possible as $C_s$ cannot be obtained analytically. 

Fig.~\ref{fig:Seg_CA} also shows that the three-phase contact angle $\theta$ 
decreases with increasing $\chi$. Although surface energies of both $\beta_1$ 
and $\beta_2$ phases decrease with increasing $\chi$, the reduction in $\sigma_2$
is greater than that of $\sigma_1$ (see Table~\ref{IE_CA}). This is a direct
consequence of the way $\chi$ affects the $f$-$c$ curves of Fig.~\ref{fig:fmiddle}:
free energy function becomes more asymmetric with increasing $\chi$, dropping to 
increasingly lower levels at the $\beta_2$ end ($c=1$). Moreover, a comparison of
equilibrium $\alpha\text{-}\beta_2$ profiles reveals lower gradient energy 
contributions as $\chi$ increases (see the bottom row of Fig.~\ref{1deq}). Huang 
\emph{et al.} have discussed a similar asymmetry in the context of interfacial
adsorption in ternary alloys~\cite{Huang}. Thus increasing $\chi$ results in a 
monotonic increase of the difference between the surface energies,
$\Delta{\sigma}=\sigma_1-\sigma_2$, which leads to a corresponding decrease in
contact angle $\theta$. At $\chi=0.75$,  $\theta$ becomes zero; any further 
increase in $\chi$ continues to increase the segregation, but no contact angle 
can be defined under this perfect wetting regime as the low-energy 
$\alpha\text{-}\beta_2$ interface completely replaces the three-phase contact.

We note that there may be challenges in determining $\chi$ accurately from 
first-principles calculations or experiments. Experimental measurements of 
surface segregation in alloy nanoparticles~\cite{Hennes2015b,Peng2015,Tang2019a}
can be a viable route to determine $\chi$ if a proper correlation between
surface energy and extent of segregation can be established. Atomistic 
simulations of alloy nanocrystals that account for interactions between the
alloy constituents with the surrounding medium can also give useful insights
on the surface segregation and its role on the stability of nanocrystals%
~\cite{Hennes2015b, Peng2015, Zhao2017, Bochicchio2016b, Wang2005}.

\subsection{Drop-on-substrate simulations}

The contact angle can play a crucial role during phase separation in confined
geometries. In some recent phase field models~\cite{Lee2011, Yu2012}, it has 
been imposed as a boundary condition over the confined domain. Our model, 
however, does not require imposition of any additional condition or 
constraint: the correct contact is angle is recovered naturally for the chosen
set of relevant free energy parameters  (\emph{viz.}, $f_0^m,f_0^p,\omega_0,
\kappa_c,\kappa_\phi$, and $\chi$).

\begin{figure*}[hbpt]
 \captionsetup[subfigure]{justification=centering}
  \begin{subfigure}[b]{0.19\linewidth}%
    \centering%
    {\includegraphics[width=0.9\textwidth,trim={5cm 0 5cm 0},clip]{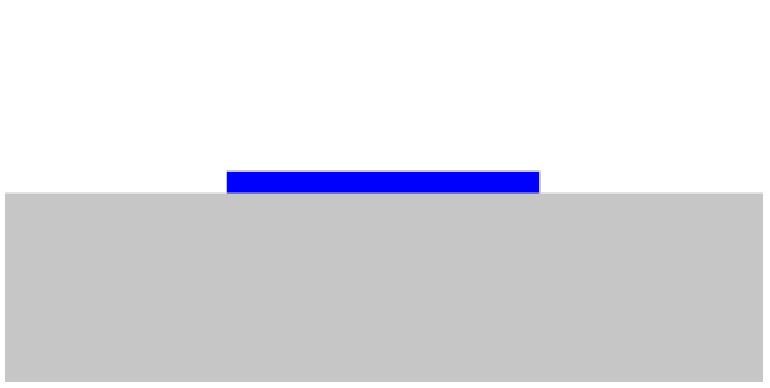}}%
    \caption{Initial profile}%
    \label{Init_rect}%
  \end{subfigure}%
  \begin{subfigure}[b]{.19\linewidth}%
    \centering
    {\includegraphics[width=0.9\textwidth,trim={12cm 0 12cm 0},clip]{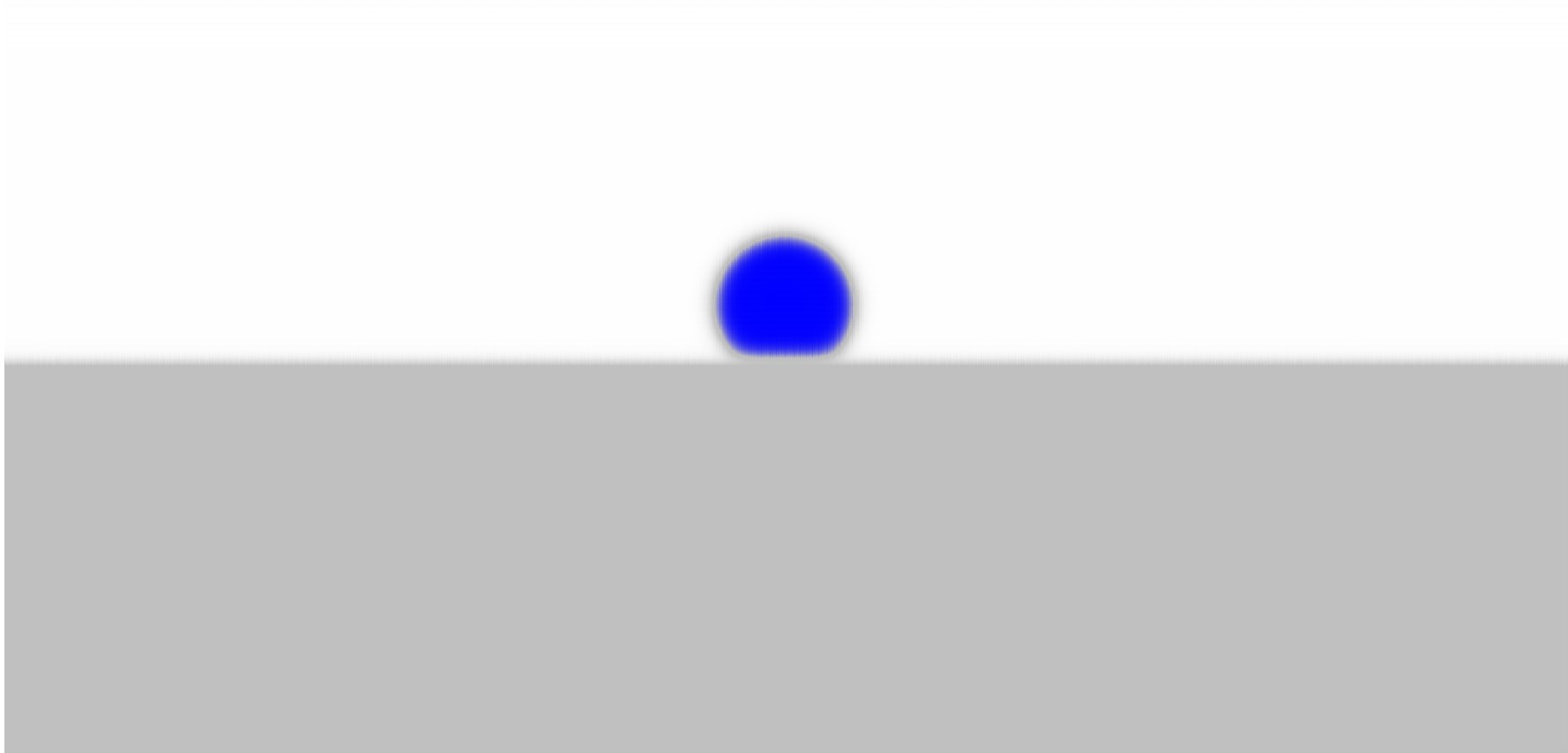}}%
    \caption{$\chi=-0.5$, $\theta=\ang{137}$}%
    \label{theta138}
  \end{subfigure}%
  \begin{subfigure}[b]{0.19\linewidth}%
    \centering
    {\includegraphics[width=0.9\textwidth,trim={12cm 0 12cm 0},clip]{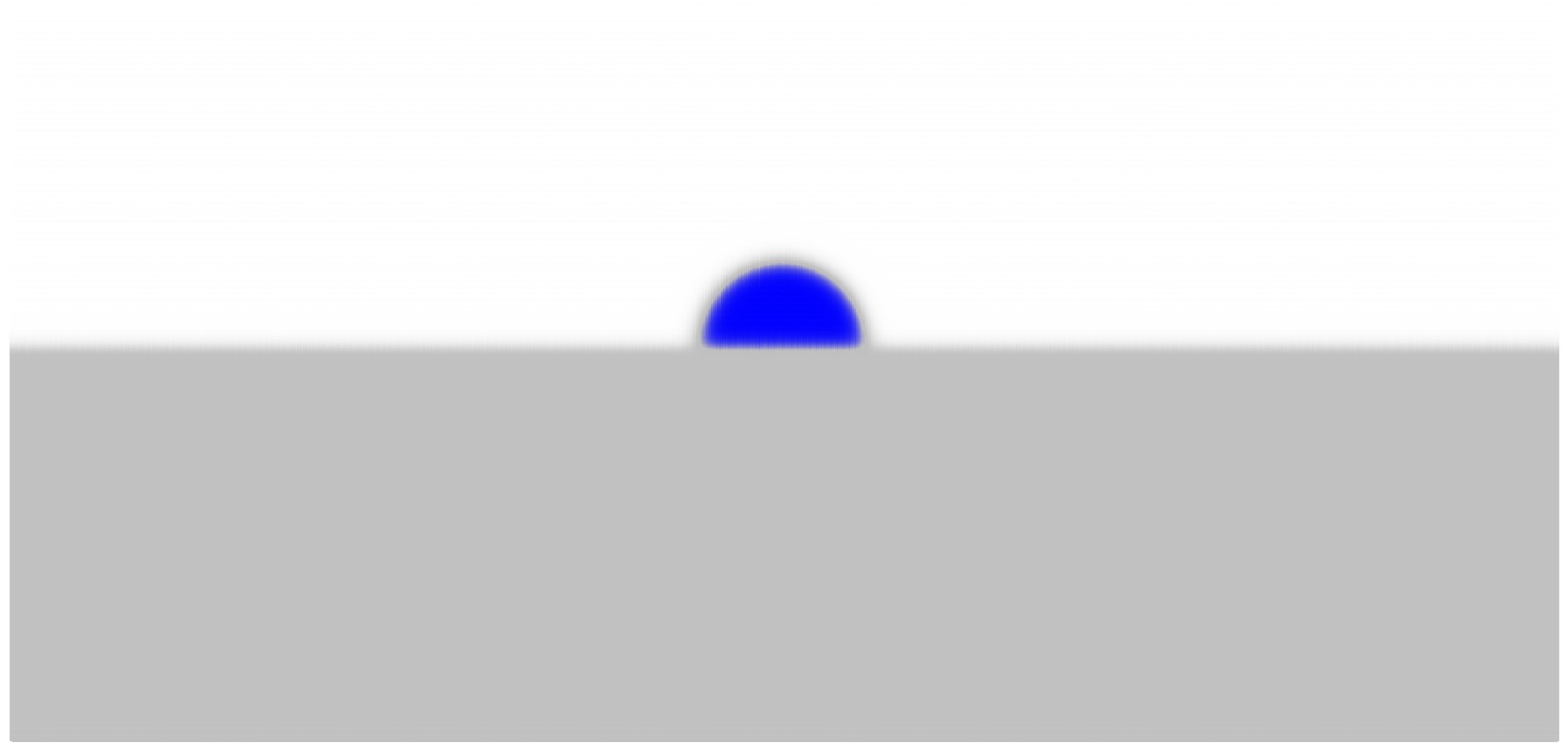}}%
    \caption{$\chi=0$, $\theta=\ang{90}$}%
    \label{theta90}
  \end{subfigure}%
  \begin{subfigure}[b]{0.19\linewidth}%
    \centering
    {\includegraphics[width=0.9\textwidth,trim={12cm 0 12cm 0},clip]{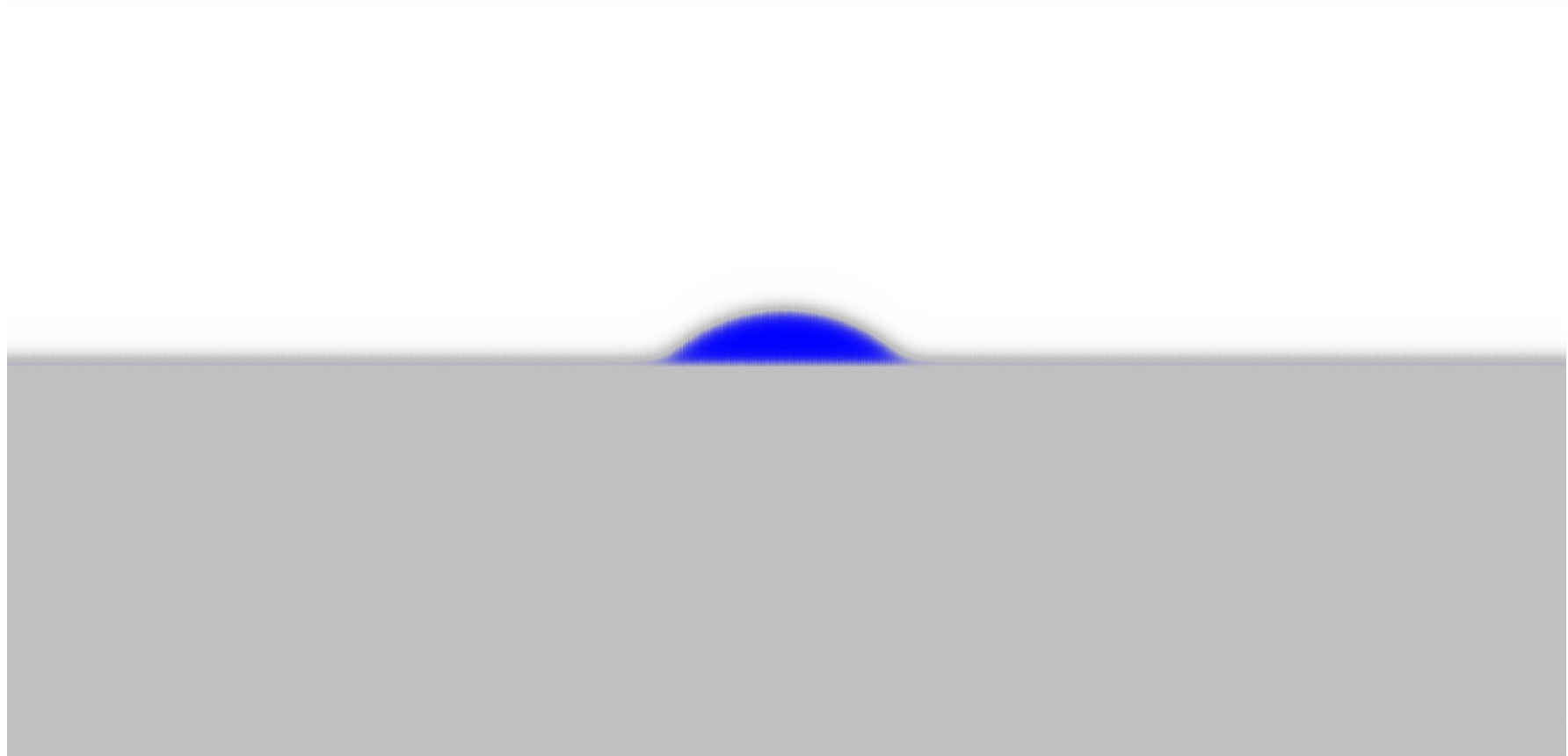}}
    \caption{$\chi=0.5$, $\theta=\ang{43}$}%
    \label{theta42}
  \end{subfigure}%
  \begin{subfigure}[b]{0.19\linewidth}%
    \centering
    {\includegraphics[width=0.9\textwidth,trim={12cm 0 12cm 0},clip]{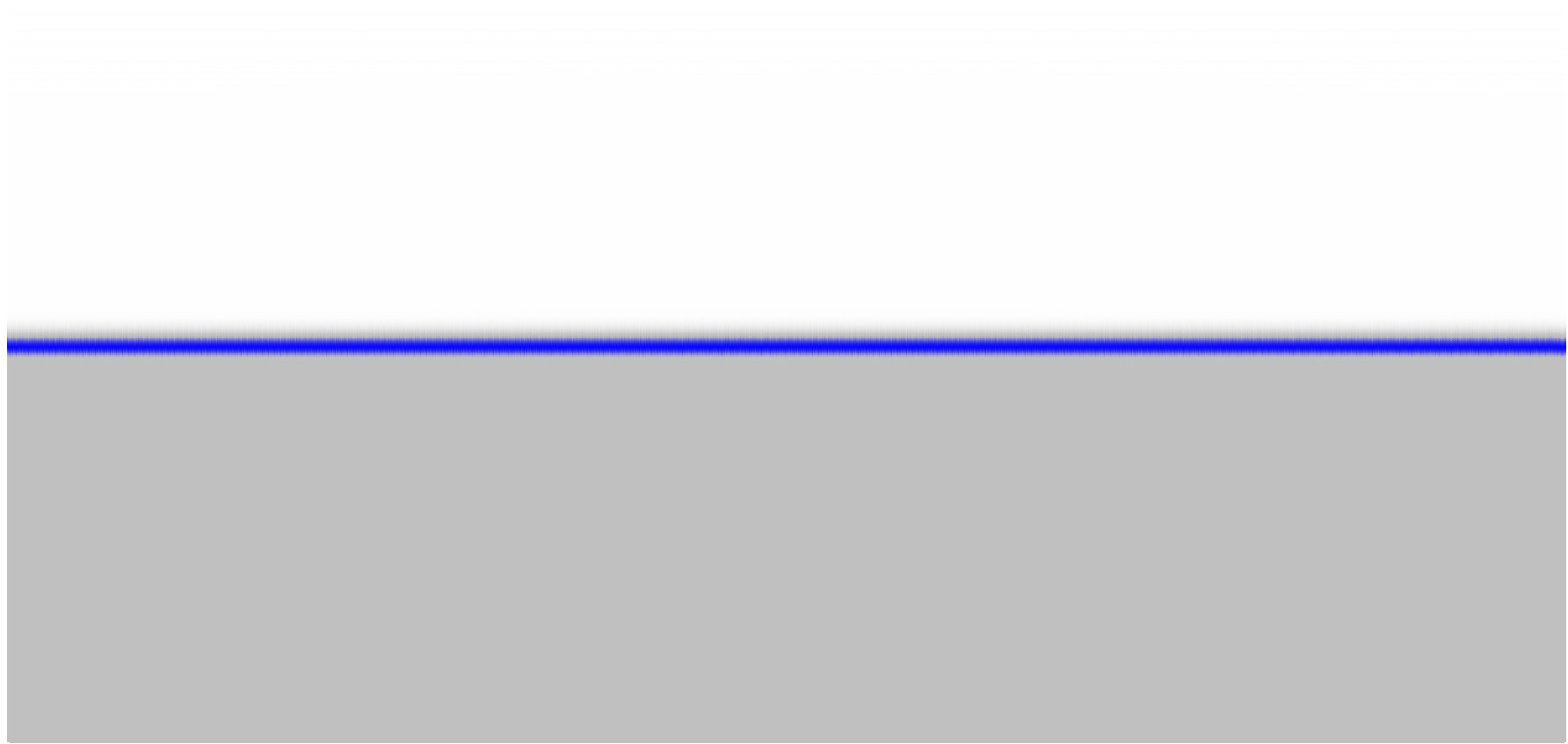}}
    \caption{$\chi=0.5$, $\theta=\ang{0}$}%
    \label{CWC}
  \end{subfigure}%
  \caption{Attainment of correct contact angle of a drop on a substrate: (a)
  initial drop (b-e) final configurations for different values of $\chi$}
  \label{result_drop}
\end{figure*}

Fig.~\ref{result_drop} demonstrates the capability of our model in capturing 
wetting behavior with the correct contact angle. We choose a drop-on-substrate
configuration where we initially place a $\beta_2$ droplet (blue) on an $\alpha$
(grey) substrate; $\beta_1$ phase (white) surrounds the droplet, thereby creating
two three-phase contact points. The droplet evolves via diffusion along the 
interfaces till the three-phase contact angle attains equilibrium. To have both
acute and obtuse contact angles in these simulations, we have chosen positive and
negative values of $\chi$, respectively. For a given size, the initial shape of
the drop does not influence the kinetic path to the final configuration. The 
contact angles measured on the final configurations using level sets for $c$ and
$\phi$ at $0.5$ agree with those predicted by Young's equation within 
$\pm\ang{2}$ (Table~\ref{IE_CA}). 

\subsection{Morphology as a function of size and contact angle}

\begin{figure}[htbp]
\captionsetup[subfigure]{justification=centering}
\centering%
\hspace*{\fill}%
\begin{subfigure}{0.1\textwidth}
\centering
    \includegraphics[scale=0.0175]{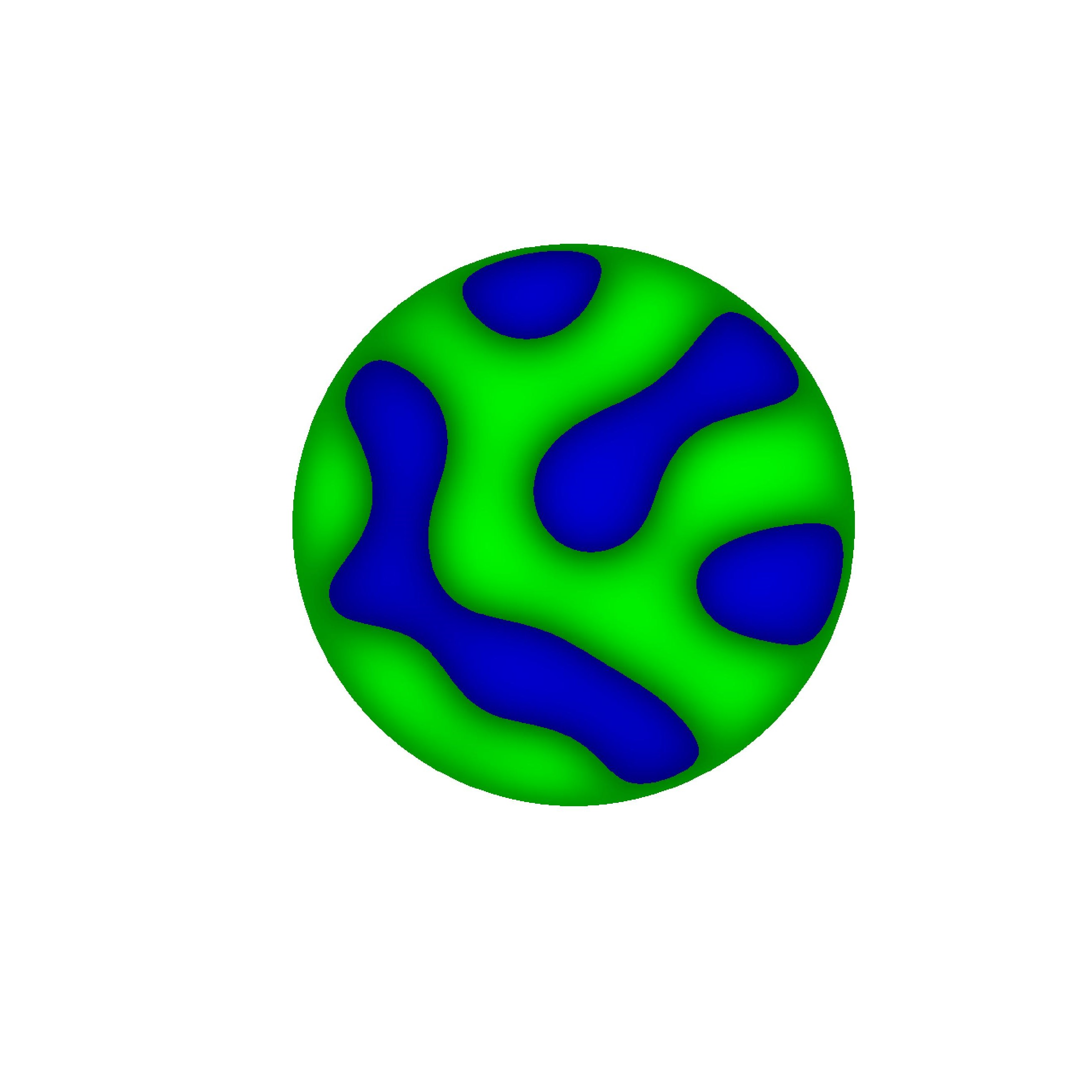}\hfil%
    \caption{}
    \label{r70a_early}
\end{subfigure}%
\begin{subfigure}{0.1\textwidth}
\centering
    \includegraphics[scale=0.0175]{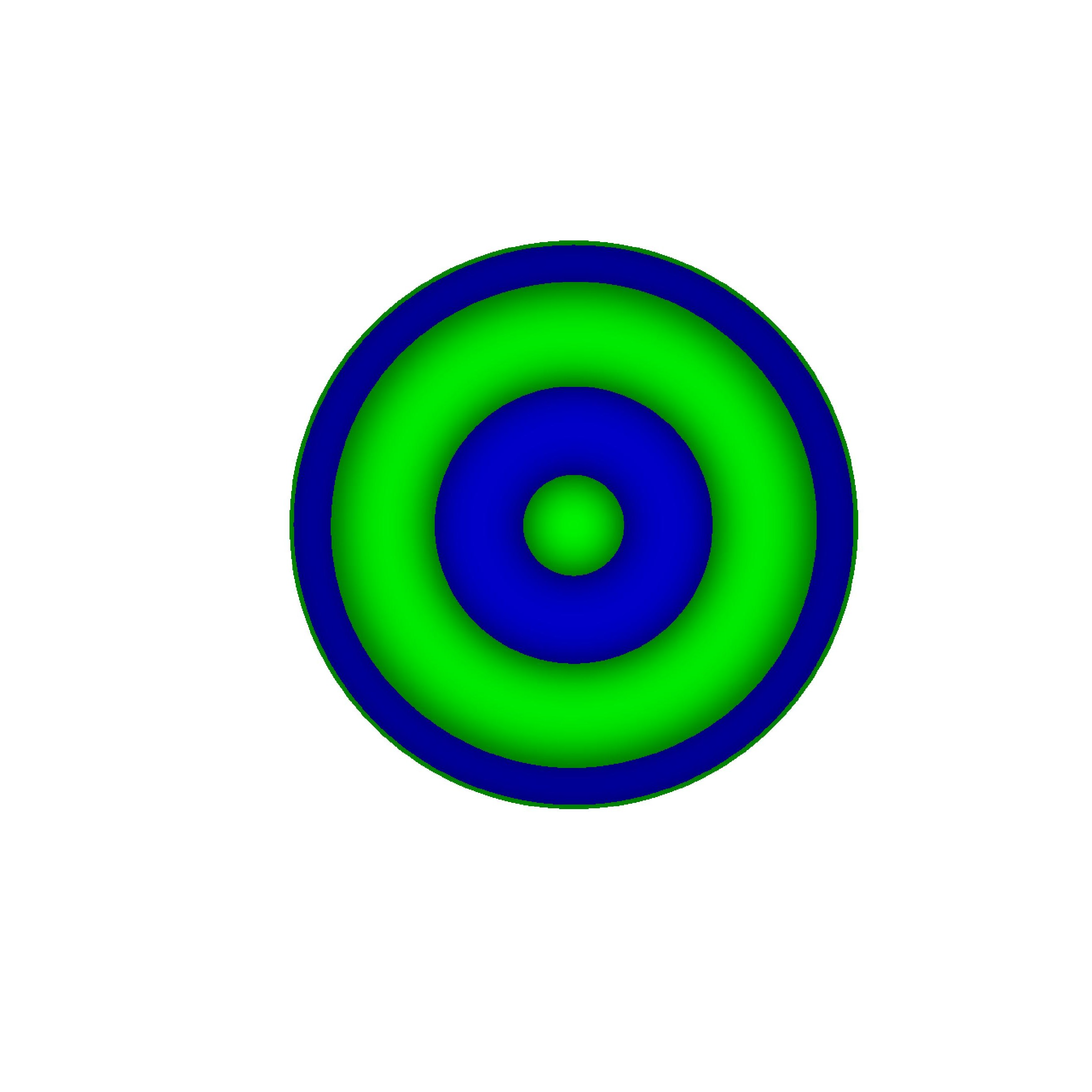}\hfil%
    \caption{}
    \label{r70b_early}
\end{subfigure}
\begin{subfigure}{0.1\textwidth}
\centering
    \includegraphics[scale=0.0175]{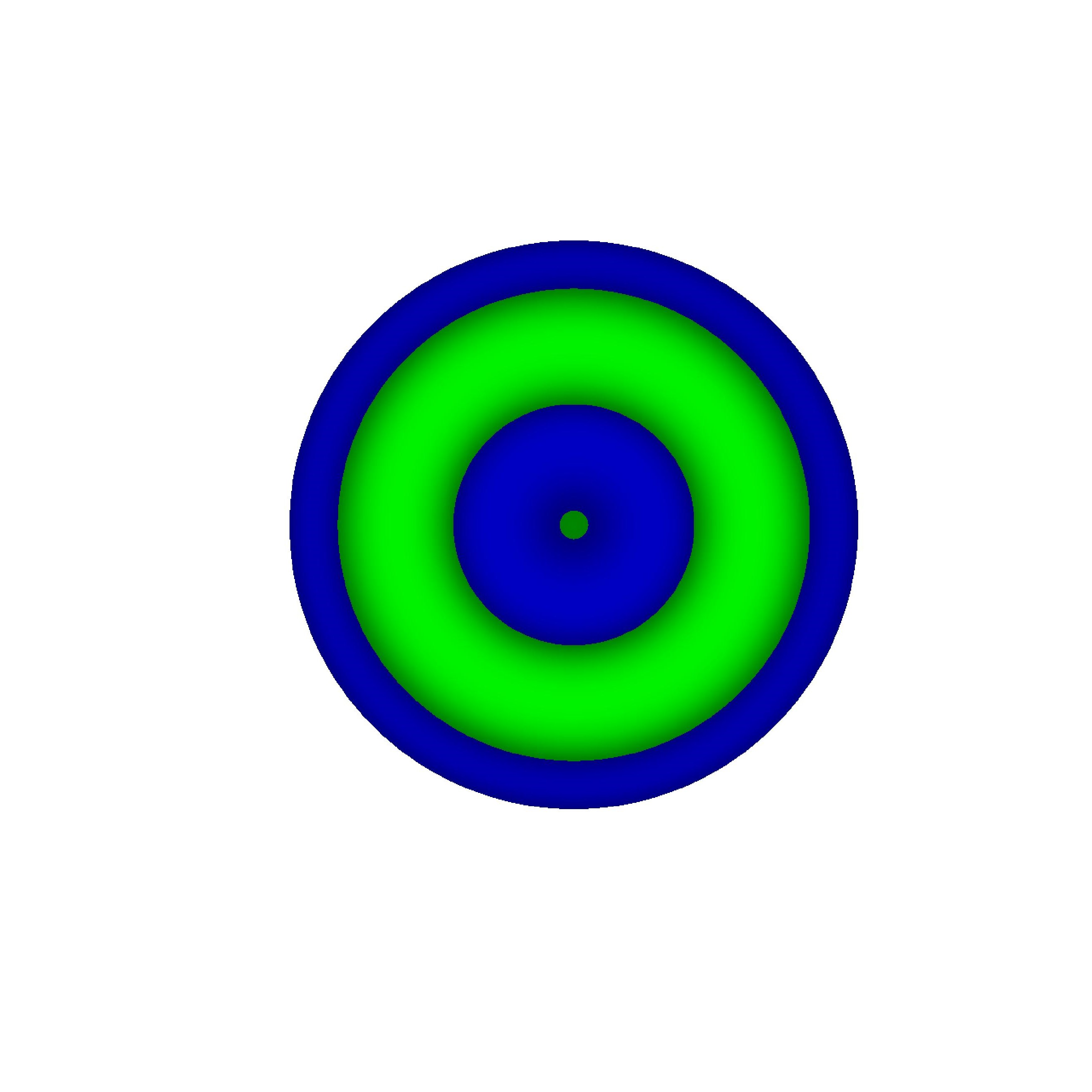}\hfil%
    \caption{}
    \label{r70c_early}
\end{subfigure}%
\begin{subfigure}{0.1\textwidth}
\centering
    \includegraphics[scale=0.0175]{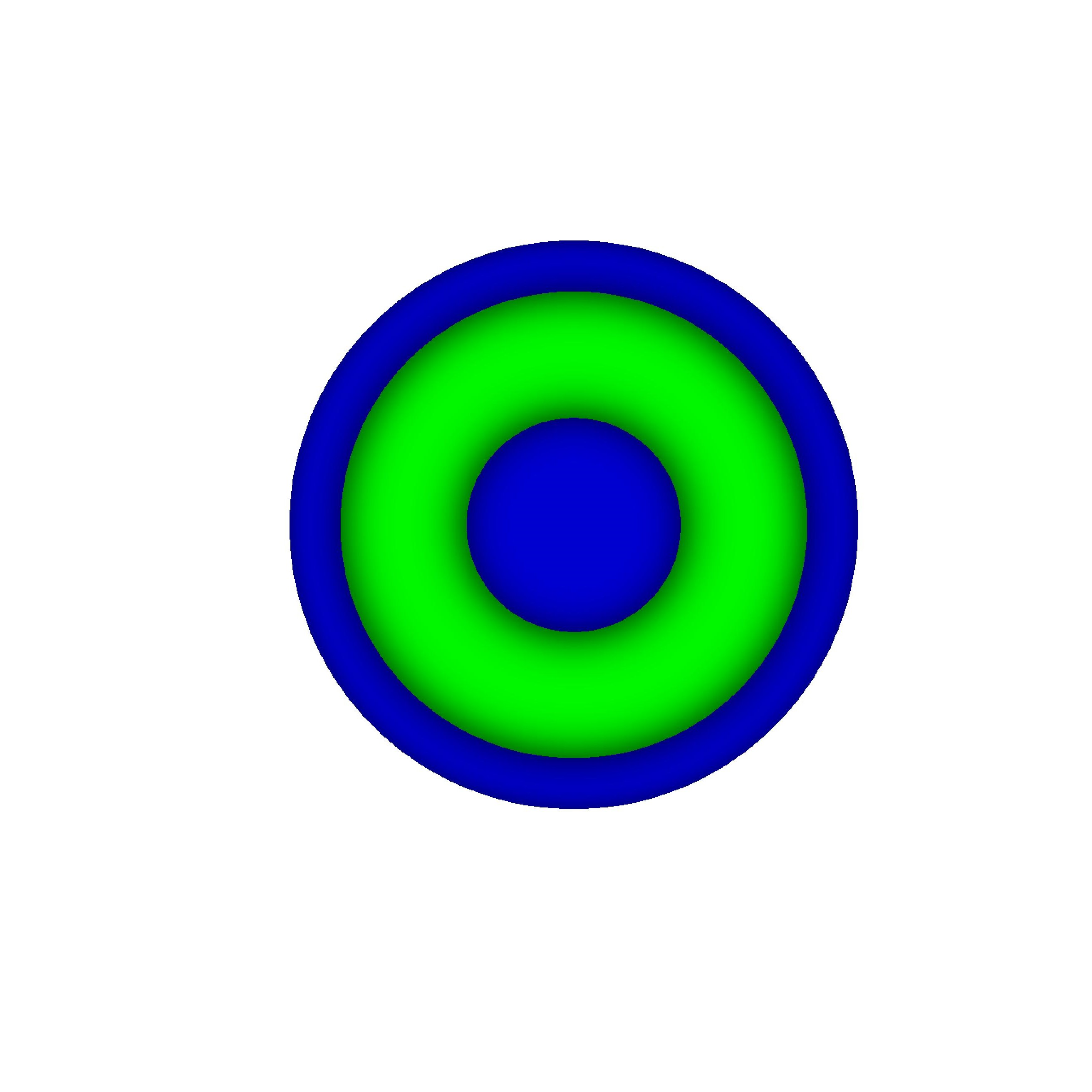}%
    \caption{}
    \label{r70d_early}
\end{subfigure}
\hspace*{\fill}%

\hspace*{\fill}%
\begin{subfigure}{0.1\textwidth}
\centering
    \includegraphics[scale=0.02]{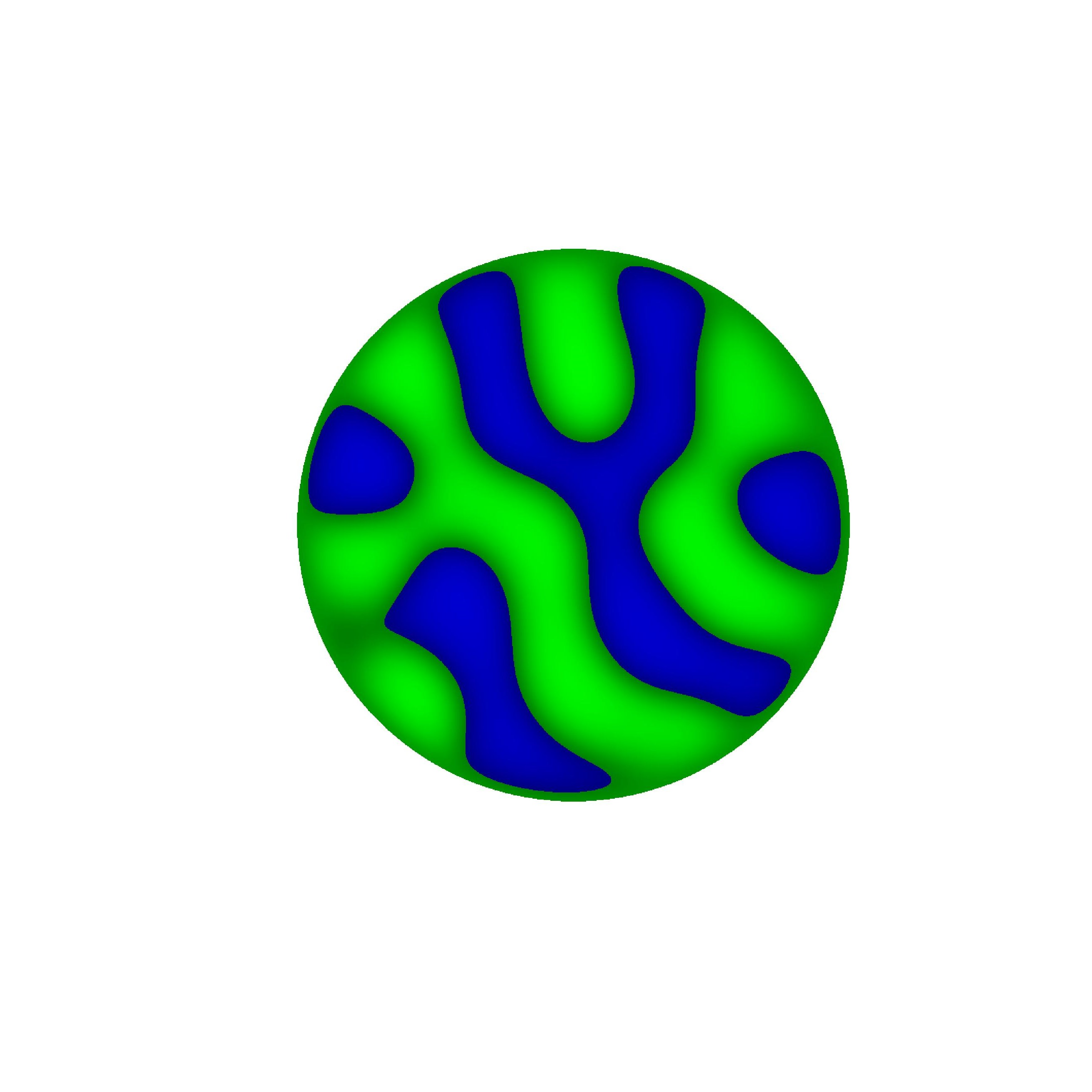}\hfil%
    \caption{}
    \label{r80a_early}
\end{subfigure}%
\begin{subfigure}{0.1\textwidth}
\centering
    \includegraphics[scale=0.0175]{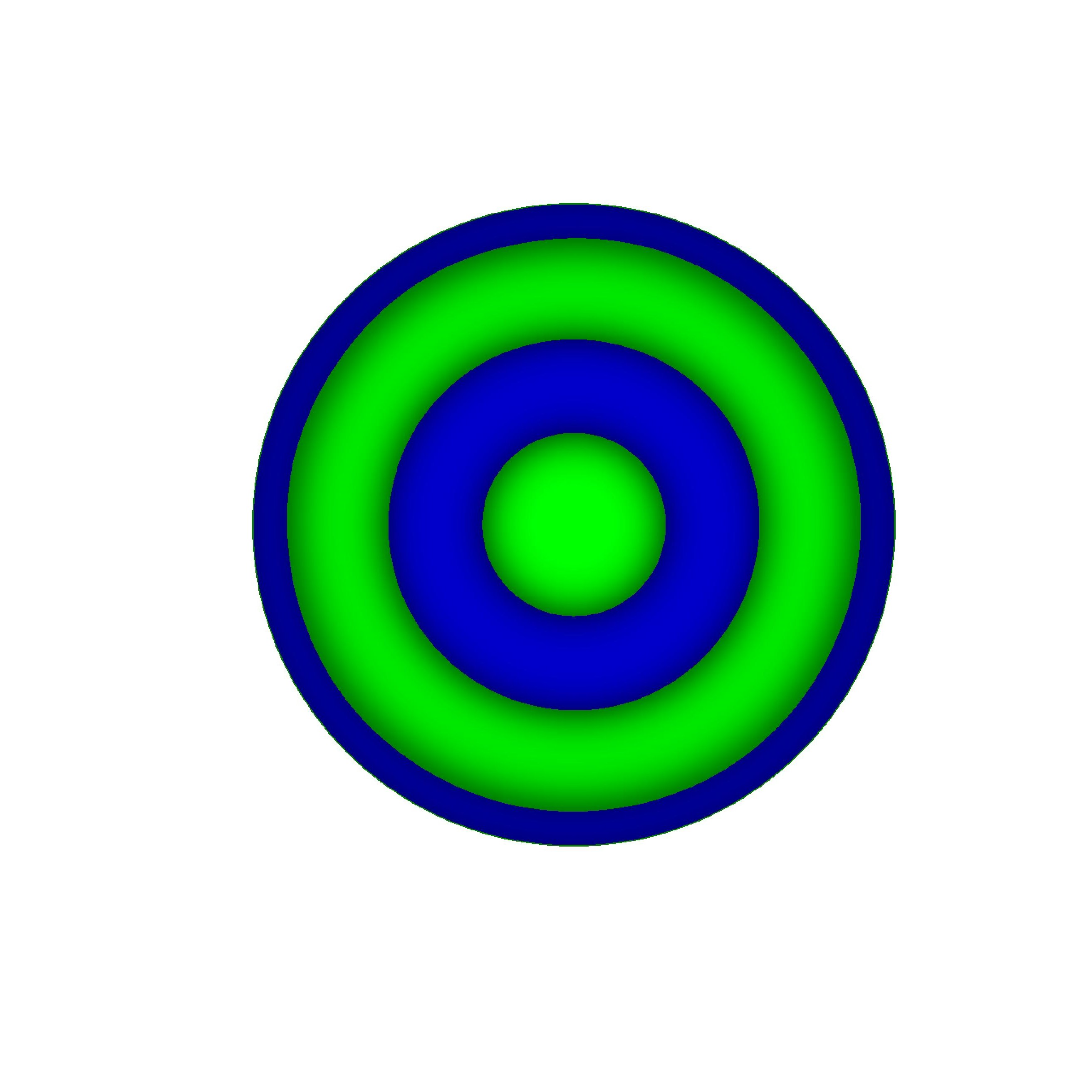}\hfil%
    \caption{}
    \label{r80b_early}
\end{subfigure}
\begin{subfigure}{0.1\textwidth}
\centering
    \includegraphics[scale=0.02]{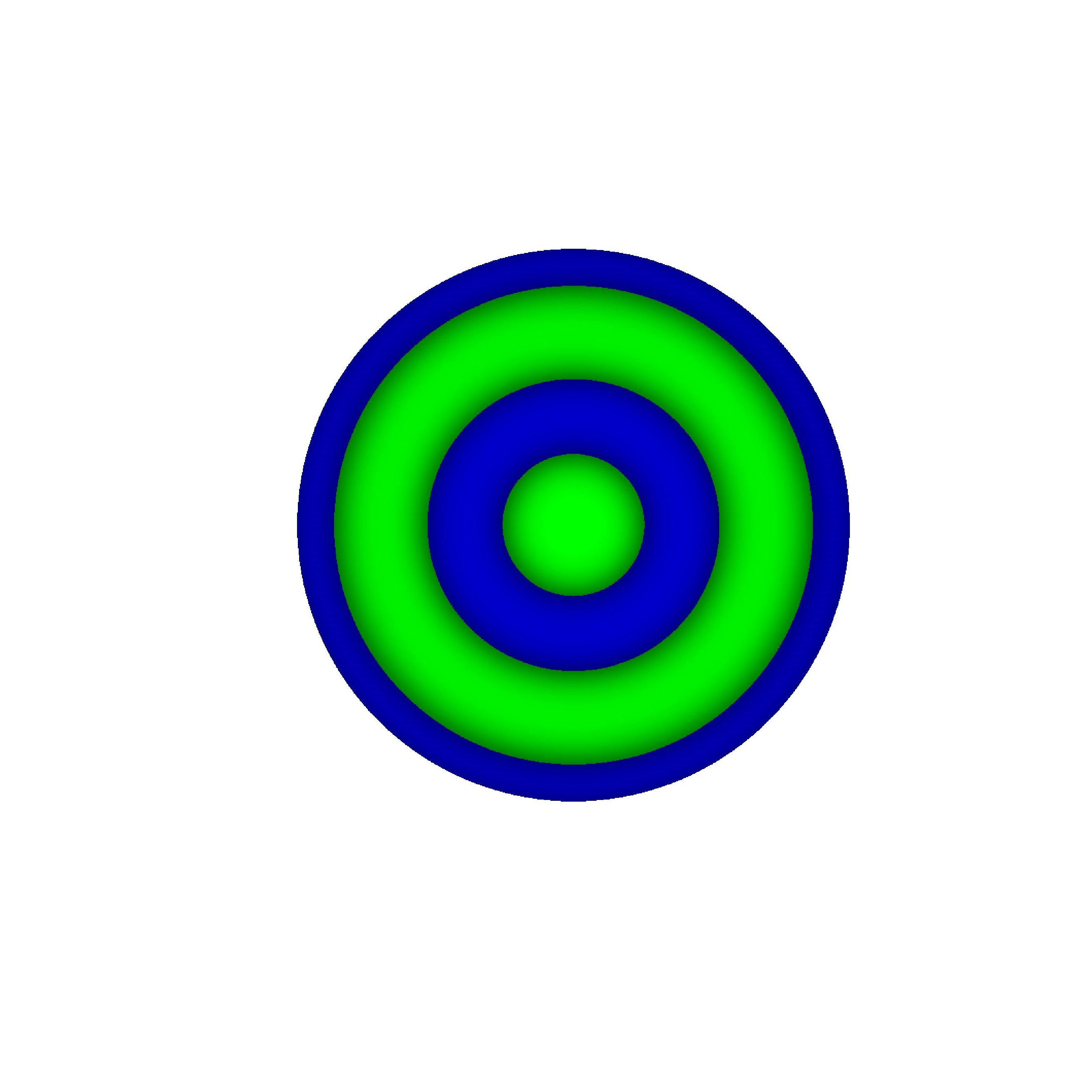}\hfil%
    \caption{}
    \label{r80c_early}
\end{subfigure}%
\begin{subfigure}{0.1\textwidth}
\centering
    \includegraphics[scale=0.02]{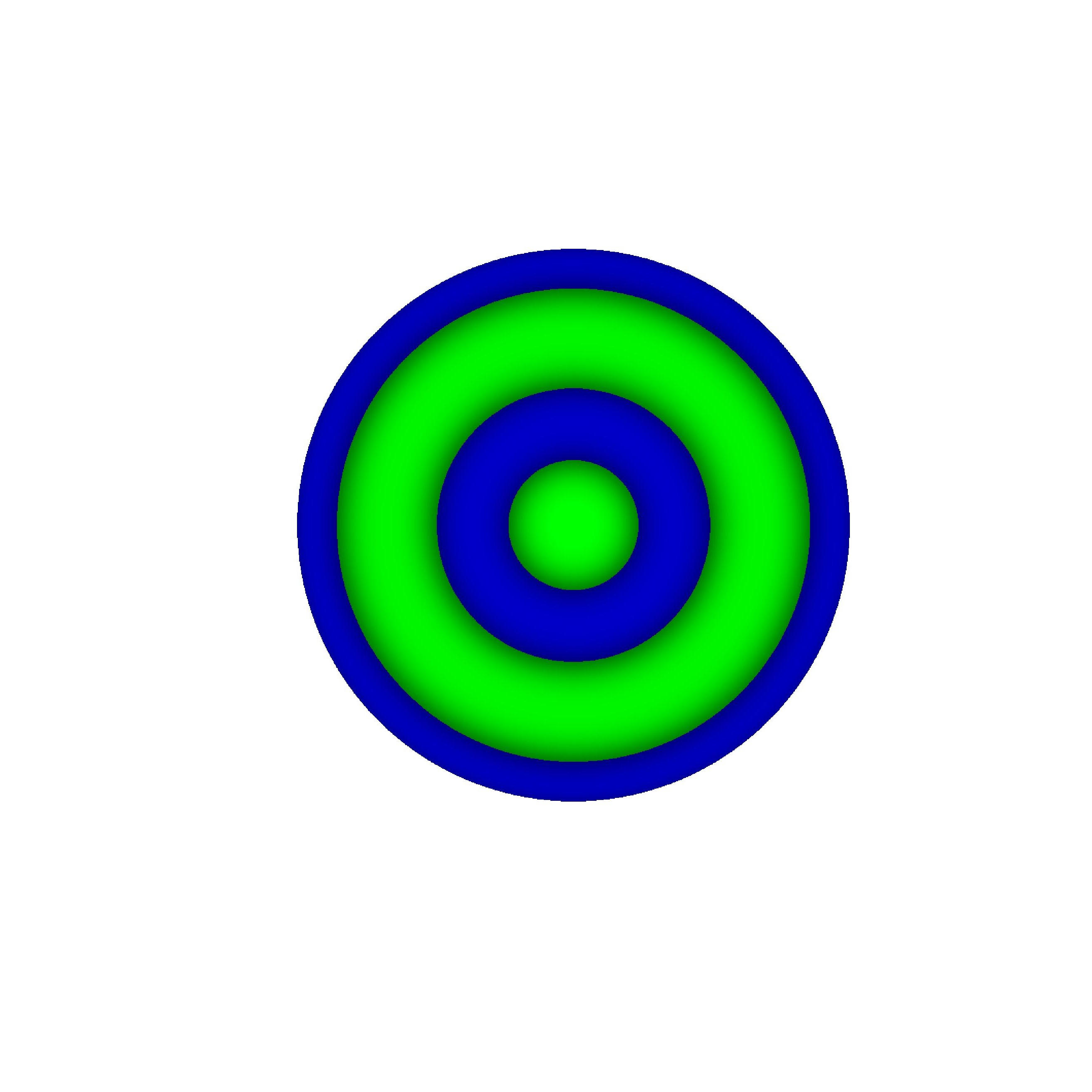}%
    \caption{}
    \label{r80d_early}
\end{subfigure}%
\hspace*{\fill}%

\hspace*{\fill}%
\begin{subfigure}{0.1\textwidth}
\centering
    \includegraphics[scale=0.0225]{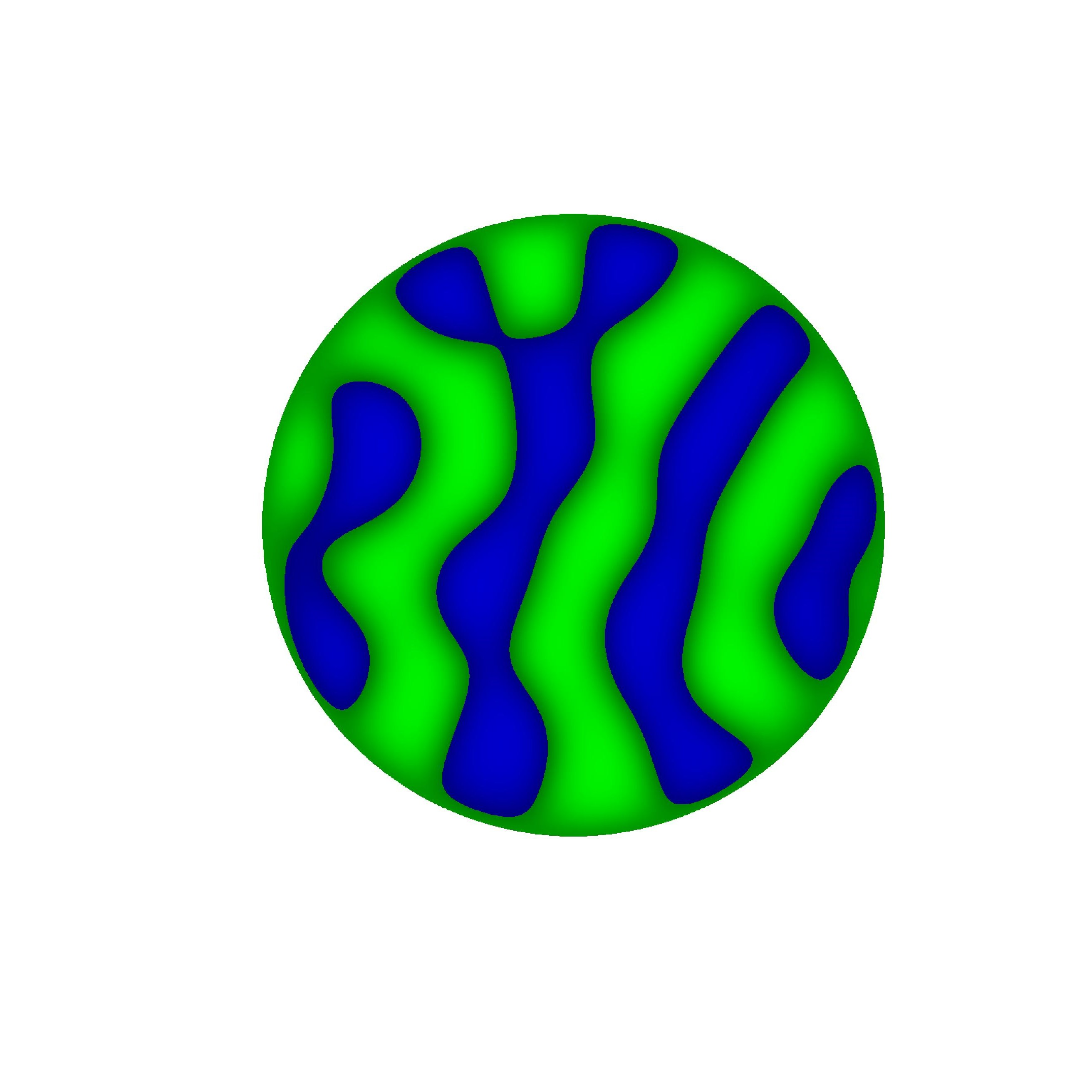}\hfil%
    \caption{}
    \label{r90a_early}
\end{subfigure}%
\begin{subfigure}{0.1\textwidth}
\centering
    \includegraphics[scale=0.0225]{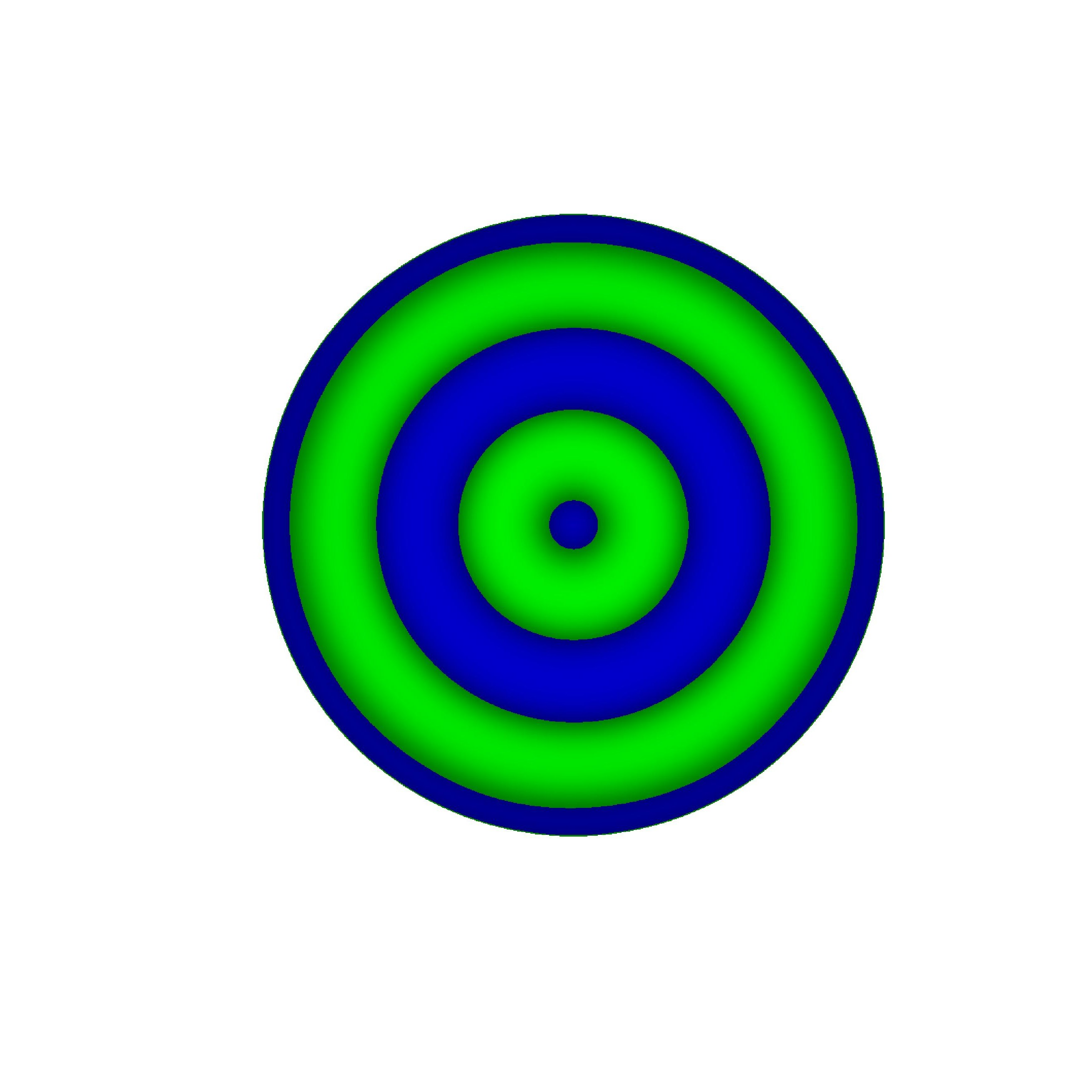}\hfil%
    \caption{}
    \label{r90b_early}
\end{subfigure}
\begin{subfigure}{0.1\textwidth}
\centering
    \includegraphics[scale=0.0225]{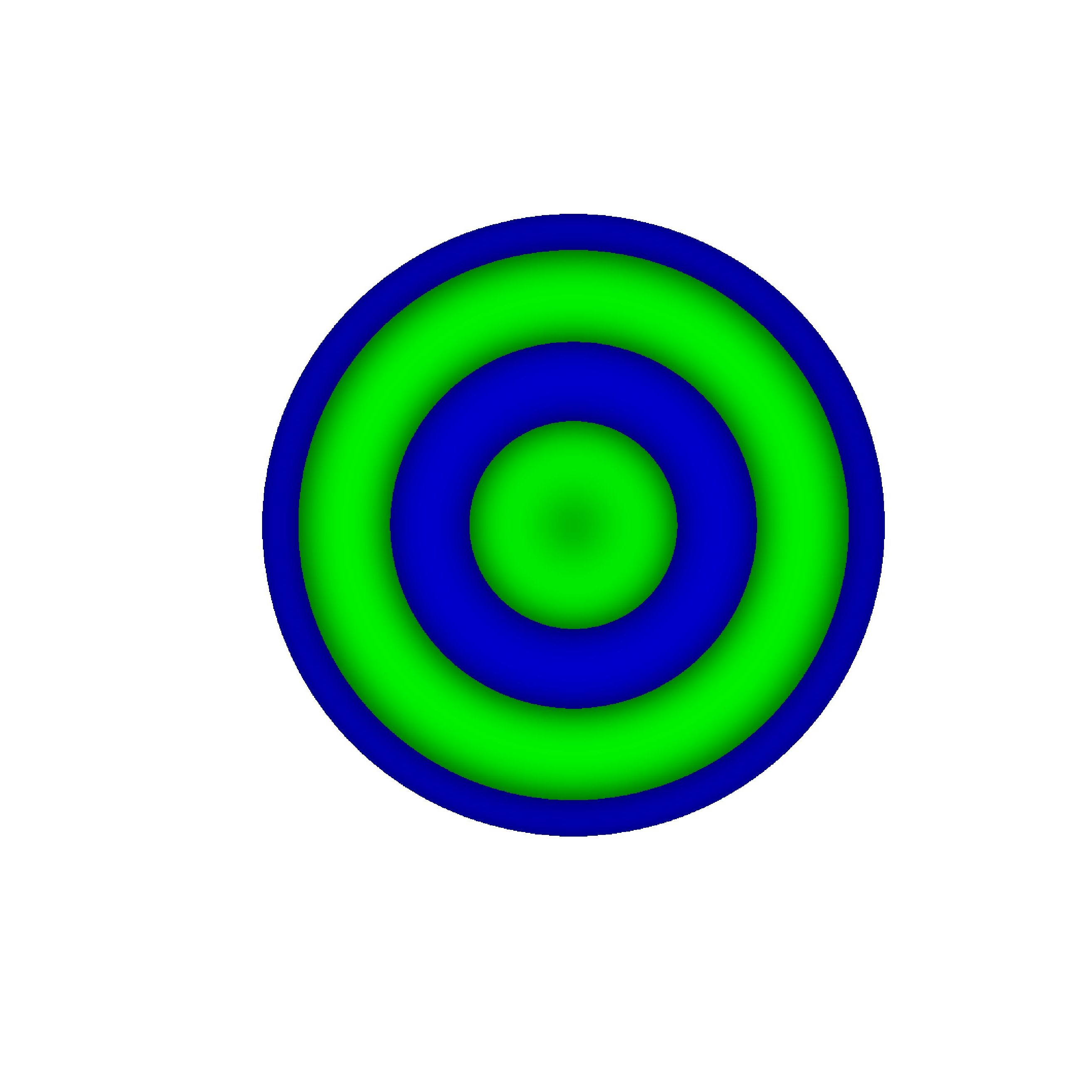}\hfil%
    \caption{}
    \label{r90c_early}
\end{subfigure}%
\begin{subfigure}{0.1\textwidth}
\centering
    \includegraphics[scale=0.0225]{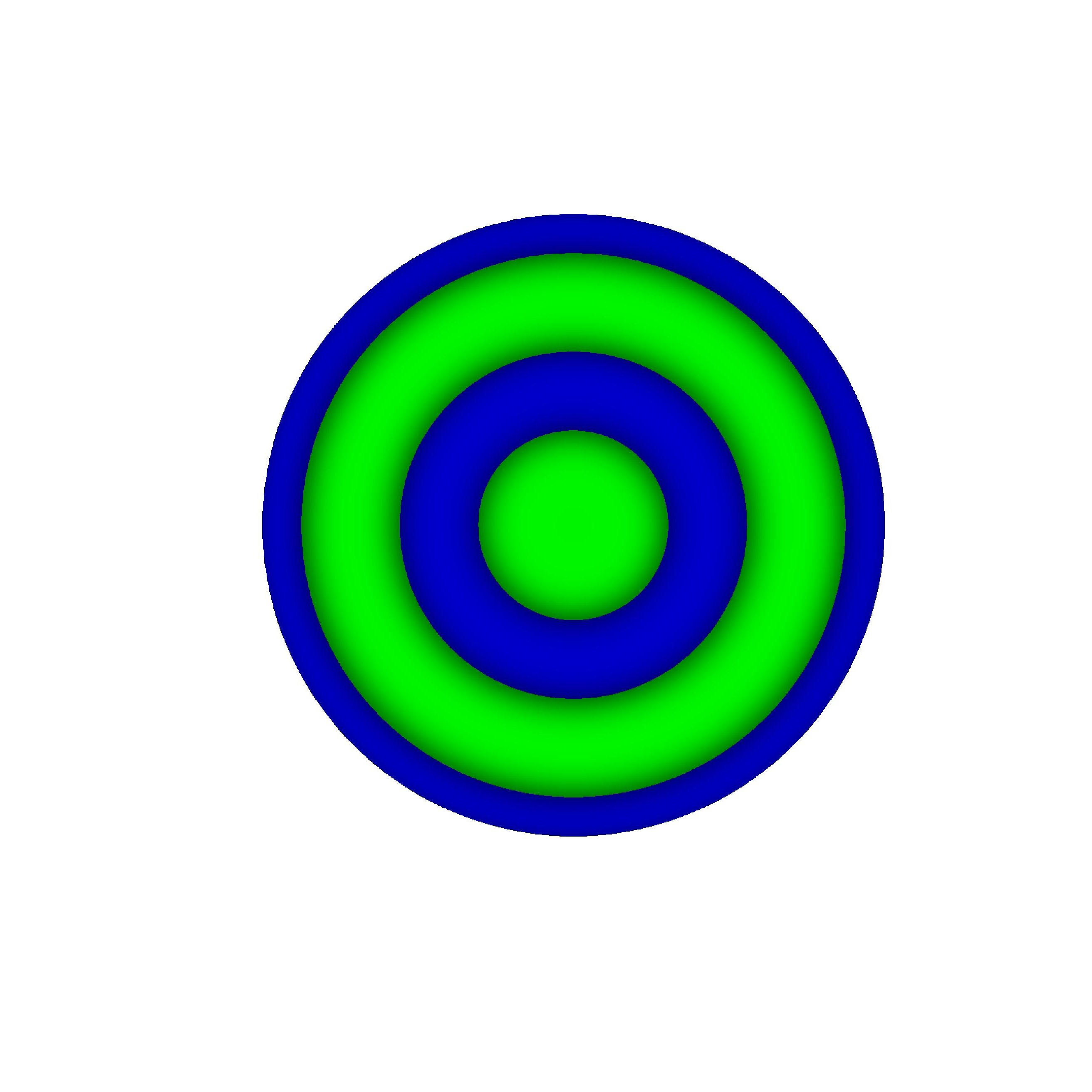}%
    \caption{}
    \label{r90d_early}
\end{subfigure}%
\hspace*{\fill}%

\caption{Early-stage time snapshots of evolution of BNP morphology. First, second, 
third and fourth columns correspond to $\chi=0$, $\chi=0.25$, $\chi=0.5$, and 
$\chi=0.75$, respectively. Particle size $d$ in the top, middle, and bottom row is 
140, 160, and 180, respectively; simulation boxes are scaled accordingly to reflect
relative sizes. Blue and green colors represent solute-rich $\beta_2$ and solute-poor
$\beta_1$ phases, respectively.}%
\label{Early_stage}
\end{figure}

BNPs primarily exhibit two distinct types of morphology: CS and Janus%
\footnote[1]{Although some other forms such as ``onion-like'', ``dumbbell'', ``ball
and cup'' morphologies have been reported%
~\cite{Xu2008, Hennes2015b, Ferrando2008, Cheng2006, Baletto2003, Cheng2008, Paz-Borbon2008},
these are mostly intermediate forms of CS and Janus structures (see, for example, 
Fig.~\ref{Early_stage} for an ``onion-like'' layered structure)}. In addition to
contact angle, factors such as particle size and shape, and alloy composition too
can influence the morphology. Since one of our major goals is to understand the role
of surface energies on SD-induced transformations in confined domains, we focus mainly
on circular particles which have a constant curvature at the surface. Such a configuration
permits a fundamental investigation of the effect of size on wetting/dewetting dynamics 
without complexities arising from curvature gradients in non-circular particles.

\begin{figure}[htbp]
\captionsetup[subfigure]{justification=centering}
\centering%
\hspace*{\fill}%
\begin{subfigure}{0.1\textwidth}
\centering
    \includegraphics[scale=0.0175]{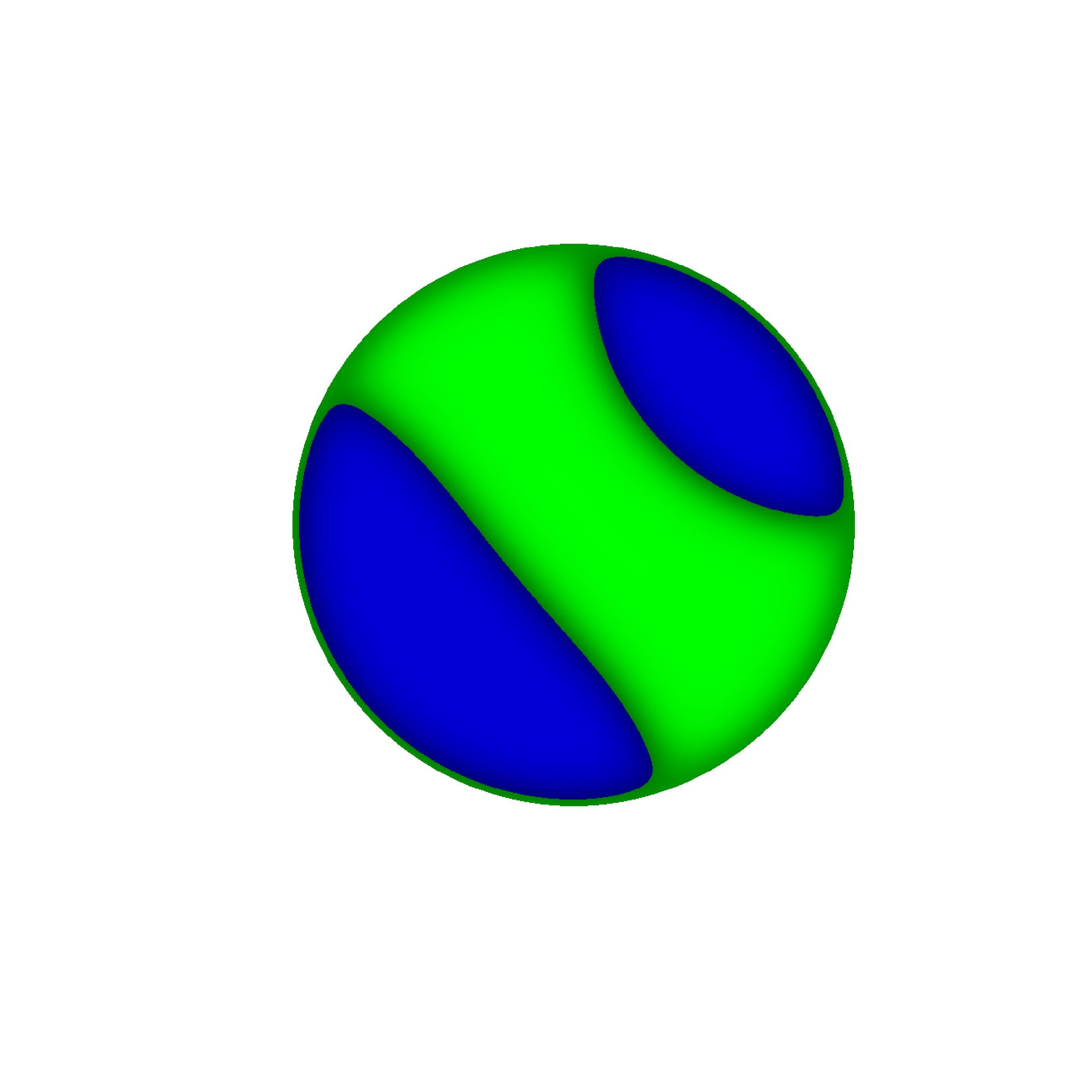}\hfil%
    \caption{}
    \label{r70a_int}
\end{subfigure}%
\begin{subfigure}{0.1\textwidth}
\centering
    \includegraphics[scale=0.0175]{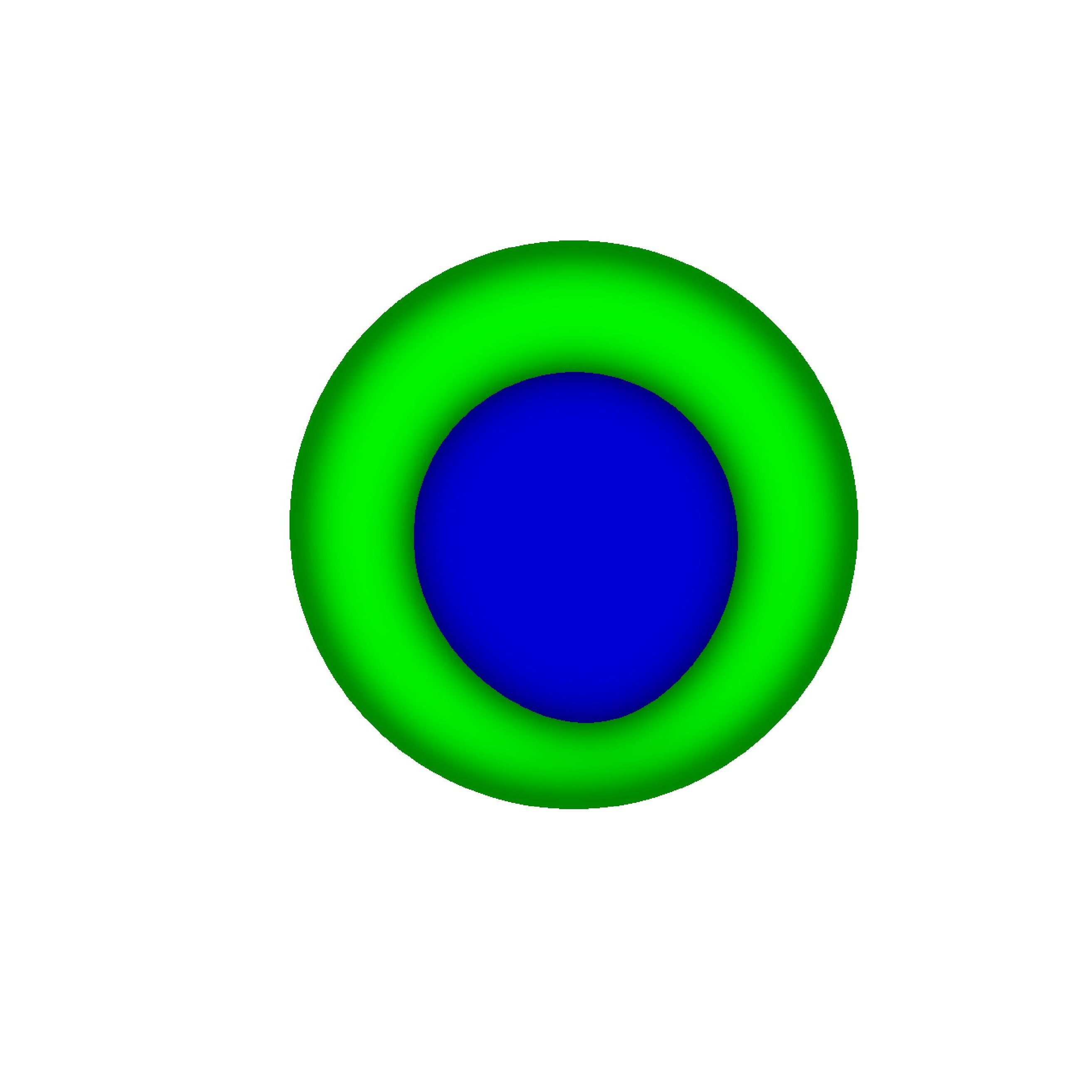}\hfil%
    \caption{}
    \label{r70b_int}
\end{subfigure}
\begin{subfigure}{0.1\textwidth}
\centering
    \includegraphics[scale=0.0175]{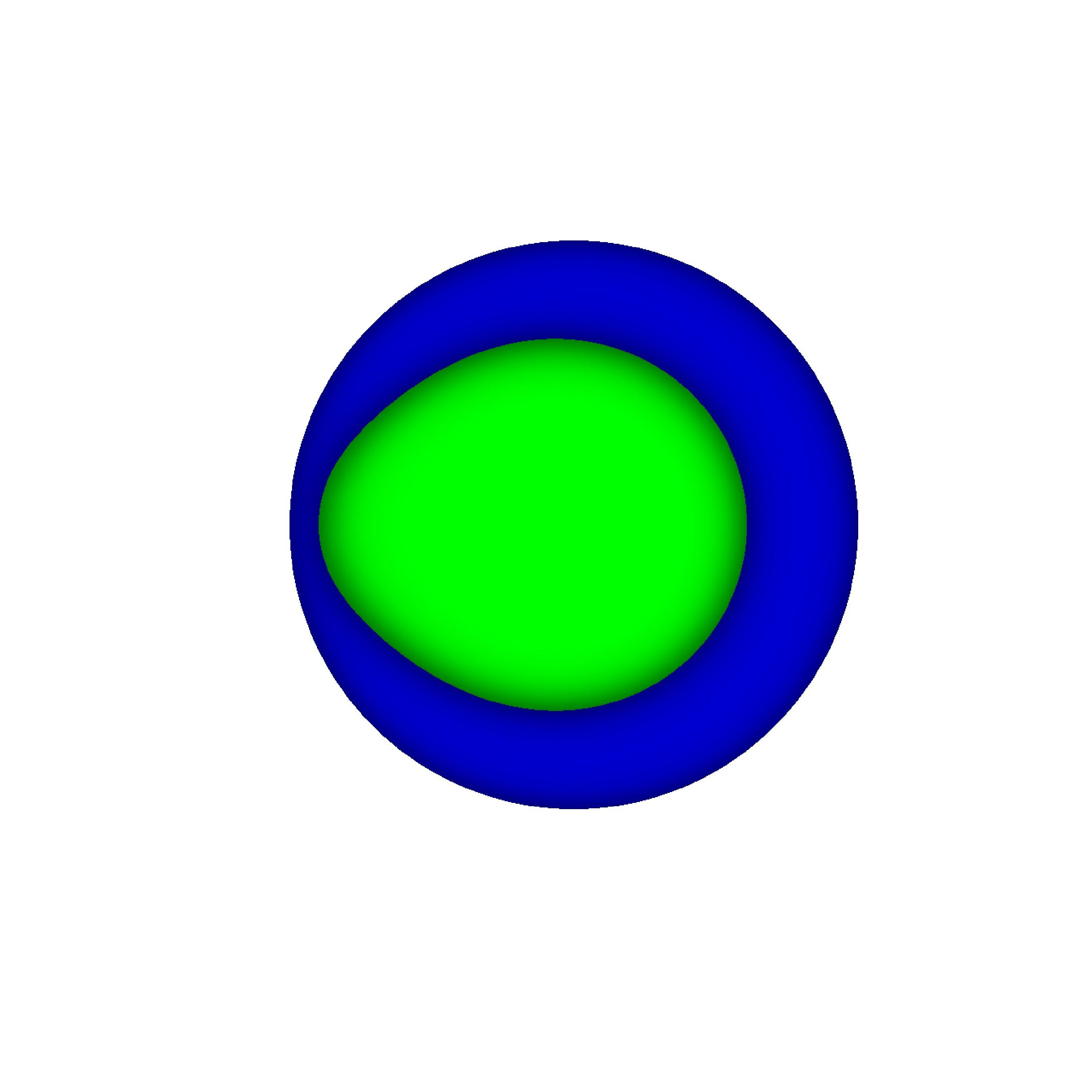}\hfil%
    \caption{}
    \label{r70c_int}
\end{subfigure}%
\begin{subfigure}{0.1\textwidth}
\centering
    \includegraphics[scale=0.0175]{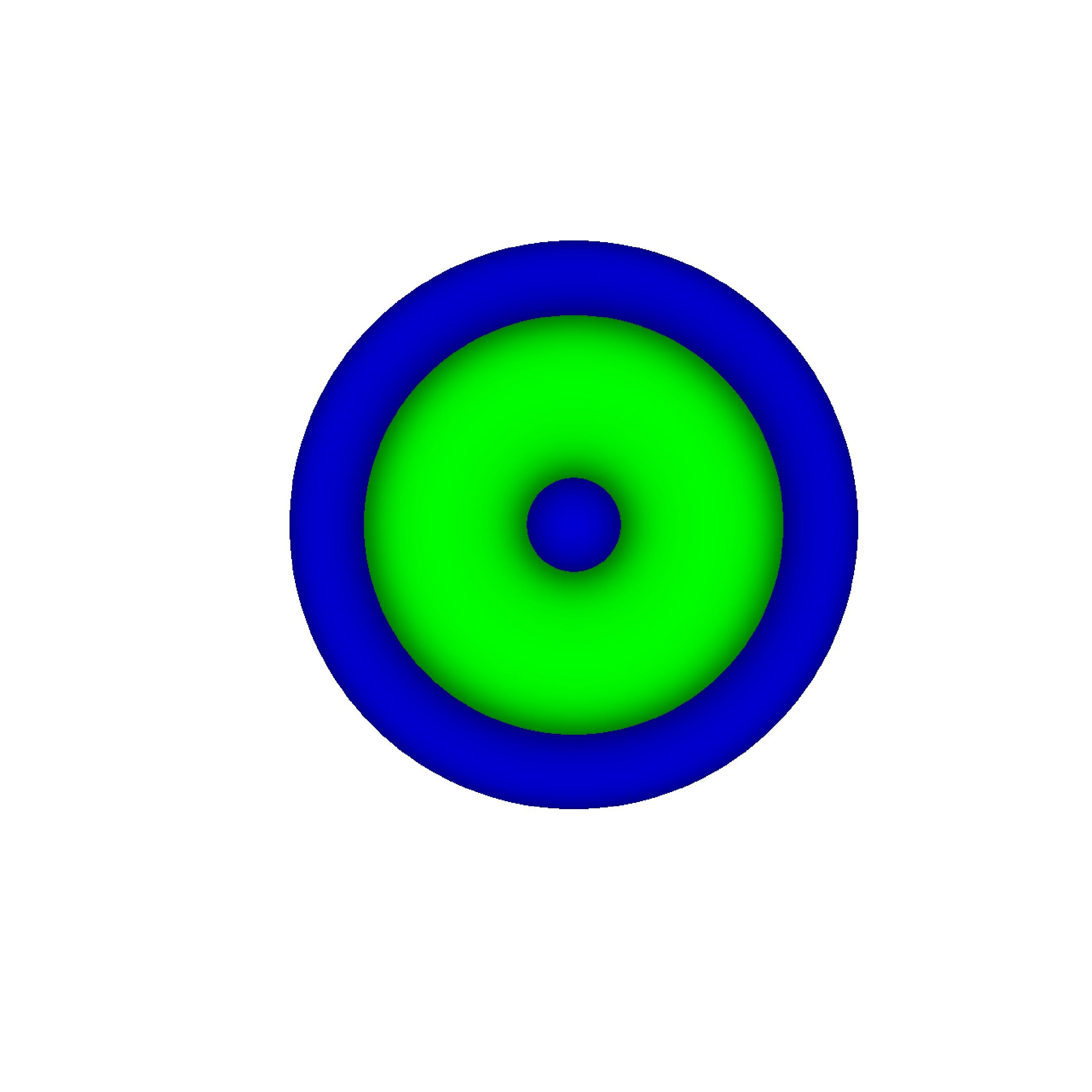}%
    \caption{}
    \label{r70d_int}
\end{subfigure}
\hspace*{\fill}%

\hspace*{\fill}%
\begin{subfigure}{0.1\textwidth}
\centering
    \includegraphics[scale=0.02]{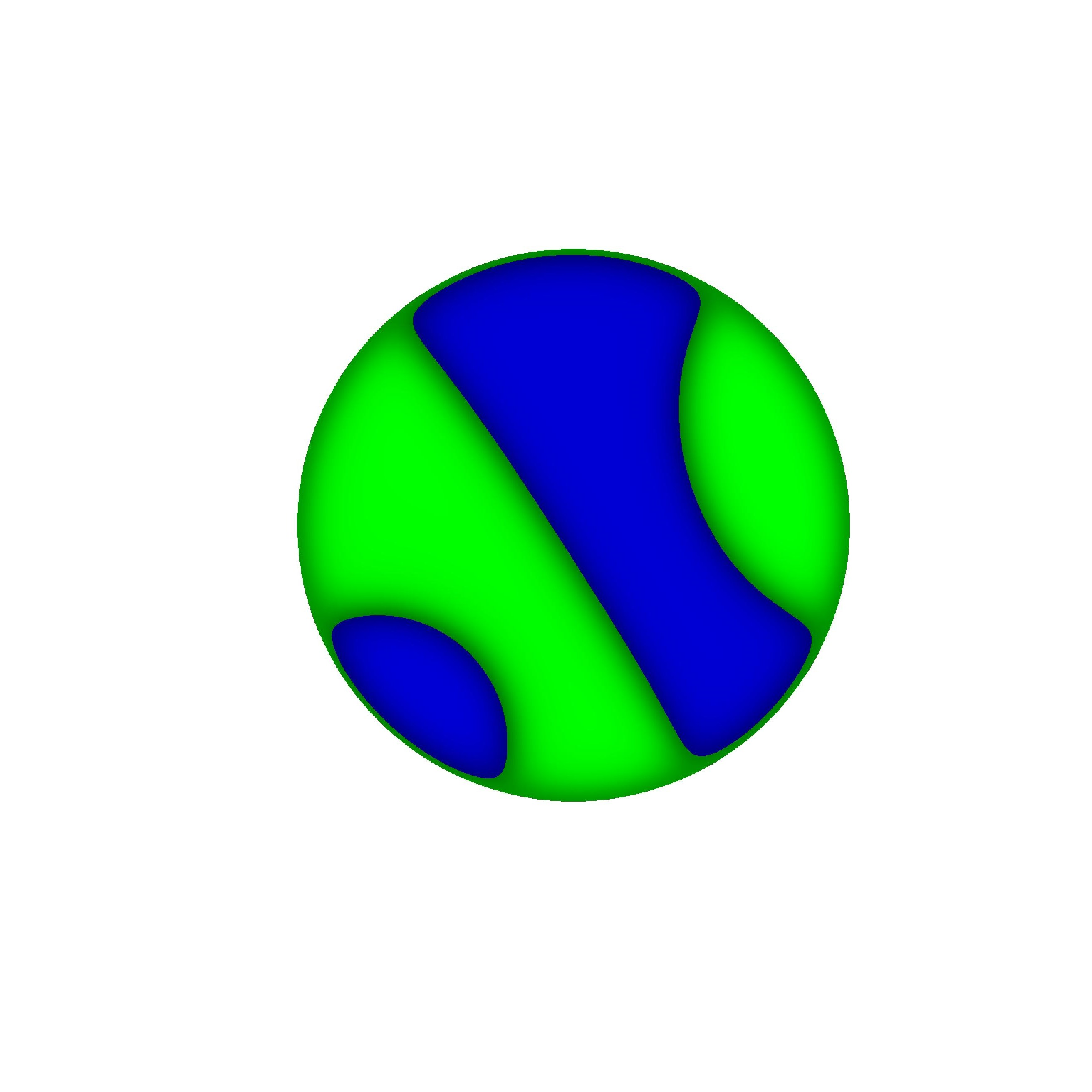}\hfil%
    \caption{}
    \label{r80a_int}
\end{subfigure}%
\begin{subfigure}{0.1\textwidth}
\centering
    \includegraphics[scale=0.0175]{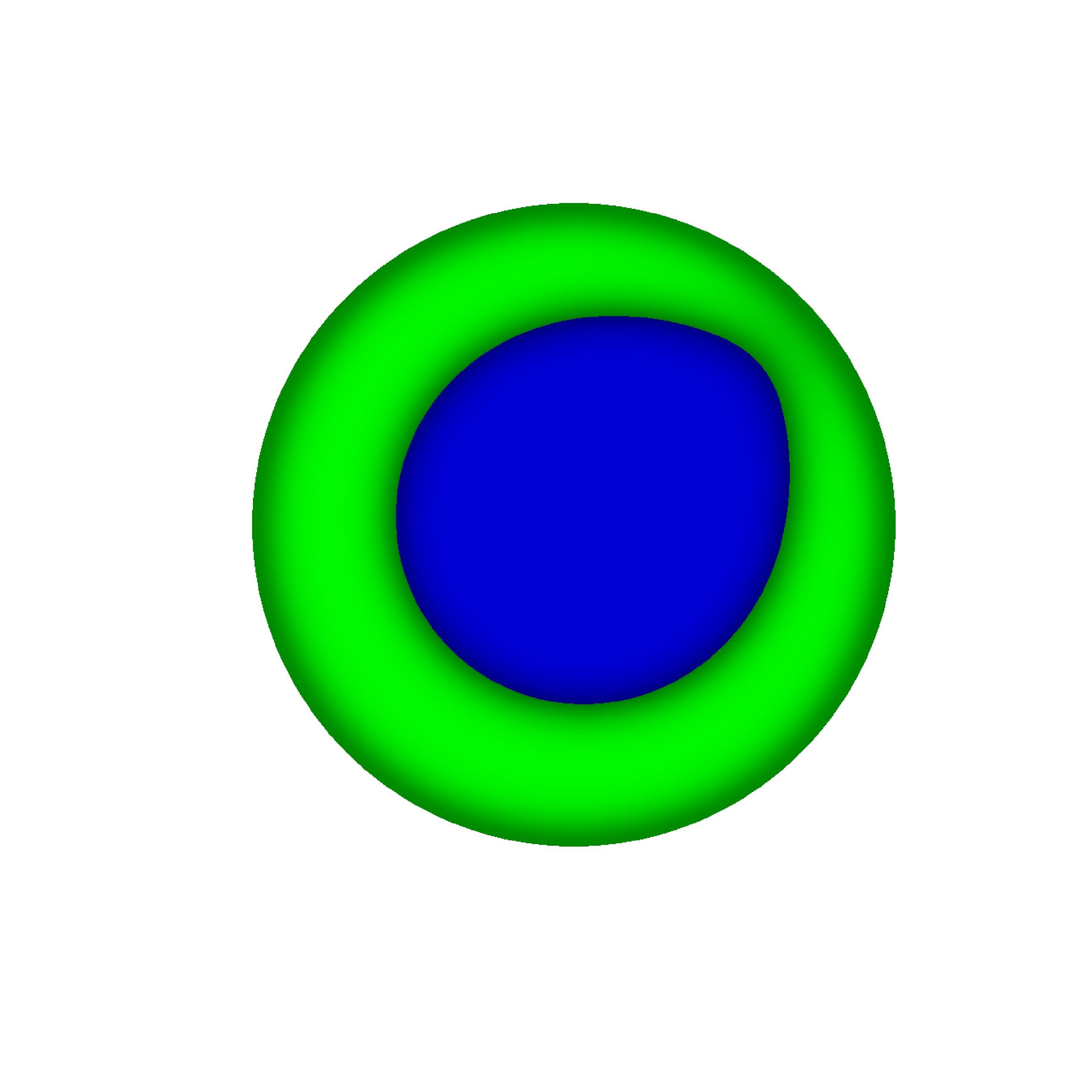}\hfil%
    \caption{}
    \label{r80b_int}
\end{subfigure}
\begin{subfigure}{0.1\textwidth}
\centering
    \includegraphics[scale=0.02]{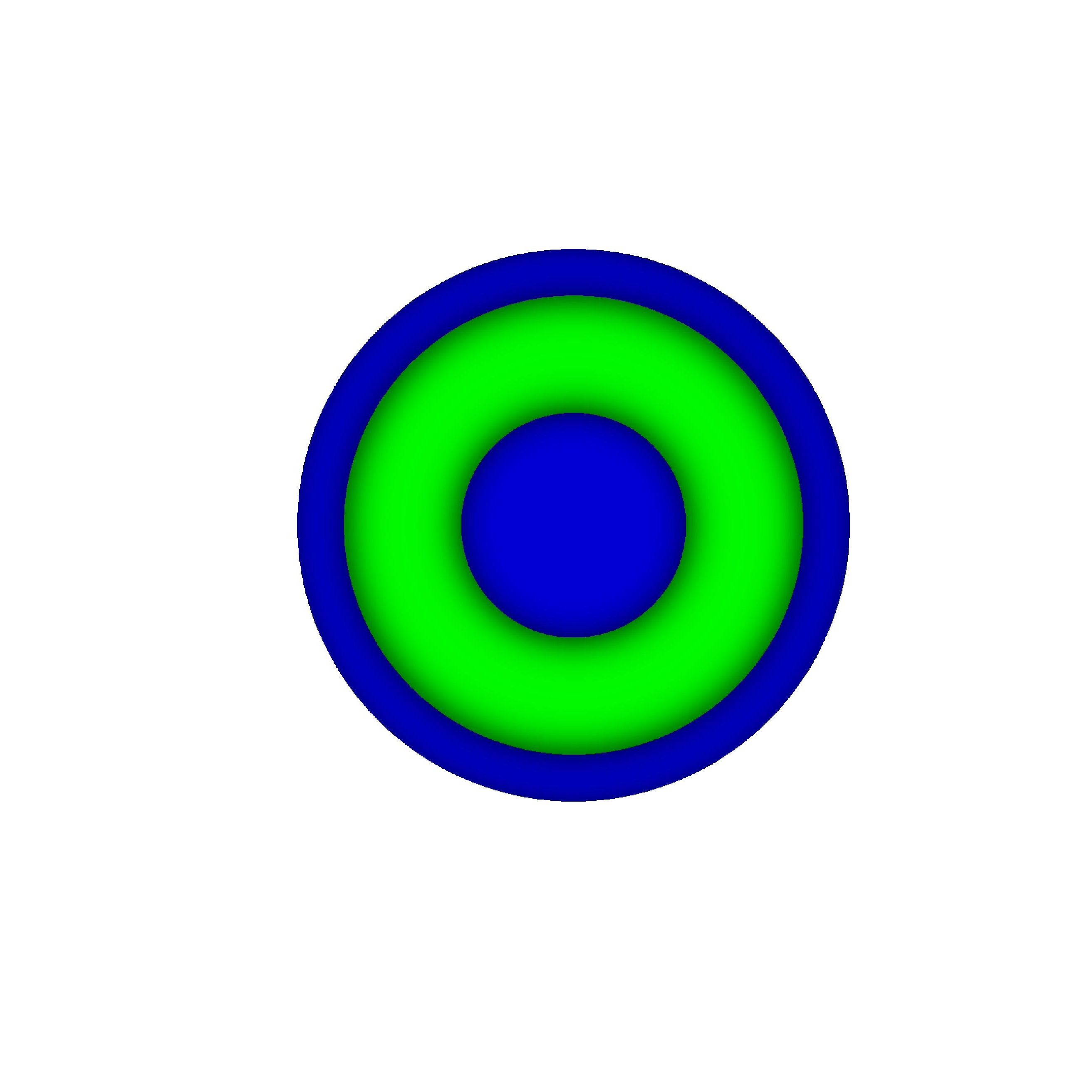}\hfil%
    \caption{}
    \label{r80c_int}
\end{subfigure}%
\begin{subfigure}{0.1\textwidth}
\centering
    \includegraphics[scale=0.02]{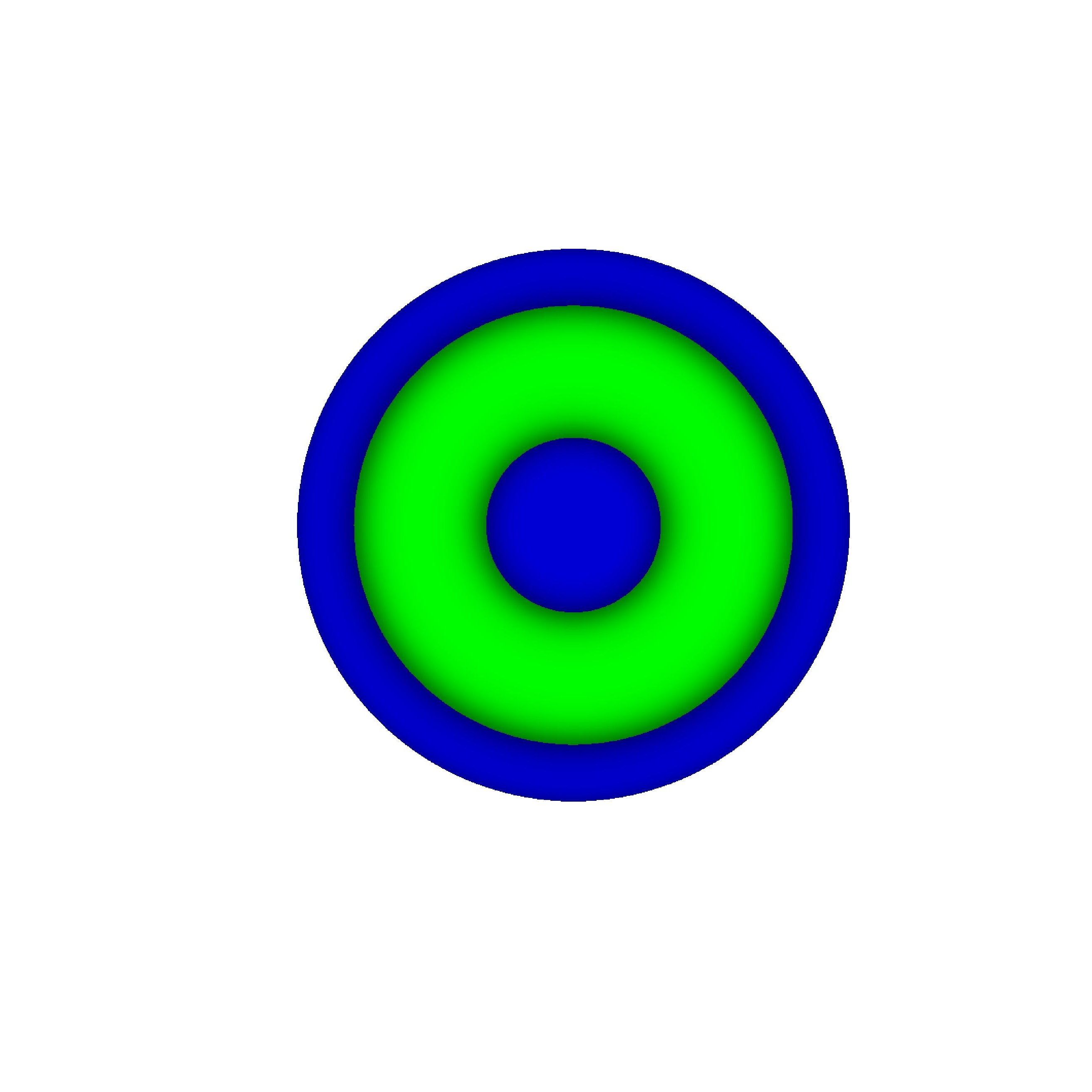}%
    \caption{}
    \label{r80d_int}
\end{subfigure}%
\hspace*{\fill}%

\hspace*{\fill}%
\begin{subfigure}{0.1\textwidth}
\centering
    \includegraphics[scale=0.0225]{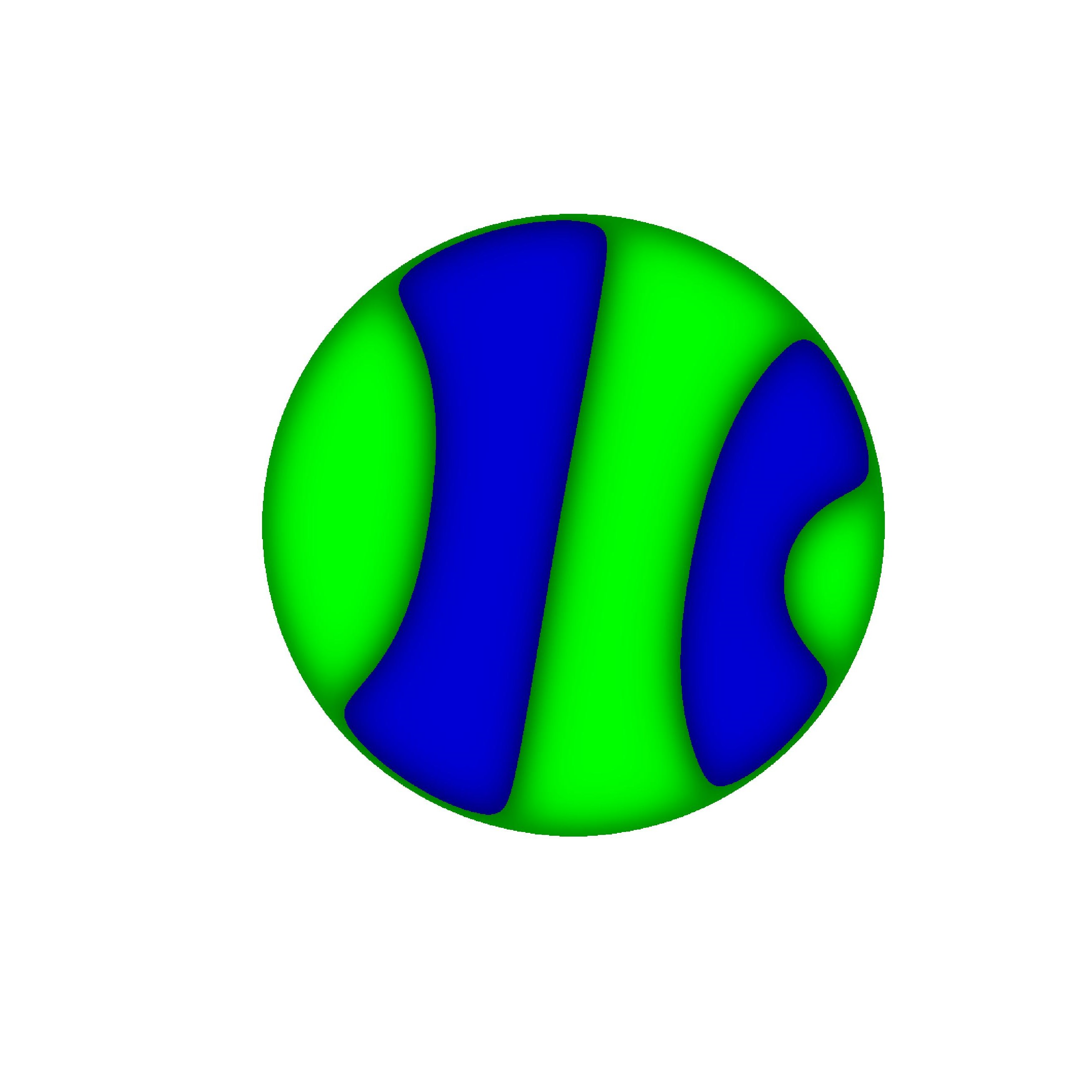}\hfil%
    \caption{}
    \label{r90a_int}
\end{subfigure}%
\begin{subfigure}{0.1\textwidth}
\centering
    \includegraphics[scale=0.0225]{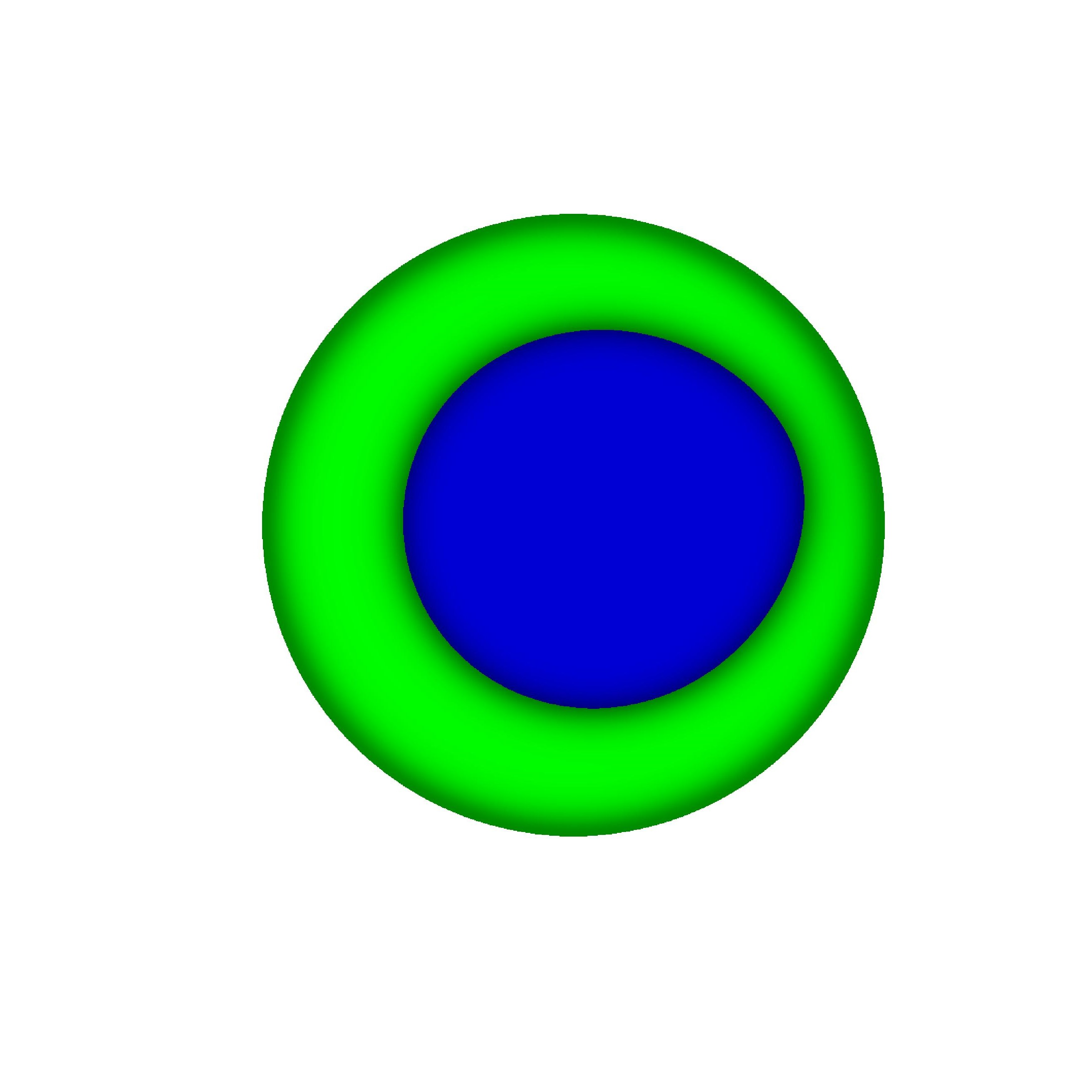}\hfil%
    \caption{}
    \label{r90b_int}
\end{subfigure}
\begin{subfigure}{0.1\textwidth}
\centering
    \includegraphics[scale=0.0225]{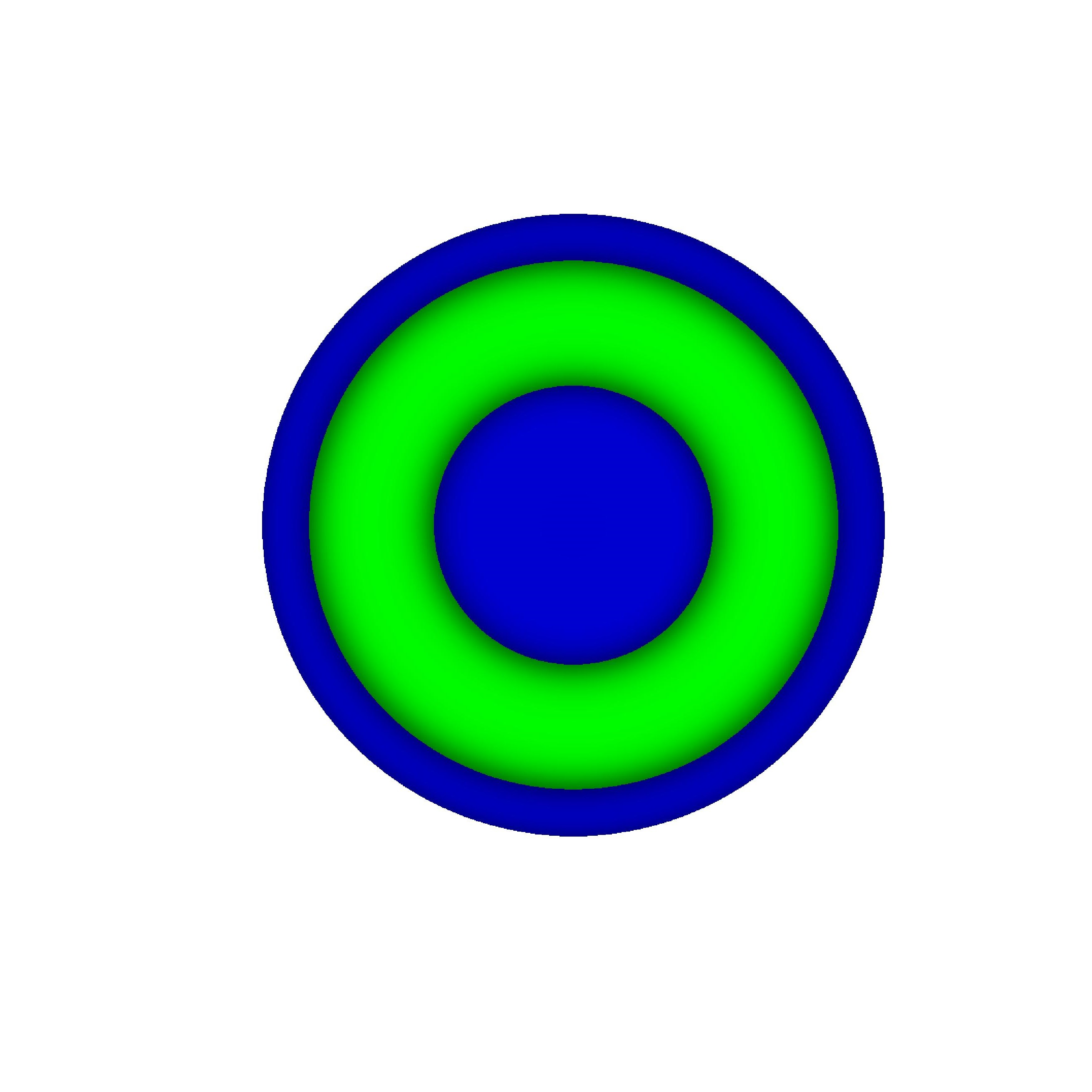}\hfil%
    \caption{}
    \label{r90c_int}
\end{subfigure}%
\begin{subfigure}{0.1\textwidth}
\centering
    \includegraphics[scale=0.0225]{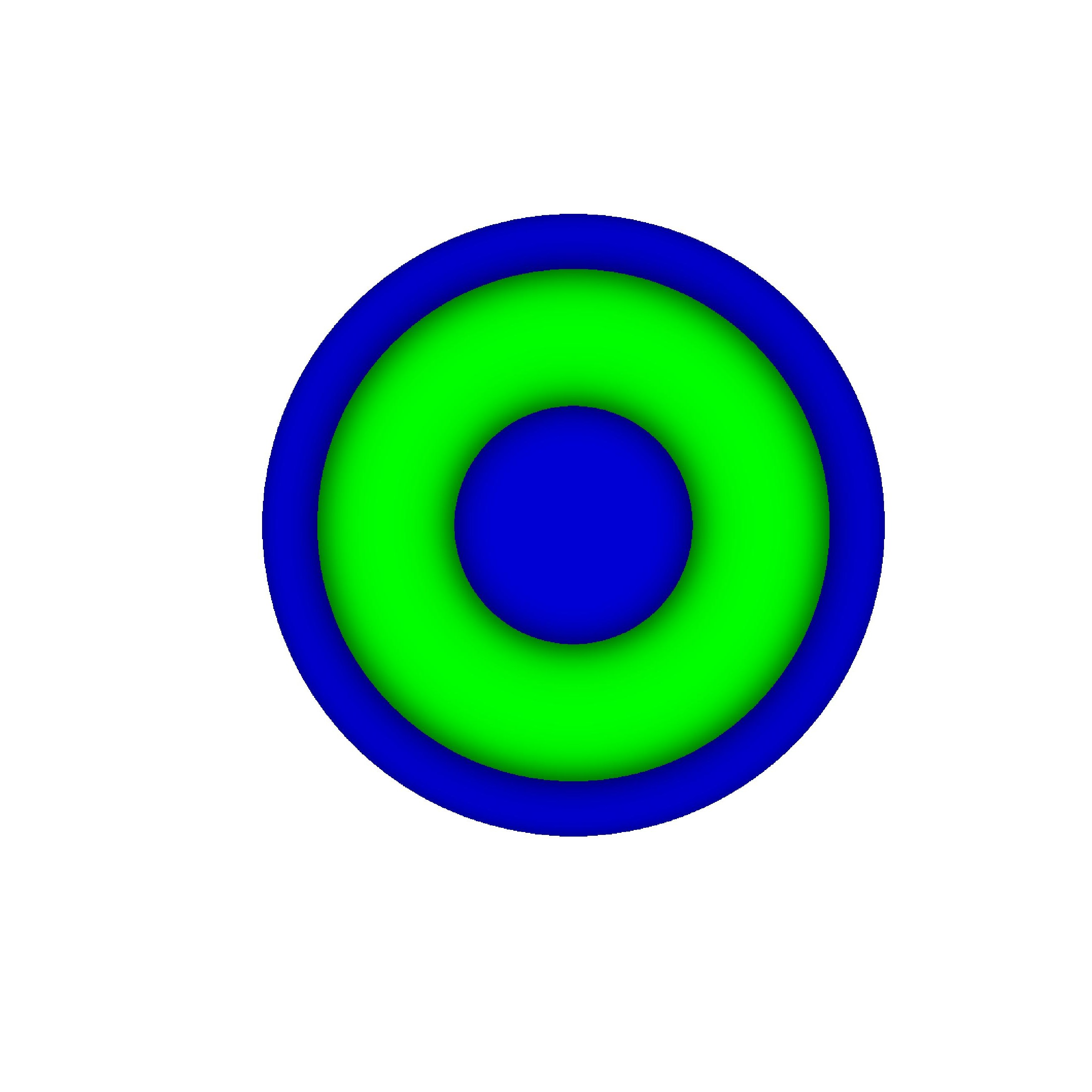}%
    \caption{}
    \label{r90d_int}
\end{subfigure}%
\hspace*{\fill}%

\caption{Intermediate-stage time snapshots of evolution of BNP morphology. 
First, second, third and fourth columns correspond to $\chi=0$, $\chi=0.25$, 
$\chi=0.5$, and $\chi=0.75$, respectively. Particle size $d$ in the top, 
middle, and bottom row is 140, 160, and 180, respectively; simulation 
boxes are scaled accordingly to reflect relative sizes. Blue and green 
colors represent solute-rich $\beta_2$ and solute-poor $\beta_1$ phases,
respectively.}
\label{Int_stage}
\end{figure}

\begin{figure}[htbp]
\captionsetup[subfigure]{justification=centering}
\centering%
\hspace*{\fill}%
\begin{subfigure}{0.1\textwidth}
\centering
    \includegraphics[scale=0.0175]{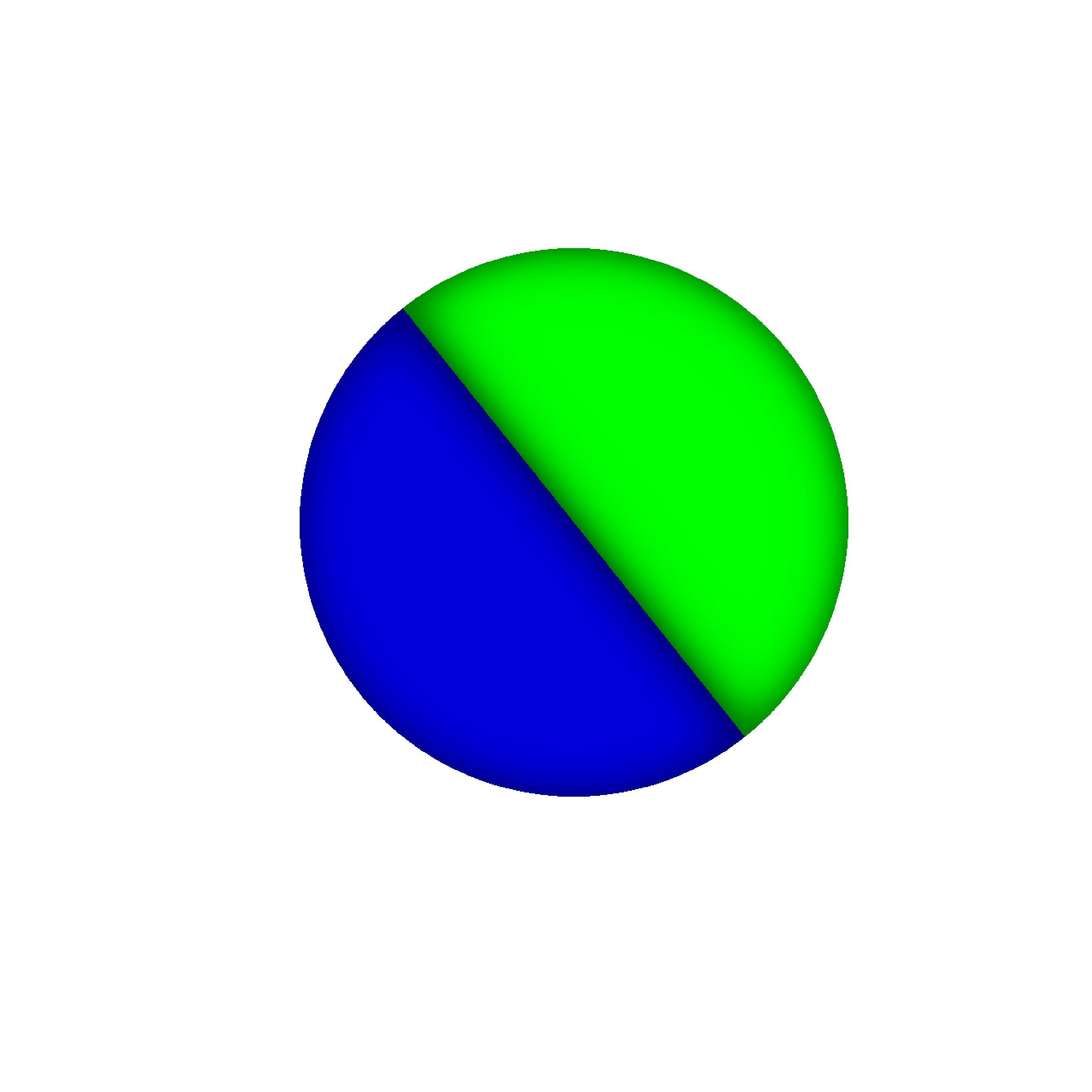}\hfil%
    \caption{}
    \label{r70a_eqm}
\end{subfigure}%
\begin{subfigure}{0.1\textwidth}
\centering
    \includegraphics[scale=0.0175]{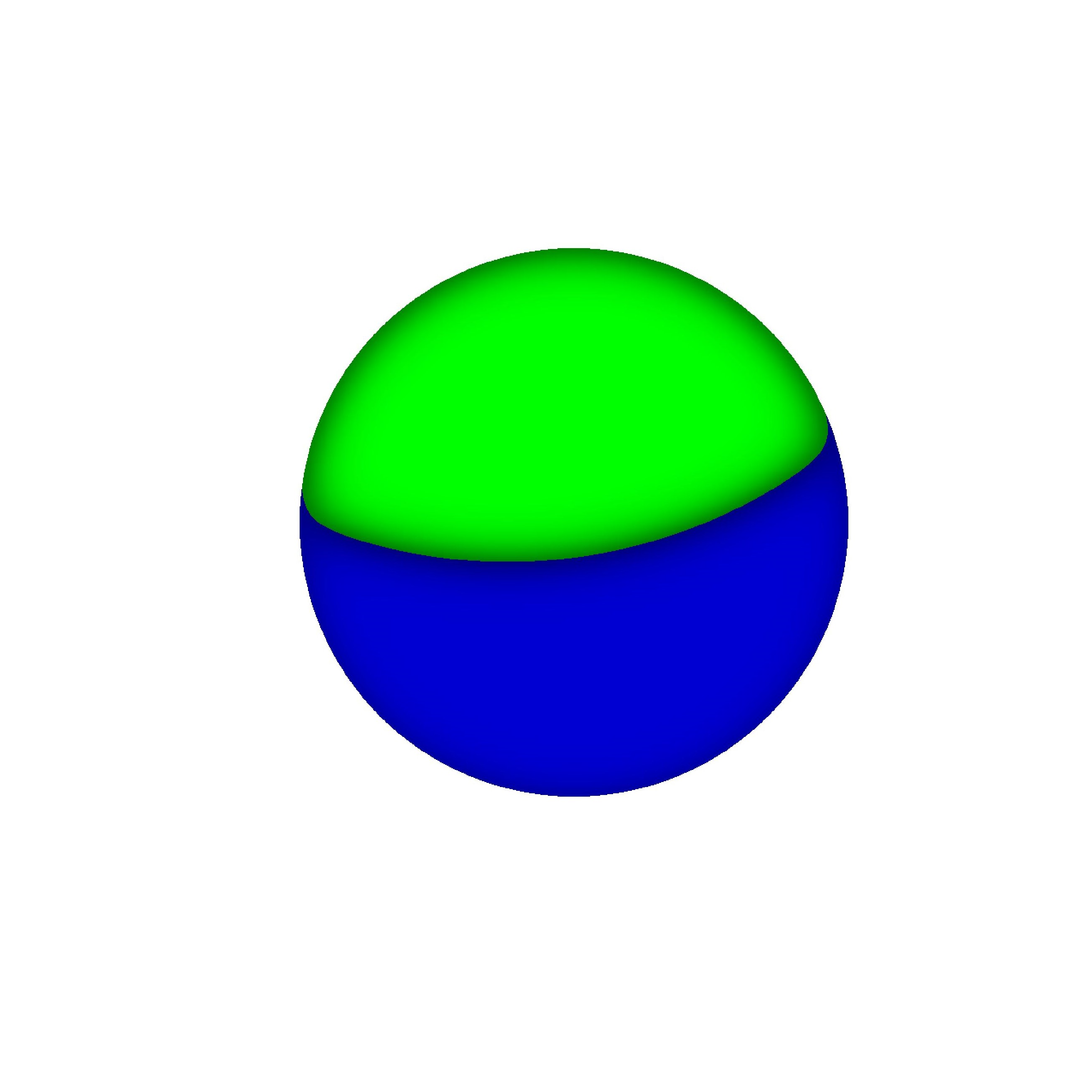}\hfil%
    \caption{}
    \label{r70b_eqm}
\end{subfigure}
\begin{subfigure}{0.1\textwidth}
\centering
    \includegraphics[scale=0.0175]{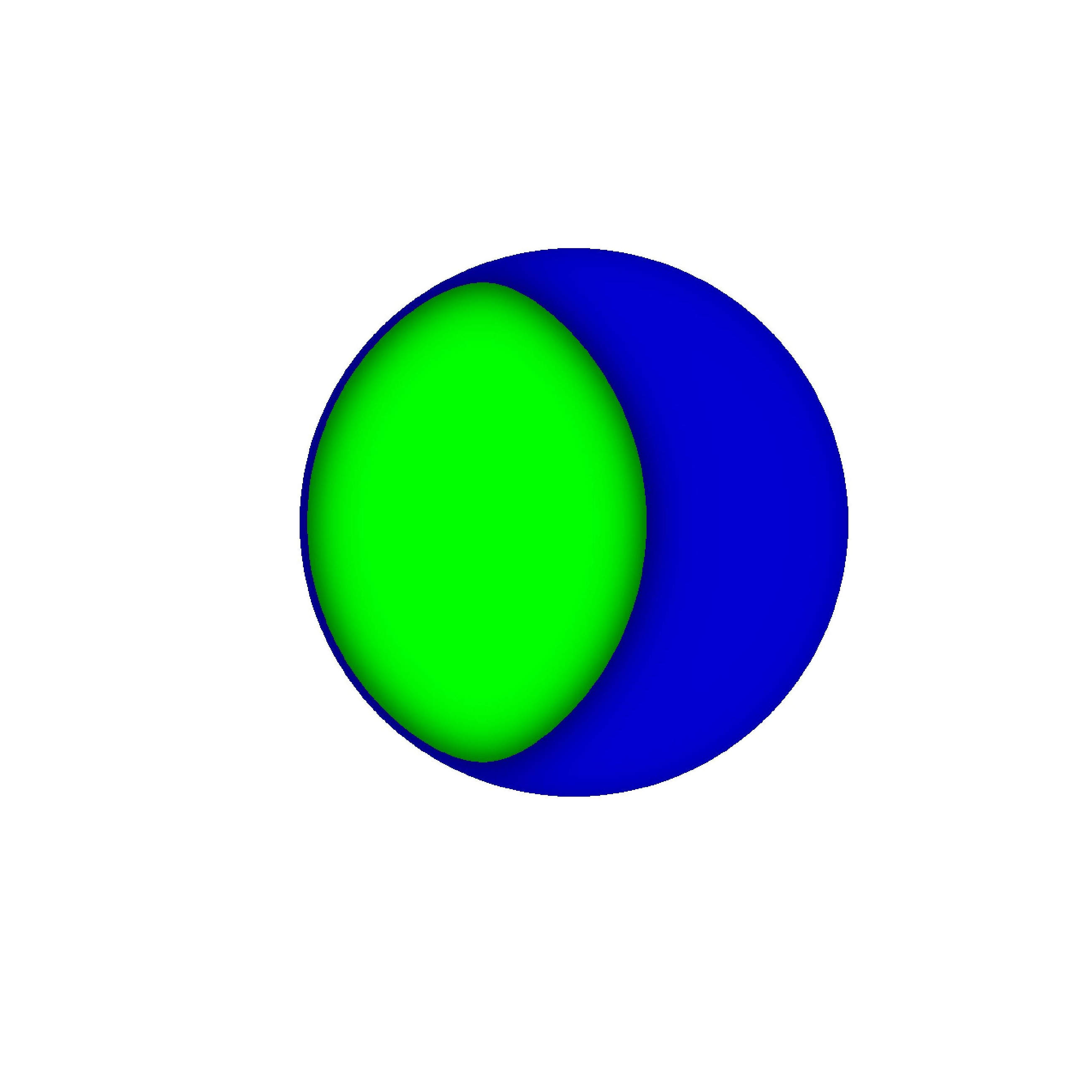}\hfil%
    \caption{}
    \label{r70c_eqm}
\end{subfigure}%
\begin{subfigure}{0.1\textwidth}
\centering
    \includegraphics[scale=0.0175]{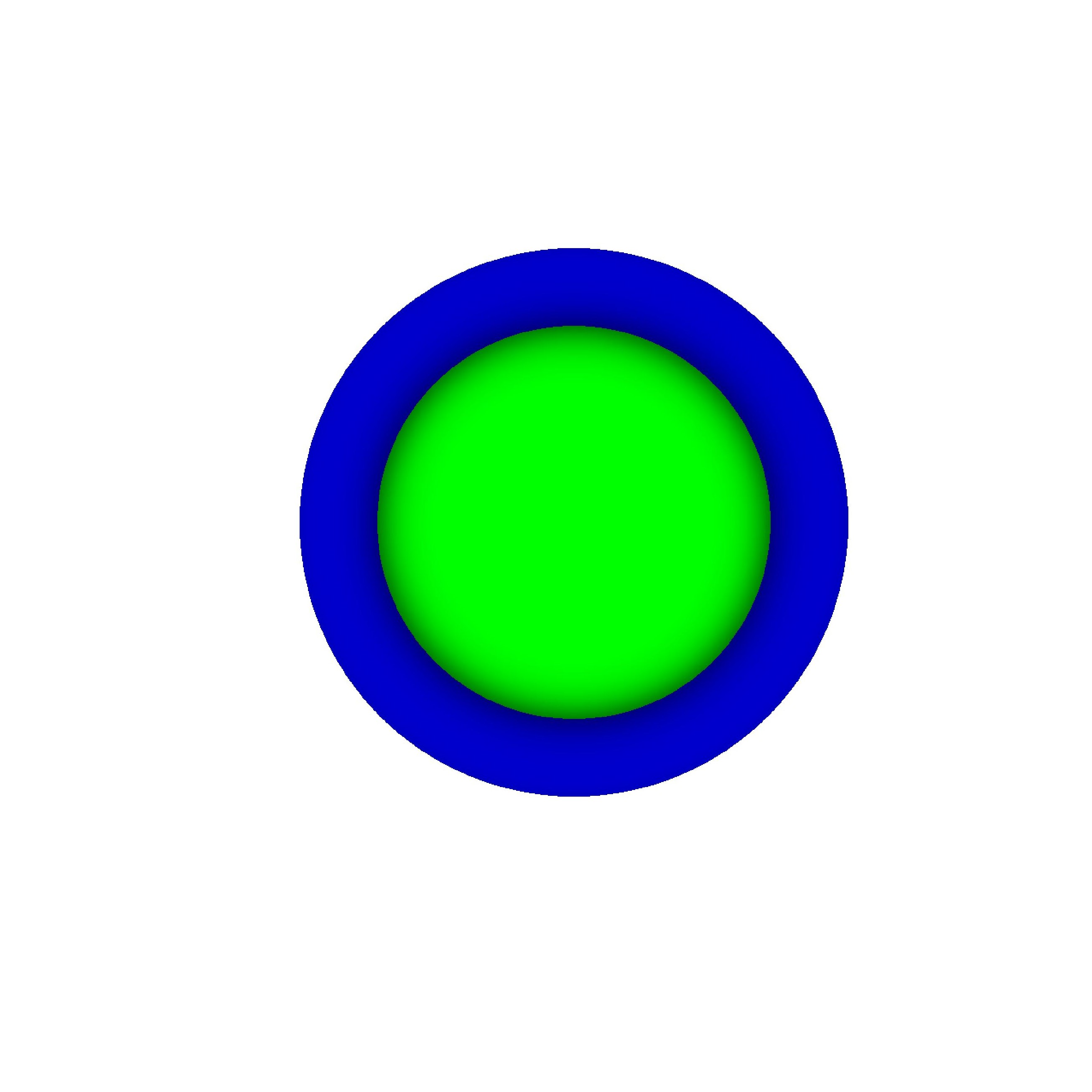}%
    \caption{}
    \label{r70d_eqm}
\end{subfigure}
\hspace*{\fill}%

\hspace*{\fill}%
\begin{subfigure}{0.1\textwidth}
\centering
    \includegraphics[scale=0.02]{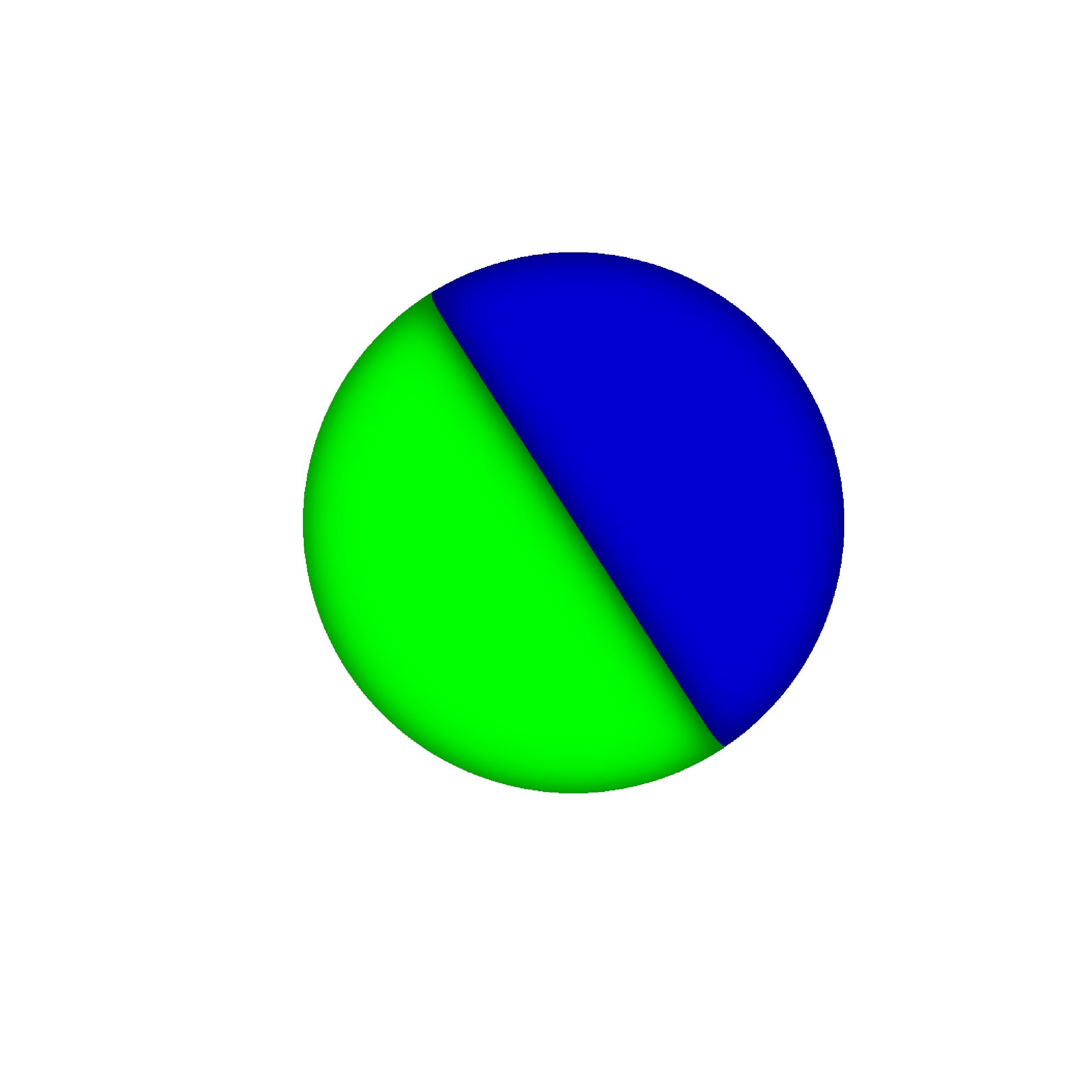}\hfil%
    \caption{}
    \label{r80a_eqm}
\end{subfigure}%
\begin{subfigure}{0.1\textwidth}
\centering
    \includegraphics[scale=0.0175]{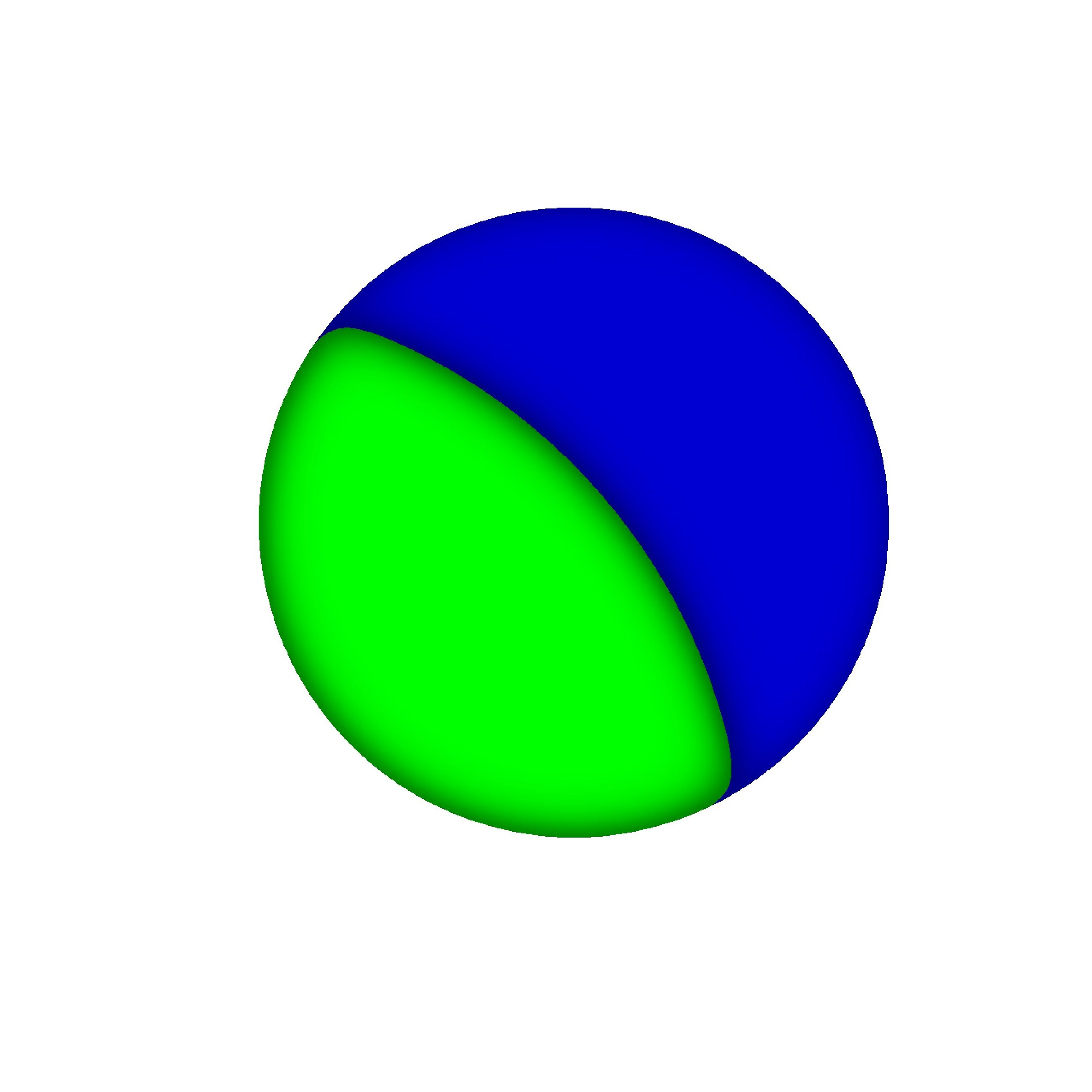}\hfil%
    \caption{}
    \label{r80b_eqm}
\end{subfigure}
\begin{subfigure}{0.1\textwidth}
\centering
    \includegraphics[scale=0.02]{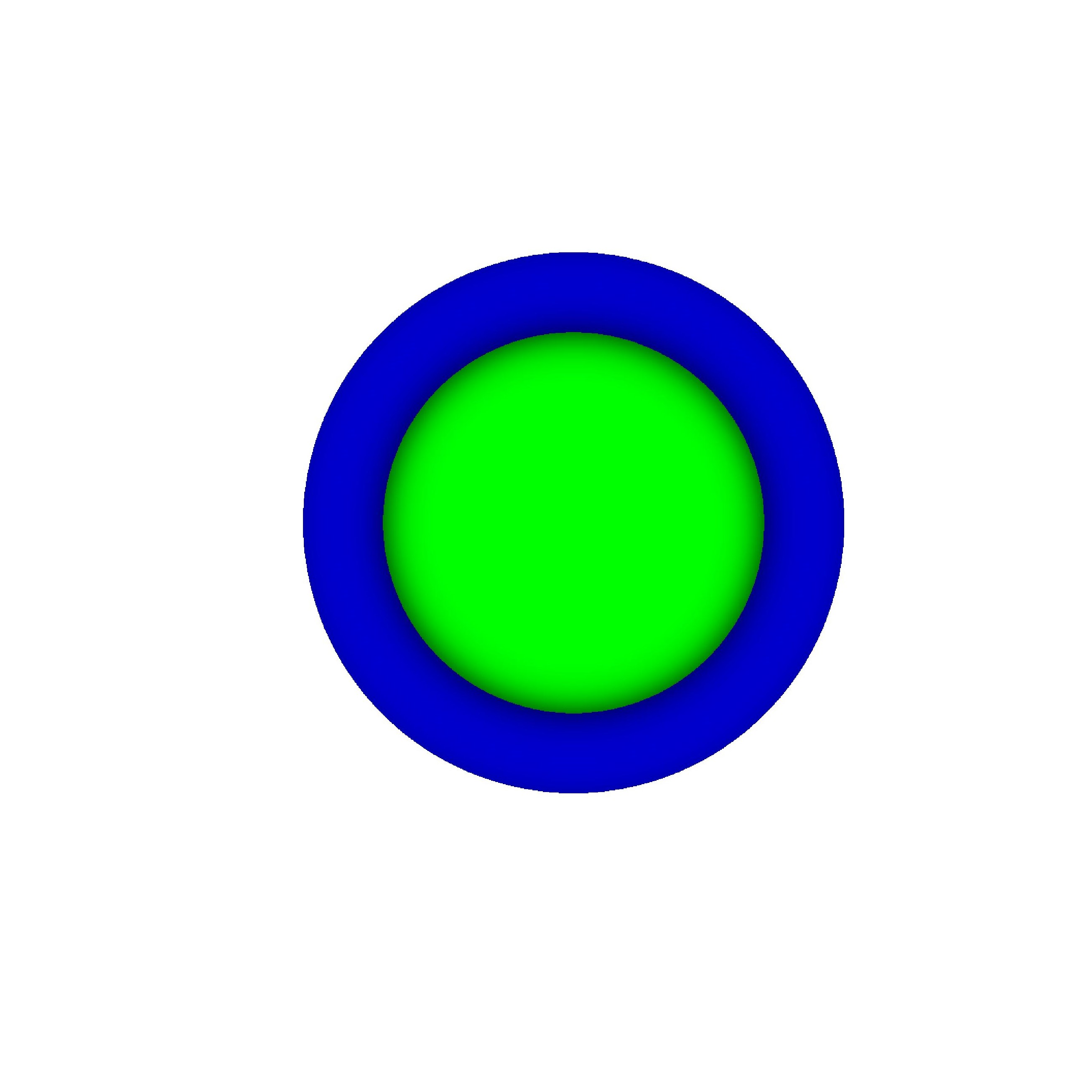}\hfil%
    \caption{}
    \label{r80c_eqm}
\end{subfigure}%
\begin{subfigure}{0.1\textwidth}
\centering
    \includegraphics[scale=0.02]{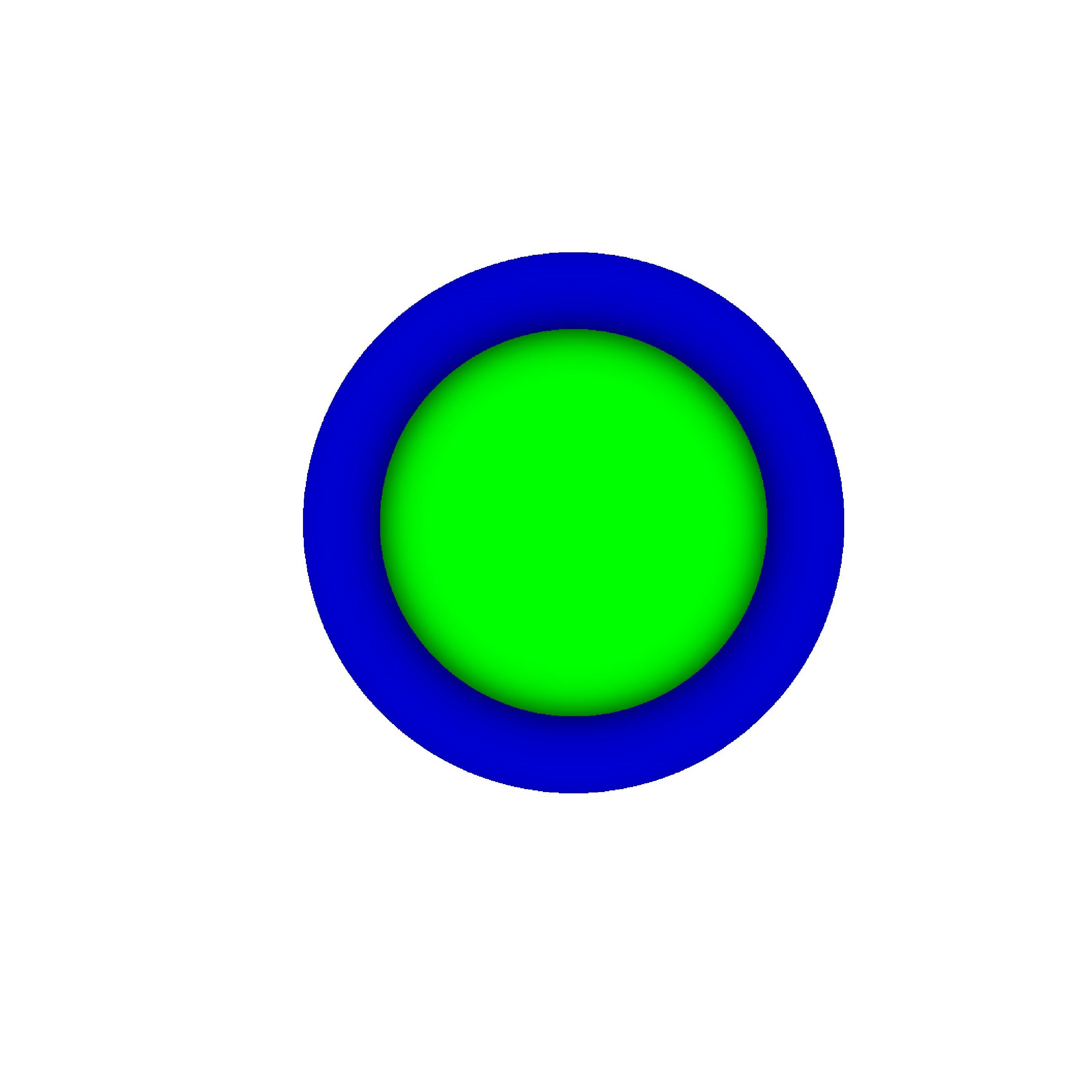}%
    \caption{}
    \label{r80d_eqm}
\end{subfigure}%
\hspace*{\fill}%

\hspace*{\fill}%
\begin{subfigure}{0.1\textwidth}
\centering
    \includegraphics[scale=0.0225]{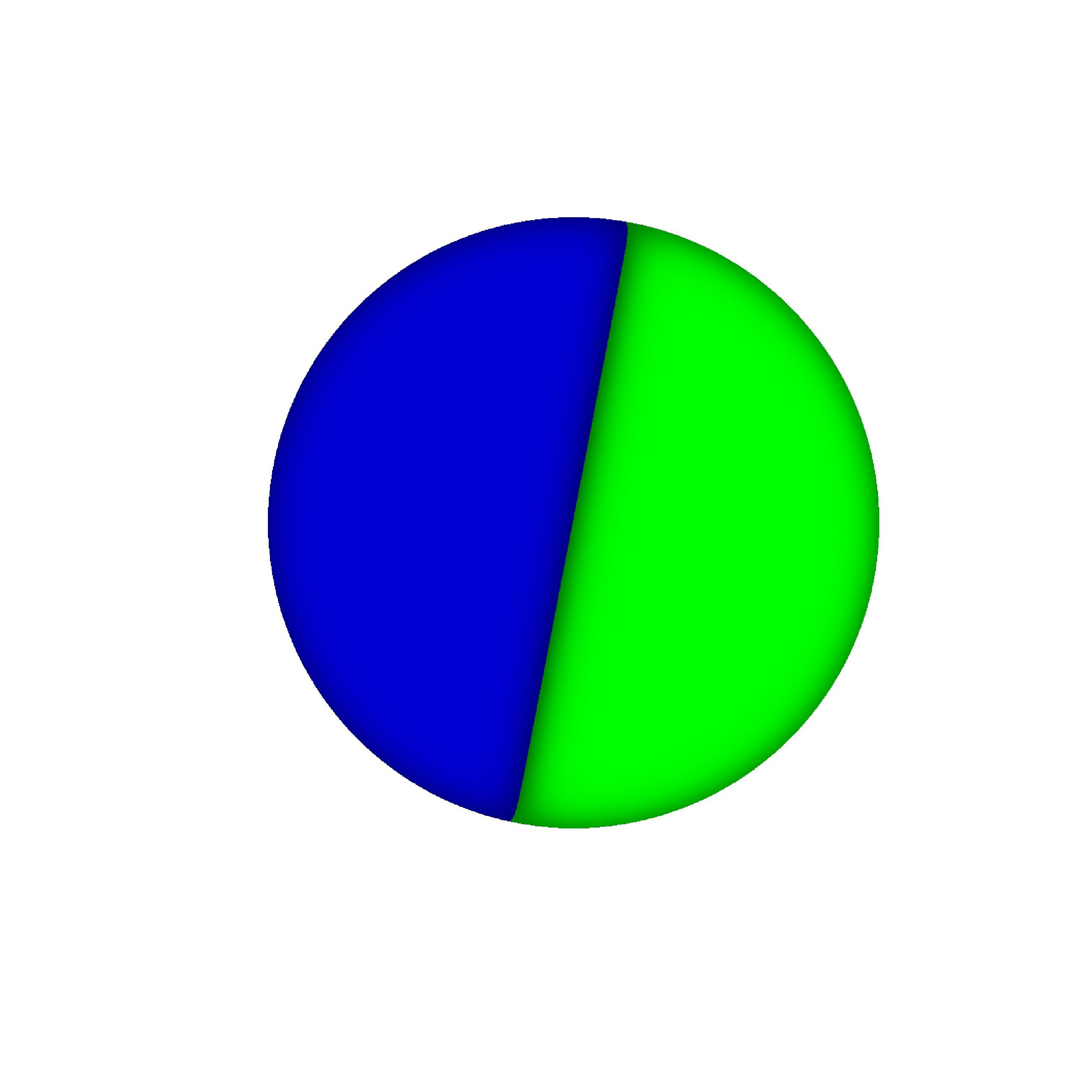}\hfil%
    \caption{}
    \label{r90a_eqm}
\end{subfigure}%
\begin{subfigure}{0.1\textwidth}
\centering
    \includegraphics[scale=0.0225]{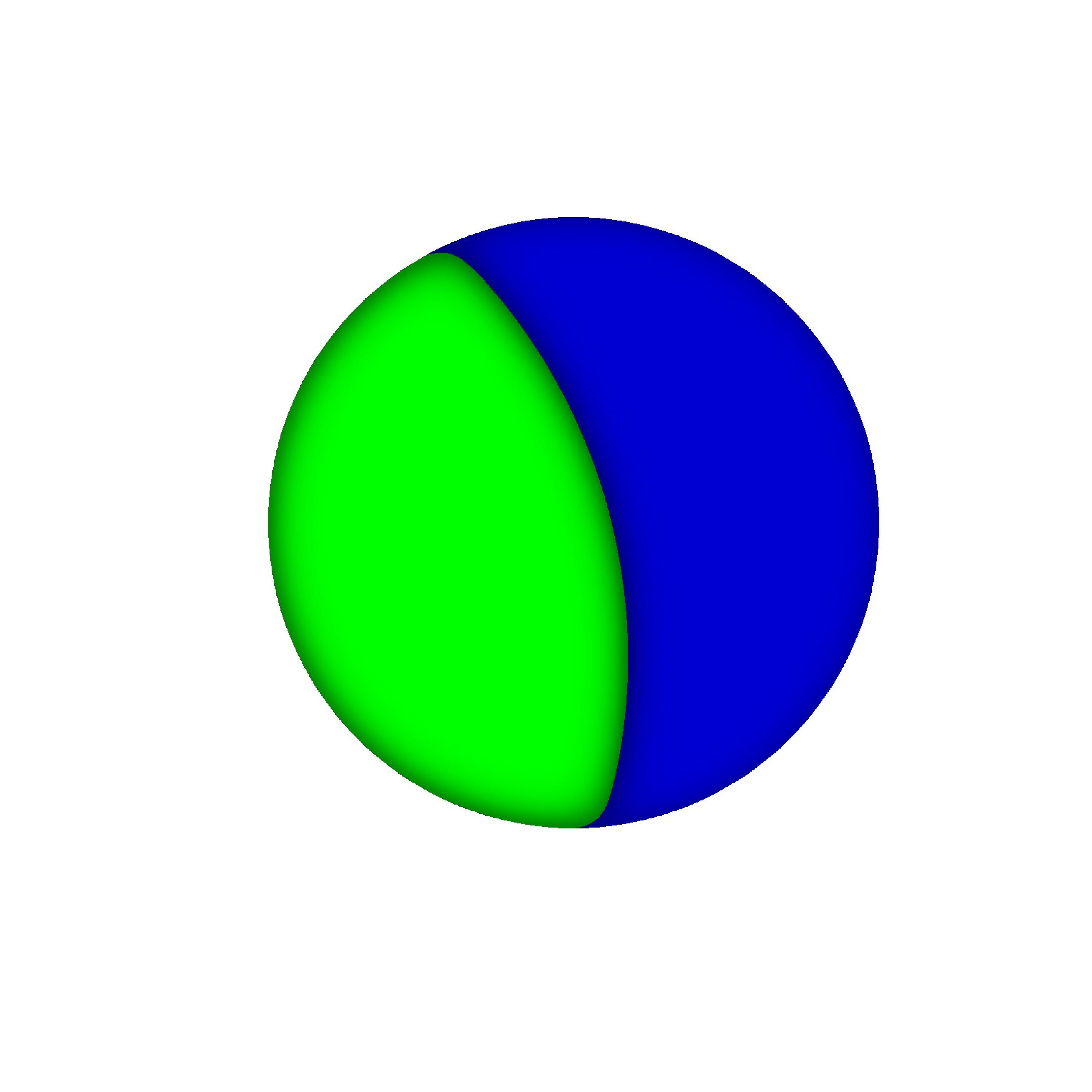}\hfil%
    \caption{}
    \label{r90b_eqm}
\end{subfigure}
\begin{subfigure}{0.1\textwidth}
\centering
    \includegraphics[scale=0.0225]{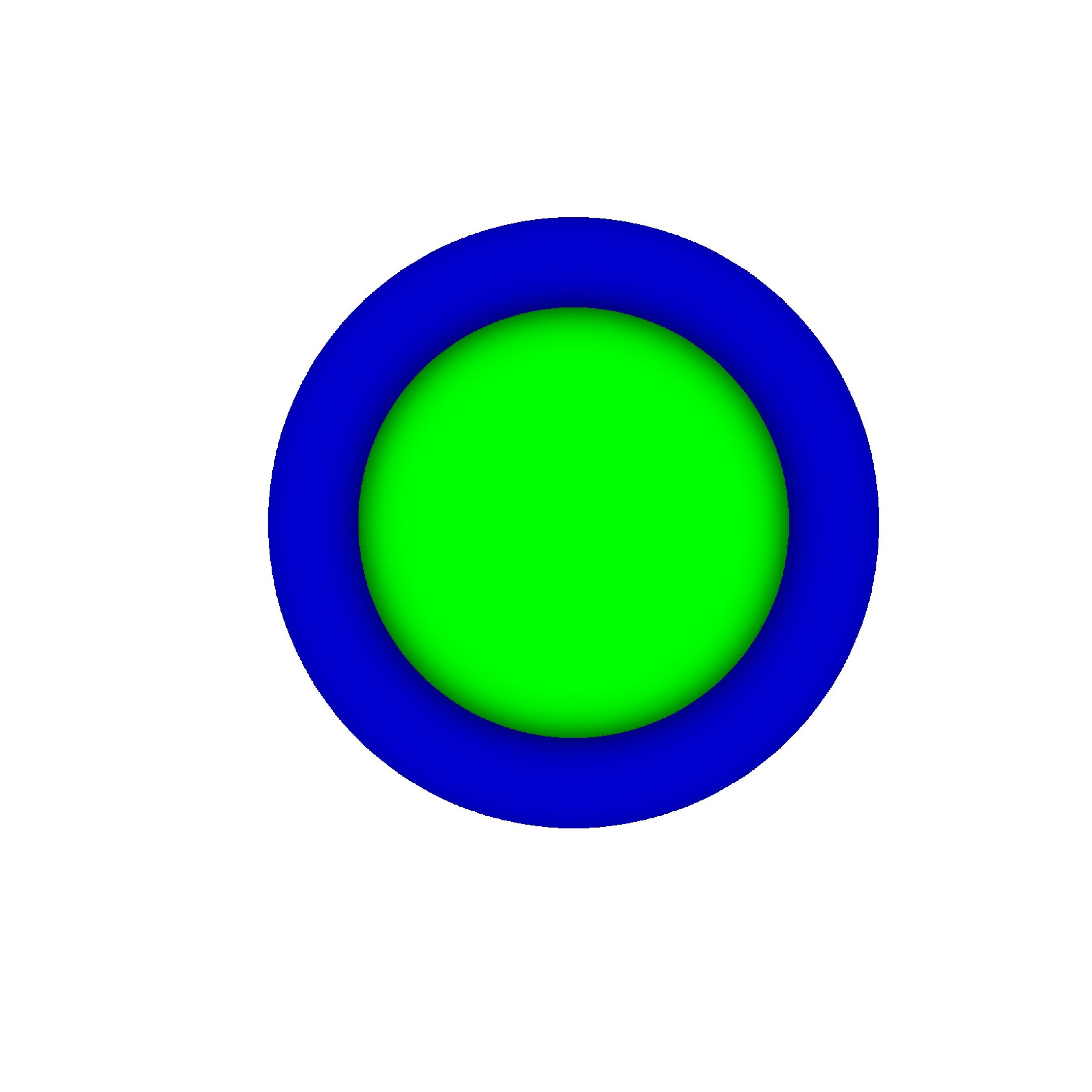}\hfil%
    \caption{}
    \label{r90c_eqm}
\end{subfigure}%
\begin{subfigure}{0.1\textwidth}
\centering
    \includegraphics[scale=0.0225]{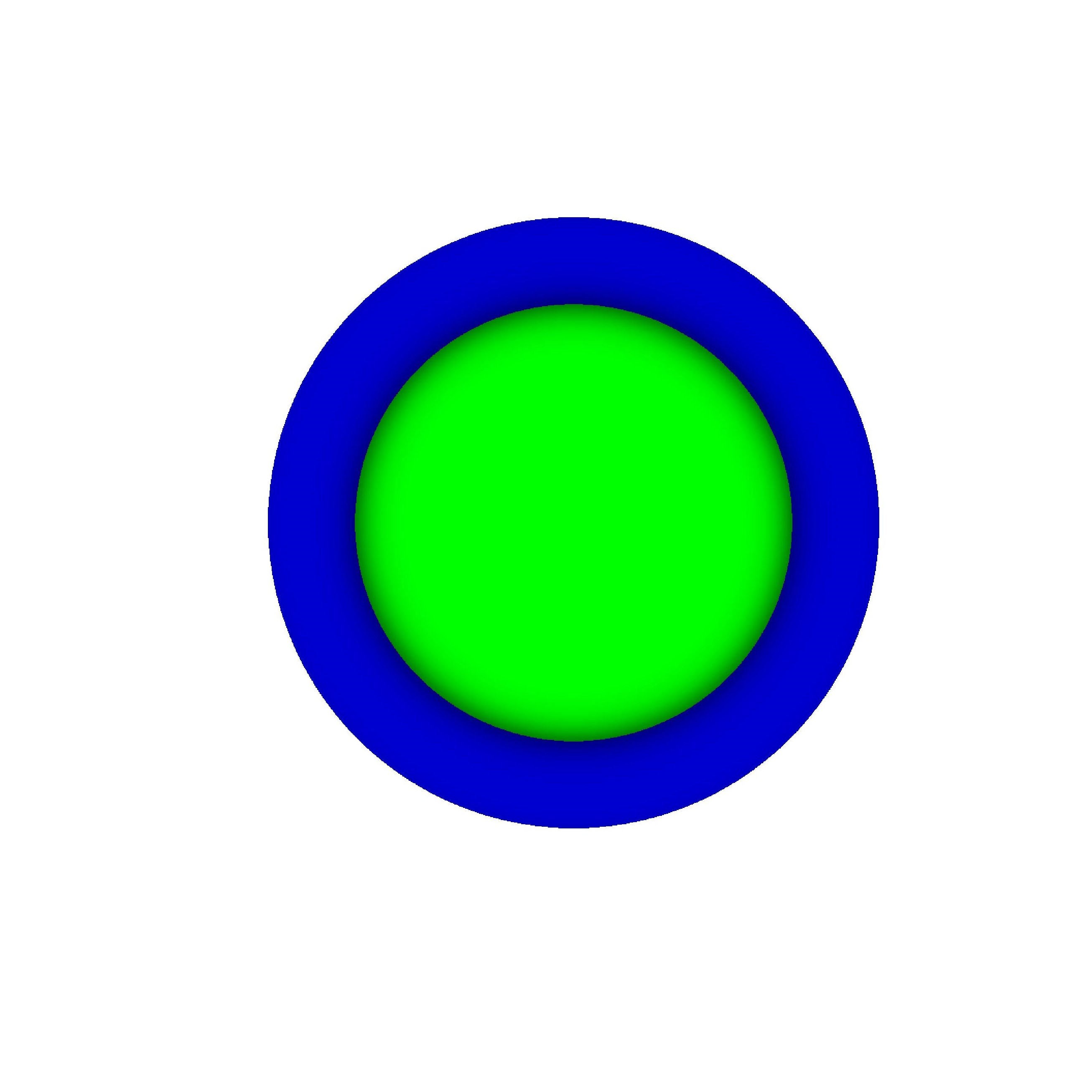}%
    \caption{}
    \label{r90d_eqm}
\end{subfigure}%
\hspace*{\fill}%

\caption{Final time snapshots of evolution of BNP morphology. First, second, third
and fourth columns correspond to $\chi=0$, $\chi=0.25$, $\chi=0.5$, and $\chi=0.75$,
respectively. Particle size $d$ in the top, middle, and bottom row is 140, 160, 
and 180, respectively;simulation boxes are scaled accordingly to reflect relative
sizes. Blue and green colors represent solute-rich $\beta_2$ and solute-poor
$\beta_1$ phases, respectively.}
\label{Eqm_stage}
\end{figure}

We first present the evolution of domains inside an isolated, initially homogeneous BNP
of equiatomic composition which has been quenched from a high temperature to a temperature
below the critical point of the miscibility gap. Figs.~\ref{Early_stage}--\ref{Eqm_stage}
present time snapshots corresponding to the early, intermediate and late stages of its 
evolution, respectively. In all these figures, $\chi$ increases (and $\theta$ decreases) 
from left to right, and particle size increases from top to bottom. The solute-rich 
$\beta_2$ and solute-poor $\beta_1$ phases in the particle are represented by blue and
green colors, respectively; matrix is colored white to render it invisible. Animations
of the evolution process for representative cases are available in the Electronic
Supplementary Information (ESI).

Simulations with $\chi=0$ (\emph{i.e.}, $\theta=\ang{90}$) exhibit similar morphological
evolution at all times irrespective of particle size and it finally leads to a straight
Janus morphology with a 2-fold symmetry about the center of the particle (see the first 
column of Fig~\ref{Eqm_stage}). First column of Fig.~\ref{Early_stage} shows that in this
case, SD initiates homogeneously within the bulk of the particle with no apparent preference 
for the surface. During later times (first column of Fig.~\ref{Int_stage}), coarsening sets
in, and eventually $\beta_1\text{-}\beta_2$ interface makes a $\ang{90}$ contact angle at
the particle surface (first column of Fig.~\ref{Eqm_stage}), as is expected from Young's
equation (Eq.~\eqref{eq-Young}). Absence of solute segregation in the equilibrium 
composition profiles in Fig.~\ref{1deq}(\subref{chi0_AB1}) implies a surface-agnostic
evolution in this case ($\chi=0$ or $\theta=\ang{90}$).

When $0<\chi\leq 0.75 \;(\ang{90}>\theta\geq\ang{0})$, early stages of evolution
at all particle sizes lead to solute segregation at the surface. Consequently, 
alternate solute-rich and solute-poor rings form due to SDSD. Subsequent coarsening
of the rings due to capillarity leads to an intermediate CS morphology, either with
a solute-rich $\beta_2$ shell and a solute-poor $\beta_1$ core, or a solute-poor
$\beta_1$ shell and a solute-rich $\beta_2$ core -- we call them CS-$2@1$ and CS-$1@2$,
respectively, where `1@2' implies $\beta_1$ shell on $\beta_2$ core and vice versa.

The circular core in these cases starts to become distorted at later times for certain
combinations of particle size and $\theta$. Specifically, we can observe this in Figs.~\ref{Int_stage}(\subref{r70b_int},~\subref{r70c_int},~\subref{r80b_int},%
~\subref{r90b_int}) where the core undergoes a symmetry-breaking shape transition
to an oval with its sharper end coming in close proximity to the particle surface. This
sets up a diffusion field along the $\beta_1$-$\beta_2$ interface which leads to the
breakdown of the intermediate CS structure and eventual formation of a Janus structure.
In the very late stages, the contact angle formed by the Janus configuration approaches
the corresponding equilibrium value, as shown in Figs.~\ref{Eqm_stage}(\subref{r70b_eqm},%
~\subref{r70c_eqm},~\subref{r80b_eqm},~\subref{r90b_eqm}). Although our simulations
provide discrete combinations of size and contact angle that exhibit these CS$\to$Janus
transitions, these are likely to take place within broader window of the $d$-$\theta$
continua.

When perfect wetting condition is satisfied ($\theta=\ang{0}$), last columns of 
Figs.~\ref{Early_stage}--\ref{Eqm_stage}), decomposition proceeds with spontaneous
solute enrichment of the particle surface which increases with time until the 
surface concentration reaches that of equilibrium $\beta_2$. In this case, the end 
product of coarsening is always CS-$2@1$, in accordance with the Cahn wetting
condition, Eq.~\eqref{eq_Cahn_wet}. Interestingly, we observe spontaneous 
enrichment of solute even when wetting is imperfect (\emph{e.g.}, $\chi = 0.5$). 
For small enough contact angles, the difference between the surface energies,
$\Delta{\sigma}$, is large, and CS$\to$Janus transition is arrested, especially at
larger particle sizes. As a result, CS-$2@1$ configuration remains stable (see,
\emph{e.g.}, Figs.~\ref{Eqm_stage}(\subref{r80c_eqm},~\subref{r90c_eqm})).

\begin{figure}[htbp]
\centering%
\begin{subfigure}{0.1\textwidth}
\centering%
    \includegraphics[scale=0.028]{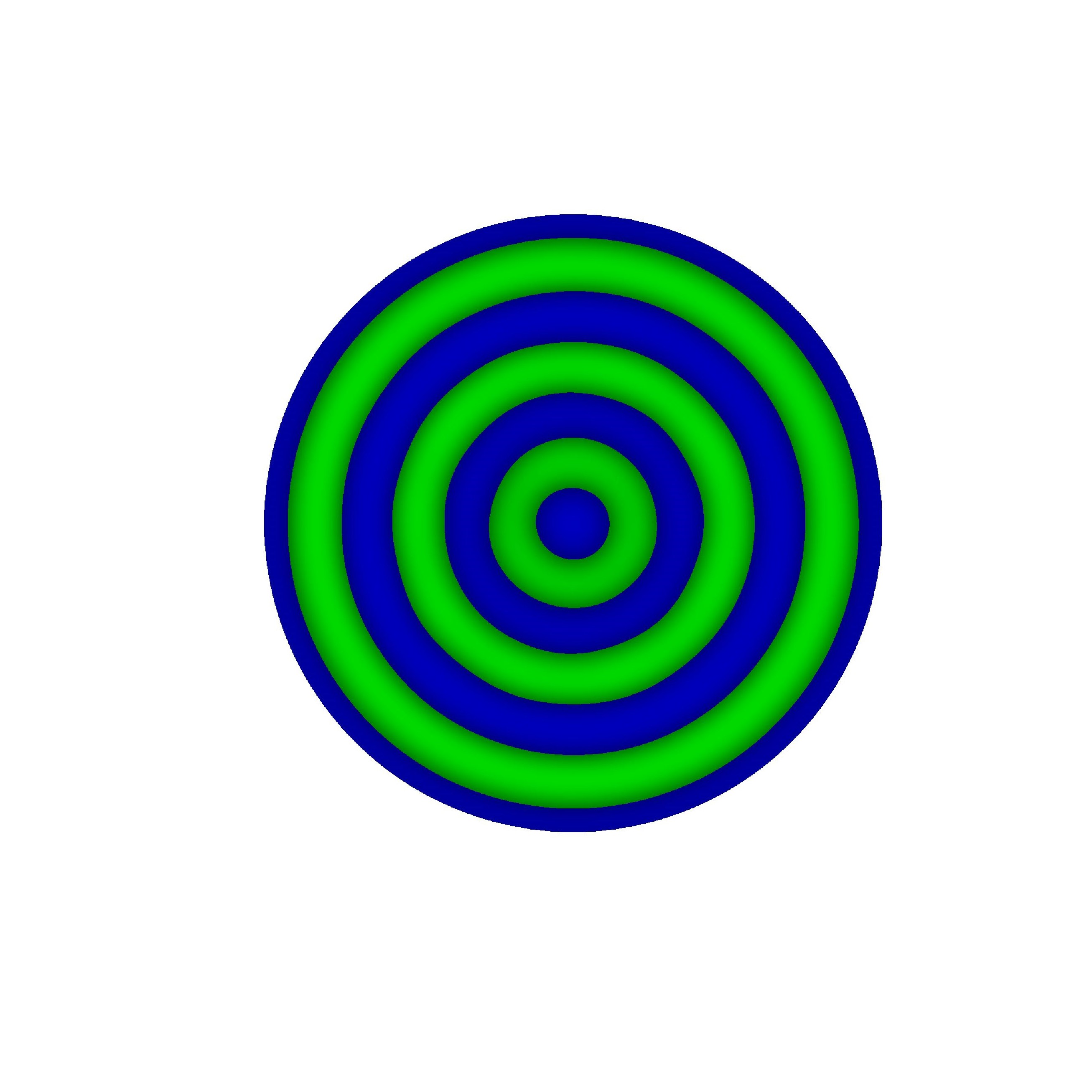}\hfil%
    \caption{}
    \label{r120a}
\end{subfigure}%
\begin{subfigure}{0.1\textwidth}
  \centering
  \includegraphics[scale=0.028]{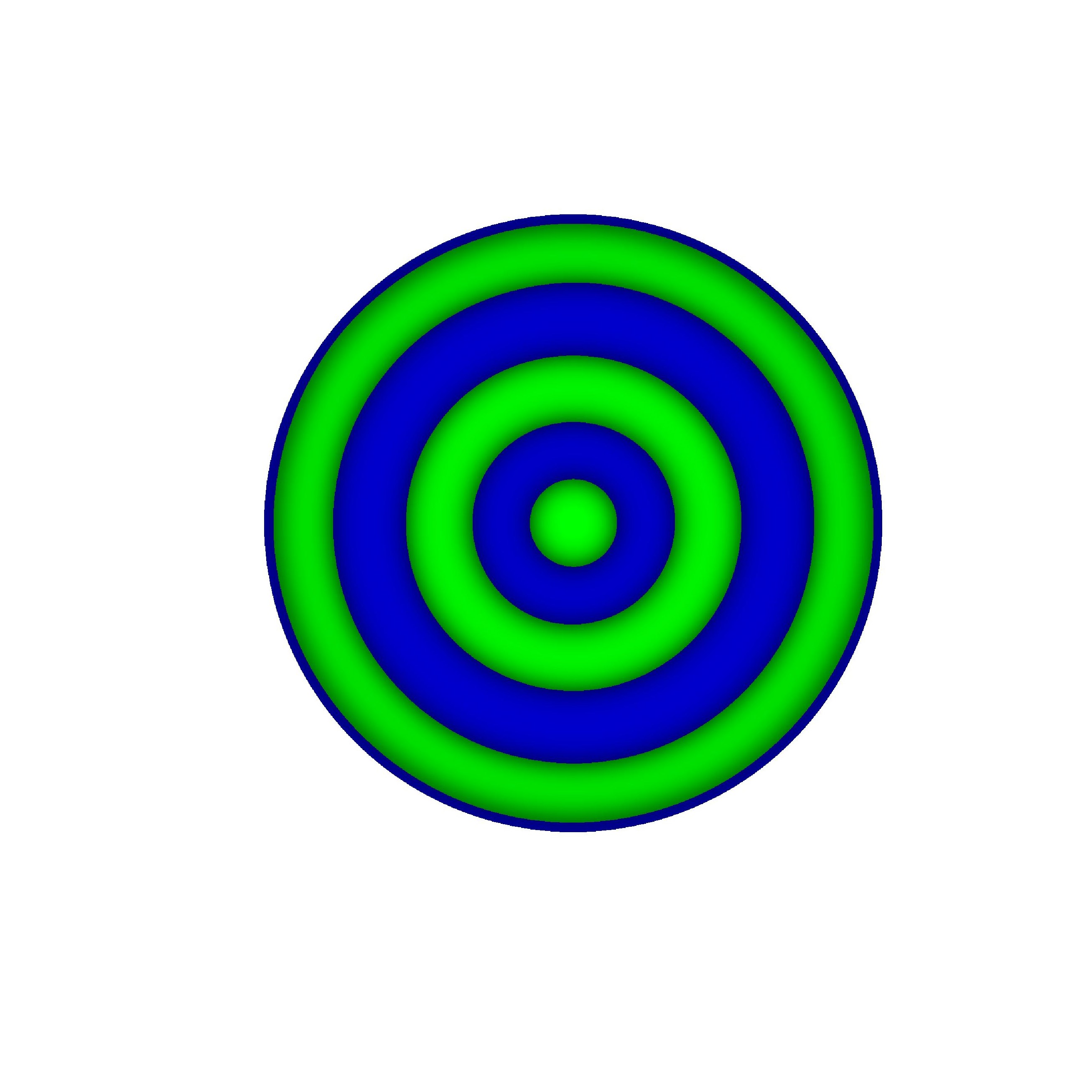}\hfil%
  \caption{}
  \label{r120b}
\end{subfigure}%
\begin{subfigure}{0.1\textwidth}
  \centering
  \includegraphics[scale=0.028]{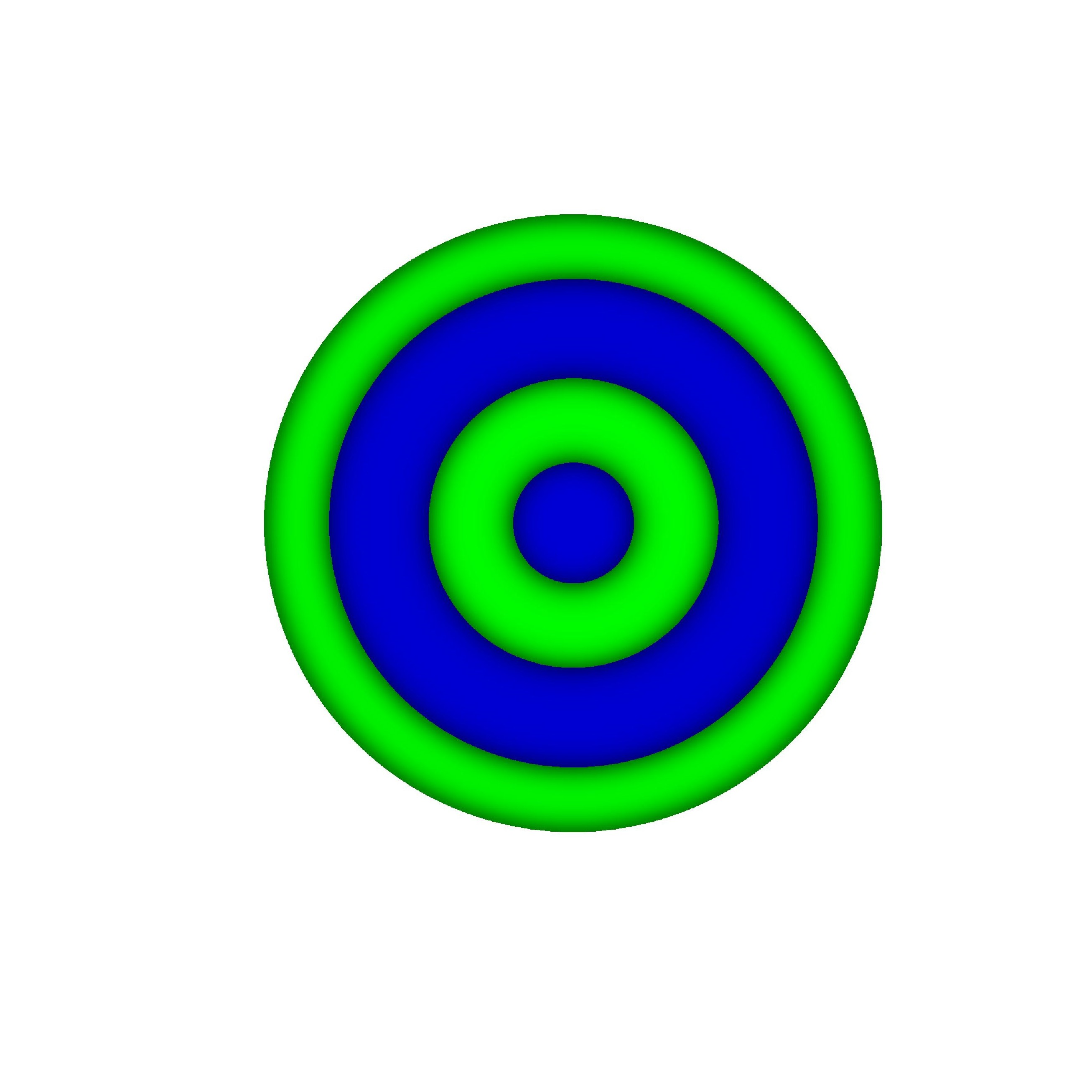}\hfil%
  \caption{}
  \label{r120c}
\end{subfigure}%
\begin{subfigure}{0.1\textwidth}
  \centering
  \includegraphics[scale=0.028]{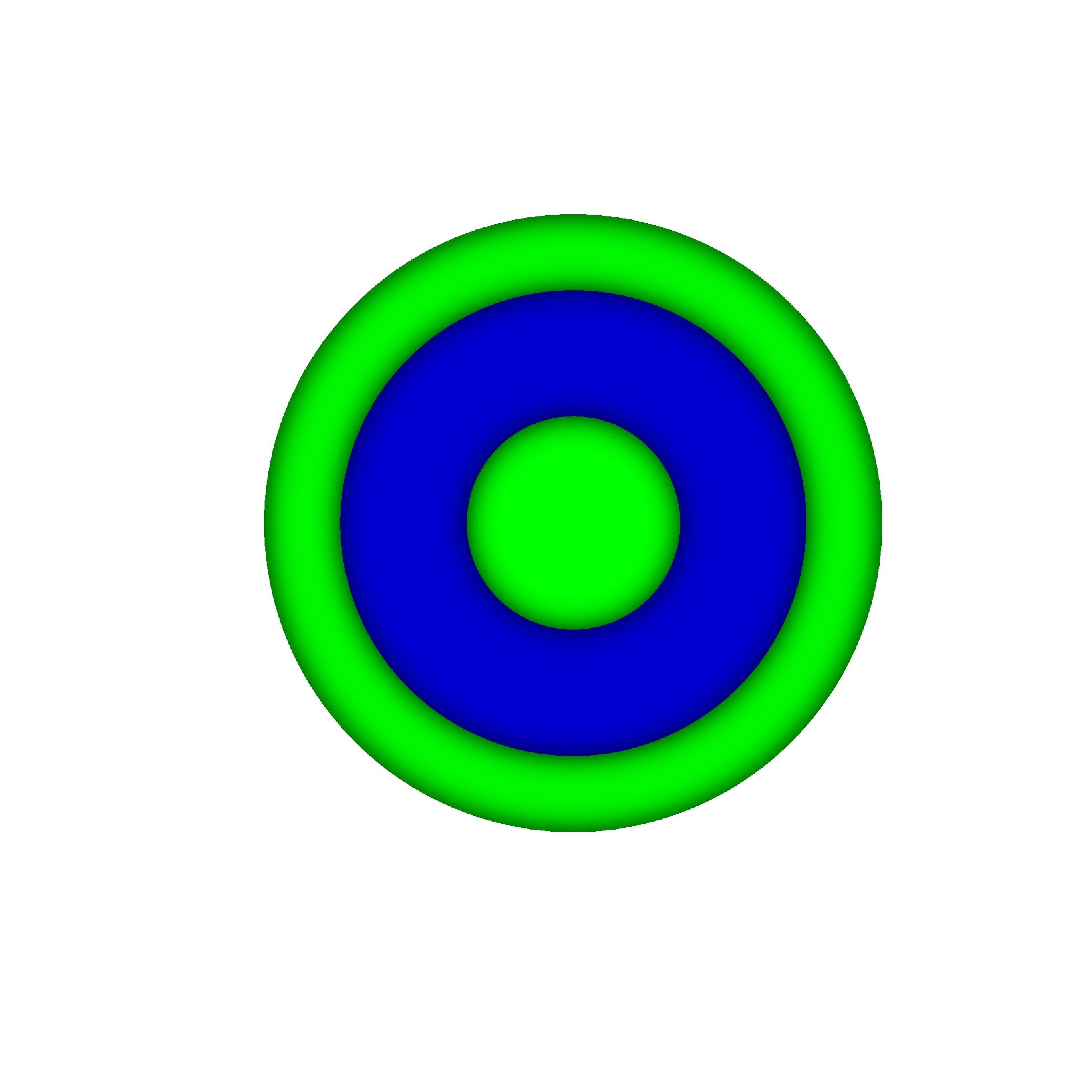}\hfil%
  \caption{}
  \label{r120d}
\end{subfigure}%
\begin{subfigure}{0.1\textwidth}
  \centering
  \includegraphics[scale=0.028]{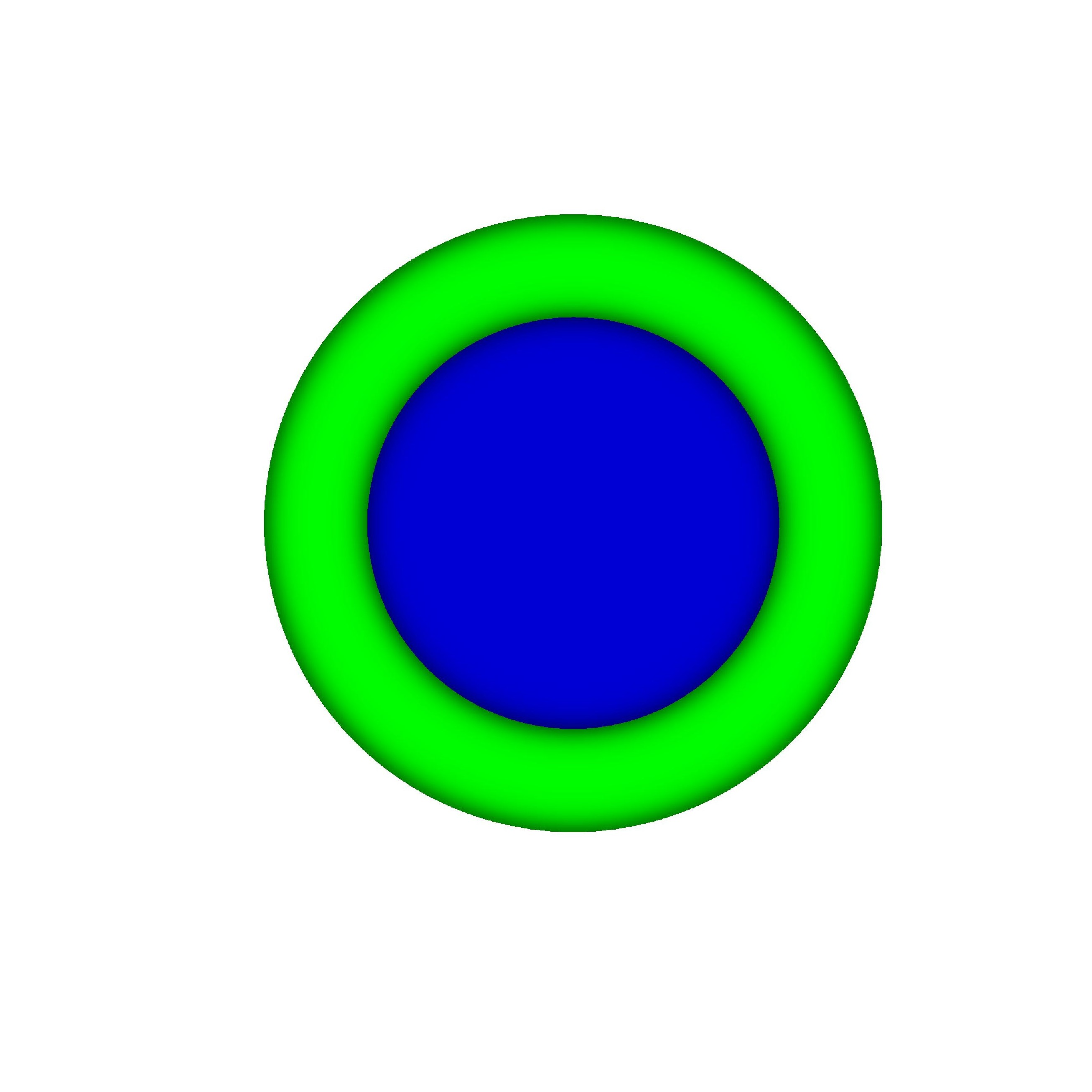}\hfil%
  \caption{}
  \label{r120e}
\end{subfigure}%

\bigskip

\begin{subfigure}{0.1\textwidth}
  \centering
  \includegraphics[scale=0.028]{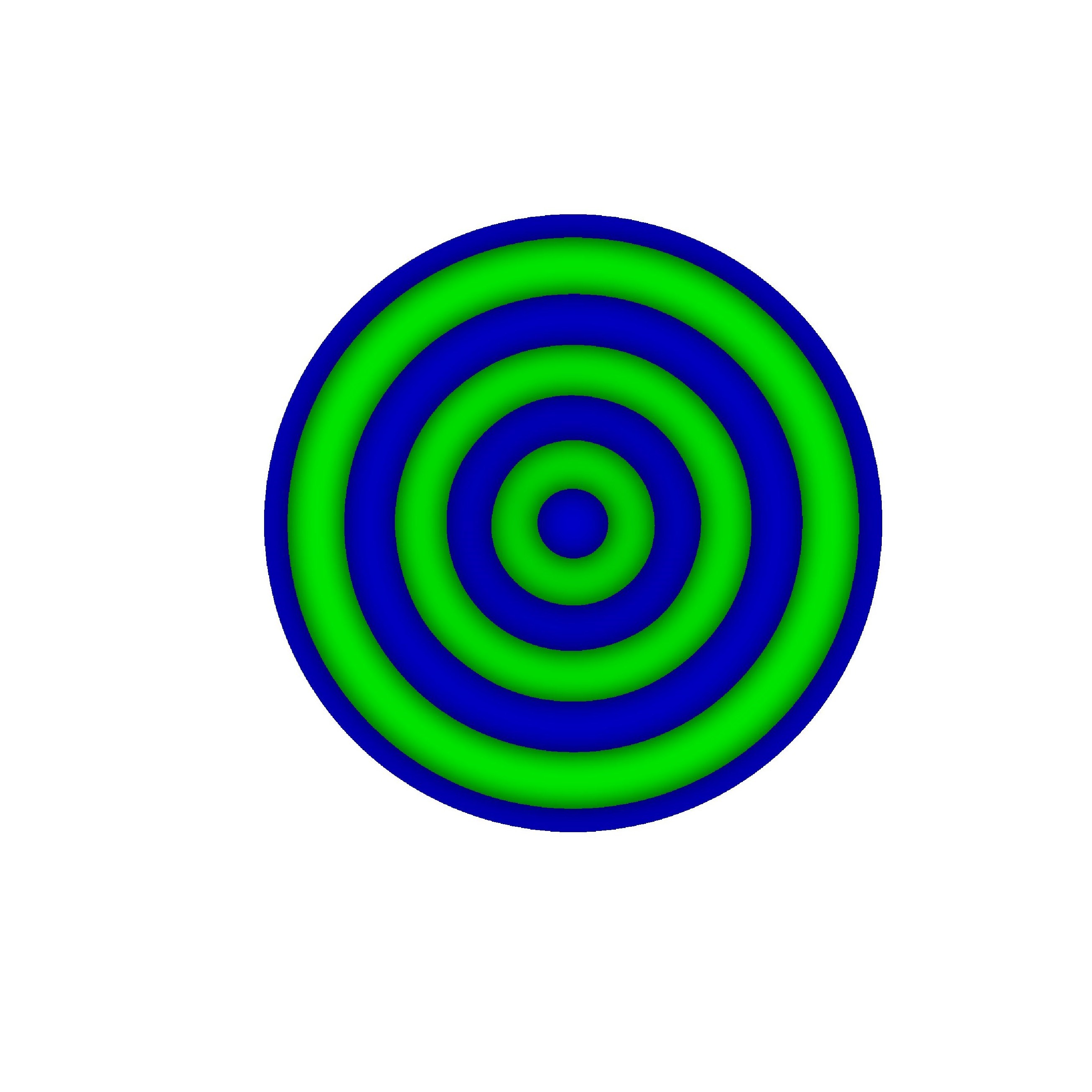}\hfil%
  \caption{}
  \label{r120chi05a}
\end{subfigure}%
\begin{subfigure}{0.1\textwidth}
  \centering
  \includegraphics[scale=0.028]{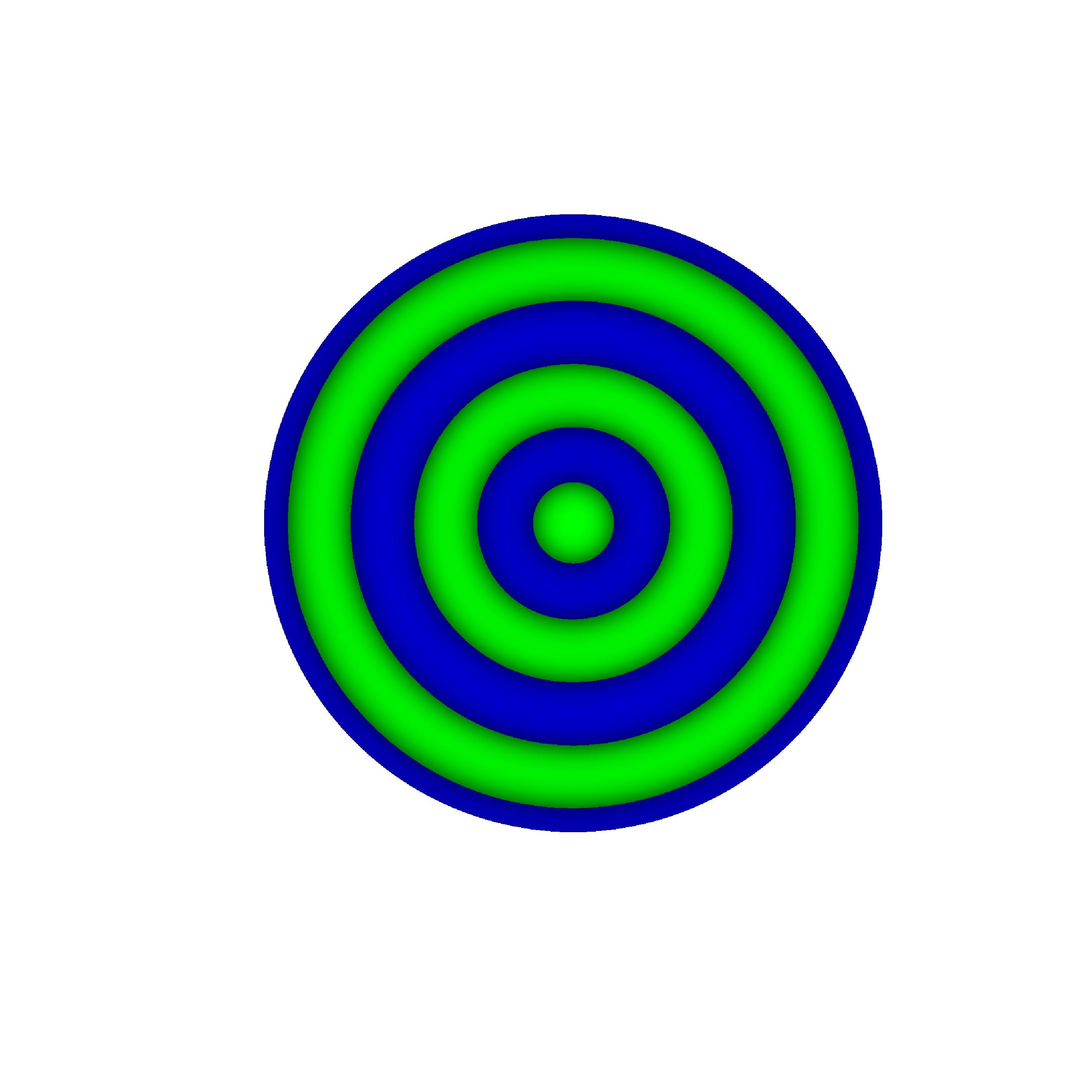}\hfil%
  \caption{}
  \label{r120chi05b}
\end{subfigure}%
\begin{subfigure}{0.1\textwidth}
  \centering
  \includegraphics[scale=0.028]{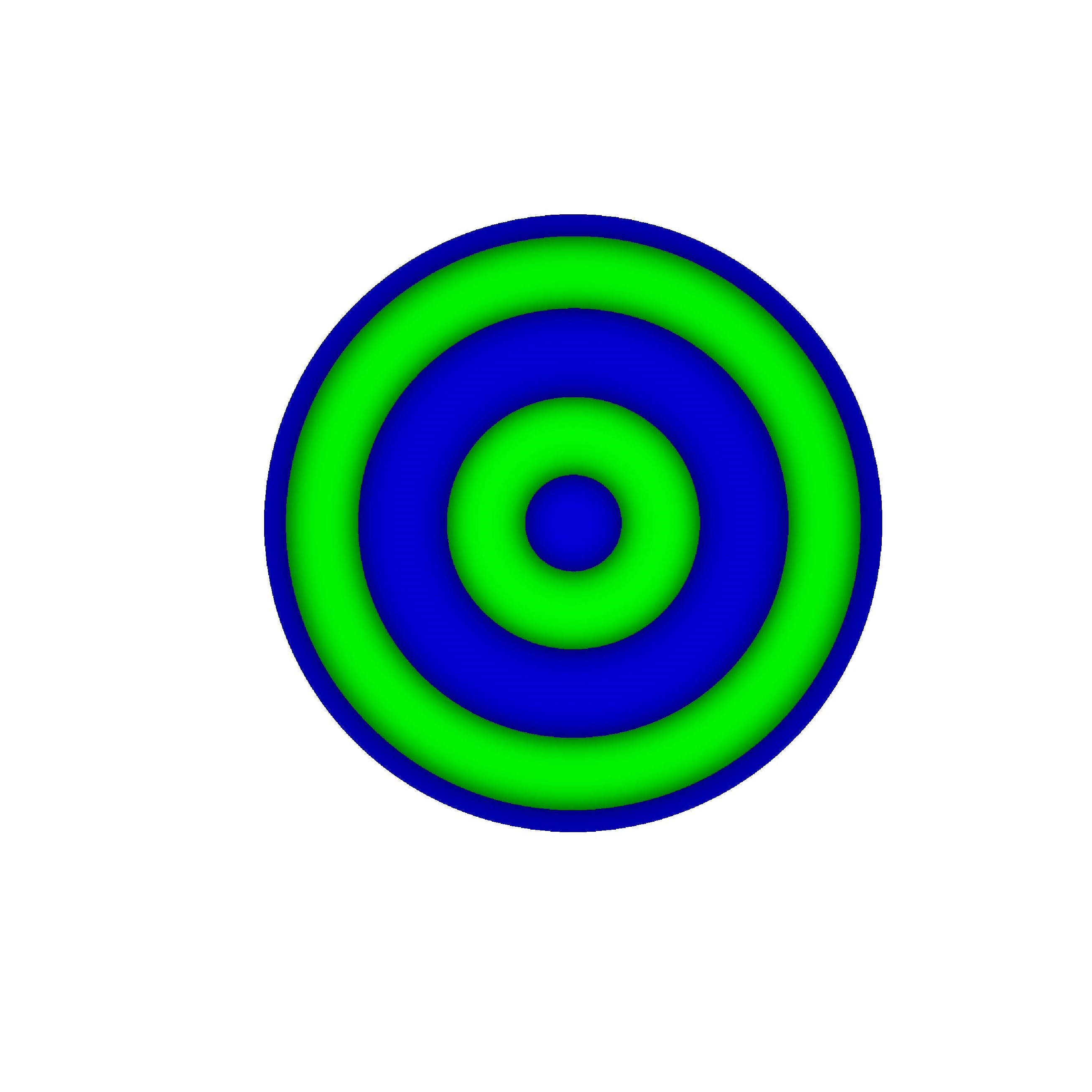}\hfil%
  \caption{}
  \label{r120chi05c}
\end{subfigure}%
\begin{subfigure}{0.1\textwidth}
  \centering
  \includegraphics[scale=0.028]{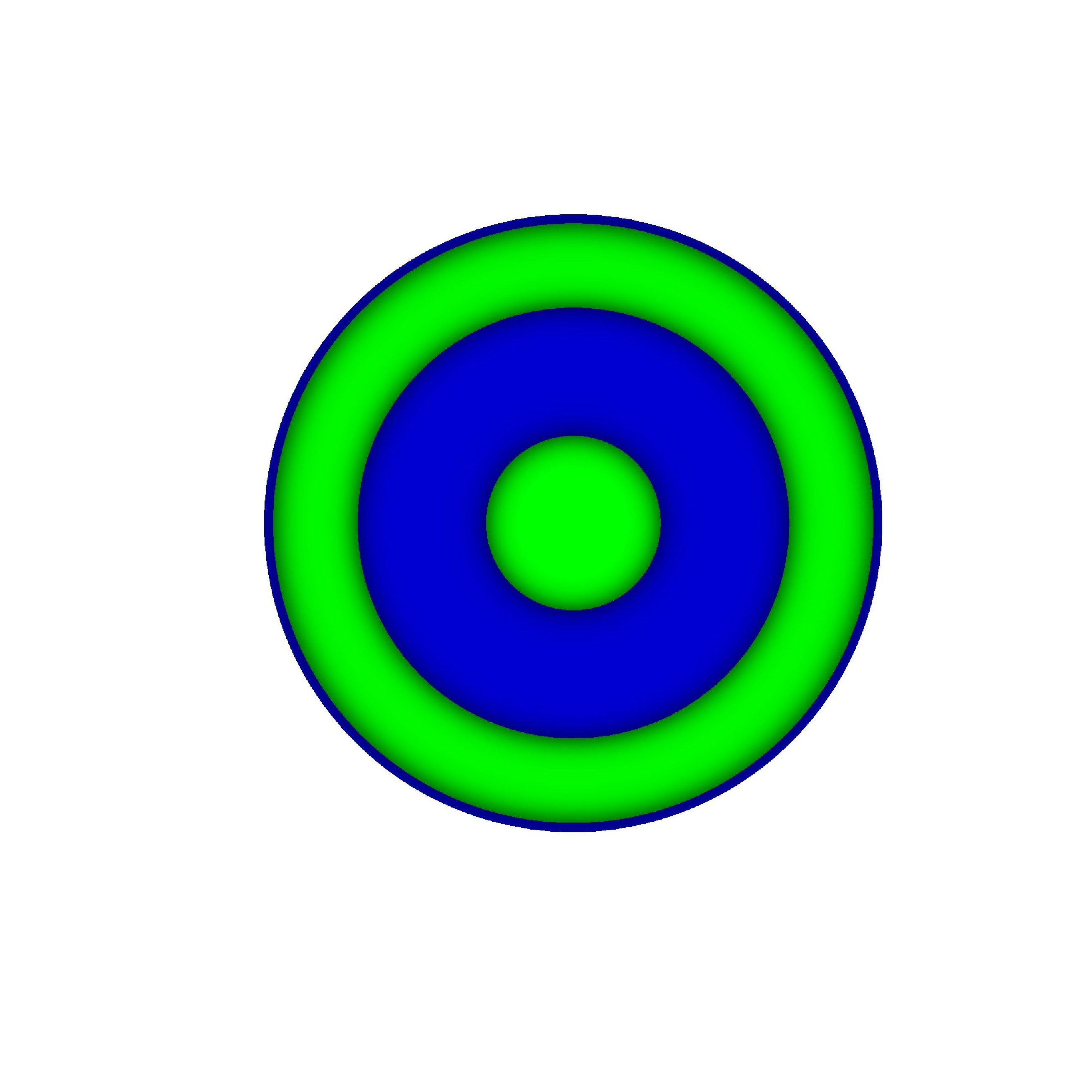}\hfil%
  \caption{}
  \label{r120chi05d}
\end{subfigure}%
\begin{subfigure}{0.1\textwidth}
  \centering
  \includegraphics[scale=0.028]{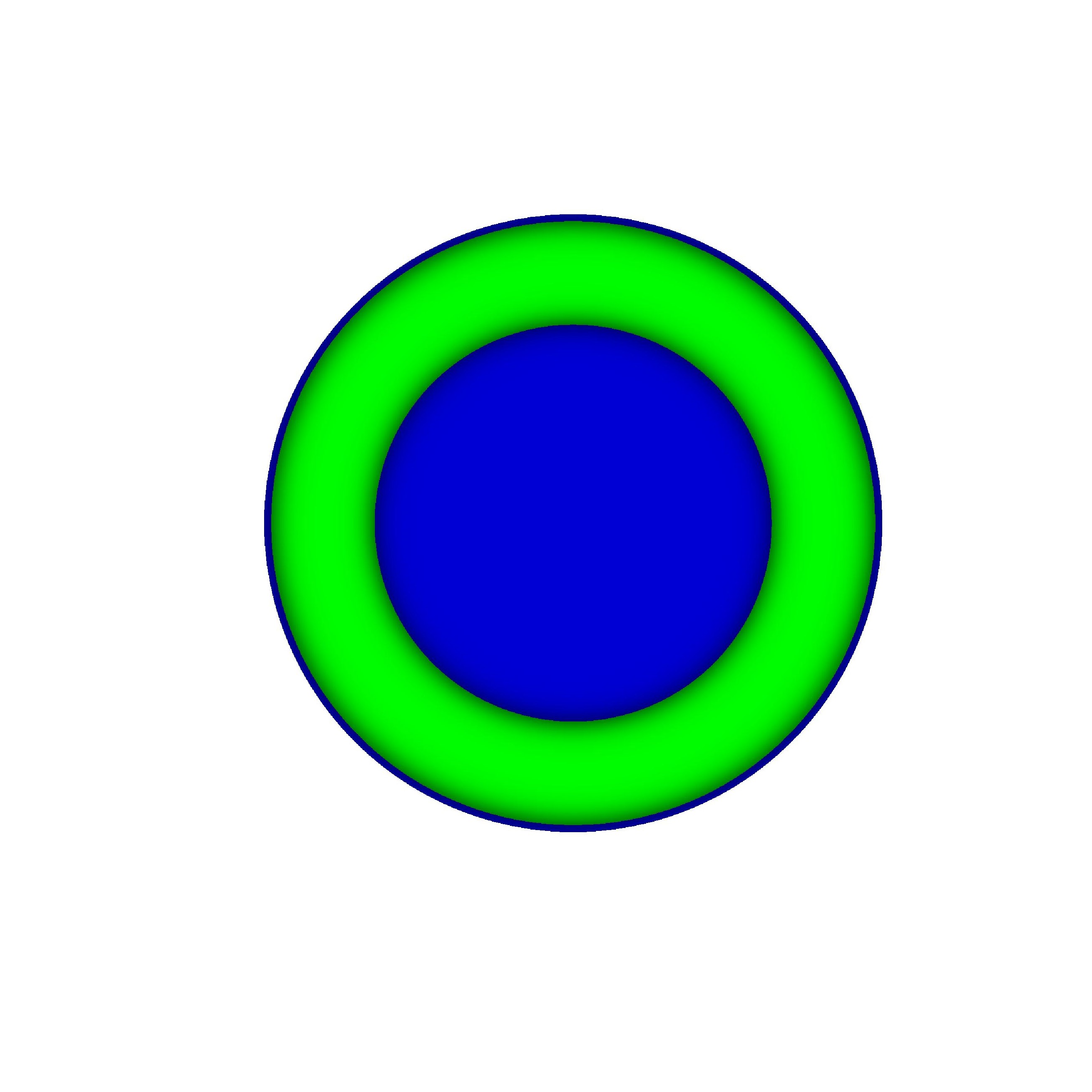}\hfil%
  \caption{}
  \label{r120echi05}
\end{subfigure}%



\caption{Time snapshots of microstructural evolution within a circular
particle of size $d=240$ showing various stages of evolution. Top row 
(a-e): $\chi=0.25$, bottom row (f-j): $\chi=0.5$. Time increases from 
left to right in a row. Blue and green colors represent solute-rich
$\beta_2$ and solute-poor $\beta_1$ phases, respectively.}%
\label{d_240}%
\end{figure}

In addition to CS-2@1 and Janus, there is yet another final configuration
possible at larger particle sizes. When $0<\chi<0.75$ and $d>180$, solute 
redistribution following SDSD leads to a CS-1@2 structure where the phase having
a larger surface energy ($\beta_1$) forms the shell. Fig.~\ref{d_240} shows the
morphological evolution in particles of size $d=240$ for $\chi=0.25$ and $0.5$
(corresponding animations are provided in ESI).
Initial layered structure formed by SDSD in these larger particles is similar to 
what is observed in smaller particles (see second column of Fig.~\ref{Early_stage} 
for comparison), although the number of rings is greater in the larger particle. 
Unlike in smaller particles, however, the core does not destabilize in the larger 
particle at later times, and the outermost solute-poor layer (green) continues to
widen, resulting in the final CS-1@2 morphology. 
Note that when the phase with a higher surface energy - here solute-poor $\beta_1$ 
- forms the shell, its outer surface is always enriched with solute, as indicated by
the $C_{mx}$ values in the top row of Fig.~\ref{fig:1d-eqm-profiles} corresponding to
$\alpha$-$\beta_1$ interface. The likelihood of ending up with this morphology 
increases with increasing particle size, except for $\chi=0.75$
(\emph{i.e.}, $\theta=\ang{0}$) which always forms CS-2@1 morphology.

\subsection{Mechanisms of morphological evolution}

As mentioned earlier, various studies have reported a wide spectrum of BNP
configurations, CS and Janus being the two most frequently observed ones.
Although most of these studies have focused on the final morphology and not
how it is attained, some mechanisms have been put forward or speculated as
possible means to achieve CS and Janus configurations. These include
coalescence of individual NPs~\cite{Yuan2008,Langlois2012,Lu2014b,Zhang2016},
thin film dewetting and re-solidification~\cite{McKeown2015a}, or strain 
effects arising from the size mismatch of the constituent atoms
~\cite{Bochicchio2013,Ferrando2015,Peng2015modeling,Peng2015strain}.
Simulations presented here establish another route, \emph{viz.}, SD of
homogeneous alloyed BNPs, by which these structures can form. Incidentally,
this appears to the operative mechanism in BNPs synthesized using solvothermal
co-reduction~\cite{Sopousek2015}, pulsed laser ablation~\cite{Malviya2016} or sputtering~\cite{Radnoczi2017,Gajdics2018}.

Our simulations show that when segregation is absent ($\chi=0$, 
Fig.~\ref{chi0_AB1}), evolution proceeds with bulk phase separation.
It eventually leads to an ideal Janus structure where the two
halves of the particle are separated by a flat interface along a
diameter (for example Fig.~\ref{r70a_eqm}). In the presence of 
segregation ($\chi\neq0$), there are two possible pathways to reach
a Janus morphology. The first one involves heterogeneous nucleation
of one of the phases on the surface of a supersaturated alloy 
particle, and its subsequent growth leads to a `side-segregated' 
morphology~\cite{Malviya2016, Langlois2012}. Alternatively, Janus
structures can form through a three-step process: (i) initially, 
SDSD produces an onion-like morphology comprising of alternate 
rings, (ii) these rings coarsen to form a CS morphology, and finally,
(iii) the CS structure is destabilized by interface-induced
disturbances and small fluctuations present in the system to form a
Janus.

Heterogeneous nucleation mechanism requires composition waves that
are \emph{large in degree, but small in extent}~\cite{cahn1961spinodal}.
To investigate this mechanism, one requires to introduce large, sustained
fluctuations. The current study, on the other hand, deals with systems 
having a miscibility gap which permits spontaneous phase separation. It
can  occur with small initial fluctuations or even without it. Most of
the systems constituting BNPs have miscibility gaps, which makes SD 
the more likely mechanism adopted by the system to form Janus.

\begin{figure}[htbp]
 \centering
 \begin{subfigure}{0.3\textwidth}
    \includegraphics[scale=0.05]{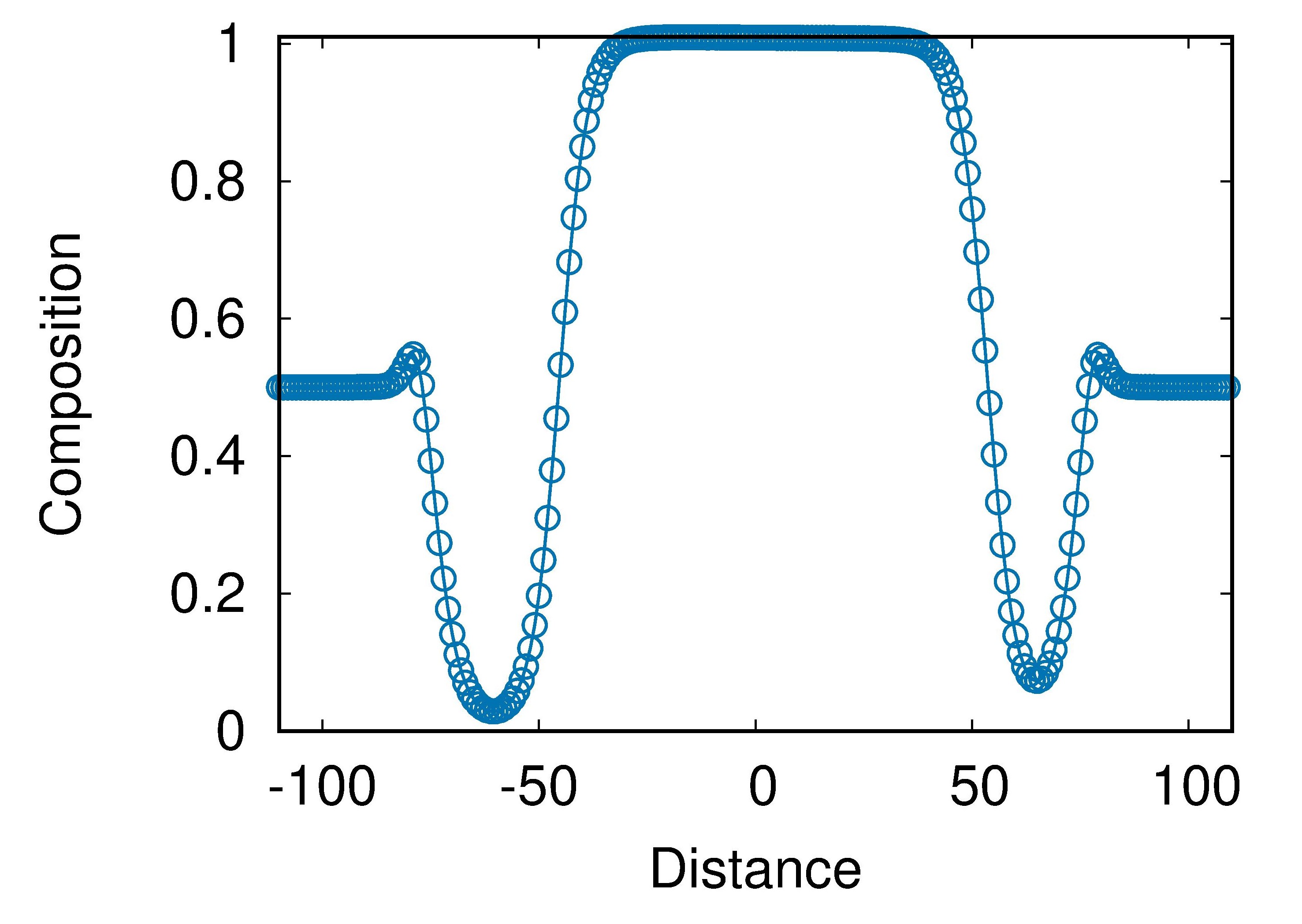}%
    \caption{}
    \label{fig:lcut_d160_chi25}
 \end{subfigure}%
 \begin{subfigure}{0.3\textwidth}
  \centering
    \includegraphics[scale=0.05]{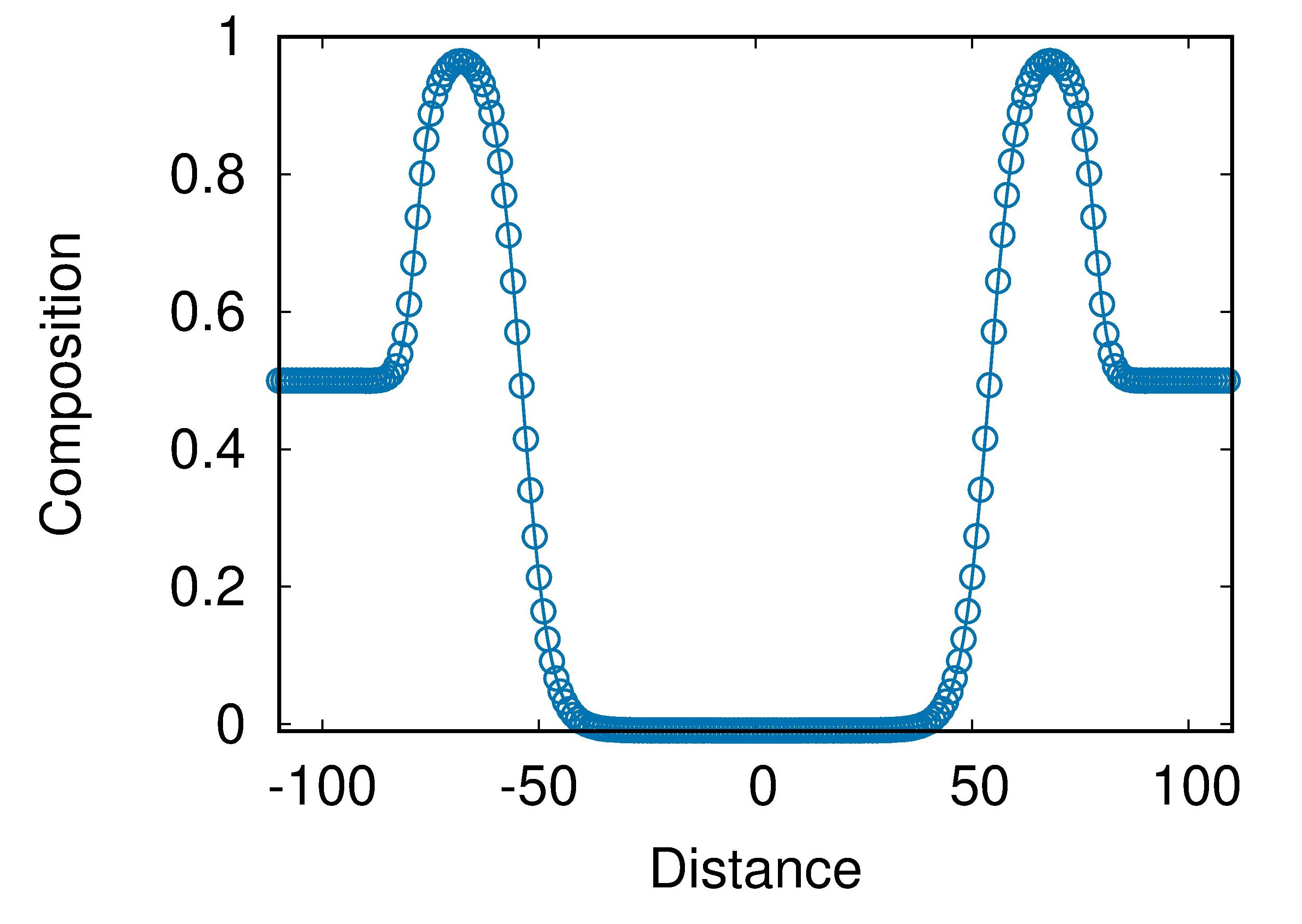}%
    \caption{}
    \label{fig:lcut_d160_chi05}
 \end{subfigure}%
 \caption{Composition profiles along a diameter of a particle with 
 $d=160$ for (a) $\chi=0.25$ and (b) $\chi=0.5$ highlighting the 
 breakdown and retention, respectively, of circular symmetry in
 these two cases.}%
 \label{fig:lcut_d160}
\end{figure}

Disruption of intermediate CS structure produced by  SDSD proceeds
through core destabilization and subsequent core movement towards the
surface of the particle. Core destabilization initiates through a
breakdown of the circular symmetry of the radial composition profiles
due to interface induced disturbances. For example,
Fig.~\ref{fig:lcut_d160_chi25} presents an intermediate-time composition
profile taken along a diameter of a particle shown in  Fig.~\ref{r80b_int}
($d=160$, $\chi=0.25$): this plot clearly illustrates the breaking down of
the symmetry of the  composition `line scan' with respect to the center of
the particle. In contrast, the profile presented in 
Fig.~\ref{fig:lcut_d160_chi05} for the same particle size but with a 
higher $\chi$ (Fig.~\ref{r80c_int}) retains its symmetry. Consequently, the
particle with a lower  $\chi$ (\emph{i.e.}, higher $\theta$) undergoes core
distortion, which evolves from a circle to an oval, and it finally results 
in a Janus configuration; the particle with a higher $\chi$ (lower $\theta$),
however, retains the symmetric CS structure. Animations depicting the 
morphological evolution for both the cases are provided as ESI.

Core destabilization depends on the contact angle ($\theta$) as well as
particle size ($d$). For systems closer to the perfect wetting condition
(\emph{i.e.}, smaller $\theta$'s), radial composition profiles have a 
stronger tendency to retain their circular symmetry and prefer a CS
structure. However, for a given contact angle $\theta$, circular rings
are more unstable in smaller particles as their relatively large higher 
surface area to volume ratio makes them more susceptible to interface-induced
disturbances. 

Surface solute segregation plays an important role in events
leading up to the formation of CS as the final configuration in
certain systems. Formation of a segregated layer initiates SDSD
with radial composition waves propagating from surface of the
particle towards its center. For systems with a larger $\chi$ and
relatively smaller $d$, transformation proceeds by spontaneous 
and monotonic solute enrichment of the segregated layer by the 
spinodal mechanism. The maximally growing wavelength of the
composition waves formed by this mechanism is dictated by the 
greater of the surface energies. It results in an outermost 
$\beta_2$ ring that grows inward, leading to a CS-2@1 structure.

\begin{figure*}[htbp]
\centering%
\begin{subfigure}{0.24\textwidth}%
\centering%
\begin{tikzpicture}
    \node[anchor=south west,inner sep=0] at (0,0) 
    {\includegraphics[scale=0.044]{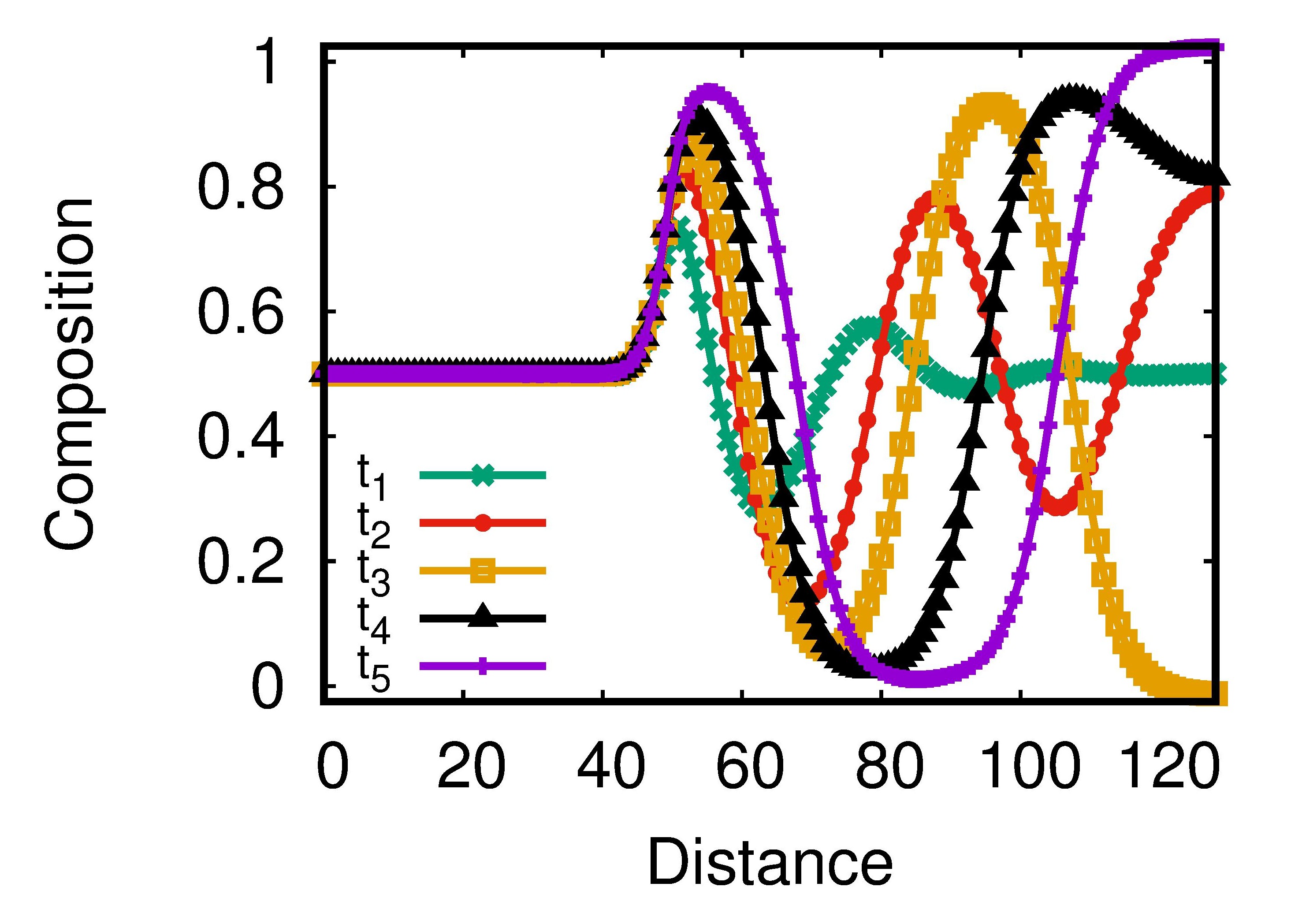}};
    \draw[red,dashed,thick,rounded corners] (2,0.8) rectangle (3,3); 
\end{tikzpicture}%
\end{subfigure}%
\begin{subfigure}{0.24\textwidth}%
\centering%
  \begin{tikzpicture}
    \node[anchor=south west,inner sep=0] at (0,0) 
    {\includegraphics[scale=0.044]{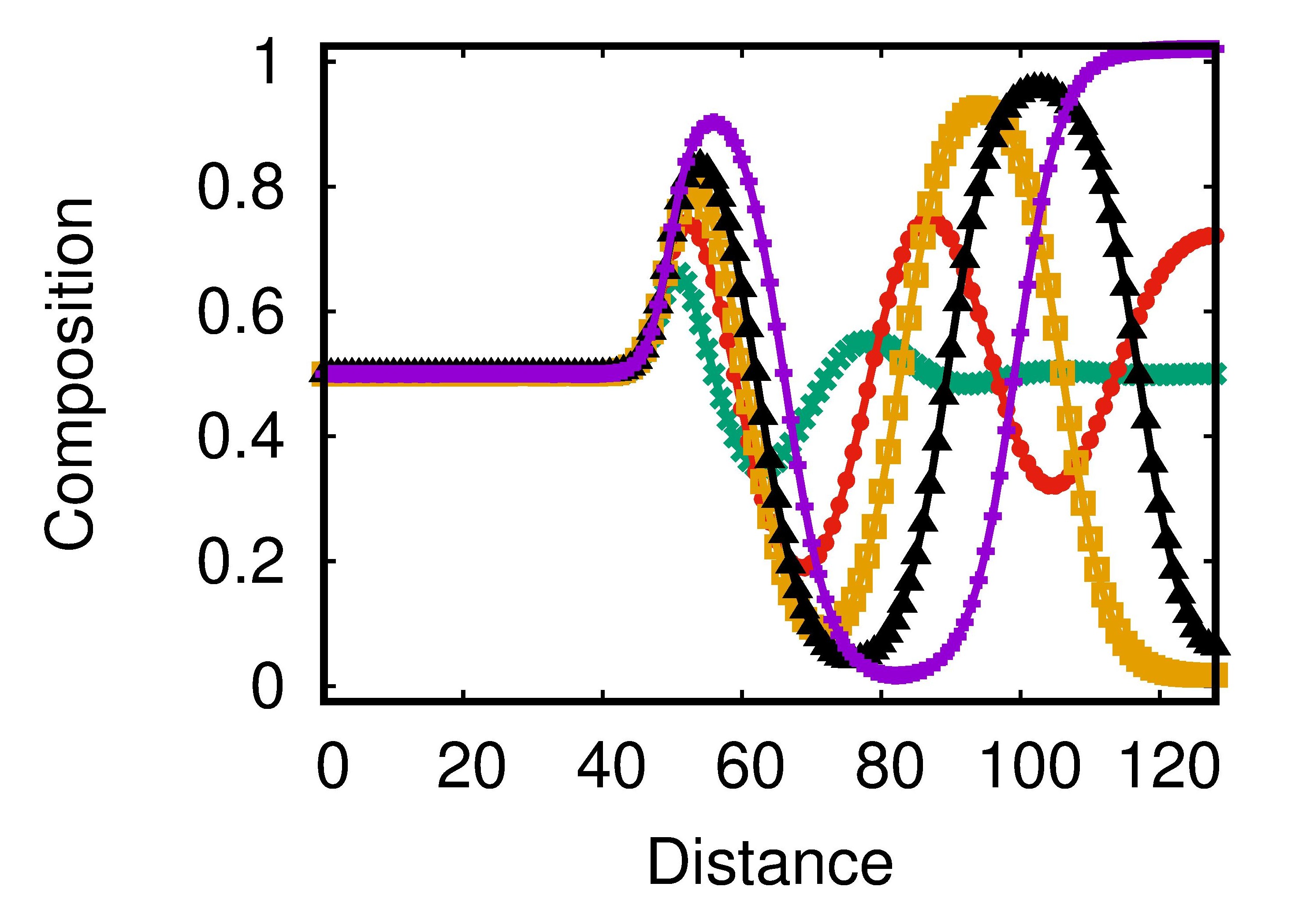}};
    \draw[red, thick,dashed,rounded corners] (2,0.8) rectangle (3.1,3); 
 \end{tikzpicture}
\end{subfigure}%
\begin{subfigure}{0.24\textwidth}%
\centering%
 \begin{tikzpicture}
    \node[anchor=south west,inner sep=0] at (0,0) 
    {\includegraphics[scale=0.044]{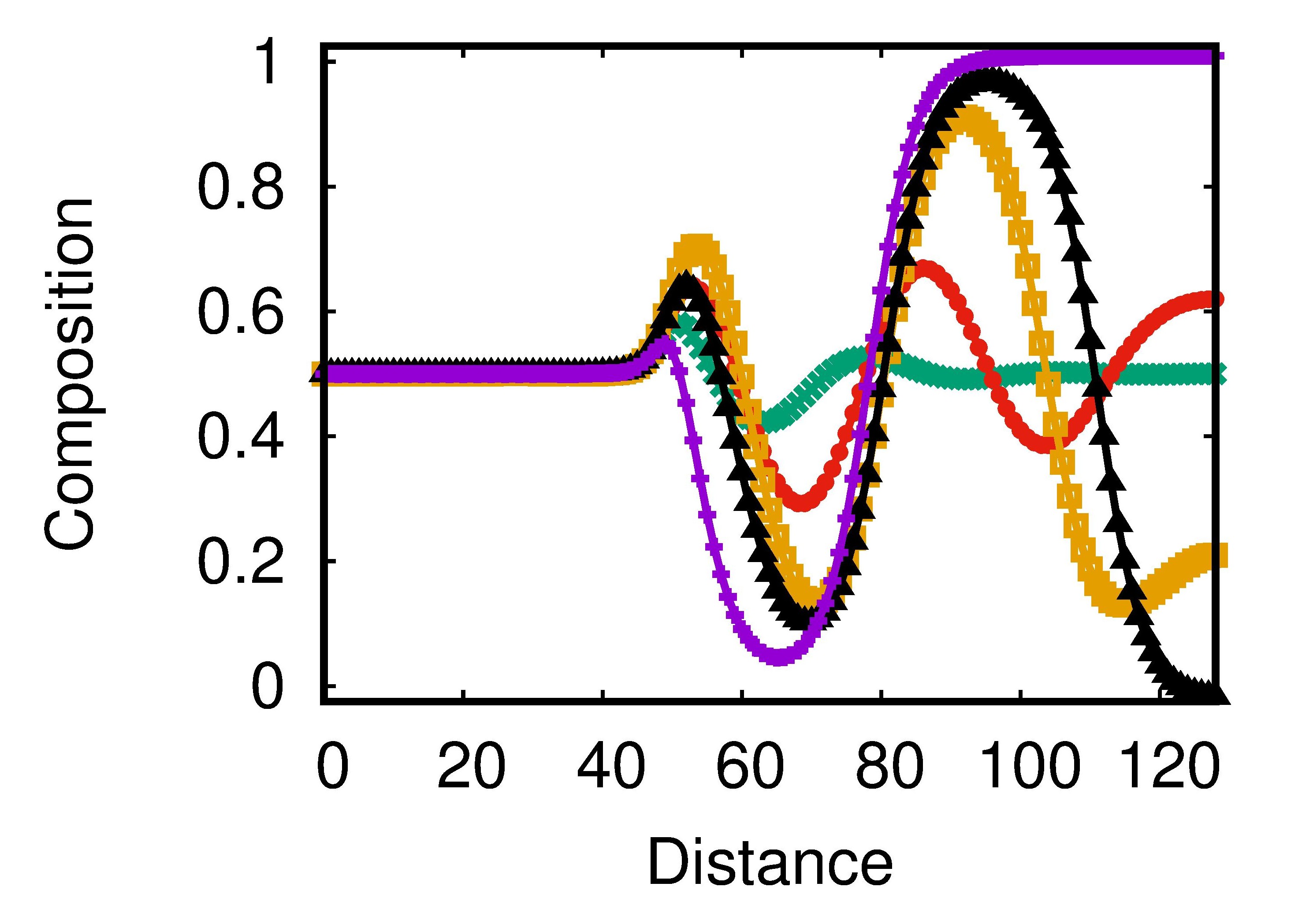}};
    \draw[red, thick,dashed,rounded corners] (2,0.8) rectangle (3,2.5); 
 \end{tikzpicture}
\end{subfigure}%
\begin{subfigure}{0.24\textwidth}%
\centering%
  \begin{tikzpicture}
    \node[anchor=south west,inner sep=0] at (0,0) {\includegraphics[scale=0.044]{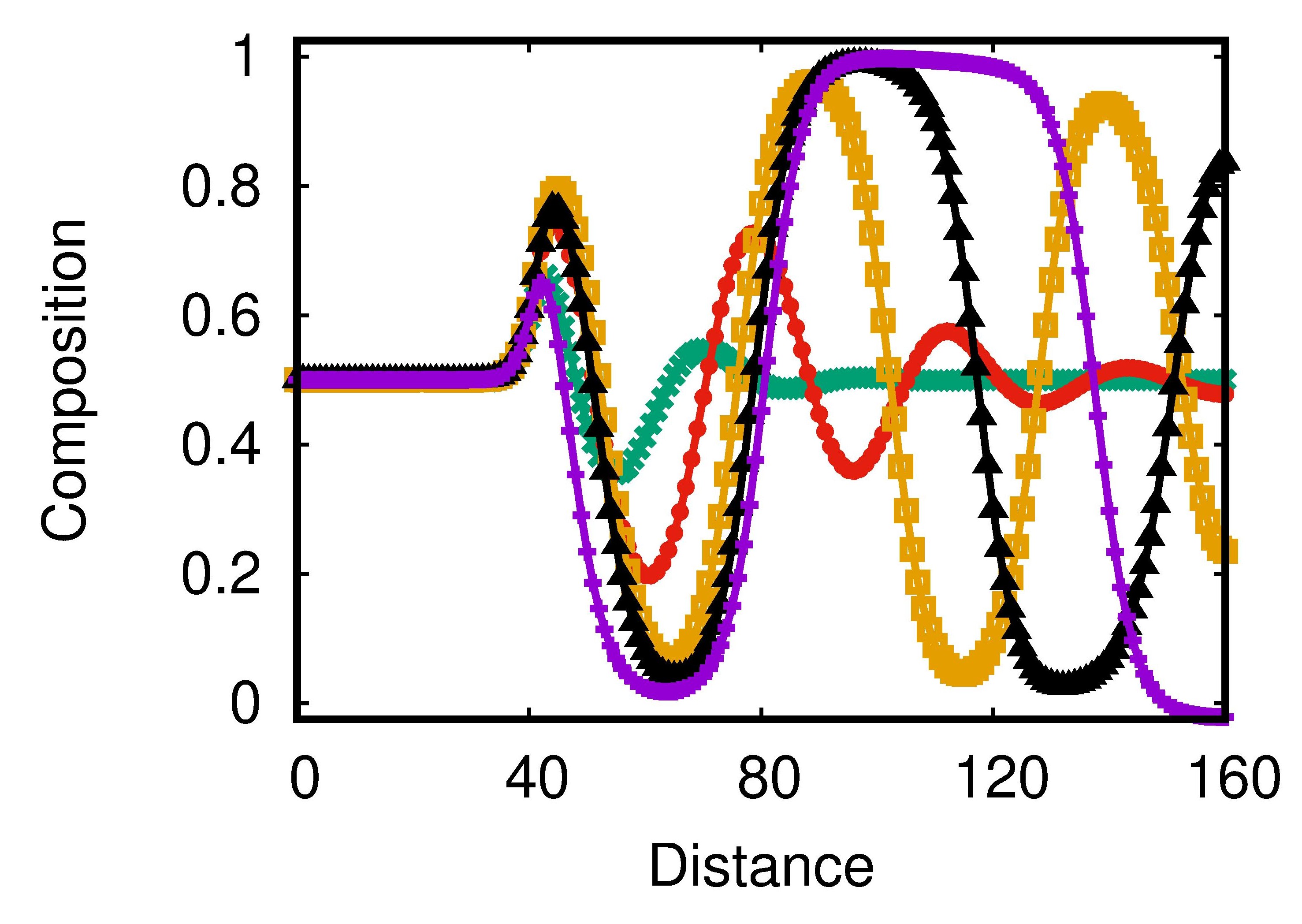}};
    \draw[red, thick,dashed,rounded corners] (1.7,0.8) rectangle (2.1,2.8); 
 \end{tikzpicture}
\end{subfigure}%
\hspace*{\fill}

\bigskip

\begin{subfigure}{0.24\textwidth}%
\centering
 \begin{tikzpicture}
    \node[anchor=south west] at (0,0) 
    {\includegraphics[scale=0.045]{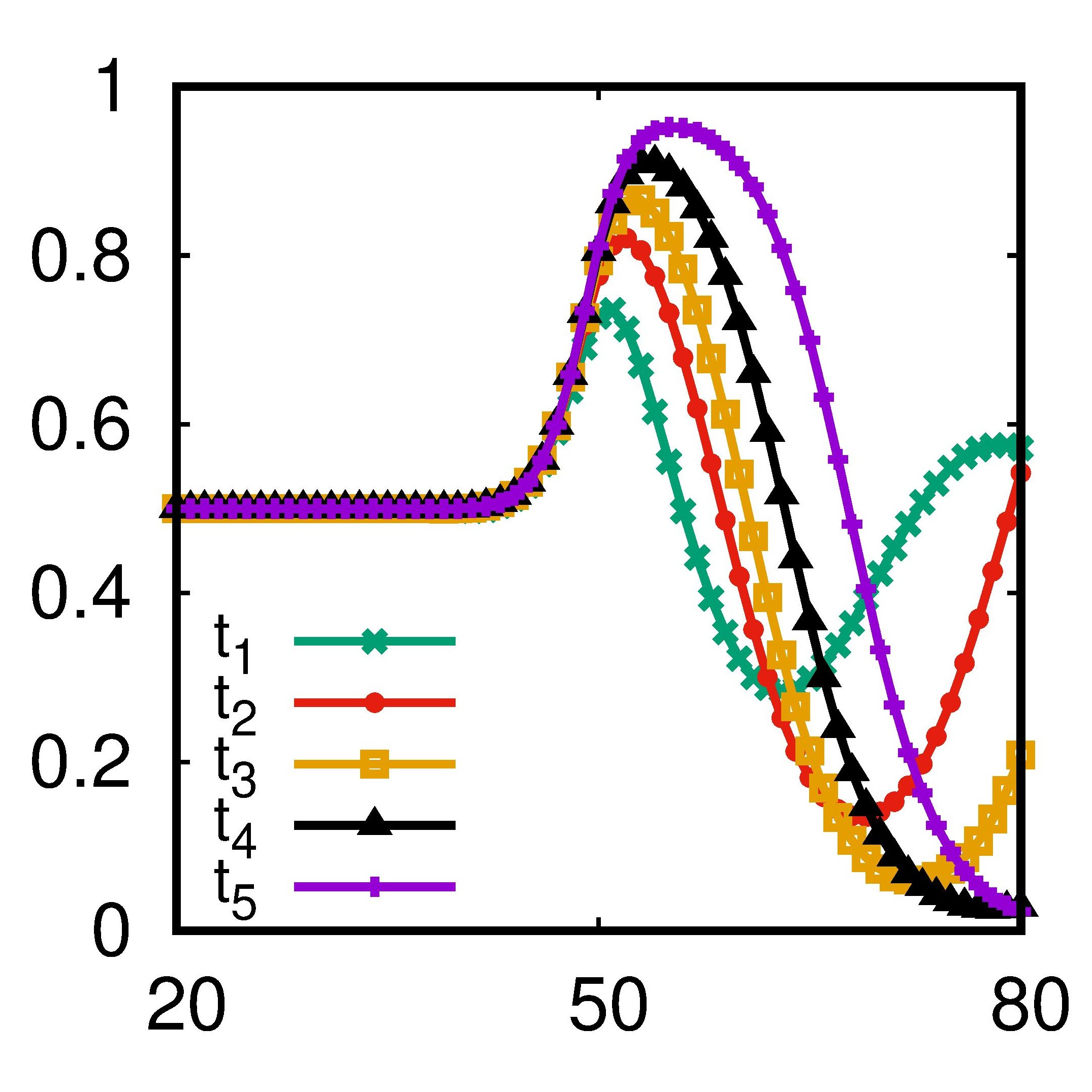}};
    \draw[->,line width=0.3mm](1.8,2.4)--(1.8,2.9);
    \node at (2.,-0.3) {};
 \end{tikzpicture}%
 \caption{}
\end{subfigure}%
\begin{subfigure}{0.24\textwidth}%
\centering
 \begin{tikzpicture}
    \node[anchor=south west] at (0,0) 
    {\includegraphics[scale=0.045]{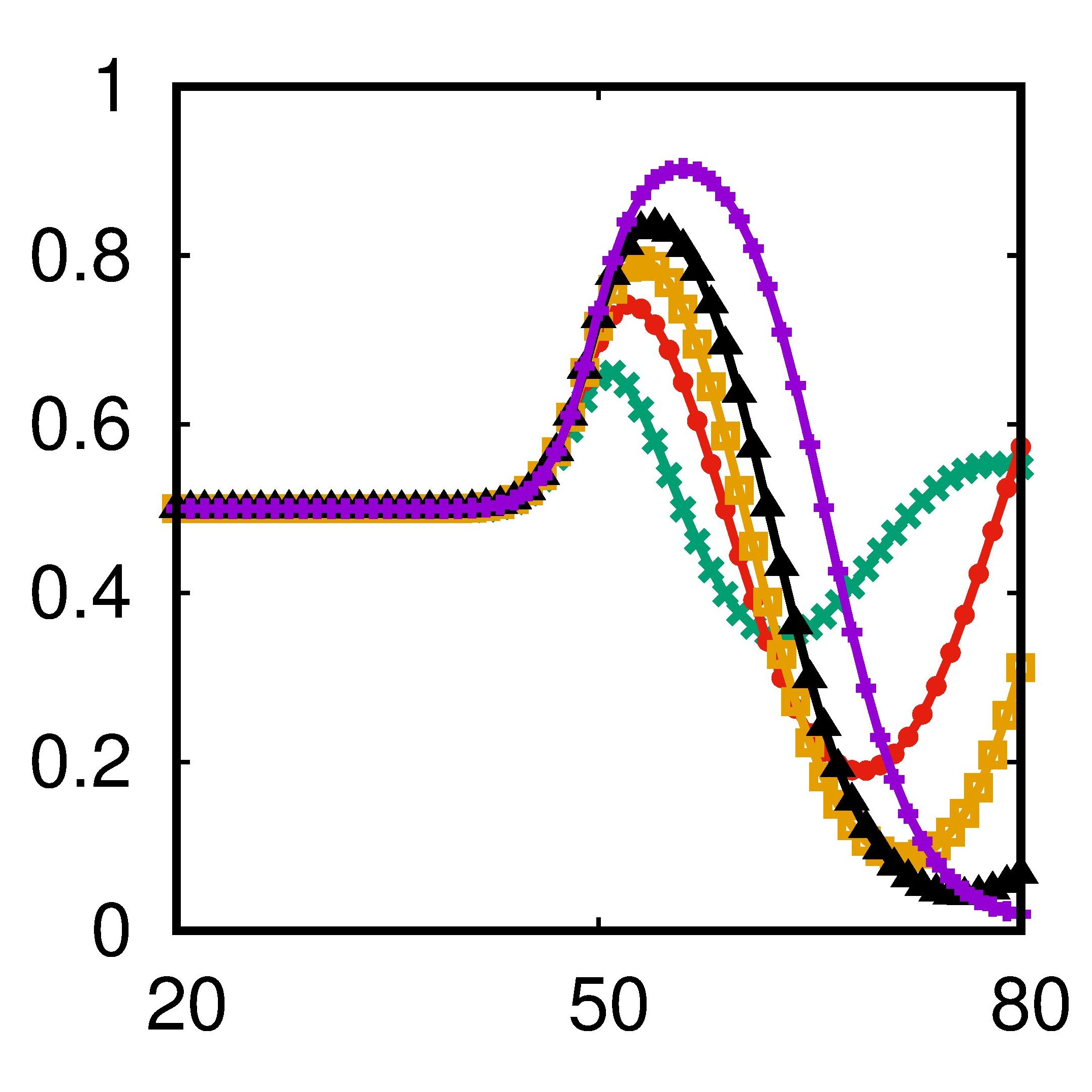}};
    \draw[->,line width=0.3mm](1.8,2.4)--(1.8,2.9);
    \node at (2.,-0.3) {};
 \end{tikzpicture}%
 \caption{}
\end{subfigure}%
\begin{subfigure}{0.24\textwidth}%
\centering
  \begin{tikzpicture}
    \node[anchor=south west] at (0,0) 
    {\includegraphics[scale=0.045]{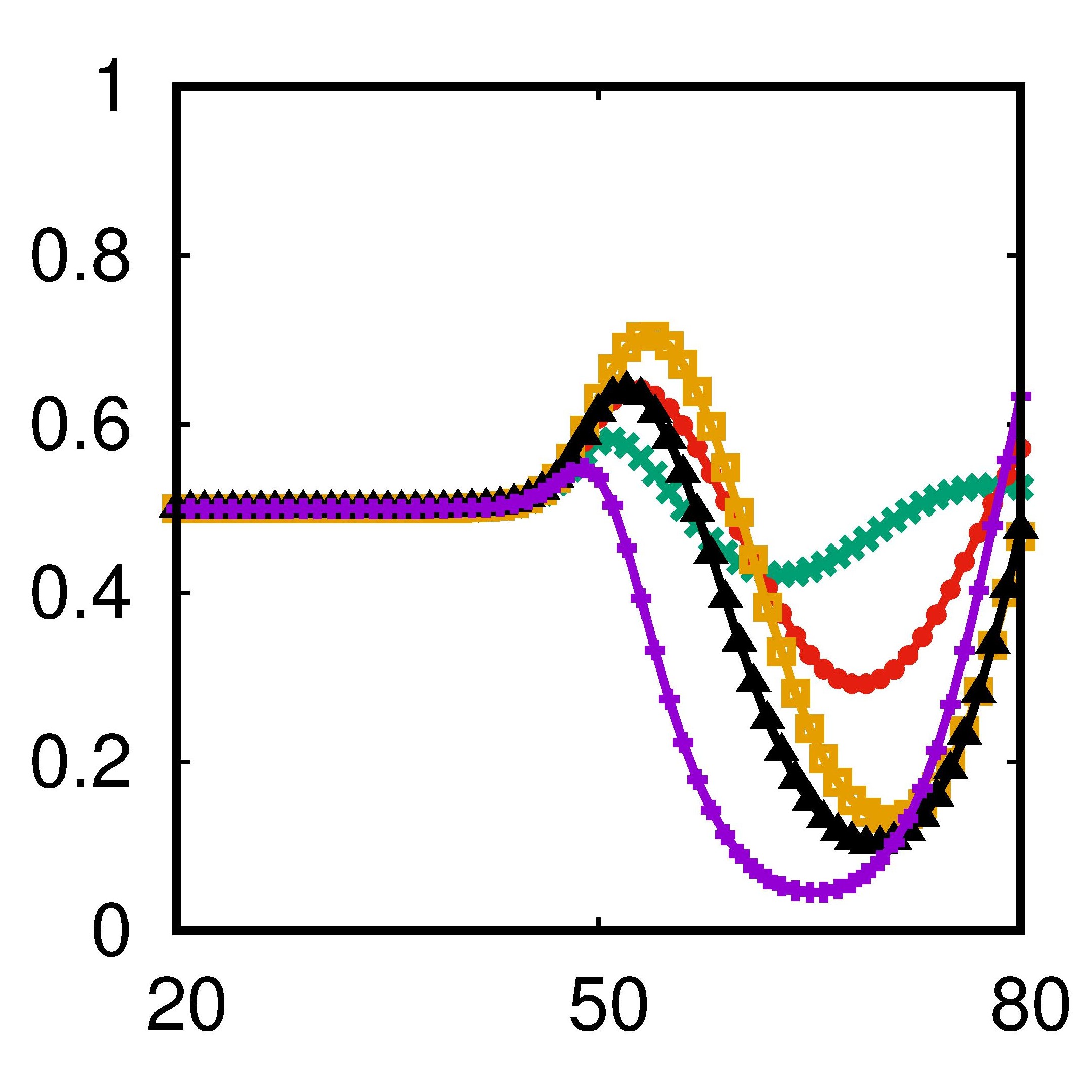}};
    \node at (2.,-0.3) {};
    \draw[->,line width=0.3mm](2.,2.6) -- (2.,3.1);
    \draw[->,line width=0.3mm](2.2,3.1) -- (2.2,2.6);
 \end{tikzpicture}%
 \caption{}
\end{subfigure}%
\begin{subfigure}{0.24\textwidth}%
\centering
  \begin{tikzpicture}
    \node[anchor=south west] at (0,0) 
    {\includegraphics[scale=0.045]{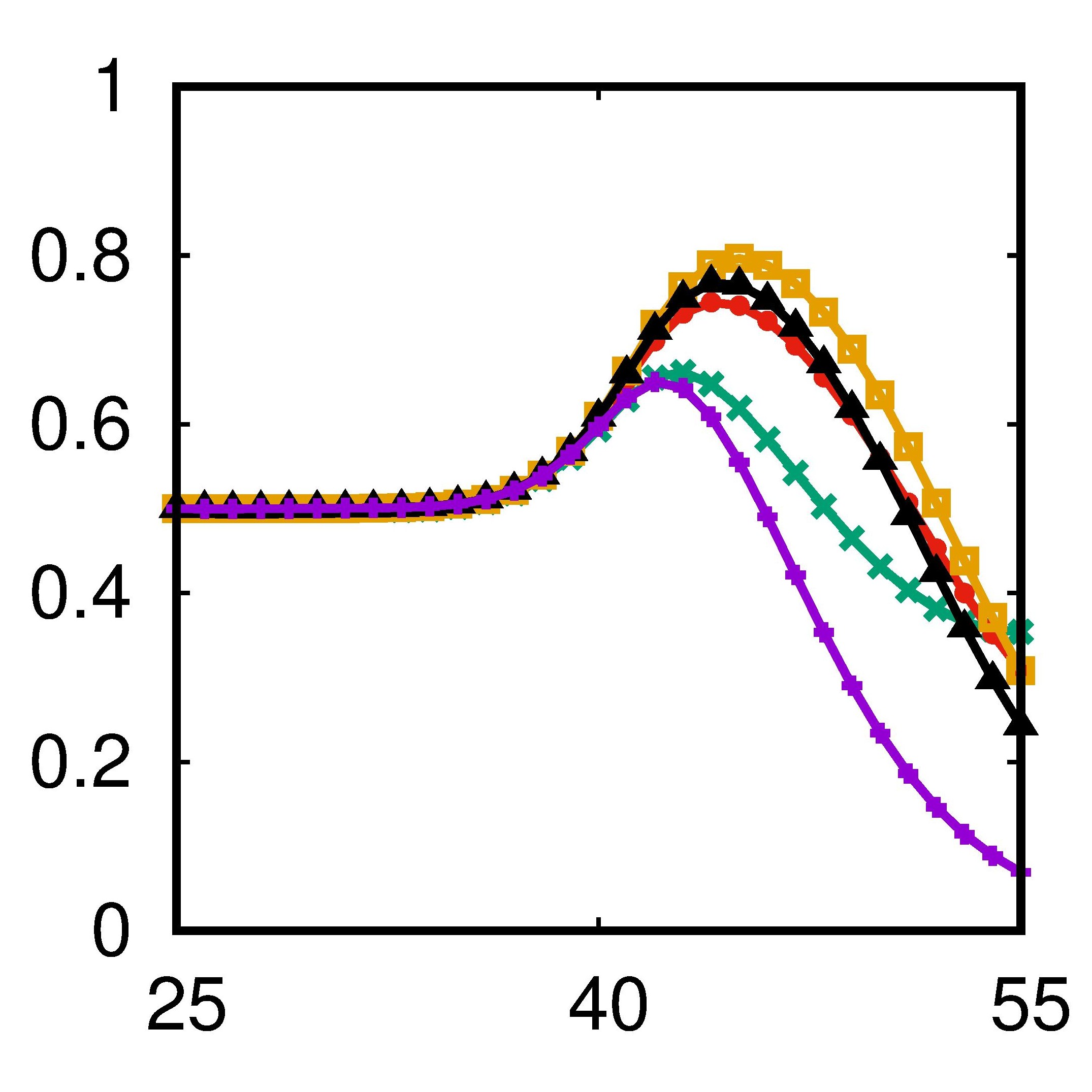}};
    \node at (2.,-0.3) {};
    \draw[->,line width=0.3mm](2.1,2.8) -- (2.1,3.2);
    \draw[->,line width=0.3mm](2.3,3.2) -- (2.3,2.8);
 \end{tikzpicture}%
 \caption{}
\end{subfigure}%
  \caption{Top row: Evolution of radial composition line profiles drawn from matrix 
  side into the particle.  Bottom row: Magnified views of the red outlined regions
  in the corresponding figures above. $t_1<t_2<t_3<t_4<t_5$. (a) $d=160,\;\chi = 0.75$,
  (b) $d=160,\;\chi=0.5$ and (c) $d=160,\;\chi=0.25$, (d) $d=240,\;\chi=0.5$.}
   \label{fig:line_cut}
\end{figure*}
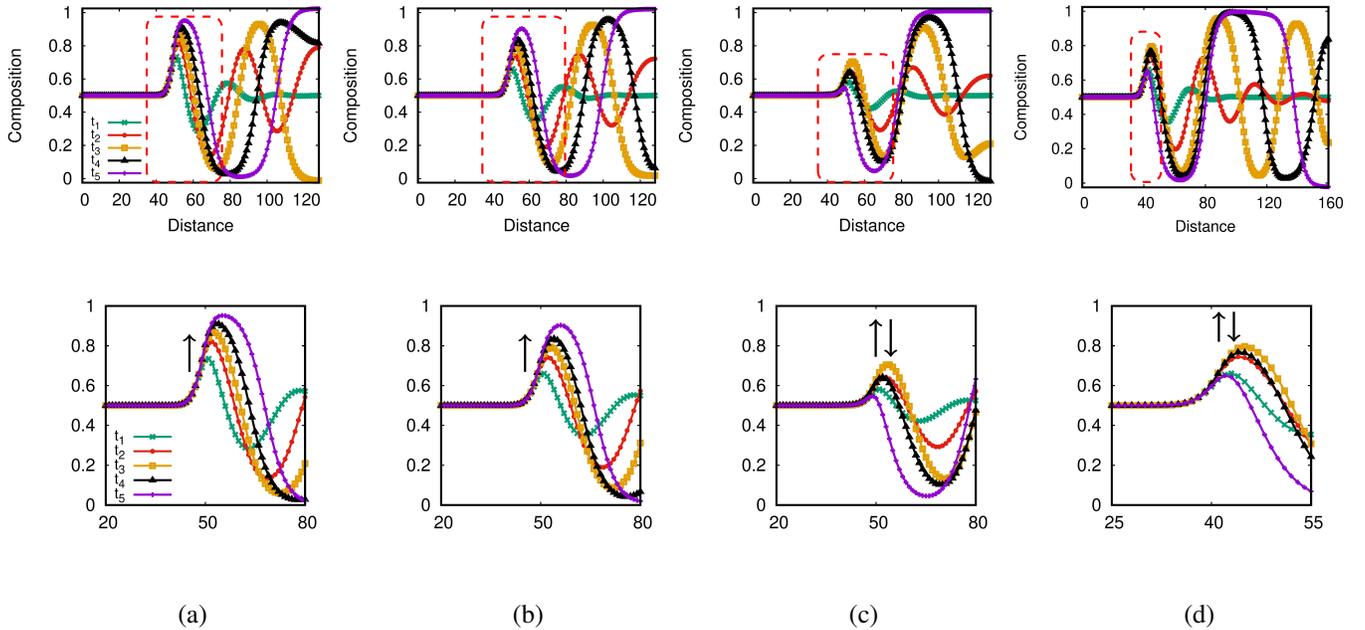

We present evolution of the radial composition profiles for two such
systems with $d=160$ in Fig.~\ref{fig:line_cut}(a) and (b) for
$\chi=0.75$ and 0.5, respectively. The upward pointing arrow in these
figures represent a monotonically increasing solute enrichment at the
matrix-particle interface. In contrast, for systems with a lower
$\chi$ and/or at larger sizes, the initial solute enrichment at the
surface is followed by a decay, eventually settling to an equilibrium 
segregation value $C_{\textrm{mx}}$. Figs.~\ref{fig:line_cut}c-d show
this non-monotonic evolution (indicated by the up-and-down arrows) of
surface composition for two different combinations of $d$ and $\chi$.

\subsubsection*{Driving force for evolution of intermediate ring structures.}
Ring structures develop within the circular BNPs as a result of surface-directed
compositional modulation during SD.  To understand the evolution and coarsening
of intermediate ring structures, we need to analyze the variation of average ring
size with time and correlate it with the driving force for diffusion, which in 
this case is the radial chemical potential gradient $-{\partial \mu}/{\partial r}$.
For statistical quantification of phase-separated microstructures, the first zero
of the pair correlation function is often taken as a measure for the average domain
size or characteristic length scale~\cite{Zhu1999}. To determine this, we first
compute the structure function $S(\mathbf{k},t)$ of the composition field 
$c(\mathbf{r})$ by taking the Fourier transform of its spatial correlation:

\begin{equation}
    S(\mathbf{k},t) =
    \frac{1}{N}\Big\langle \sum\limits_{r}\sum\limits_{r^{\prime}} 
    e^{-\mathrm{i}\mathbf{k}\cdot\mathbf{r}} \left[c(\mathbf{r}+\mathbf{r}^{\prime}, t)
    c(\mathbf{r}^{\prime}, t) - \langle c \rangle^2\right] \Big\rangle.
\end{equation}

\begin{figure}[htbp]
 \centering
    \includegraphics[scale=0.2]{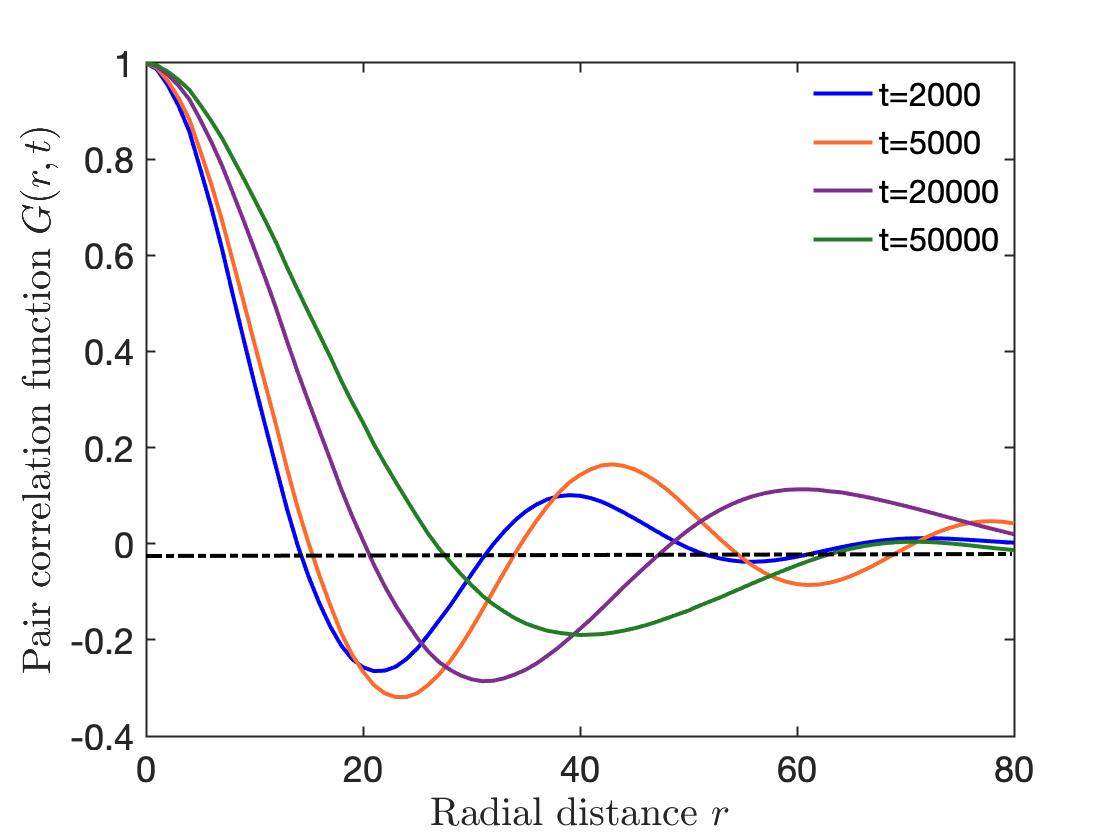}%
 \caption{Evolution of circularly-averaged pair correlation function
 $G(r)$ with time for $\chi=0.5$}
 \label{fig:PairCorr}
\end{figure}

Here $\langle\ \rangle$ denotes the average over the entire domain, $N$ is the
total number of grid points in the domain, and $\mathrm{i}=\sqrt{-1}$. The
inverse Fourier transform of $S(\mathbf{k},t)$ is the pair correlation 
function $G(\mathbf{r},t)$:

\begin{equation}
    G(\mathbf{r},t) = \sum\limits_{k} e^{\mathrm{i}\mathbf{k}\cdot \mathbf{r}}S(\mathbf{k},t)
\end{equation}
For the isotropic systems considered here, we use circular averaging of the pair
correlation function and determine the location of its first zero to obtain the
average ring size $\bar{r}$ at a given time. Evolution of circularly averaged
$G(r,t)$ for $\chi=0.5$ is shown in Fig.~\ref{fig:PairCorr}. Except at the very 
early stages, $\bar{r}$ increases continuously. indicating coarsening of rings. 
 
\begin{figure}[htbp]
 \centering
    \includegraphics[scale=0.05]{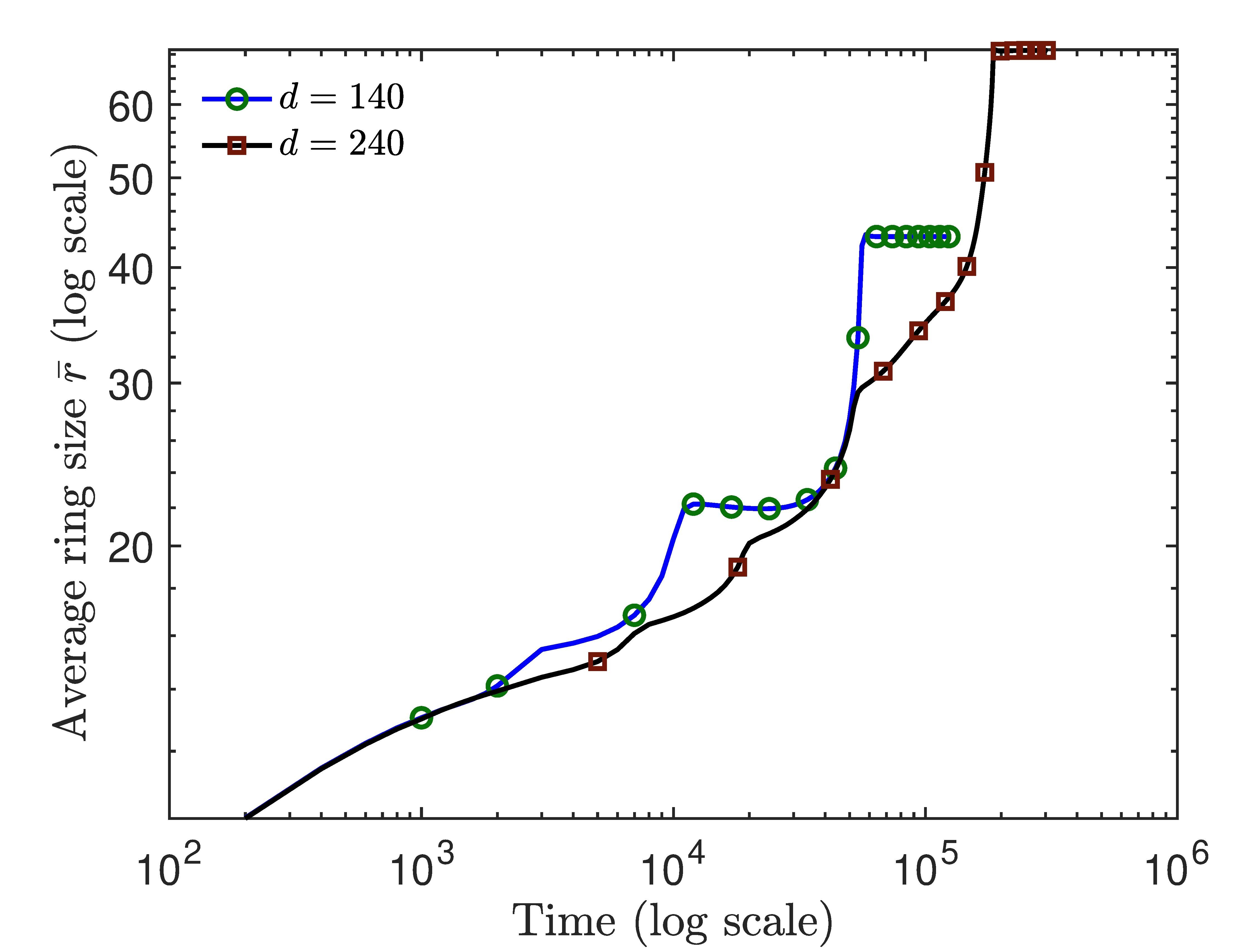}
    \caption{Evolution of average ring size with time for $\chi=0.5$. To improve clarity,
    every third data point is marked with a circle in (a). The lines are drawn through
    the data points as a guide to the eye.}%
    \label{fig:coarsening}
    \end{figure}

Variation of $\bar{r}$ with time, plotted in Fig.~\ref{fig:coarsening} in a log-
log scale, reveals a step-like coarsening behavior of these rings: the plateaus
correspond to a slow kinetic regime where adjacent rings have similar widths, 
whereas sharp changes near the end of plateaus mark an extremely fast regime in 
which a shrinking intermediate ring disappears due to a large gradient in
concentration. The plot of $\bar{r}$ vs. $t$ on a logarithmic scale does not 
conform to the conventional power-law relation~\cite{Martin1999} of the type
$\bar{r}^n \propto t$. Moreover, the plot indicates sequential, step-like 
coarsening of rings. We note that long range diffusion of solute is restricted
here due to the formation of alternating solute-rich and solute-poor rings.

\begin{figure}[htbp]
 \centering
 \includegraphics[scale=0.14]{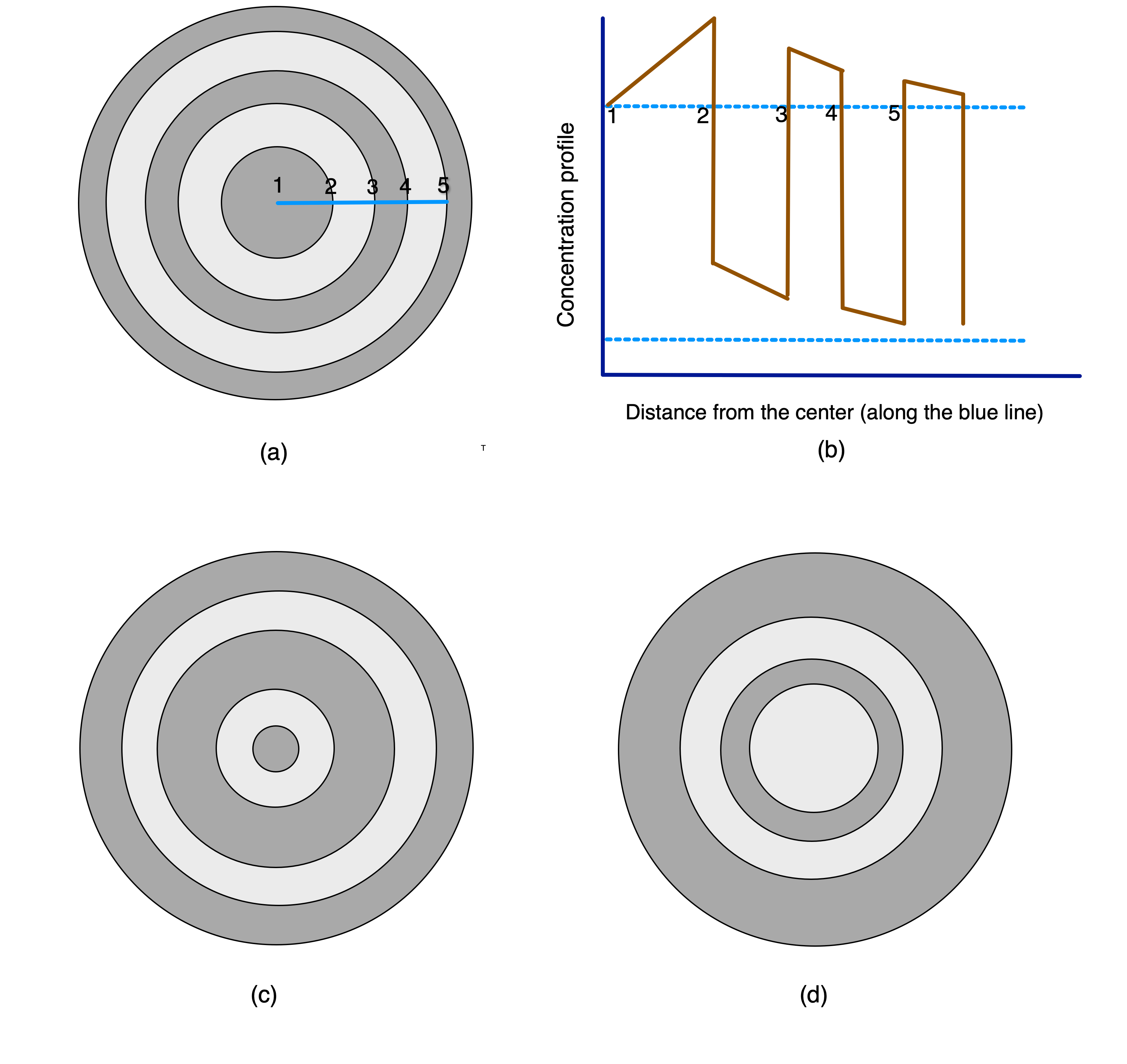}%
 \caption{Schematic illustration of the step-like process 
 of coarsening of rings. (a) Initial configuration with five compositionally 
 modulated rings, (b) radial composition line profile, (c,d) successive
 disappearance and coarsening of rings.}
\label{scheme}
\end{figure}

The step-like nature of coarsening dynamics within a particle is illustrated
schematically  in Fig.~\ref{scheme} where a circular particle initially contains
a set of compositionally modulated rings. For this initial configuration, 
composition variation is shown schematically along a radial line that intersects
the rings at points marked as `2, 3, 4, 5'; `1' denotes the center of the
particle. Note that the composition at these points deviate from their respective
equilibrium values indicated by the blue dashed lines due to Gibbs-Thomson 
effect~\cite{Porter1992}. Since this deviation $\Delta c \propto\sigma K$ increases with increasing
interface curvature $K$ (i.e., with decreasing radius of curvature), 
it gradually reduces with increasing radial 
distance as we move outward. Thus, it sets up a diffusion flux within each ring as
shown schematically in the composition line profile (Fig.~\ref{scheme}b). The
consequent local diffusion process leads to successive coarsening and disappearance
events wherein the inner rings dissolve and outer ones coarsen in a step-like 
fashion, depicted schematically in Figs.~\ref{scheme}(c) and (d). When the particles are reasonably large in size, $K$ dictates the direction of dissolution and coarsening from the center to the surface.  

However, when the particle size is small, for any non-zero contact angle, 
the outer ring can also disappear if it has higher surface energy. For example, in a particle with size $d=140$ and $\chi=0.25$, the outermost solute-poor $\beta_1$ ring dissolves while the solute-poor $\beta_1$ core grows in size as shown in the animation in ESI. In this case, the surface energy $\sigma$ in the Gibbs-Thomson relation dominates over interfacial curvature $K$, thus reversing the direction of coarsening.  
 
\begin{figure}[htbp]
    \centering
    \includegraphics[scale=0.2]{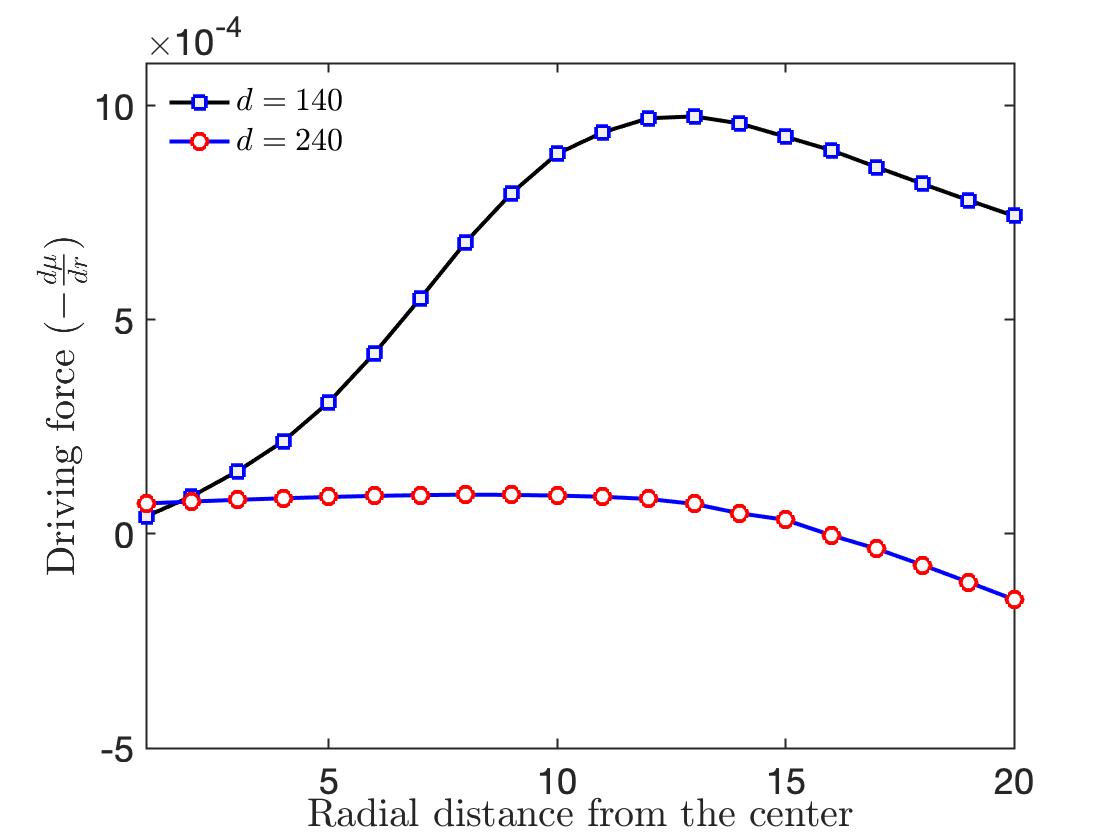}%
    \caption{Variation of driving force for diffusion as function of radial 
    distance within a single ring}%
    \label{fig:drivingforce}%
\end{figure}

Fig.~\ref{fig:drivingforce} compares the radial variation of the driving force for
solute transport $-{\partial \mu}/{\partial r}$ within a ring for two particle sizes
$d=140, 240$ when $\chi=0.5$. It is evident from the figure that driving force for
diffusion reduces considerably with increasing particle size. Hence, larger particles
are more likely to exhibit metastable `kinetically trapped' configurations.


Note that with increasing size of the particle, bulk spinodal may also start
in the interior and proceed towards the surface interfering with the radial
concentration waves propagating from the surface. Here we have restricted the
study to smaller sizes where bulk spinodal is negligible. This allows us to
study the effect of surface-induced concentration modulations on the morphological
evolution to CS/Janus structures without interference from bulk SD. When the
decomposition is surface-agnostic (\emph{i.e.}, when $\sigma_1=\sigma_2$), the
equilibrium Janus does not go through an intermediate CS morphology; instead, it
forms through the coarsening of domains produced by bulk SD. For all other $\chi$ 
values, the time taken to reach the equilibrium morphology decreases with increasing $\chi$.   
    
When the contact angle is non-zero (here $0<=\chi<0.75$), the equilibrium 
configuration (based on Young's equation) is Janus. However, we observe metastable
CS structures (CS-2@1 or CS-1@2) depending on the particle size and and the contact 
angle. In all these cases, the sequence of transformation is as follows: first a 
solute-rich layer forms at the matrix-particle interface leading to the formation of
a solute-poor layer adjacent to it. Subsequently, SDSD leads to concentric solute-rich
and solute-poor rings. Later, since the innermost ring has the highest curvature, 
disappearance of the rings initiate from the centre of the particle. While the 
early-stage diffusional waves propagate from surface to centre of the particle, the
late-stage dynamics propagates in the opposite direction. Evidently, the diffusional
redistribution of outermost solute-lean layer during coarsening becomes more sluggish
with the increase in particle size. As a result, CS morphology becomes relatively more
stable with increase in particle size -- CS 2@1, which requires more solute
redistribution across the outermost solute-lean layer, is attained at intermediate
sizes ($d < 200$), whereas CS 1@2, requiring less redistribution, becomes more
favourable at larger sizes. 

\subsubsection*{Effect of particle shape on morphological evolution} 

\begin{figure}[htbp]
\captionsetup[subfigure]{justification=centering}
\centering%
\begin{subfigure}{0.075\textwidth}%
\centering%
    \includegraphics[scale=0.02]{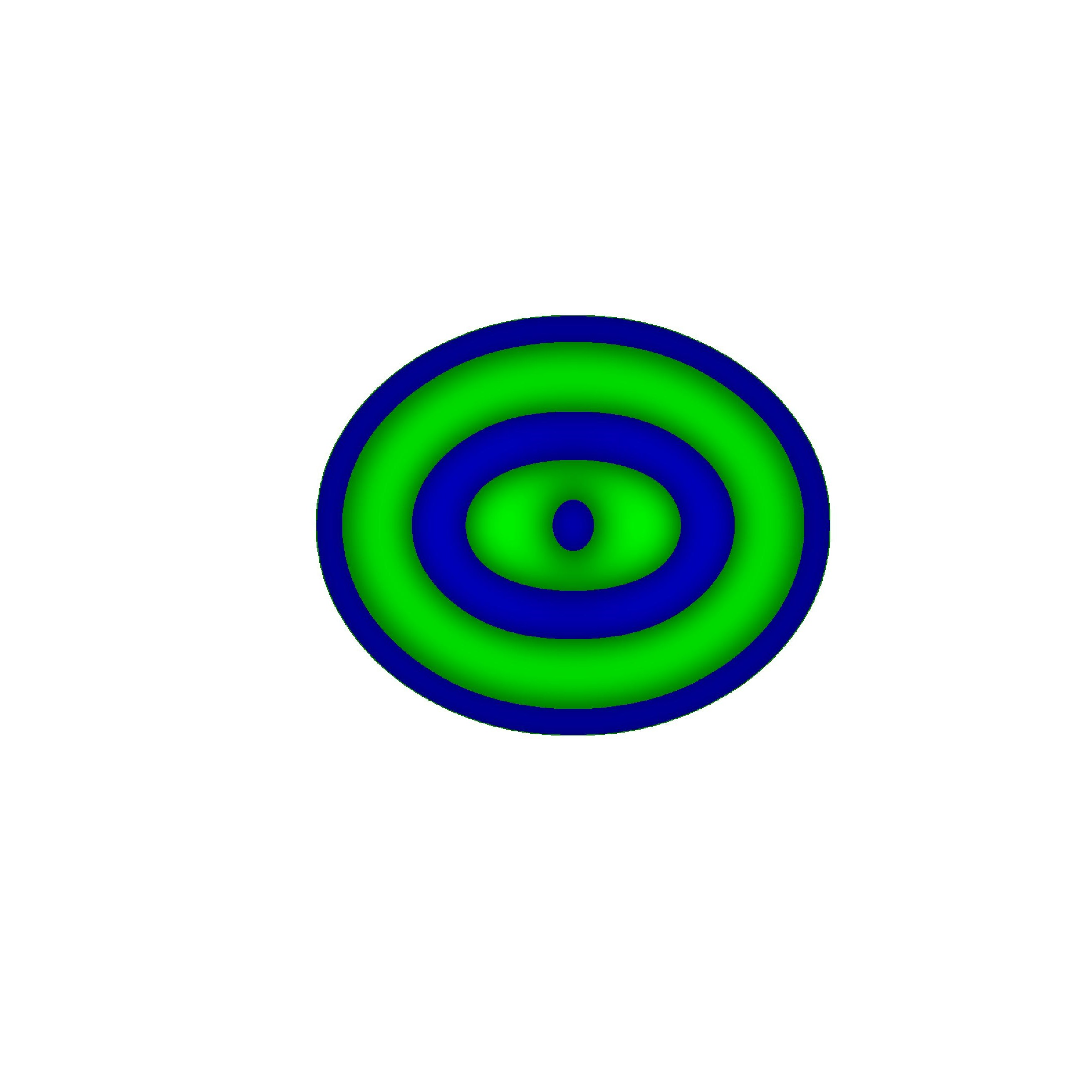}\hfil%
    \caption{}%
    \label{Ell_chi025_01}%
\end{subfigure}%
\begin{subfigure}{0.075\textwidth}
\centering%
    \includegraphics[scale=0.02]{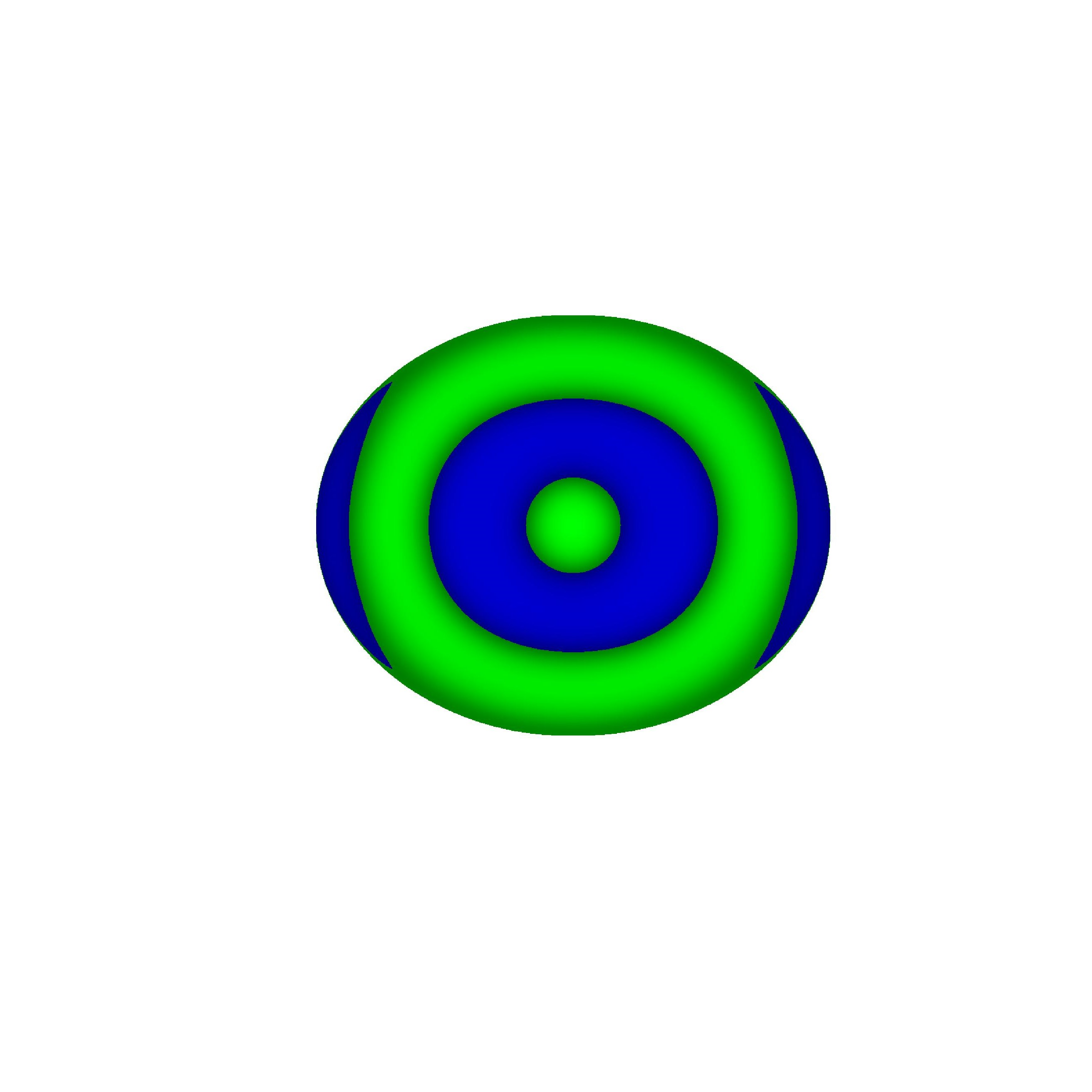}\hfil%
    \caption{}%
    \label{Ell_chi025_1}%
\end{subfigure}%
\begin{subfigure}{0.075\textwidth}%
\centering%
    \includegraphics[scale=0.02]{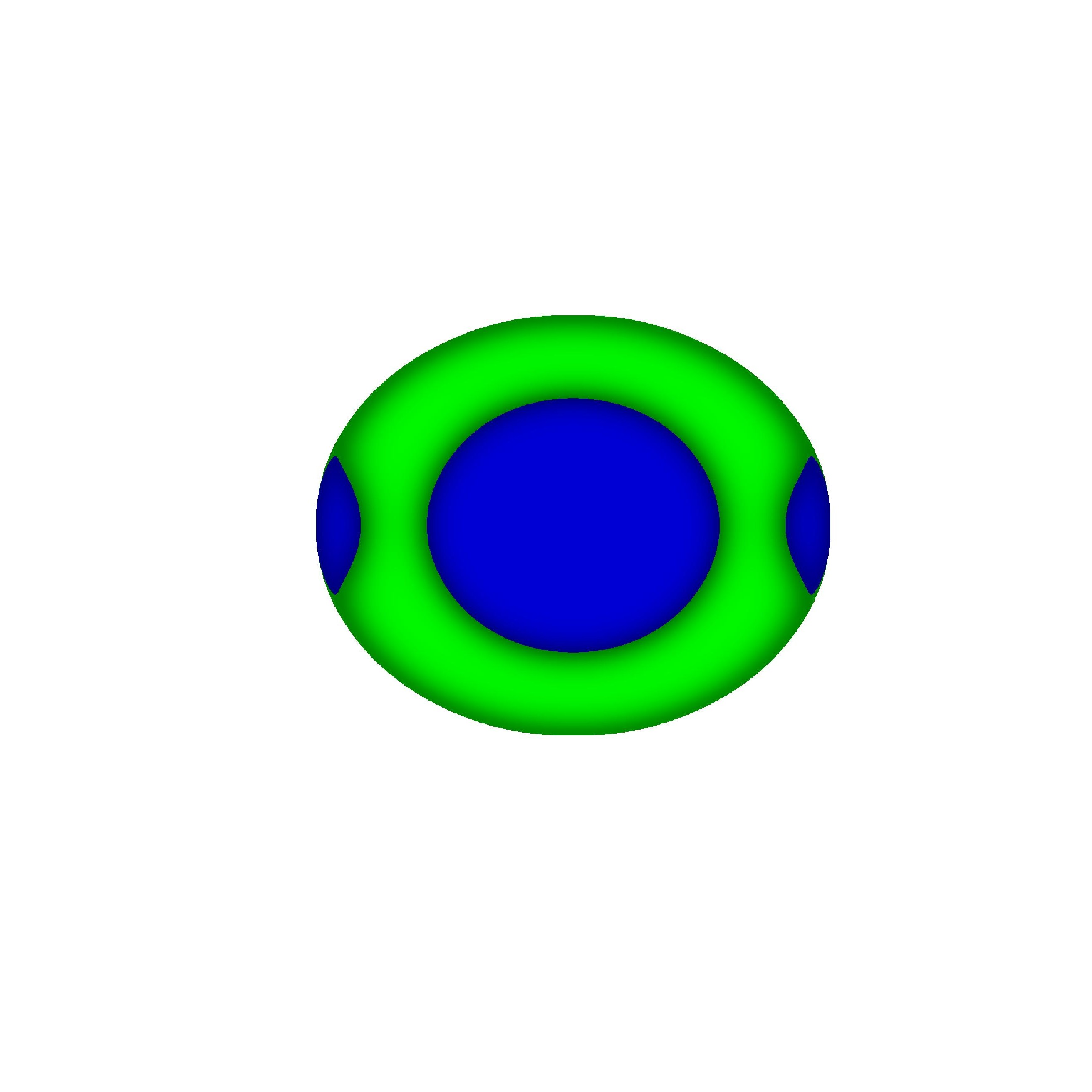}\hfil%
    \caption{}%
    \label{Ell_chi025_2}%
\end{subfigure}%
\begin{subfigure}{0.075\textwidth}%
\centering%
    \includegraphics[scale=0.02]{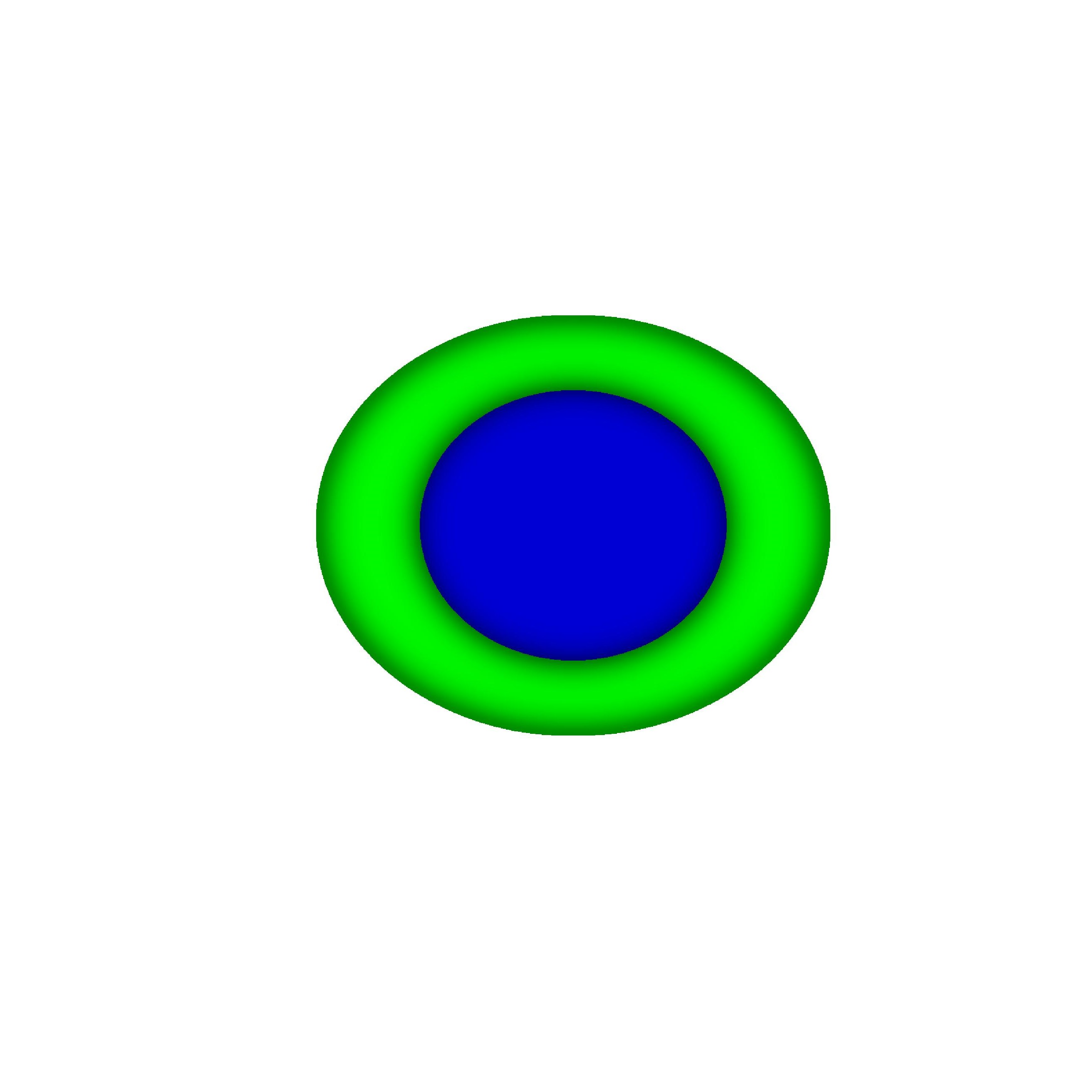}\hfil%
    \caption{}%
    \label{Ell_chi025_4}%
\end{subfigure}%
\begin{subfigure}{0.075\textwidth}%
\centering%
    \includegraphics[scale=0.02]{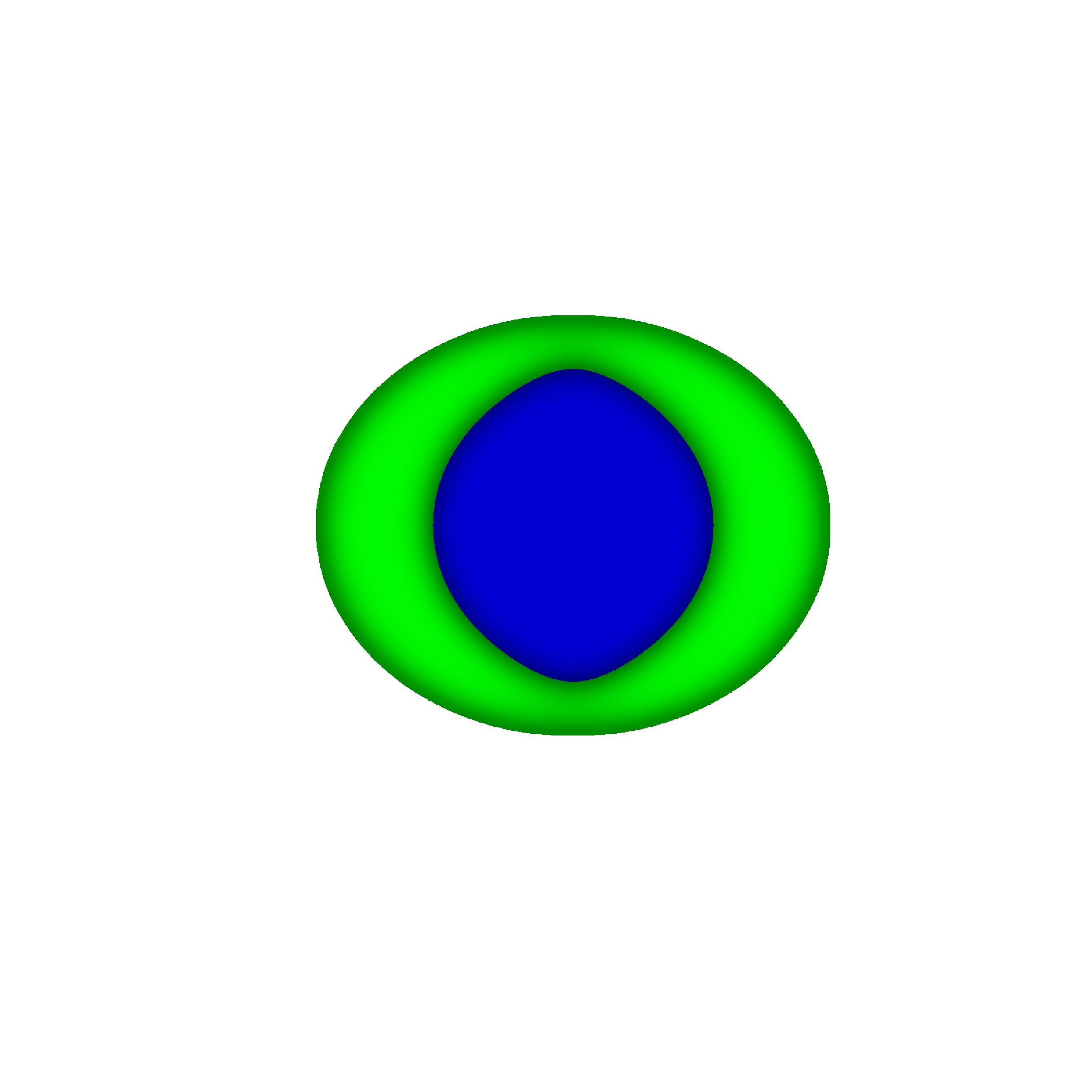}\hfil%
    \caption{}%
    \label{Ell_chi025_5}%
\end{subfigure}%
\begin{subfigure}{0.075\textwidth}%
\centering%
    \includegraphics[scale=0.02]{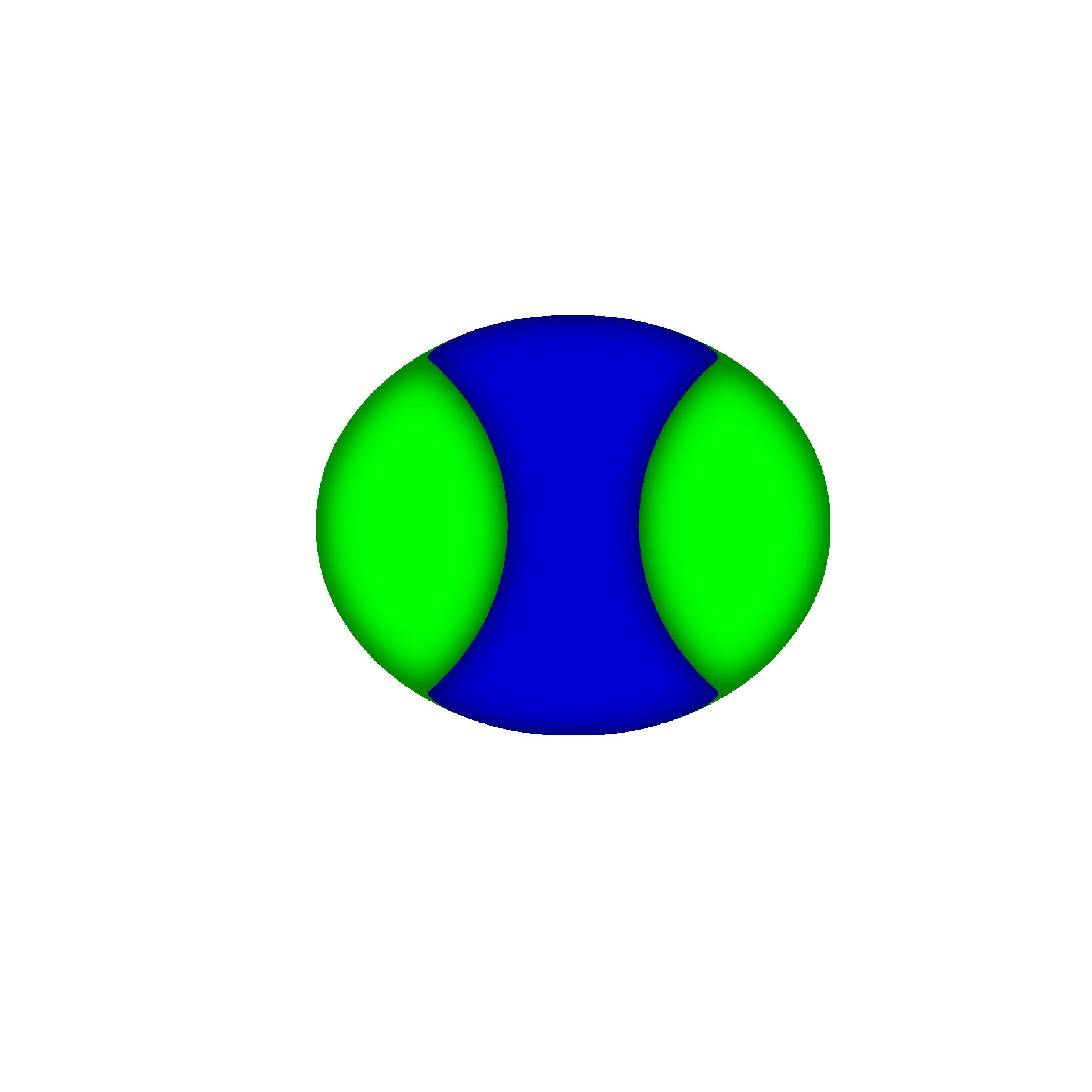}\hfil%
    \caption{}%
    \label{Ell_chi025_6}%
\end{subfigure}%

\begin{subfigure}{0.075\textwidth}
\centering%
     \includegraphics[scale=0.02]{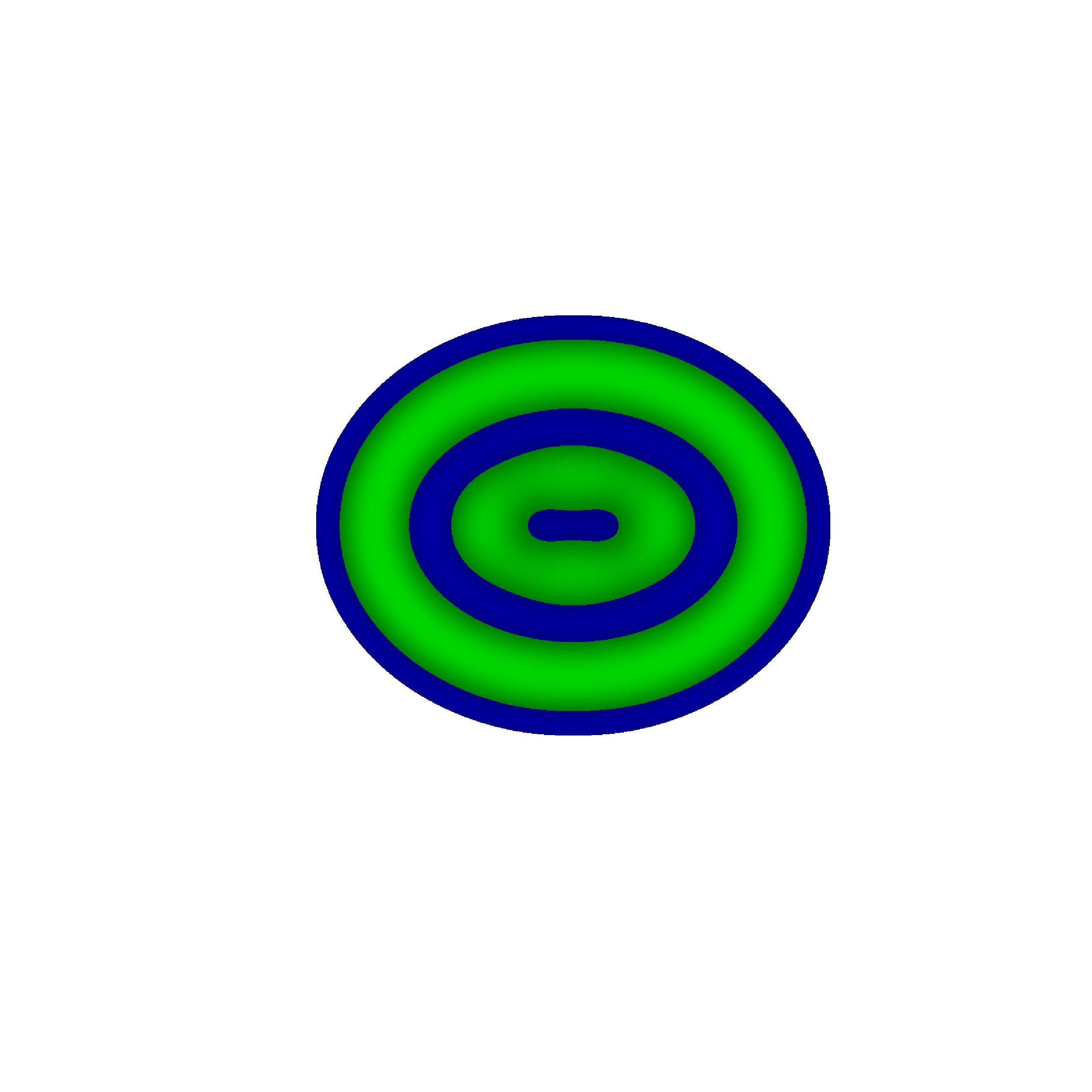}\hfil%
    \caption{}%
    \label{Ell_chi05_01}%
\end{subfigure}%
\begin{subfigure}{0.075\textwidth}%
\centering%
     \includegraphics[scale=0.02]{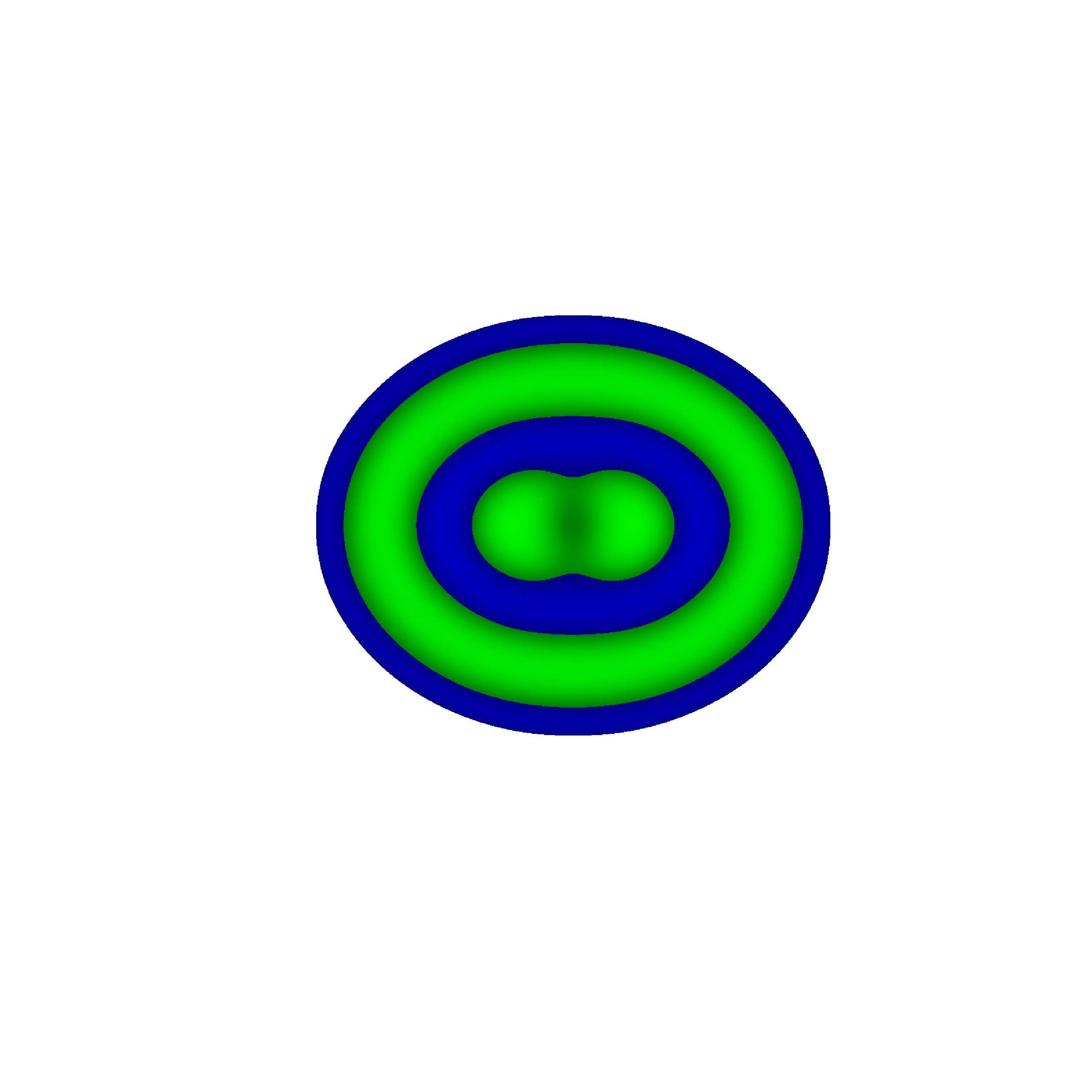}\hfil%
    \caption{}%
    \label{Ell_chi05_1}%
\end{subfigure}%
\begin{subfigure}{0.075\textwidth}
\centering%
    \includegraphics[scale=0.02]{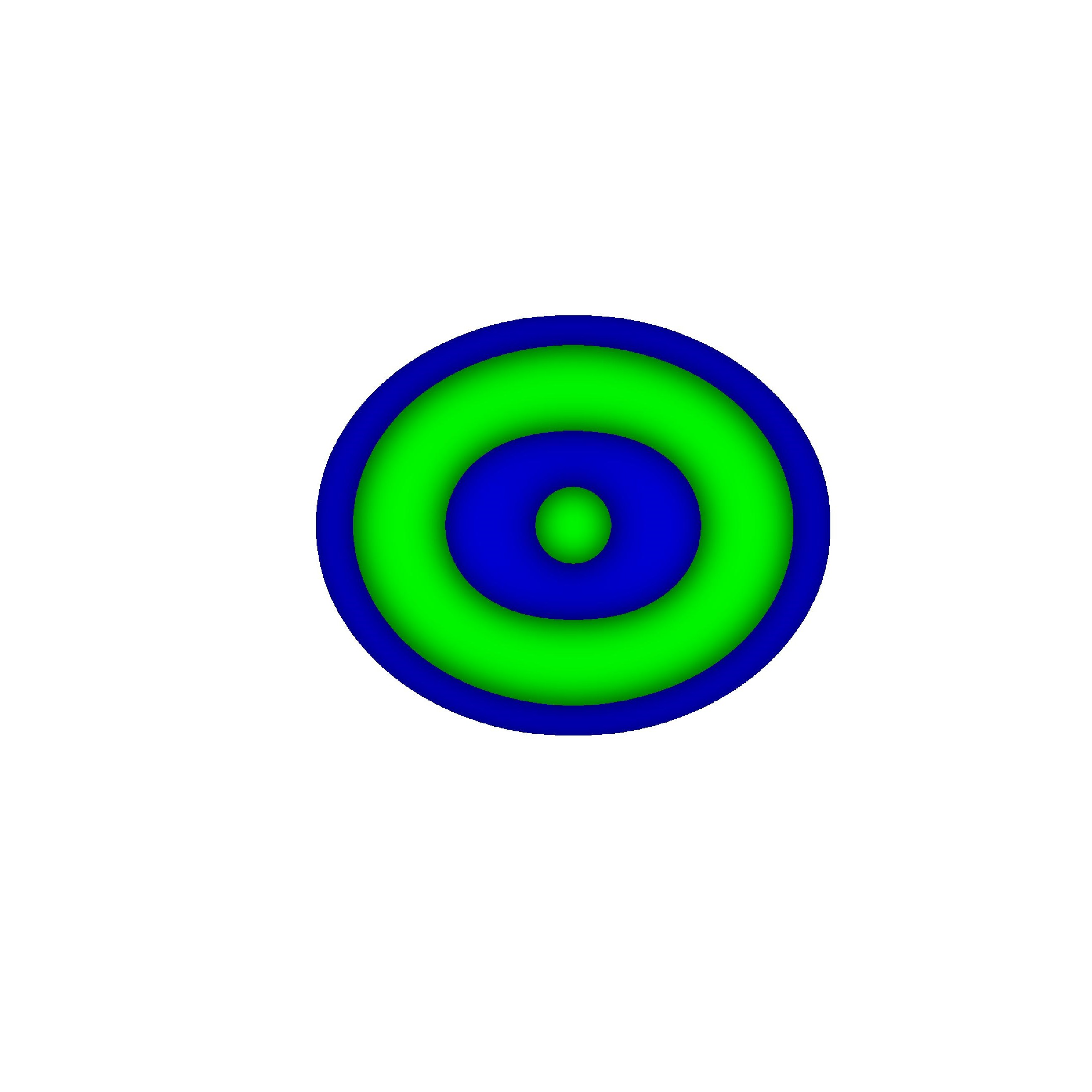}\hfil%
    \caption{}%
    \label{Ell_chi05_2}%
\end{subfigure}%
\begin{subfigure}{0.075\textwidth}
\centering%
    \includegraphics[scale=0.02]{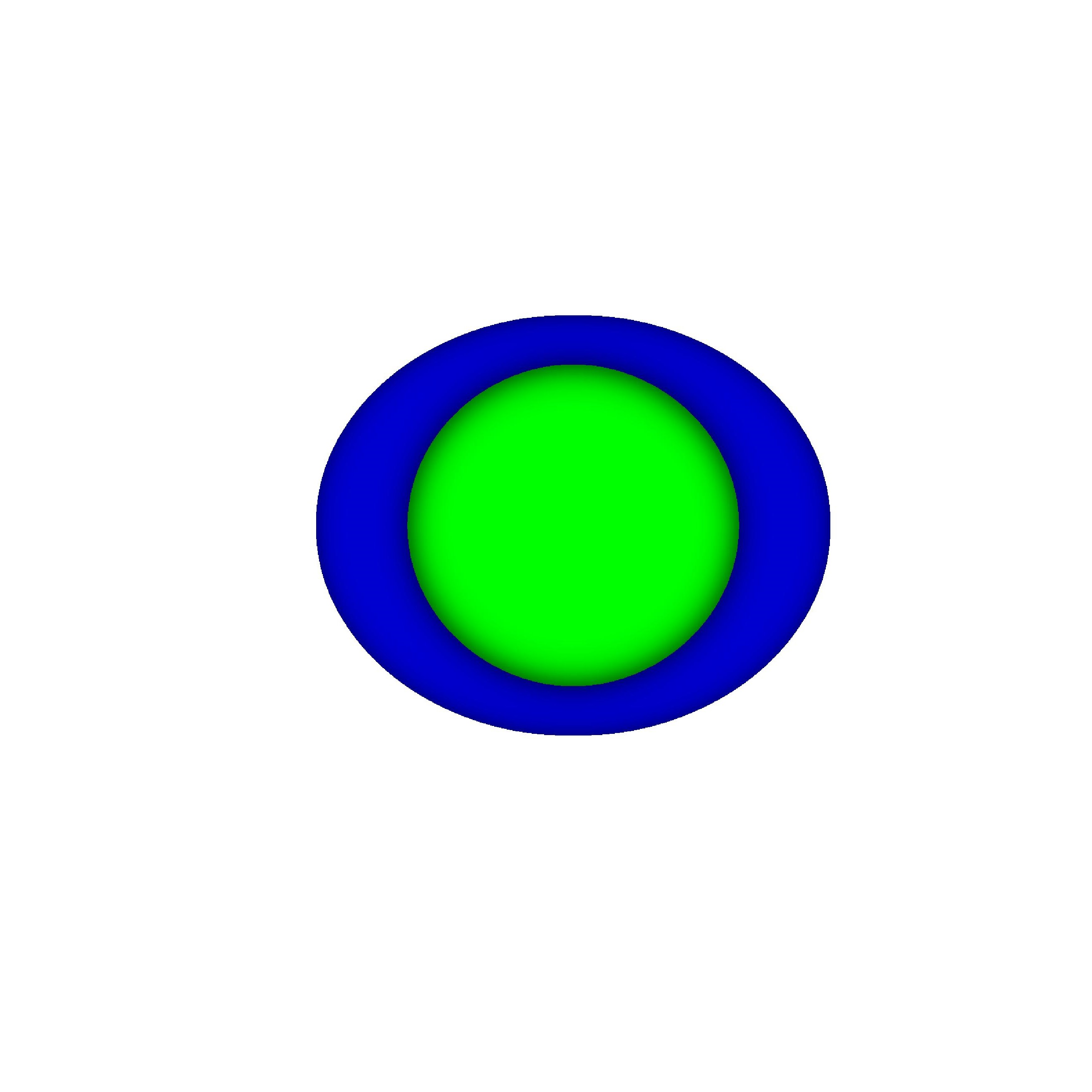}\hfil%
    \caption{}%
    \label{Ell_chi05_3}%
\end{subfigure}%
\begin{subfigure}{0.075\textwidth}%
\centering%
    \includegraphics[scale=0.02]{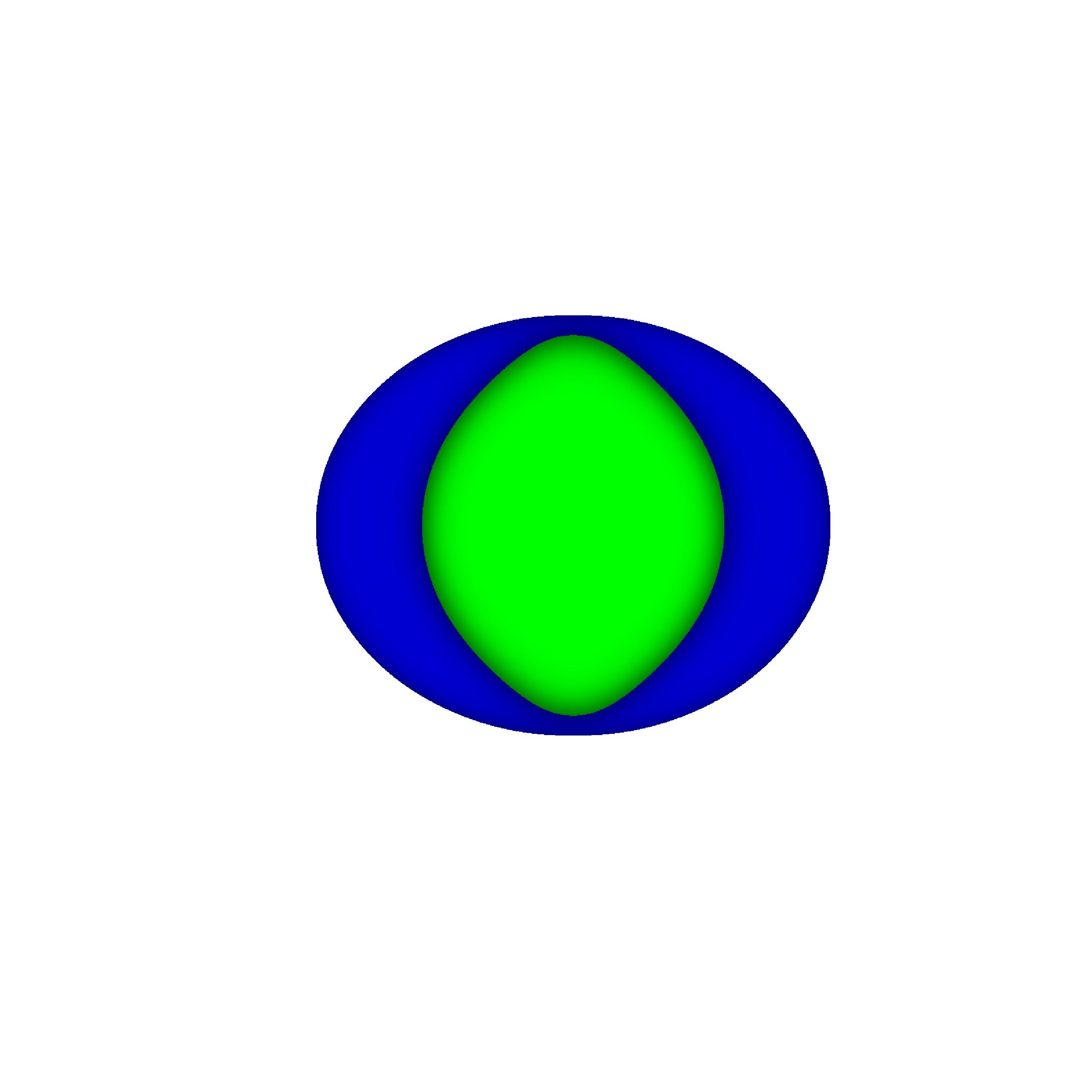}\hfil%
    \caption{}%
    \label{Ell_chi05_4}%
\end{subfigure}%
\begin{subfigure}{0.075\textwidth}
\centering%
    \includegraphics[scale=0.02]{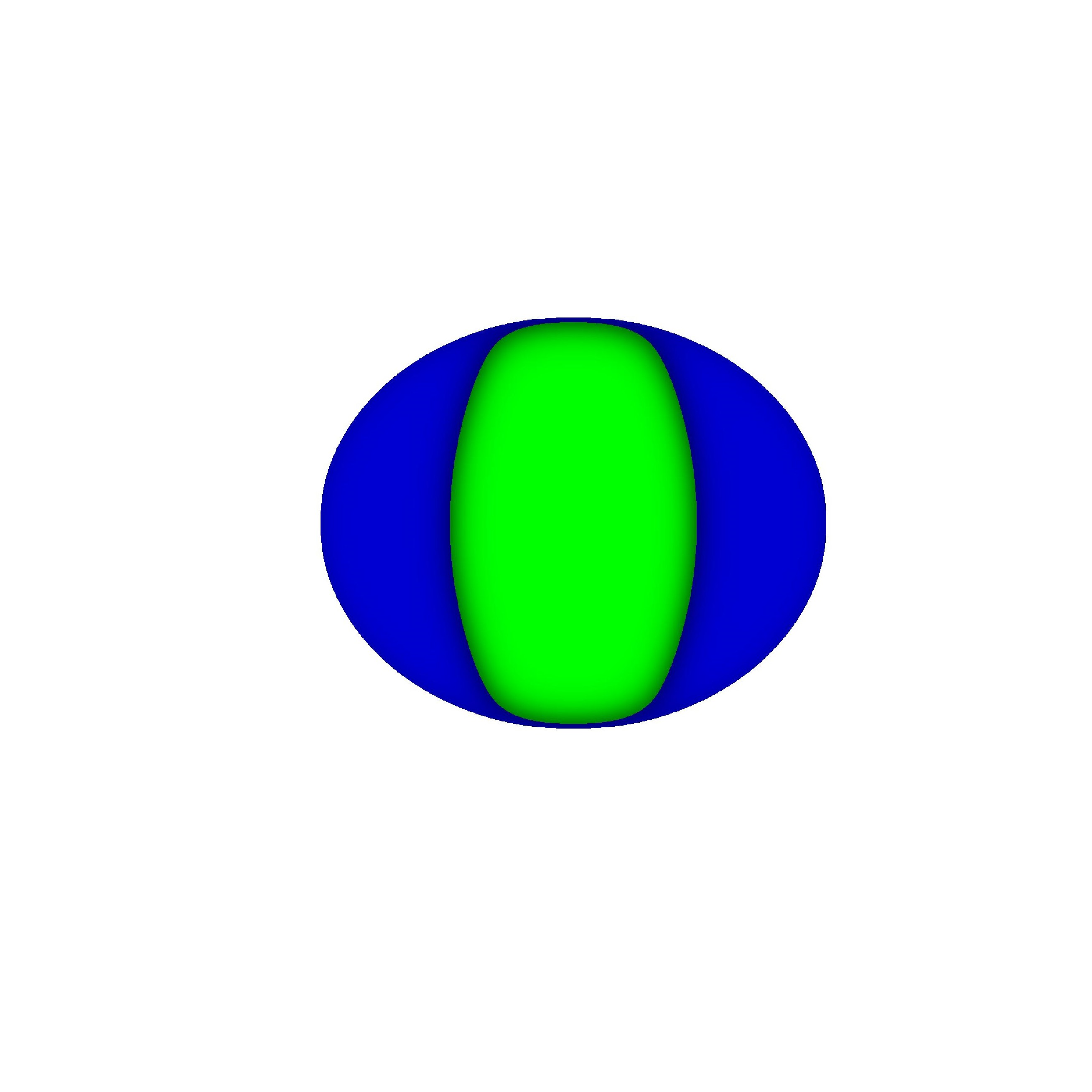}\hfil%
    \caption{}%
    \label{Ell_chi05_6}%
\end{subfigure}%

\begin{subfigure}{0.075\textwidth}%
\centering%
    \includegraphics[scale=0.02]{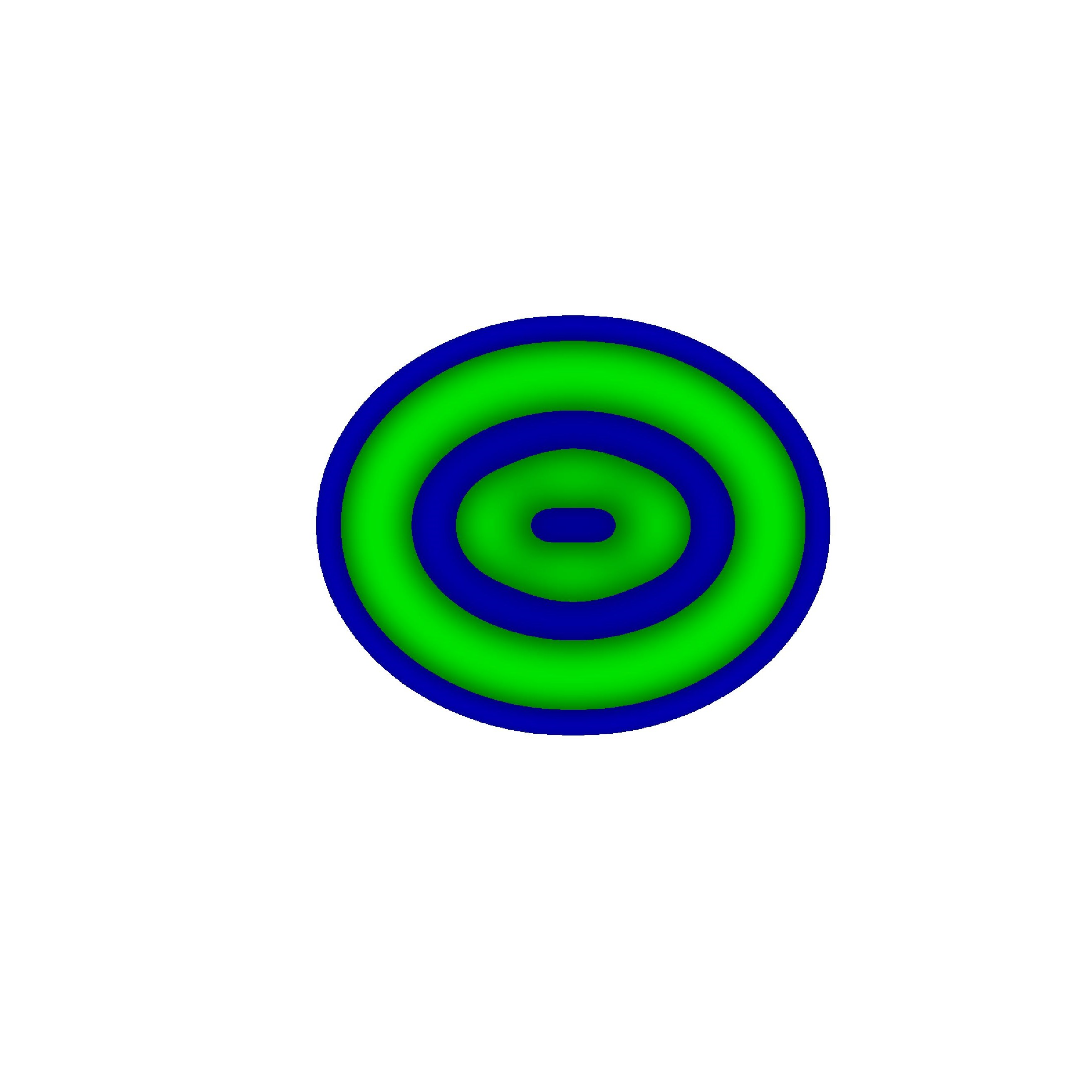}\hfil%
    \caption{}%
    \label{Ell_chi075_01}%
\end{subfigure}%
\begin{subfigure}{0.075\textwidth}%
\centering%
    \includegraphics[scale=0.02]{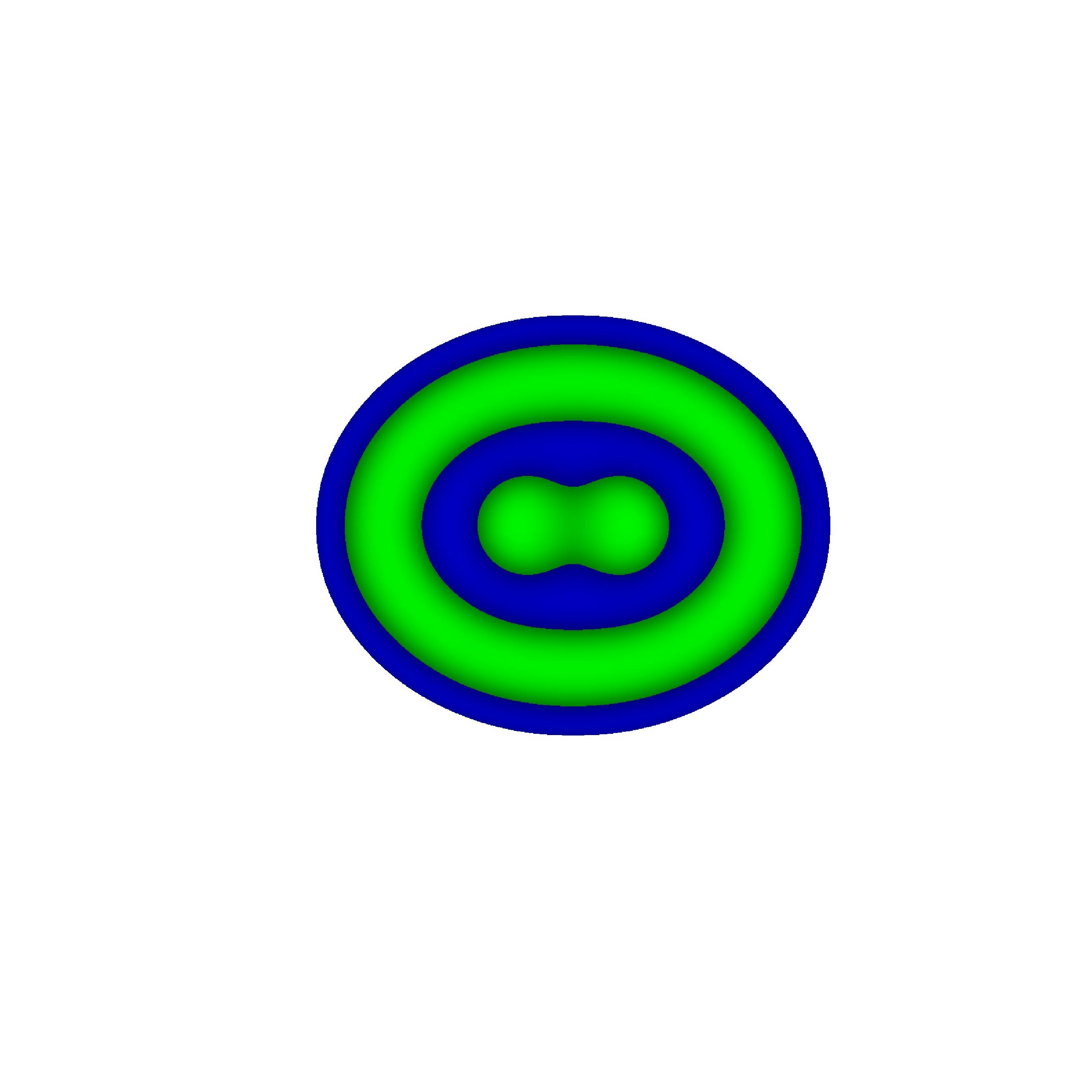}\hfil%
    \caption{}%
    \label{Ell_chi075_1}%
\end{subfigure}%
\begin{subfigure}{0.075\textwidth}%
\centering%
    \includegraphics[scale=0.02]{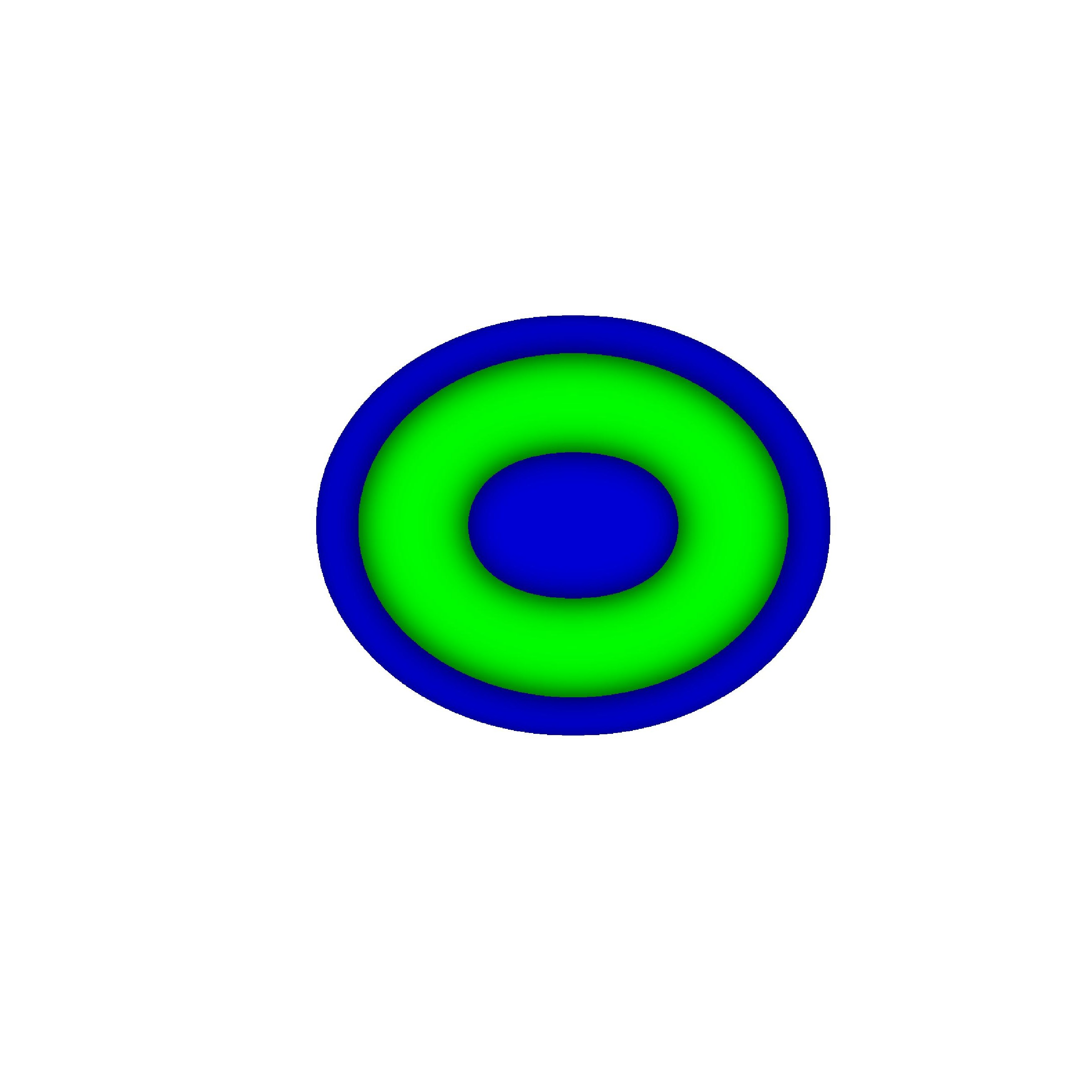}\hfil%
    \caption{}%
    \label{Ell_chi075_2}%
\end{subfigure}%
\begin{subfigure}{0.075\textwidth}%
\centering%
    \includegraphics[scale=0.02]{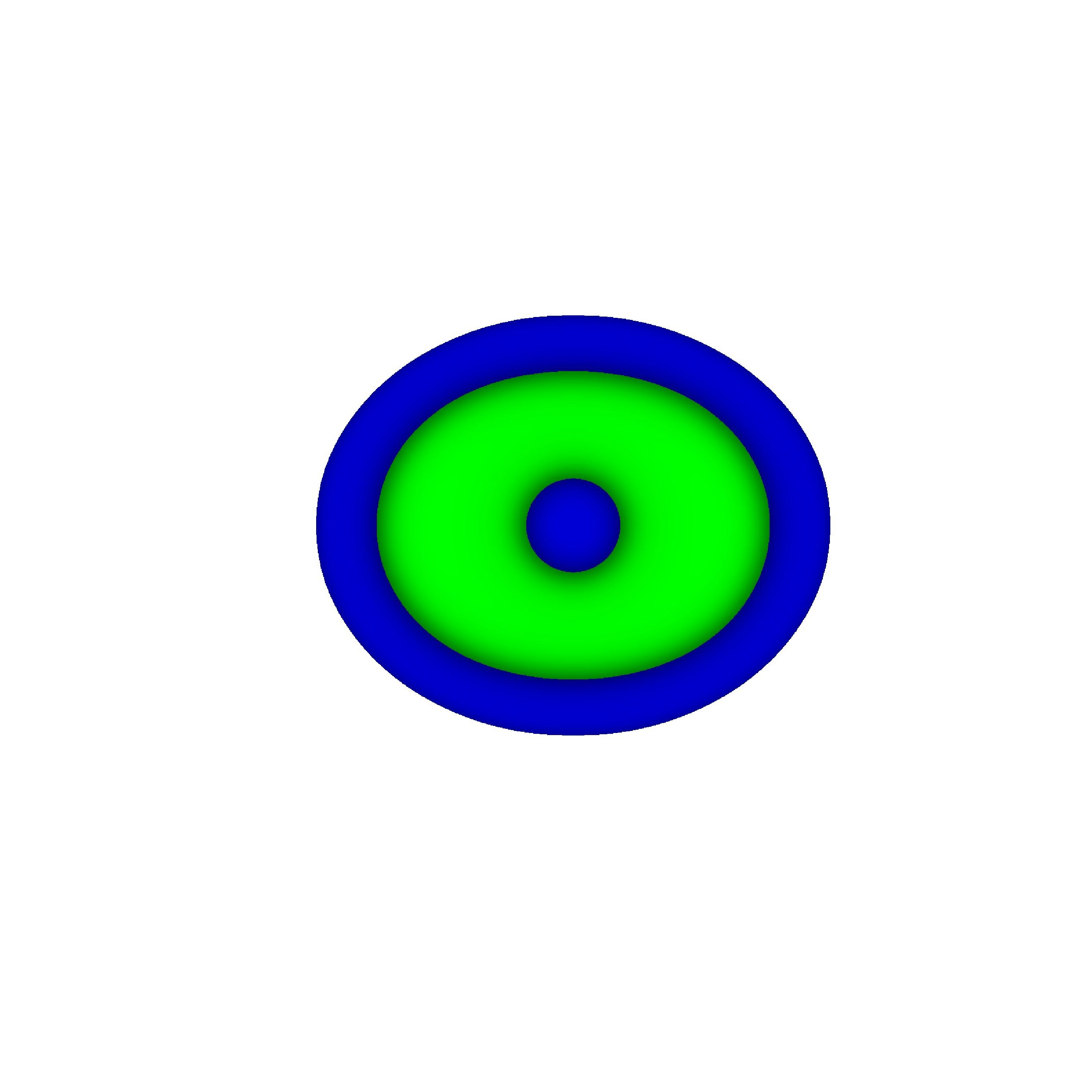}\hfil%
    \caption{}%
    \label{Ell_chi075_3}%
\end{subfigure}%
\begin{subfigure}{0.075\textwidth}%
\centering%
    \includegraphics[scale=0.02]{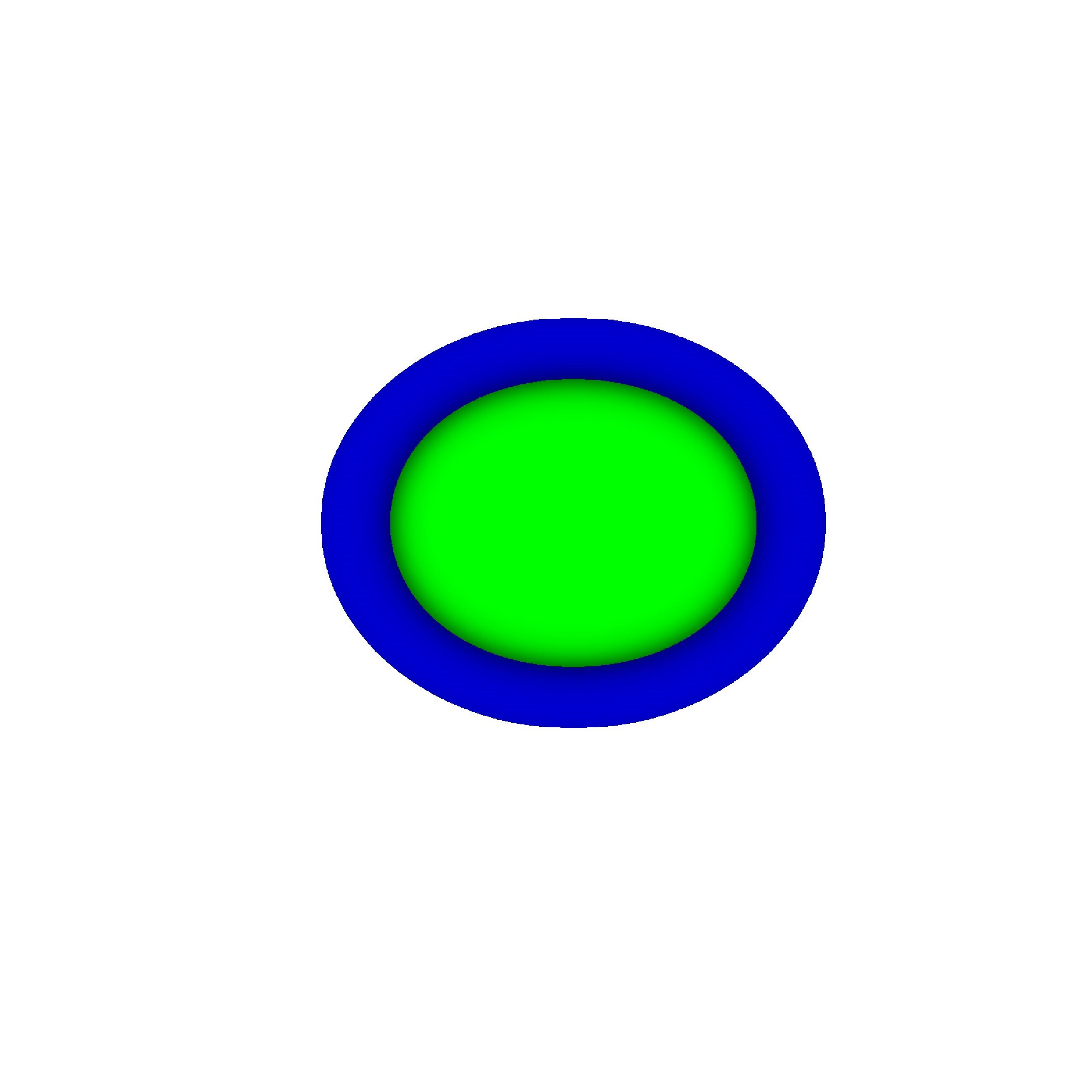}\hfil%
    \caption{}%
    \label{Ell_chi075_4}%
\end{subfigure}%
\begin{subfigure}{0.075\textwidth}%
\centering%
    \includegraphics[scale=0.02]{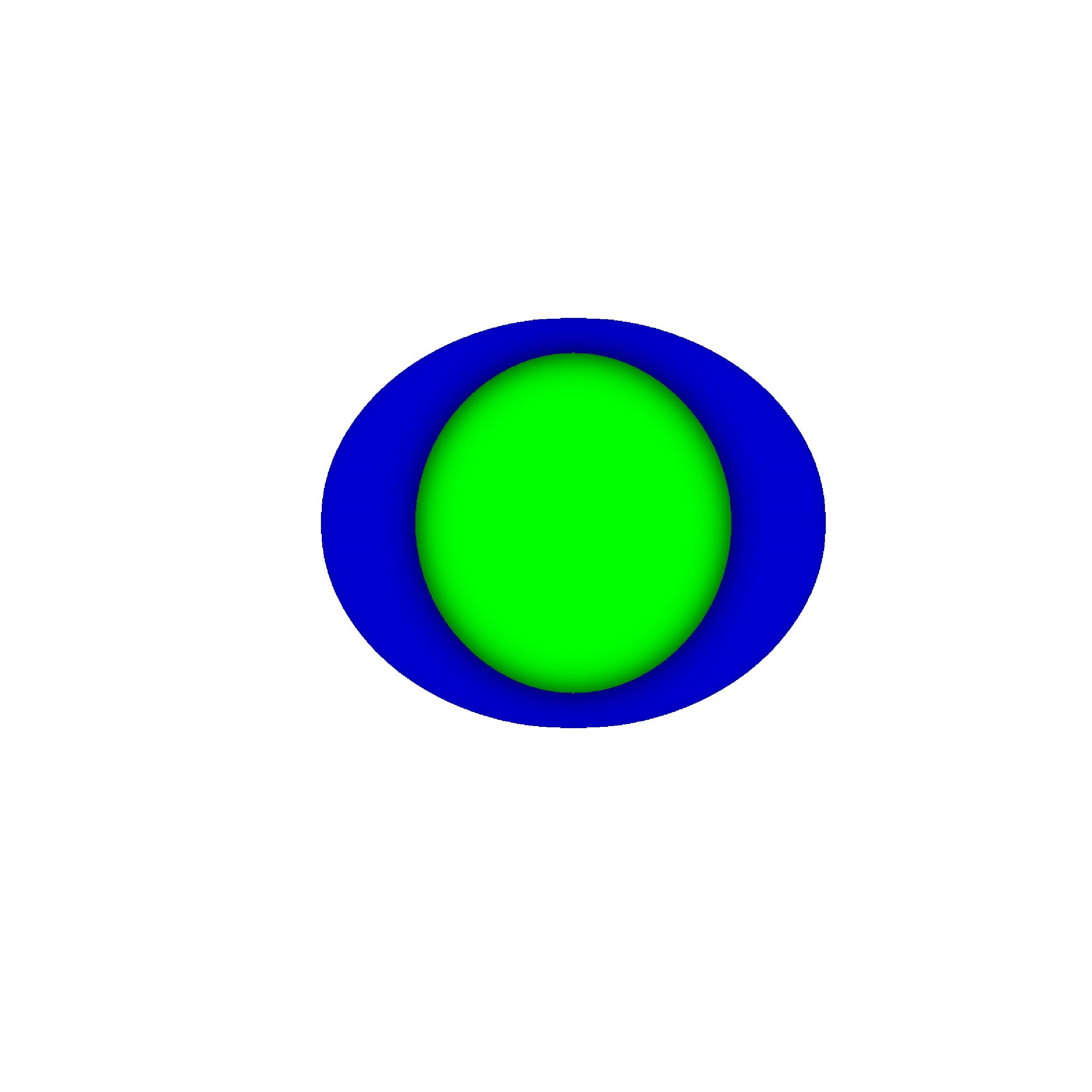}\hfil%
    \caption{}%
    \label{Ell_chi075_5}%
\end{subfigure}%
\caption{Time snapshots of microstructural evolution (time increasing 
from left to right) within an elliptic particle of aspect ratio 1.2 
and area same as that of a circular particle with $d=160$. Top row:
$\chi=0.25$, middle row: $\chi=0.5$, and bottom row: $\chi=0.75$.
Blue and green colors represent solute-rich $\beta_2$ and solute-poor 
$\beta_1$ phases, respectively.}%
\label{Ellipse}
\end{figure}

Particle shape can also modify the evolution significantly -- the more the 
deviation from circular symmetry, the more anisotropic is the decomposition path.
In Fig.~\ref{Ellipse}, we demonstrate this using an elliptical particle whose
area is equivalent to circular particle of $d=160$ with different contact angles
(ESI contains the corresponding animations).

When $\chi=0.25$, initial decomposition is similar
to circular particle and leads to an intermediate CS-$1@2$ microstructure 
(Fig.~\ref{Ell_chi025_4}). 

Since the  radius of curvature of the elliptic particle is position-dependent,
Gibbs-Thomson effect associated with the each layer creates an additional
composition gradient along its perimeter. Simultaneously, the elongated core 
wants to become circular to reduce the interfacial energy. This sets up two 
opposing currents: (a) flux of solute solute towards the surface along a radial
direction parallel to the major axis, and  (b) flux of solute along the surface 
(tangential) from regions of higher curvature (poles) to those with lower 
curvature (equatorial points on the minor axis). This leads to a nonuniform 
shell thickness and an eventual pinch off of the thinnest sections at the 
equator (Fig.~\ref{Ell_chi025_5}). Further solute redistribution pushes the
solute-poor layer towards the poles and formation of a characteristic 
\emph{pole-segregated}, trimorphic structure (one of the phases at two poles
separated the other phase in the equator).

When $\chi=0.5$, the initial SDSD pattern remains the same. However, due to 
relatively larger capillarity, the solute-rich shell remains intact till later
times (Fig.~\ref{Ell_chi05_3}), while a solute-lean core forms via coarsening.
Eventually, the continuity of the shell disrupts, forming a pole-segregated
structure. Unlike the previous $\chi=0.25$ case, here the poles are solute-rich
and the interfaces between equatorial solute-poor phase and the polar phase is
convex inwards. For complete wetting (bottom row), the initial layered structures
are similar. However, coarsening produces a continuous solute-rich shell 
surrounding a solute-poor core. Unlike the partial wetting case, the CS structure
is retained at the later stages even though the shell thins around the equator
and thickens around the poles. 

\section{Conclusions}

\begin{figure}[htbp]
  \centering
  \includegraphics[scale=0.05]{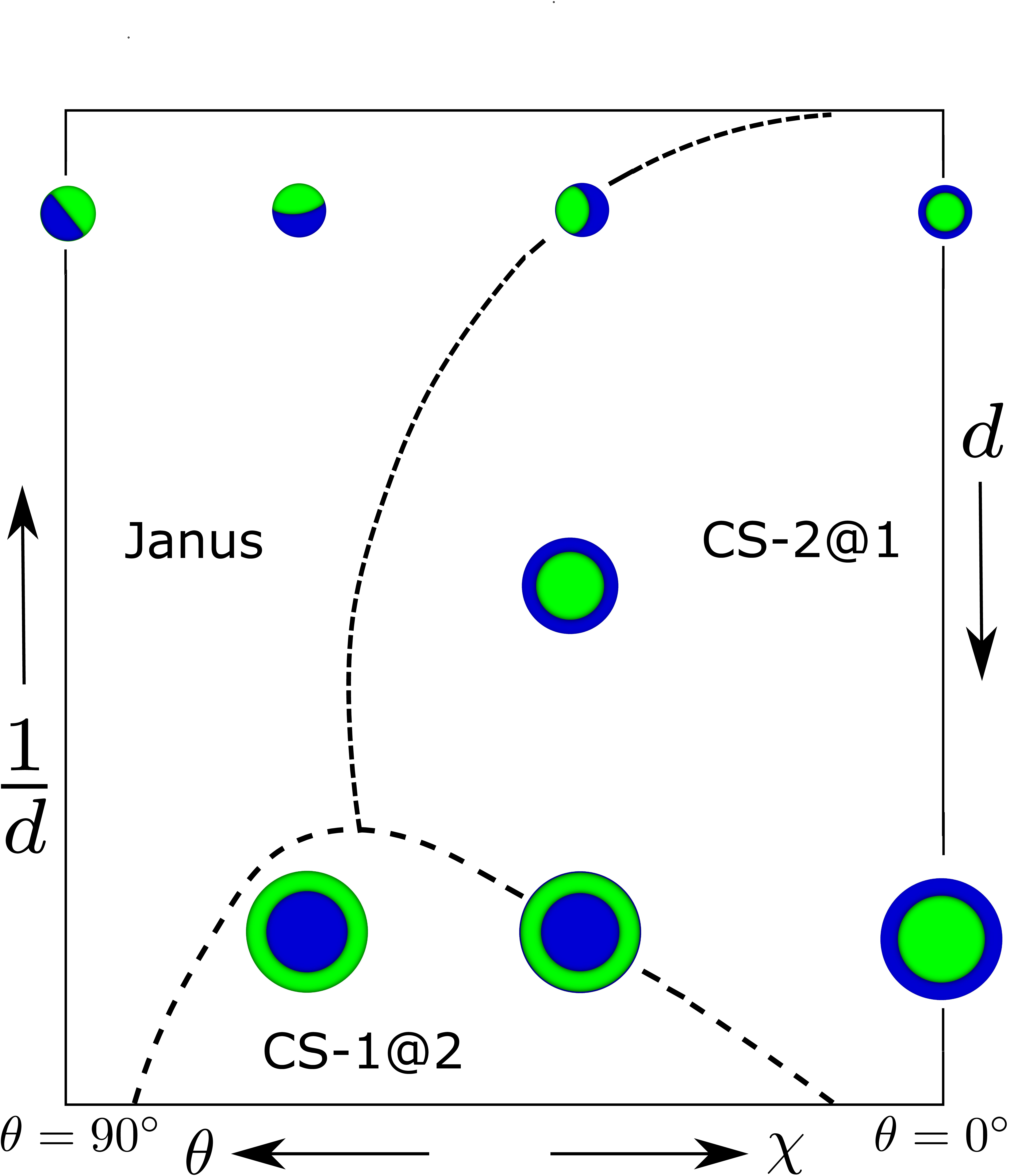}
  \caption{Morphology map in the particle size ($d$)
  -- contact angle ($\theta$) space}
  \label{fig:map}
\end{figure}

Through a systematic study, we have shown how interfacial energy and particle
size interact to influence the development of CS/Janus morphology in BNPs.
Fig.~\ref{fig:map} summarizes these findings through a morphological stability
map in the space of segregation parameter $\chi$ and particle size $d$. Since
this is not merely an equilibrium diagram but incorporates kinetic factors as 
well, it has a wider applicability.

The key outcomes of our study are:

\begin{enumerate}

    \item A novel phase field method for studying morphological evolution during
    phase transformations in finite systems has been developed. We have applied
    it here to understand kinetic pathways to various configurations in BNPs.
    
    \item The interplay of three interfacial energies, captured
    through the contact angle, and particle size, critically influences the 
    development of CS, Janus and intermediate structures in BNPs. In particular,
    smaller size and higher contact angle tend to promote Janus, while the opposite conditions promote CS.
    
    \item We show that discontinuous, step-like coarsening of intermediate ring
    structures govern the final morphology. We find that the diffusional driving
    force for coarsening decreases dramatically with increasing particle size,
    leading to kinetically trapped configurations (e.g., inverse-CS) at larger sizes.
    
    \item For a given particle size, lower contact angle enhances 
    the tendency to form metastable CS morphologies even when complete wetting
    condition given by Eq.~\eqref{eq_Cahn_wet} is not satisfied.   

   \item Our simulations with elliptical particles also highlight the possibility
   of attaining novel morphologies due to non-uniform particle curvature.
\end{enumerate}

\section{Acknowledgement}

P.P. and S.B. gratefully acknowledge the computational support from DST Grant No. EMR/2016/006007.

\bibliographystyle{rsc}

\bibliography{npmerged}

\end{document}